\documentclass[12pt,preprint]{aastex62}

\shorttitle{Tip Error}
\shortauthors{Madore, Freedman, Owens \& Jang}

\begin{document}


\title{ Quantifying Uncertainties on the  \\Tip of the Red Giant Branch Method \\}


\author{ Barry F. Madore}
\affil{The Observatories \\ Carnegie Institution for Science \\ 813
Santa Barbara St., Pasadena, CA ~~91101}
\affil{Dept. of Astronomy \& Astrophysics, University of Chicago, Chicago, IL}
\email{barry.f.madore@gmail.com}
\author{ Wendy L. Freedman}
\affil{Dept. of Astronomy \& Astrophysics, University of Chicago, Chicago, IL}
\email{wfreedman@uchicago.edu}
\author{ Kayla A. Owens}
\affil{Dept. of Astronomy \& Astrophysics, University of Chicago, Chicago, IL}
\email{kowens@uchicago.edu}
\author{ In Sung Jang}
\affil{Dept. of Astronomy \& Astrophysics, University of Chicago, Chicago, IL}
\email{hanlbomi@gmail.com}

%% are identified by the \altaffilmark after each name.  Specify  alternate

\begin{abstract} 
We present an extensive grid of numerical simulations quantifying the uncertainties in measurements of the Tip of the Red Giant Branch (TRGB). These simulations incorporate a luminosity function composed of 2 magnitudes of red giant branch (RGB) stars leading up to the tip, with asymptotic giant branch (AGB) stars contributing exclusively to the luminosity function for at least a magnitude above the RGB tip. 
We quantify the sensitivity of the TRGB detection and measurement to three important error sources:  
(1) the sample size of stars near the tip, 
(2) the photometric measurement uncertainties at the tip, and
(3) the degree of self-crowding of the RGB population. 
The self-crowding creates a population of supra-TRGB stars due to the blending of one or more RGB stars just below the tip. This last population is ultimately difficult, though still possible, to disentangle from true AGB stars.

In the analysis given here, the precepts and general methodology as used in the {\it Chicago-Carnegie Hubble Program} (CCHP) has been followed. However, in the Appendix, we introduce and test a set of new tip detection kernels which internally incorporate self-consistent smoothing. These are generalizations of the two-step model used by the CCHP (smoothing followed by Sobel-filter tip detection), where the new kernels are based on successive binomial-coefficient approximations to the Derivative-of-a-Gaussian (DoG) edge detector, as is commonly used in modern digital image processing.  
\end{abstract}

\section{Introduction}
Over a century ago, Harlow Shapley (1918, 1919, 1930) used blue-sensitive photographic plates to measure (by eye) the mean apparent magnitudes of the 25 brightest stars in galactic globular clusters (his Table 1, 1919), in order to go on to (incorrectly) build a case for his version of an {\it Island Universe} cosmology (see Berendzen, Hart \& Seeley 1976). 
With the availability of newly-developed, red-sensitive photographic plates, Walter Baade (1944) serendipitously resolved the brightest red giant stars (which, to his surprise, suddenly appeared at approximately the same red-band magnitudes) in several dwarf elliptical companions galaxies to the Andromeda galaxy, M31. 
That unanticipated discovery precipitated a revision in the size and age of the Universe by a factor of two. 
Four decades later, and armed with some of the first available panoramic linear charge-coupled devices (CCDs) Mould, Kristian \& Da Costa (1984) revisited one of Baade's original dwarf galaxies, NGC 205. 
They produced full color-magnitude diagrams revealing a broad swath of RGB stars all of which cumulatively defined a constant I-magnitude plateau in the CMD, later to be named the Tip of the Red Giant Branch, or simply known by its initialism, the TRGB. 
They also had earlier observed NGC 147 (Mould, Kristian \& Da Costa, G.S. 1983) finding the same feature. 
But perhaps more interestingly they observed TRGB stars in the halo of the Local Group spiral galaxy, M33. 
By good fortune, at about the same time M33 had been the subject of two different investigations into Cepheid distance moduli to this galaxy: one by Sandage \& Carlson (1983) coming in high, with a value of $(m-M)_o =$ 25.23~mag, and another by Madore et al. (1985) nearly a full magnitude closer at $(m-M)_o =$ 24.25 $\pm$~0.15~mag. 
The TRGB distance fell in the mid-range, at $(m-M)_o =$ 24.8 $\pm$~0.2~mag, right between the other two extremes. 
However, not all of the early cross-comparisons of TRGB and Cepheid distance scales were in conflict. 
For example, Freedman (1988a) used the first CCD camera available on the CFHT and measure the TRGB in the halo of the Local Group dwarf irregular galaxy, IC~1613. 
She found a true distance modulus of $(m-M)_o $ = 24.2~mag, which did agree the Cepheid-based distance modulus of $(m-M)_o $ = 24.3~mag (Freedman 1988b). 

The TRGB Method finally came of age with the publication of two papers:
The first was the calibration paper by Da Costa \& Armandroff (1990) who were inspired to undertake an I-band CCD survey a sample of 8 southern Milky Way globular clusters. 
In doing so they demonstrated that, while the mean colors of the giant branches were rank-ordered by the mean metallicities of the parent globular clusters (as previous known to Frogel, Cohen \& Persson 1983 from pioneering their studies of red giant branch stars in globular clusters in the near-infrared) the brightest of those RGB stars had a remarkably stable absolute magnitude, in the I band, independent of color. The second paper was that of Lee et al. (1996). It laid out, in one place, most of the key issues concerning systematics involving reddening, metallicity, star formation history and host galaxy type, etc. It also introduced the widely adopted Sobel filter for precisely deriving the magnitude at which the discontinuity in the RGB luminosity function occurs, as well as its uncertainty,  while exploring a range of smoothing kernels. 
This was carried out in the context of anticipating a re-furbished Hubble Space Telescope, and applying the TRGB method widely to the extragalactic distance scale. The authors demonstrated its ground-based application to 10 galaxies spanning a wide range of Hubble types, metallicities and absolute magnitudes, and found overall consistency in the TRGB, Cepheid and RR Lyrae distance scales at the level of 0.1~mag.
The success of the TRGB method might be measured by its subsequent adoption: over the intervening three decades, more than 500 TRGB distances to nearby galaxies have been published (for instance NED lists over 900 references to TRGB distance determinations to 302 distinct galaxies (https://ned.ipac.caltech.edu)
%/cgi$-$bin/INFatt?dom$=$z&id$=$4338)$ 
and EDD lists 588 galaxies that they have derived uniformly processed TRGB distances to %$(https://edd.ifa.hawaii.edu/dsecond.php)$. 
%20Detection%20FPCV-2-1.pdf. 
And recently, the TRGB method has been extended to the calibration of Type Ia supernovae and determination of the Hubble constant (Freedman et al. 2019; Freedman 2021).

For theoretical discussions of the evolution of stars up to and including degenerate helium core flash, we recommend that readers consult the monographs by Cassisi \& Salaris (2013), Salaris \& Cassisi (2005) and Lamers \& Levesque (2017).
For updated discussions of modeling, with special reference to near- and mid-infrared applications of the TRGB method see Serenelli et al. (2017), McQuinn et al. (2019) and Durbin et al. (2020).

\section{Motivation} 

A quick census of the published determinations of the apparent magnitude of the tip of the red giant branch (TRGB) in even the nearest of galaxies (NED-D August 2022 version) immediately reveals a wide range of quoted uncertainties\footnote{That said, the errors presented in NED are in no way homogenized. NED presents the data as published, and in many cases the original authors make no distinction between statistical and systematic error, or combinations of the two. }. 
The published errors, for the tip measurement in a given galaxy, can vary by as much as a factor of ten; as in the case of M31 (0.05 to 0.57~mag uncertainties quoted) and M33 (0.03 to 0.30~mag), but more typically they range over a factor of 3 to 6 as, for example, in the published values for the nearest galaxies: the LMC (0.04 to 0.25~mag), IC~1613 (0.05 to 0.20~mag) and NGC~6822 (0.06 to 0.19~mag).
On the other hand, some of the reported statistical uncertainties on the tip determination can go as low as 0.01~mag (e.g., Lee \& Jang 2012 for M101; or even smaller than that in the case of Conn et al. 2011 for Andromeda I and II). 
In an appendix to Cioni et al. (2000) those authors rightly note that many of the methods used, (counter-intuitively) do not in any way scale with population size of stars detected and measured at the tip. They should.
For a given photometric error, population size certainly needs to be a part of the calculation of the statistical uncertainty on the mean of the TRGB distance.\footnote{However, see Mendez et al. (2002) and/or Makarov et al. (2006) for extensive discussions specifically concerning the Maximum Likelihood technique and its error sensitivity to photometry and sample size.} 
Upon closer examination of any given paper, it is not always clear what exactly the source of the quoted uncertainty is or even how it was actually calculated. 
In this paper we attempt to bring some clarity to the situation.

In earlier papers (Madore \& Freedman 1995; Madore, Mager \& Freedman 2009) we presented computer simulations of the TRGB in its use as an extragalactic distance indicator. In the first paper there is an often quoted and paraphrased conclusion that ``at least 100 stars in the first magnitude interval below the tip are needed to secure a distance modulus to better than $\pm$0.1 mag". At that time, the method was still in its infancy and small number statistics were a major concern (especially when the early focus was on applying the method to sparsely populated individual galactic globular clusters, or very small fields of view in the halos of very nearby galaxies, say). The field has matured, the demand for higher precision has prevailed, and the numbers of stars measured in extragalactic halo fields has gone into the thousands, while at the same time other sources of uncertainty in determining the precision of the TRGB have become clear. We feel that it is time now to explore parameter space a bit more thoroughly. In the following we consider, in turn, a total of three independent, major sources of uncertainty: 
 
(1) The formal way in which the statistical uncertainty in the tip magnitude can be quantified, specifically in terms of its sensitivity to numbers of stars at the tip, and its independent sensitivity to individual photometric errors of those same tip-defining stars. 
(2) The effects of having an asymptotic giant branch (AGB) population of stars contributing to  the one-magnitude interval directly above the TRGB.
And finally, (3) The explicit modelling of the mutual (line-of-sight) crowding of all stars along the RGB, and the inevitable production of a new, but totally spurious, population of (crowded) stars, systematically brighter than the TRGB.

We use a modified Sobel edge-detection filter (see Appendix I) for measurement of the TRGB, which is largely consistent with our GLOESS-smoothed, Sobel-filtered analysis used in the Carnegie Chicago Hubble Program (Hatt et al. 2017; Jang et al. 2018; Hoyt et al. 2018; Madore et al. 2018). In Appendix I we derive and tabulate a complete series of new digital filters that are derived from successive discrete approximations of the first derivative of a Gaussian, using the binomial theorem as the gradient detector. 
We also adopt the weighted (noise-suppression) versions of these kernels 
as first introduced and applied to a simple Sobel filter in Madore, Mager \& Freedman (2009) and much later adopted and utilized by Gorski et al. (2018).   

\section{The Underlying Model}

The basic model adopted here for the intrinsic luminosity function, above and below the TRGB, now consists of three distinct input populations: (a) a red giant branch (RGB) population with a power-law increase in numbers with increasing (fainter) magnitudes, (b) a bright-end truncation/discontinuity of the RGB luminosity function, defining the tip, and (c) an asymptotic giant branch (AGB) population, stretching at least one magnitude above and brighter than the TRGB. 
We model the luminosity function from one magnitude above to two magnitudes below the TRGB (but note that only the first magnitude below the tip is shown in the figures) assuming a flat luminosity function for the AGB down to the TRGB\footnote{The referee has argued that a variety of shapes to the AGB luminosity are apparent in published CMDs including data above the TRGB, and that a flat AGB LF may not be representative. We agree with that statement, but as shown in the Appendix D the shape of the AGB luminosity function, be it falling rising or flat, has no impact on the ability of the Sobel filter to detect the TRGB in an unbiased manner.}, at which point there is a discontinuous offset to the RGB population. The RGB then assumes a steeply-rising luminosity function of the form $log[N(m)] = 0.3 \times [I - I_{TRGB}] + a$. 
In these first simulations the relative RGB-to-AGB  normalization is six to one, such that there are 17 AGB stars in total in the one-magnitude interval seen above the TRGB, for every 100 RGB stars in the one-magnitude interval fainter than (i.e., below) the tip. For the purpose of this simulation, the AGB luminosity function is assumed to be flat in the one-magnitude interval above the tip and zero elsewhere. References to the literature justifying the values for the parameters alluded to above are to be found in the first paragraph of Section 5.1.1.

\begin{figure*} 
\centering 
\includegraphics[width=14.0cm,angle=-0]{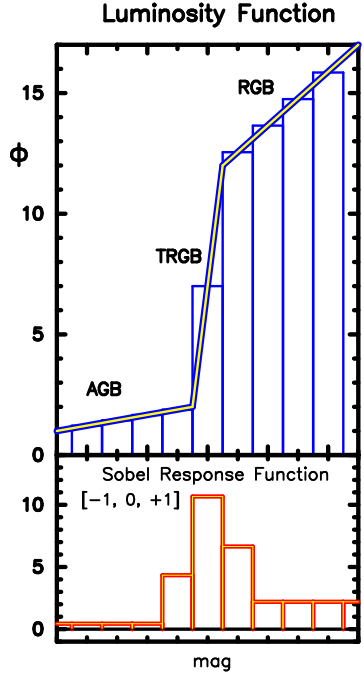} 
\caption{\small Magnified view of the idealized toy model of the RGB + AGB luminosity function, centered on the discontinuity in the RGB luminosity function at the TRGB. Solid yellow segments, from left to right show the AGB LF, the discontinuity, and the RGB LF. The blue histogram is a binned version of straight lines used as digital input to the differencing (Sobel) kernel: [-1,0,+1]. The digital output of the Sobel Response Function is shown in red histogram form in the lower portion of the figure. The maximum of the Sobel filter marks the position of the discontinuity. See text for a step-by-step description of the tip detection.}  
\end{figure*}

Here we first examine an idealization in the form of a toy model that captures the essential ingredients of the detailed simulations that follow, and try to emphasize how the various components contribute (or not) to the determination of the magnitude and location of the TRGB discontinuity.The toy model is shown in Figure 1, a luminosity function centered on the TRGB. This a plot of logarithm of numbers of stars per magnitude bin as a function of magnitude. The luminosity function is composed of an AGB population, represented by a dispersionless straight line sloping upward from left to right, stopping one bin short of the location of the discontinuity defining the TRGB. The number of AGB stars in that final bin is N. One bin beyond that magnitude the luminosity function is defined by red giant branch stars whose slope is independent of, and different from, the AGB slope. The RGB luminosity function is normalized at the tip with a value that is six times the value of the AGB population (i.e., 6N stars) at its starting point one bin brighter than the bin marking the discontinuity. The bin between the two terminal points defines the tip, and its value is the average of the two adjacent luminosity functions (i.e., 3N stars).

The upper panel of Figure 1 shows the input luminosity function binned into 9 histogram-like segments with bin No. 5 centered on the position of the discontinuity, marking the luminosity (in magnitudes) of TRGB. 
The lower panel shows the result of running a Sobel filter [-1,0,+1] across the binned luminosity function. 
The output of the Sobel filter is the discretely-sample first derivative of the function being sampled. Moving from left to right the output of the Sobel filter is constant, as expected, given the constant slope of the input AGB luminosity, i.e., at the first position the output of the Sobel function is [-1$\times$(N-3$\epsilon$) + 0$\times$2$\epsilon$ + 1$\times$(N-1$\epsilon$) =] 2$\epsilon$, where $\epsilon$ is the width of the binning. 
At the second position the output is [-1$\times$(N-2$\epsilon$ ) + 0$\times$(N-$\epsilon$) + 1$\times$N) =] again (the slope of the pure AGB) 2$\epsilon$.
These first steps do not sample the discontinuity and therefore contain no information about its position or presence.
At step No. 3, the right-most element [+1] is the first to sample the discontinuity and reports 
a increased value of the filter's output, [-1$\times$(N + $\epsilon$) + 0$\times$N + 1 $\times$ (6/2)N = ] 2N - $\epsilon$. 
The next step over continues to report larger values of its response function, where the [+1] element now sees the undiluted height of the RGB luminosity function 6N, and differences that against the response of the left-most element [-1] of the Sobel filter contributing a value of -N, with the central element of the Sobel filter always reporting a null value regardless of the function's value. 
The output is [-1$\times$N + 0$\times$(6/2)N + 1$\times$6N = ] +5N. 
The central element simply keeps track of the bin around which the derivative is being measured and reported. 
Moving the filter one more bin to the right 
reports a value of [-1$\times$(6/2)N + 0 $\times$ 6N + 1 $\times$ (6N+$\delta$) = ] +3N + $\delta$.
One step more away from the discontinuity gives [-1$\times$6N + 0$\times$(6N+$\delta$) +1$\times$(6N+2$\delta$) = ] +2$\delta$, the slope of the RGB. All subsequent steps to the right continue to report the constant slope of +2$\delta$. The maximum value of 5N for the Response Function is found at step No. 4 and marks the magnitide at which the TRGB is to be found. 

\section{A Few Preliminaries}
It is worth making explicit what exactly the criteria are for a successful experiment to be run, that aims for a detection and measurement of the position of the discontinuity marking the TRGB in magnitude space. 
It is then also important to list the real-world parameters over which we have some control in optimally undertaking the observations and subsequently analyzing the results.

Generally speaking, there are two obvious performance indices in TRGB edge detection that we are concerned with here:  accuracy and precision. 
However, the latter (which can also be classified as ``bias'') can be further broken down into (a) false-positive ``detections" of the TRGB, (b) non-detections and (c) systematic bias attributed to the edge-detector itself. 
Each of these are discussed in turn, below. 
And in the subsection following this we discuss what control we have, 
at the observational design level, in mitigating each of these kinds of errors.

\subsection{\it Accuracy}

(i) {\it False Positives:} In the presence of random noise in the output of our TRGB edge-detection response-function, there comes a point at which (a) Fluctuations in the number of detected stars (from bin to bin) and/or (b) Poisson noise in the photometry of the individual stars themselves will produce (spurious) features in the tip-detection/response-function output. These noise-induced features, if large enough, can be both qualitatively and quantitatively indistinguishable from the expected signal (i.e., being positive deflections in the response function, that have a similar width and relative height when compared to the expected/true signal.) In the controlled simulations, discussed below, we quantify when and where this situation starts to become a serious problem. 

(ii) {\it Non-Detections:} Again, in the presence of excessive photometric noise, particularly for small population sizes (or in a combination of the two) it is possible for the true signal to become so weak that it is
not detected at any significant thresholding level, with respect to the ambient response-function noise.
{\bf}{ This situation is fatal; but the circumstances under which it is likely to occur can be anticipated and identified using these simulations as a guide} (see, for example, Figures 4 and 7).

(iii) {\it Potential Bias in the Tip Detection Algorithm:} Given the unequal count rates of stars contributing to the luminosity functions above and below the TRGB, it might be thought that even a symmetric response-function kernel might return an asymmetric (i.e., biased) answer, given that more RGB stars are moving across the TRGB discontinuity to intrinsically brighter magnitudes than there are bright AGB stars moving in the opposite direction (across the TRGB discontinuity) to fainter magnitudes. 
We investigate this potential source of systematic error (detector bias) throughout the simulations studied below.
\subsection{Precision} 
We are endeavoring to measure (a) the tip magnitude, (b) its statistical uncertainty (its precision) and (c) provide any estimate of bias (its accuracy) inherent in the methodology explored here. A number of factors contribute to the outcome. 
Some of these factors can be controlled in advance while setting up the experiment/observation, and some of them can be ameliorated later in the data analysis stage. For instance, the source-count population, the amount of crowding, and the signal-to-noise ratio in the photometry can each be controlled with foreknowledge of the approximate surface brightness of the region being targeted, knowing the approximate distance, and adjusting the total exposure time (or size of the telescope), within 
allowable and practical limits. 
The type of kernel employed in measuring the first derivative of the 
luminosity function at and around the tip, and the amount of smoothing of the data chosen to be applied to the data, in advance of the kernel response function application, can both be controlled to optimize 
the output of the detector once the data have been obtained. 
We consider each of these parameters in turn.

\section{RGB + AGB Computer Simulations}

In this series of simulations, we explore changing a number of parameters while holding others fixed. These include the photometric errors and overall population size (Section 4.1), different smoothing sizes (Section 4.2 \& 4.3), and amount of crowding/blending (Section 6). In Section 5, we illustrate how the width of the smoothing function does not carry information  on the uncertainty of the tip measurement.

{\bf}{ Here we explore the systematics of changing the photometric errors at the tip (from one simulation to the next) while holding the population size and smoothing fixed.}

With Figure 2 we start at one extreme: A very densely-populated luminosity function (about 120,000 stars in total) having minimal (0.01 mag) smoothing and very high-precision photometry, as shown in the first (upper left) panel. 
We then work (left to right and top to bottom) through the observed effects of progressively increasing the photometric errors (at the tip) from $\pm$0.02, to 0.05, 0.10, 0.15 and 0.20~mag, respectively (corresponding to signal-to-noise ratios of 50, 20, 10, 6 \& 5). 

In Figures 3 through 5 {\bf} {we then re-run parallel simulations, progressively dropping the total population of stars by about a factor of ten each time}: starting with about 120,000 RGB stars in the one-magnitude interval below the tip (in Figure 2), and ending with a simulation having only 127 RGB stars in that same one-magnitude interval (in Figure 5).

{\bf}{ In each of the next sections} (each also containing four figures and six main sub-panels) {\bf} {we explore the effects of changing the smoothing} (going up from 0.01 to 0.05, and finally 0.10~mag) in Figures 6 through 9, at fixed population sizes per figure and increasing photometric errors through each of the sub-panels, as in the previous section.

{\bf} {We then close out in Section (4.1.9) holding the smoothing at a fixed value (at 0.10 mag) and assessing the effects of changing the population size in Figures 10 through 13, while changing the photometric errors in the sub-panels within those figures.}

This extensive grid of plots is provided both for their use as predictors in planning future observations, and for their use as a guide in understanding the luminosity functions and edge-detector output once they are acquired. To put this into perspective for the 12 galaxies observed by Freedman et al. (2019) in their determination of a TRGB-based value of the Hubble constant, they detected an average of 4,000 RGB stars in the one-magnitude interval below the TRGB (with anywhere from 1,000 to 20,000 RGB stars in individual cases, depending on the distance modulus of the host galaxy and how far into the halo any given exposure was taken.) As for the typical photometric errors at the tip, the exposures were scaled to the approximately known distances and they all have uncertainties at the tip of about $\pm$0.10 in F814W (I-band). This would roughly correspond to the middle-right panels of Figures 6, 7 \& 8.

To help navigate the various simulations we provide a guide to their ordered content in Table 1.

\begin{deluxetable}{rcccccccccc}
\tablecaption{Guide to Simulations: Figs 2-13 and Panels a through f}
\tablehead{\colhead{RGB Stars} & \colhead{} & \colhead{Smoothing} &\colhead{} & \colhead{} & \colhead{} &  \colhead{Error} &\colhead{at} &  \colhead{TRGB} &\colhead{} }
%\colhead{} &\colhead{Number}\colhead{} &\colhead{0.01}&\colhead{0.05}&\colhead{0.10}&\colhead{}&\colhead{0.00}&\colhead{0.02}&\colhead{0.05}& \colhead{0.10}&\colhead{0.15}&\colhead{0.20}} 
\startdata
120,000 & Fig. 2 & Fig. 6 & Fig. 10 & &a & b & c & d & e & f \\
11,331 & Fig. 3 & Fig. 7 & Fig. 11 & &a & b & c & d & e & f  \\
1,240 & Fig. 4 & Fig. 8 & Fig. 12 && a & b & c & d & e & f  \\
124 & Fig. 5 & Fig. 9 & Fig. 13 && a & b & c & d & e & f  \\
%\hline
\enddata
\end{deluxetable}
%\vfill\eject

\subsection{Low Degree of Smoothing}

\subsubsection{A Range of Photometric Errors: {\bf}{120,000 RGB Stars}, Fixed Smoothing {\bf}{$\pm$0.01~mag}}
We start this detailed discussion with a high-definition simulation of the luminosity function beginning one magnitude above, and ending one magnitude below the TRGB, where the discontinuity is set to $M =$ 0.00~mag across of the simulations in this paper. This would correspond to $M_I = $ -4.05~mag, which closely matches the value currently adopted by the CCHP (Freedman 2021).
The first magnitude interval, above the tip, is populated uniformly as a function of magnitude by AGB stars. For examples of published flat AGB luminosity functions above the tip see Beaton et al. (2019), their Figure 4, Hoyt et al. (2018), their Figure 6, and Nikolaev \& Weinberg (2000), the inset histogram  to their Figure 4, and their description of it being {\it ``The off-bar LF shows only a mild increase in the source counts at the location of TRGB, but has the same, {\bf}{roughly constant profile at Ks brighter than 12 mag, due to the AGB population}, visible in the other two luminosity functions}."  At the TRGB discontinuity the RGB population turns on at an initial rate (of stars per magnitude bin) 6x times greater than the AGB density above the tip (see Wu et al. 2023, where they calculate a variant of this contrast ratio R (using bins 0.5 mag wide, above and below the tip) for a large number of GHOSTS galaxies,  finding that it ranges from R = 4 to 7 as seen by the annotations in their Figure 5). Thereafter the binned number density of RGB stars increases with a logarithmic slope of +0.3 (Mendez et al. 2002; Makarov et al. 2006).

The upper left panel in  Figure 2 shows our highest-fidelity, and most optimistic realization, consisting of 120,000 RGB stars and some 20,000 AGB stars. 
The bin size is 0.01~mag, giving a typical RGB population of 1,200 stars per bin, leading to an expected 2-sigma scatter of $\pm$70 stars per bin (or $\pm$6\% one sigma, as can be seen in the plot).  
The solid line passing through the data is a GLOESS fit with a Gaussian smoothing window of 0.01~mag, making it a close approximation, at this fine binning/smoothing, to a spline fit through the individual data points. 
The vertical line at  M = 0.0~mag marks the exact position of the TRGB that is equidistantly flanked, in the lower panel, by two dashed lines (barely visible in this panel) that are $\pm$0.01~mag apart, showing the highest attainable resolution of the data and the response function.

Below the luminosity function, in the lower part of the panel, is the first-derivative response function as applied to the discretely-sampled and (minimally) smoothed luminosity data above it.
We use the MF5 edge detector described in Appendix A, which samples the luminosity function at 11 optimally-weighted points symmetrically placed around the output bin. 
Two versions of the output function are shown: The thin solid line is the Raw Response Function (RRF) of the MF5 filter, while the thick black line is the (inversely) Noise-Weighted Response Function (NWRF), as described in Appendix A. 
The two response functions have been scaled to agree at the their respective (close-to-peak) values at the center of the plot where the true/input value resides. 
As is evident from a casual inspection of the various plots, noise suppression results in much reduced fluctuations everywhere across the magnitude range probed by the tip detectors, without any obvious degradation (or improvement) of the sought-after signal at the TRGB discontinuity. 
We do point out however, that the width of the untreated TRGB detection is both asymmetric and wider than the noise-suppressed response, where the latter has the expected width of $\pm$ 0.01 mag, which in turn is the sampling limit of the data. 
The solid line marks the exact position of the TRGB and the two flanking solid lines are again separated by $\pm$0.01 mag for visual reference.

The GLOESS fit to the luminosity function faithfully tracks the discontinuity input at 0.0~mag, in the upper panel, and the response function, in the lower panel, {\bf}{peaks precisely at the midpoint of the M = 0.0~mag discontinuity, in all cases.}
At the resolution of the data and the detector output (0.01~mag in both cases) there is no measurable bias in the first derivative response function being used to detect the TRGB.

We do, however, want to emphasize that there is no pressing need for smoothing the data when in this high-population, high-precision-photometry portion of parameter space; the {\bf} {Noise-Weighting is sufficient in suppressing spurious signals, while simultaneously sharpening the edge-detector response.}

In the second panel of this same figure (top right) we begin to explore the effects of adding photometric errors to the individually ``observed" stars contributing to the simulated luminosity function. 
All of the other parameters (in this instance, population size, smoothing and the detection kernel) used in the 12 sub-panels of Figure 2 are kept unchanged. 

In this second simulation, randomly-generated photometric errors, having a Gaussian sigma of $\pm$0.02 mag and a mean of zero, have been applied randomly to each of the sampled stars, which were then re-binned at 0.01~mag intervals, re-plotted and re-analyzed.

The only effect obvious to the eye is the rounding of the originally sharp shoulders of the luminosity function immediately above and below the magnitude of the TRGB discontinuity. 
The dashed vertical lines in the upper panel mark the {\it one-sigma} ``smoothing radius" inflicted on the discontinuity by the degradation of the photometry. In the sub-panel below the luminosity function we again show the MF5 response function, in both the raw (thin solid line) and the noise-suppressed (solid black line) forms. 
Again the RRF is considerably noisier overall, and it is noticeably wider (with, noise-induced, broad wings) at the discontinuity. 
The noise-corrected response function still has the band-width-limited natural width of $\pm$0.01 mag.

The same general trends continue as we increase the photometric errors (from $\pm$0.05 to $\pm$0.20~mag) in the remaining four (lower) panels; that is, the raw response is always broader than the noise-suppressed response width, which is stable and effectively unresolved at the $\pm$0.01~mag level right up to and including the largest tested photometric error of $\pm$0.20~mag.  
What is progressively different is the decreasing signal-to-noise ratio of both response functions as compared to the baseline noise, {\it at the fixed baseline width of the discontinuity-sampling kernel} (MF5 in this case). 
As the observed slope of the luminosity function at the TRGB discontinuity softens with increased photometric errors, the power in the first derivative across a fixed magnitude interval drops, while the Poisson population-sampling noise in the baseline luminosity functions, on either side of the tip, remains largely unchanged. 
We emphasize here that the lower, response-function plots have been sequentially re-scaled for clarity, roughly normalized by the peak of TRGB response. 

\medskip
{\bf}{ Summary 1 --}  For very large populations of stars defining the luminosity function around the Tip of the Red Giant Branch, the Raw Response Function and the Noise-Weighted Response 
Function are each found to be unbiased indicators of the position of the discontinuity in the luminosity function marking the position of the TRGB. The RRF is found to slowly but systematically increase in width with increasing photometric errors. 
A slight skewing of the RRF distribution function towards fainter magnitudes may also be a generic feature of added noise affecting the wings. The NWRF is unresolved in all of the instances, regardless of the input photometric errors. 
As the power in the response functions fall (with increasing photometric errors) the noise on either side and surrounding the discontinuity begin to encroach upon and become competitive in amplitude with the declining response at the true position. 
This degradation is noticeable at a photometric error of $\pm$0.10~mag, and becomes problematic thereafter, for higher values of the photometric errors. 
In all cases, however, the noise-suppression is effective in damping down this background noise by about a factor of two compared to the raw response value (see the last three panels for the worst-case examples). 
At the two largest values of the photometric errors ($\pm$0.15 and $\pm$0.20~mag in the bottom two panels) noise-induced spikes in the response function become sufficiently large with respect to the declining response at the known/true position, that {\it false positive detections start to become a problem} especially downstream of the true tip. 
Noise-suppression helps to damp these fluctuations down, but does not eliminate all of the false positives in the regime of large photometric-errors ($>$0.15~mag).

\begin{figure*} \centering
\includegraphics[width=8.5cm,angle=-0]{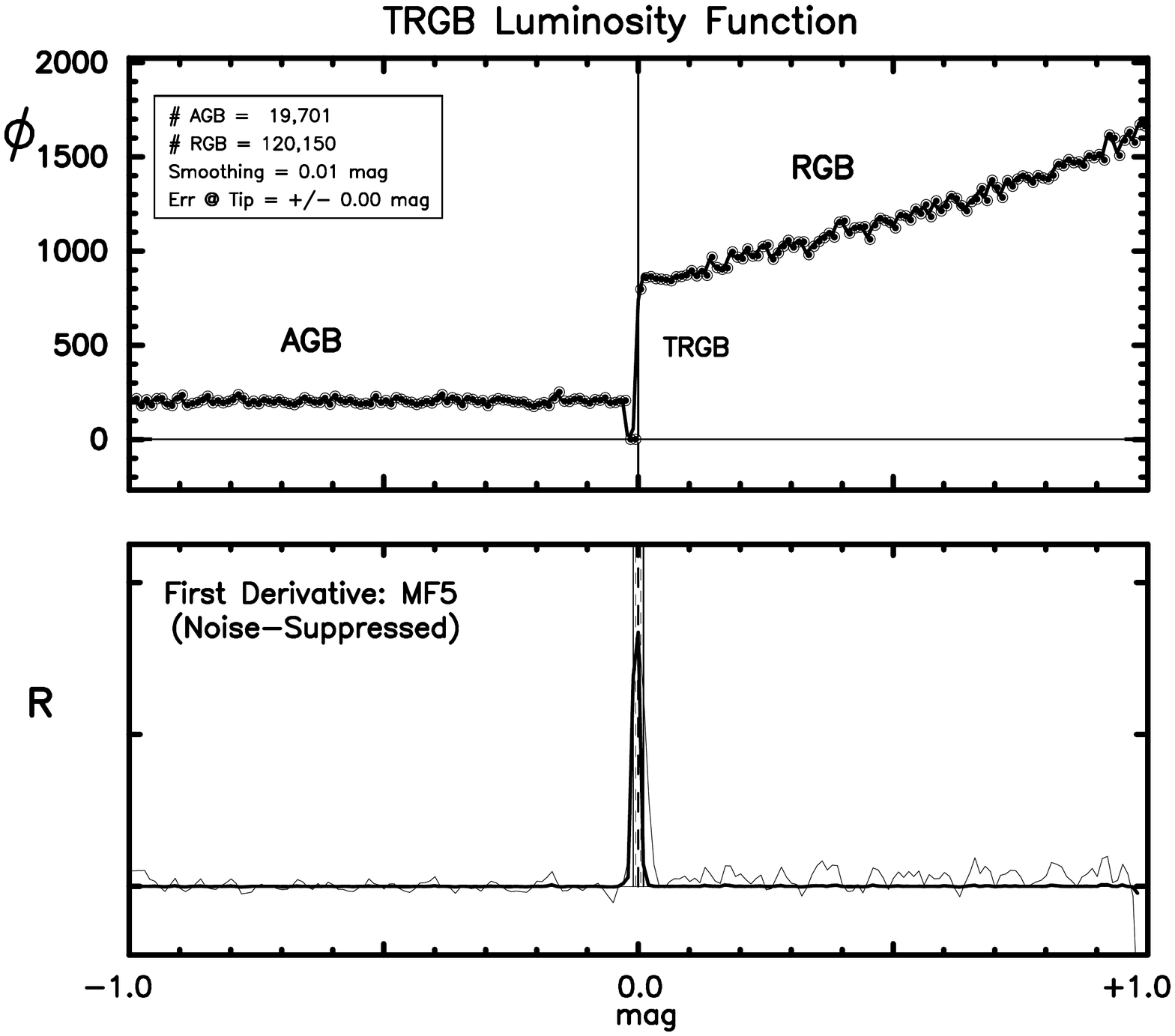}
\includegraphics[width=8.5cm,angle=-0]{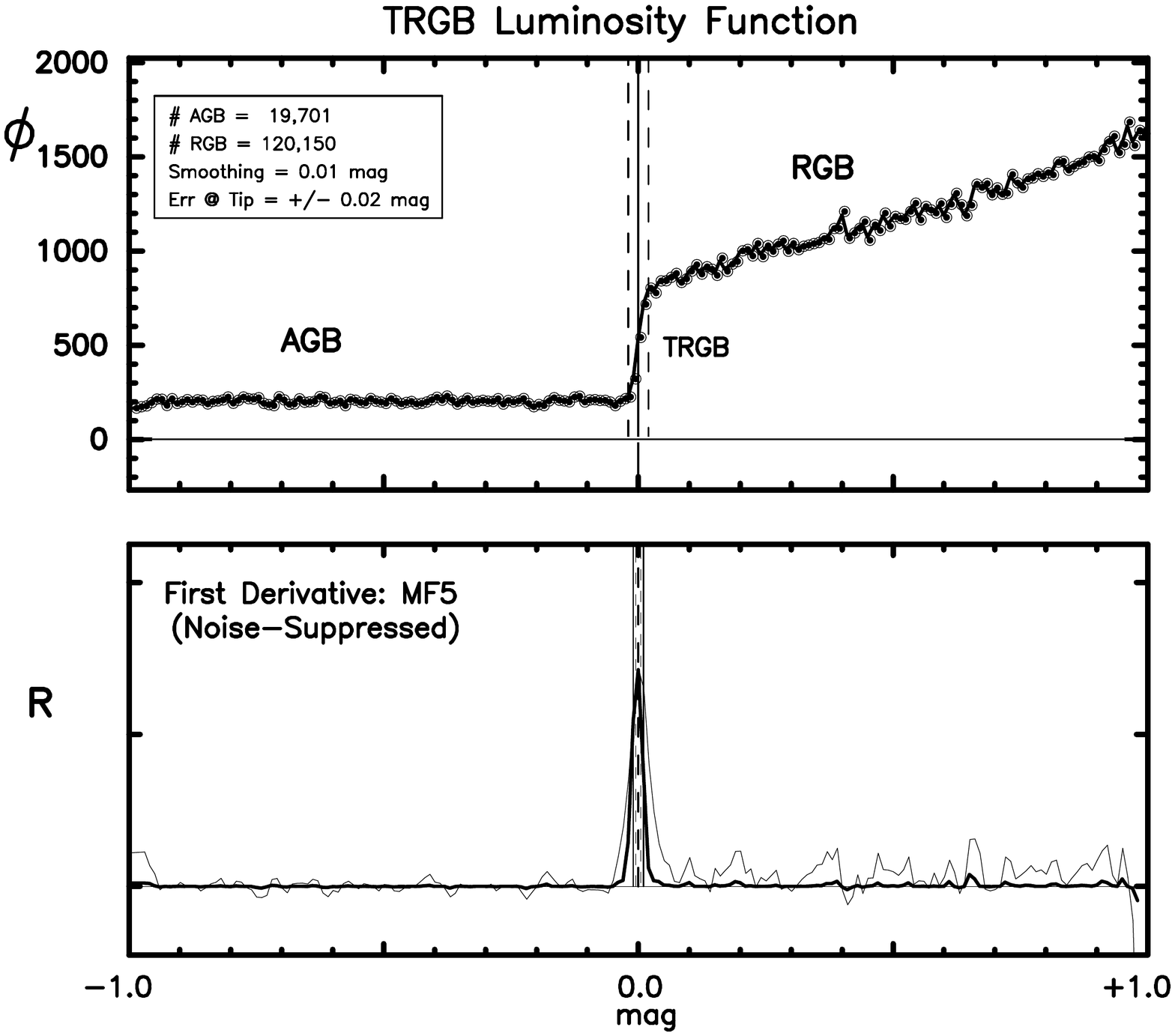}
\includegraphics[width=8.5cm,angle=-0]{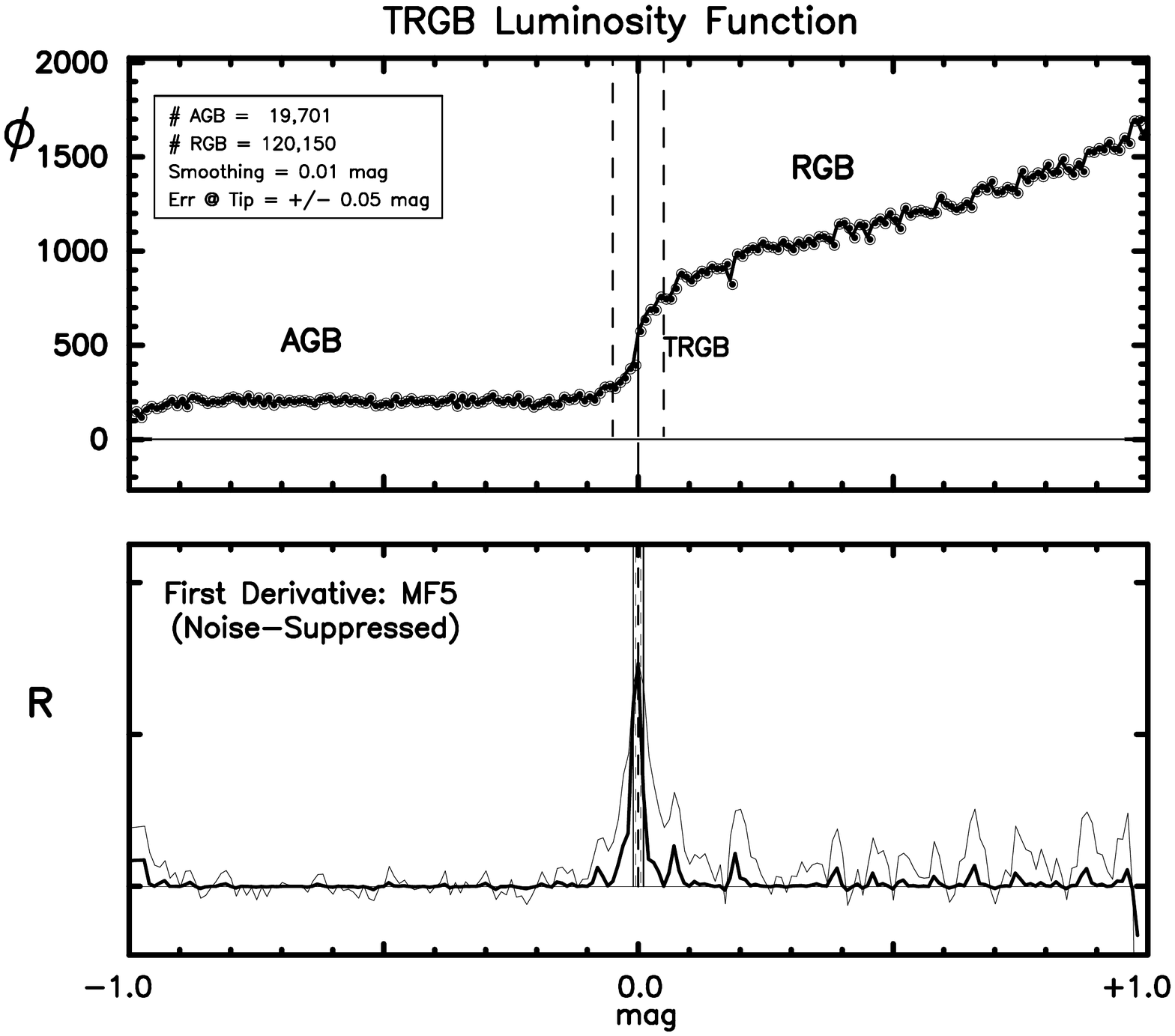}
\includegraphics[width=8.5cm,angle=-0]{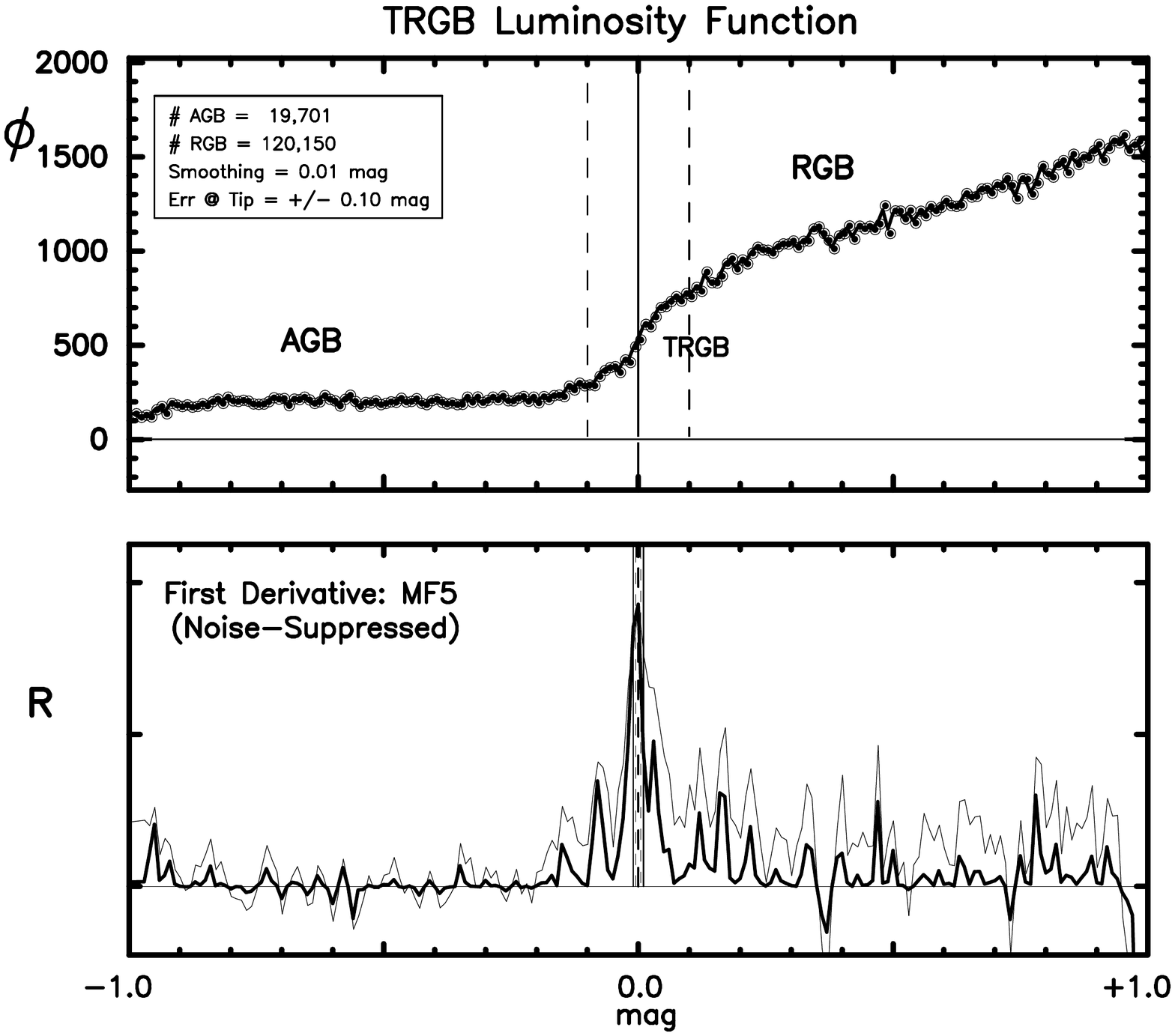}
\includegraphics[width=8.5cm,angle=-0]{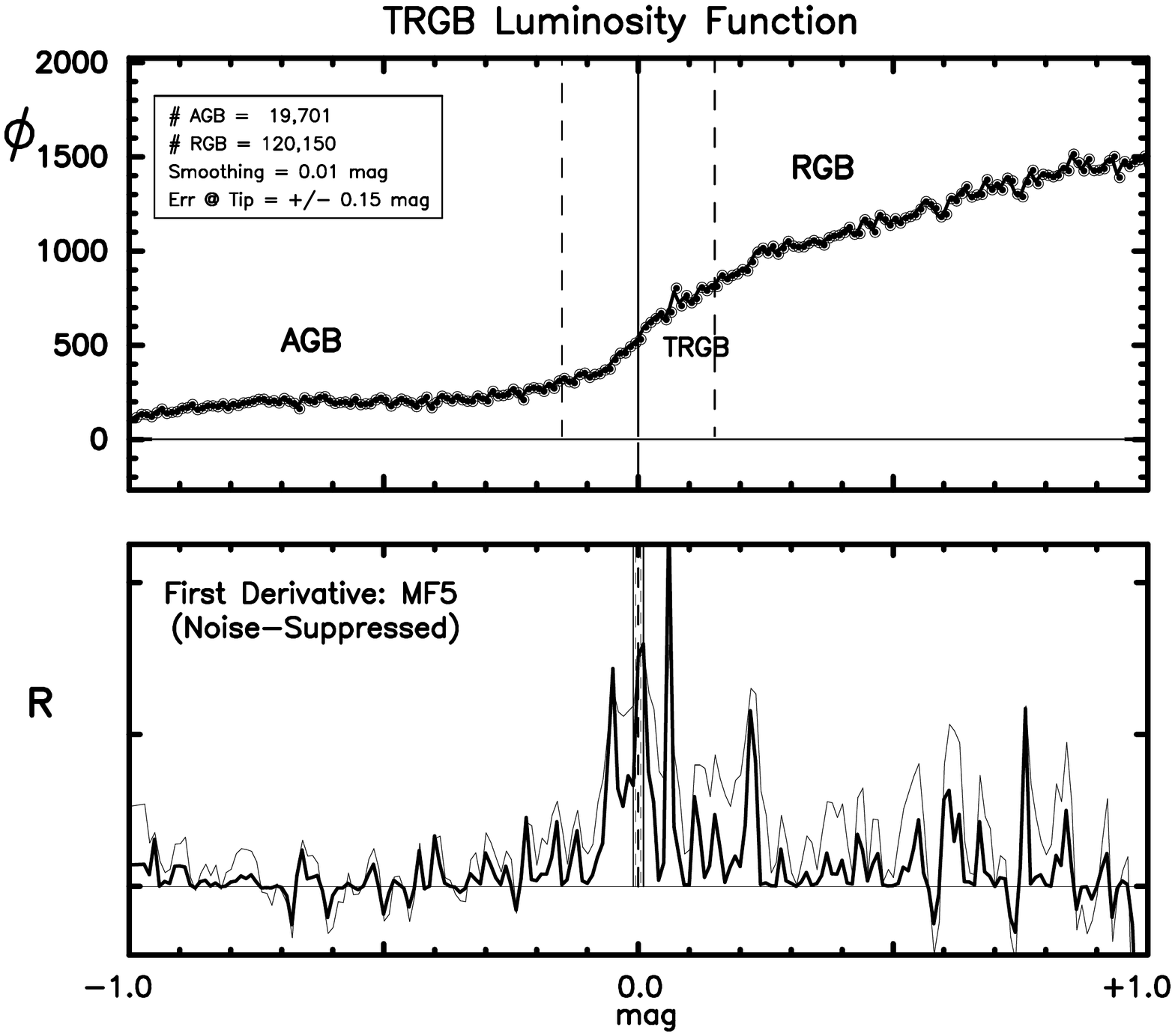}
\includegraphics[width=8.5cm,angle=-0]{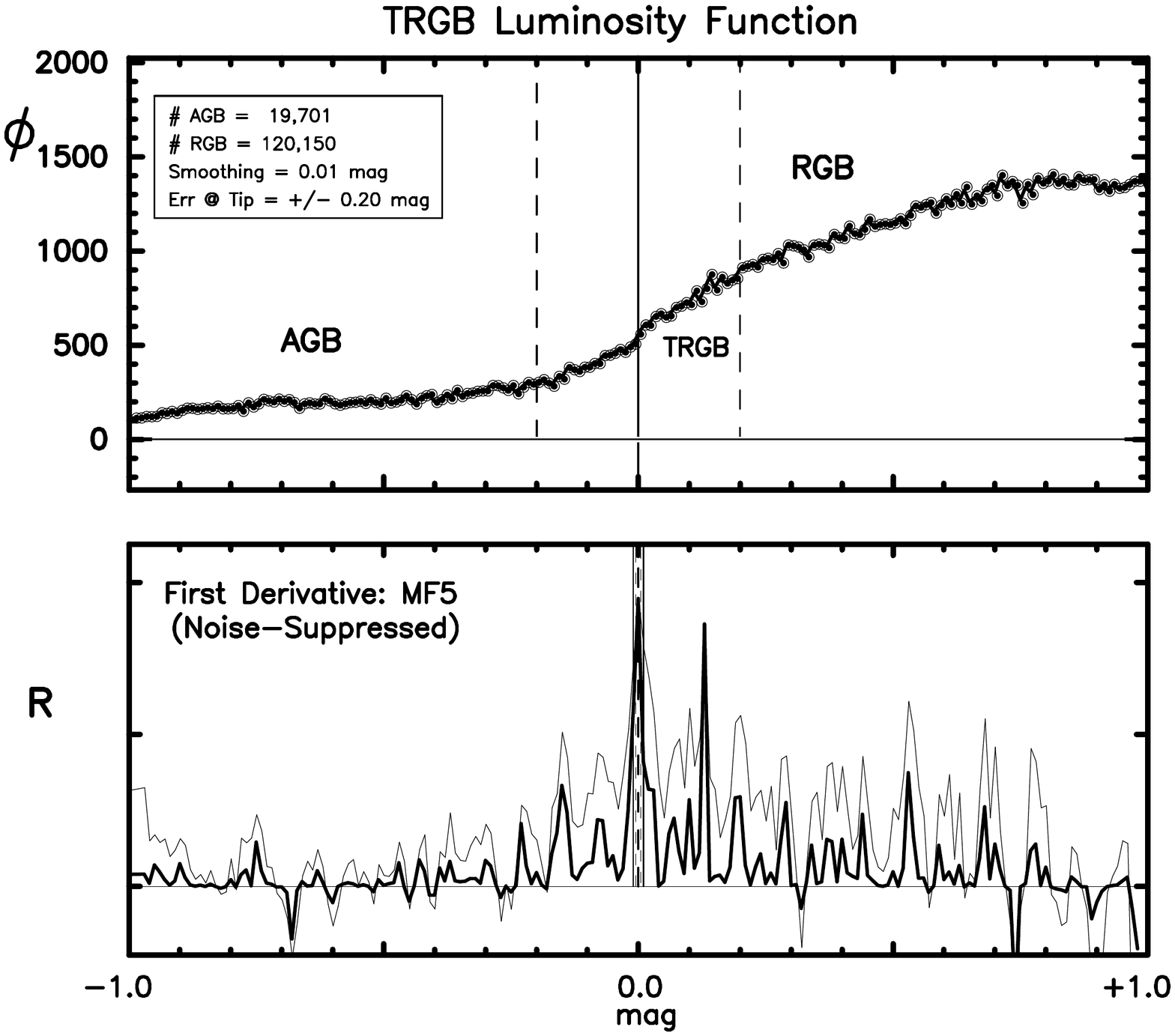} \caption{\small Six sub-panels illustrating the effect of increasing the photometric noise, (from 0.00 to $\pm$0.20~mag) at fixed smoothing ($\pm$0.01 mag) and extremely large populations of RGB stars (about 120,000). The lower portions of each of the six sub-panels shows the first-derivative edge-detector output in both its uncorrected (thin black lines) and it noise-weighted (thicker black line) form. See text for a detailed discussion of the trends.} 
\end{figure*}
\vfill\eject
\subsubsection{ A Range of Photometric Errors: {\bf}{11,000 RGB Stars}, Fixed Smoothing {\bf}{0.01~mag}}

At this iteration we drop the total population of stars contributing to the luminosity function by about factor of ten (down to 11,000 RGB and 2,000 AGB stars), keeping the smoothing at a very low level ($\pm$0.01~mag) as above, while again assessing the effects of increased photometric errors.

It is important to note at this point that the effects of decreased population size and increased photometric errors are causally independent of each other in the plotted luminosity functions. 
At fixed precision in the photometry, downsizing the population size can only {\it decrease} the number of stars in any given bin and thereby {\it increase} the relative error ($\sqrt N/N$) in that bin.
The increased scatter in all of the panels of Figure 3 as compared to Figure 2 is a direct result of the decreased number statistics and can be seen repeated and progressively amplified later on in Figures 4 and 5 as the population size decreases further. 

What may not be immediately obvious is why the ``photometric redistribution'' of the data across bins at a given population size has virtually no affect on the noise amplitude in the luminosity functions, seen on either side of the discontinuity. 
The reason for this is that while this form of smoothing redistributes data laterally, it does not significantly change the local mean value of N  in any given bin (i.e., photometric redistribution conserves total counts within its smoothing radius).  
That means, of course, that $\sqrt N/N$ is also ``conserved". 
{\bf}{Photometric blurring of individual data bin does not reduce} $\sqrt N$ {\bf}{population noise in the RGB continuum;} however, because of the strong asymmetry, inherent in the jump in the luminosity function at the TRGB, more RGB stars migrate to higher luminosities (and boost the apparent AGB population) than the other way around. 
Accordingly, photometric errors erode the tip and systematically decrease the slope of the transition marking the rise from the AGB to RGB populations, decreasing the contrast between the AGB and the tip, {\bf}{but still not moving the position of the discontinuity.}  

The small degree of ($\pm$0.01~mag) smoothing in these simulations tracks not only the population fluctuations from bin to bin, but also the precisely-defined, sharp rise marking the TRGB.
As the photometric errors increase and the transition widens and flattens the population, the power in the response function crossing the ever-widening transition region starts to drop. From a photometric error of $\pm$0.05~mag onward (middle left panel) it is approaching the noise level of the RGB population noise. 
In this simulation there are 3-4 noise spikes ``downstream'' of the true tip that are of similar power, rendering the identification of the true tip ambiguous.
At a photometric error of $\pm$0.15 mag (lower left panel) the number density of false peaks is overwhelming  and even noise in the AGB population starts to contribute to an ``upstream''  ambiguity. At this level of smoothing, population size and photometric error, the tip cannot be extracted from the noise.

\medskip
{\bf}{ Summary 2 --}  For an RGB population of approximately 10,000 stars an unambiguous detection of the tip can be assured with a photometric error of $\pm$0.05~mag or less. 
At a photometric error of $\pm$0.10~mag the first-detected discontinuity is the true one with false positives rapidly developing at fainter magnitudes, downstream. However, at a photometric error 
of $\pm$0.15~mag and beyond false positives overwhelm the signal in power and in number, both below and above the true tip.

\begin{figure*} \centering
\includegraphics[width=8.0cm,angle=-0]{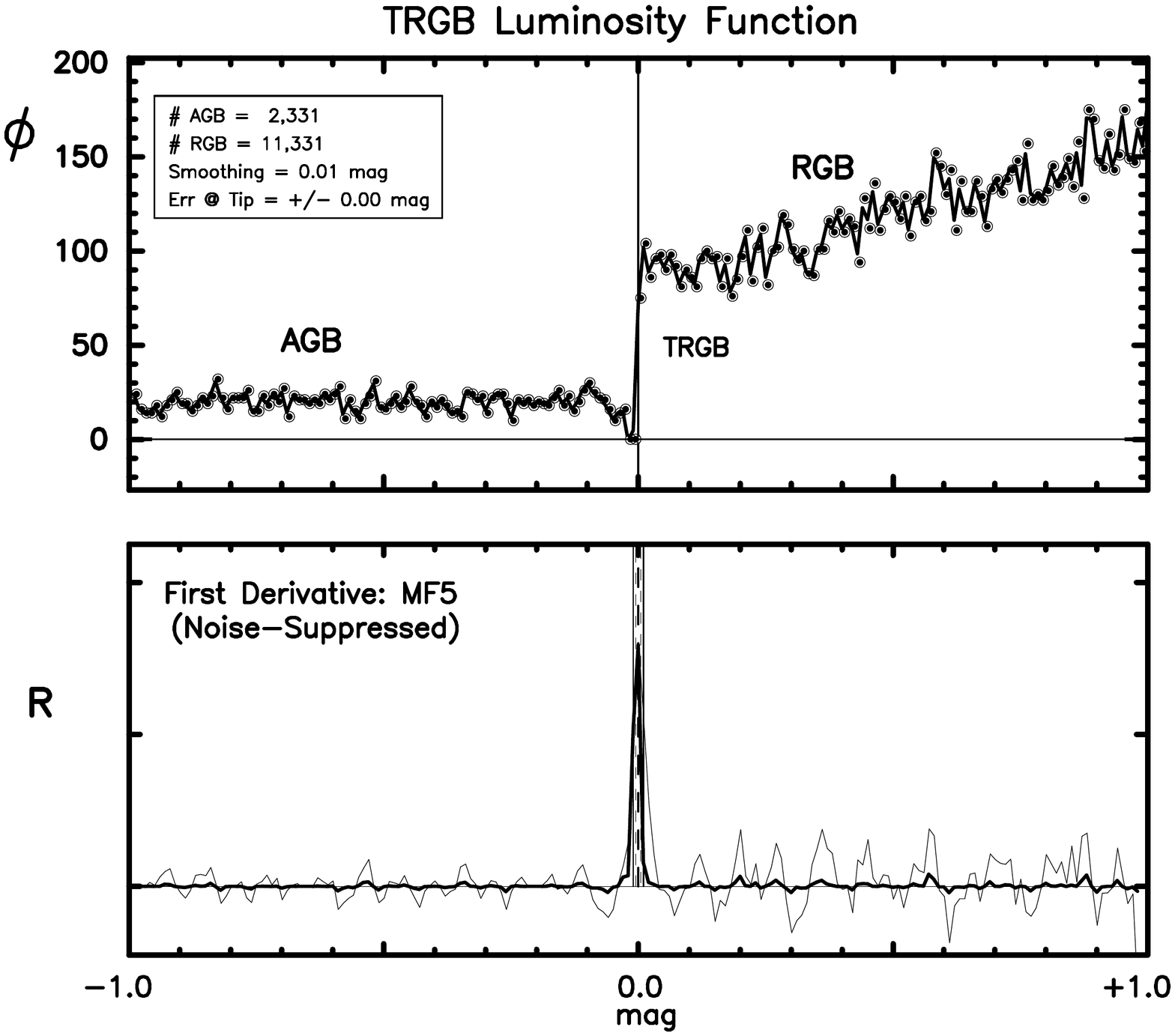}
\includegraphics[width=8.0cm,angle=-0]{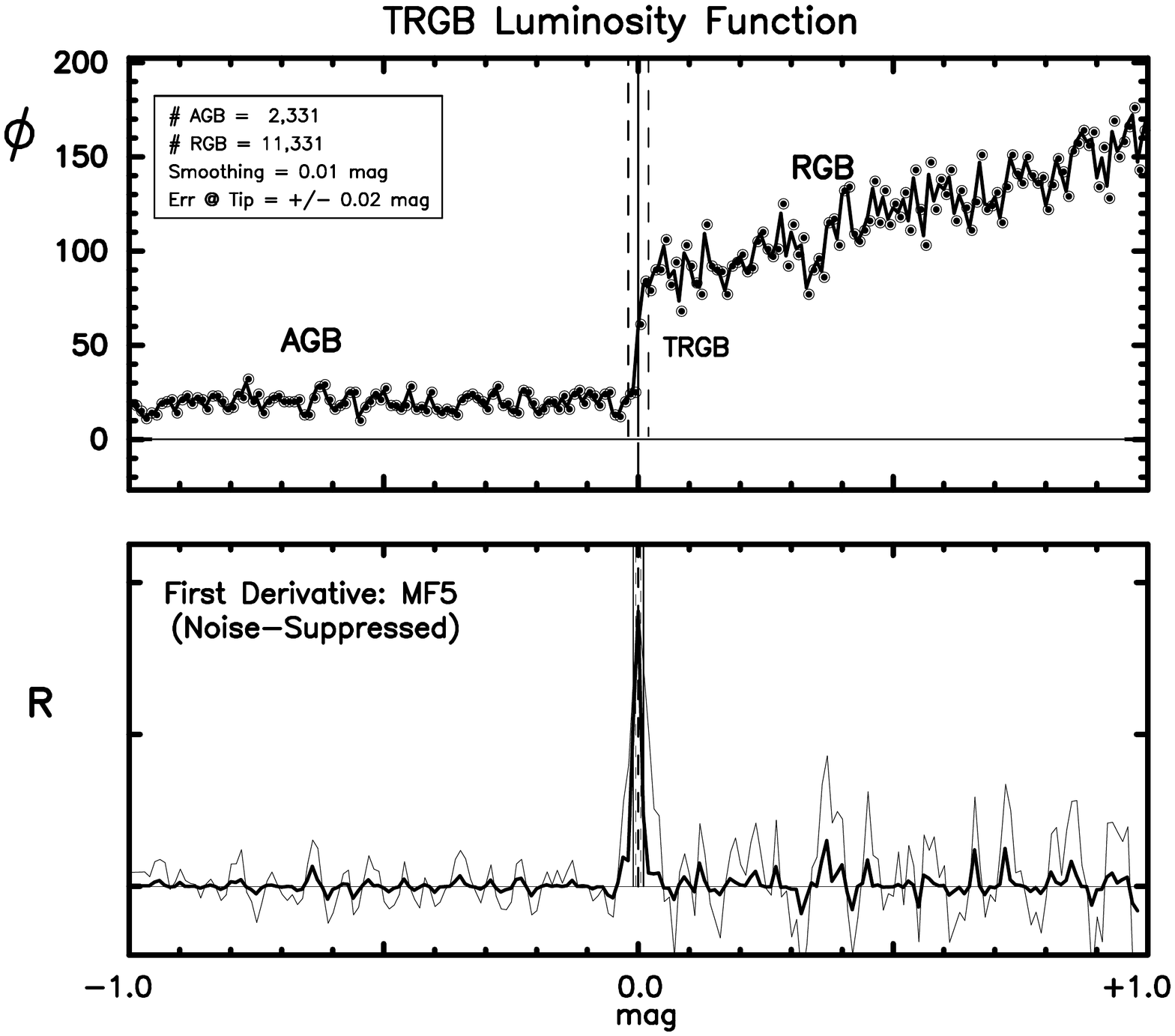}
\includegraphics[width=8.0cm,angle=-0]{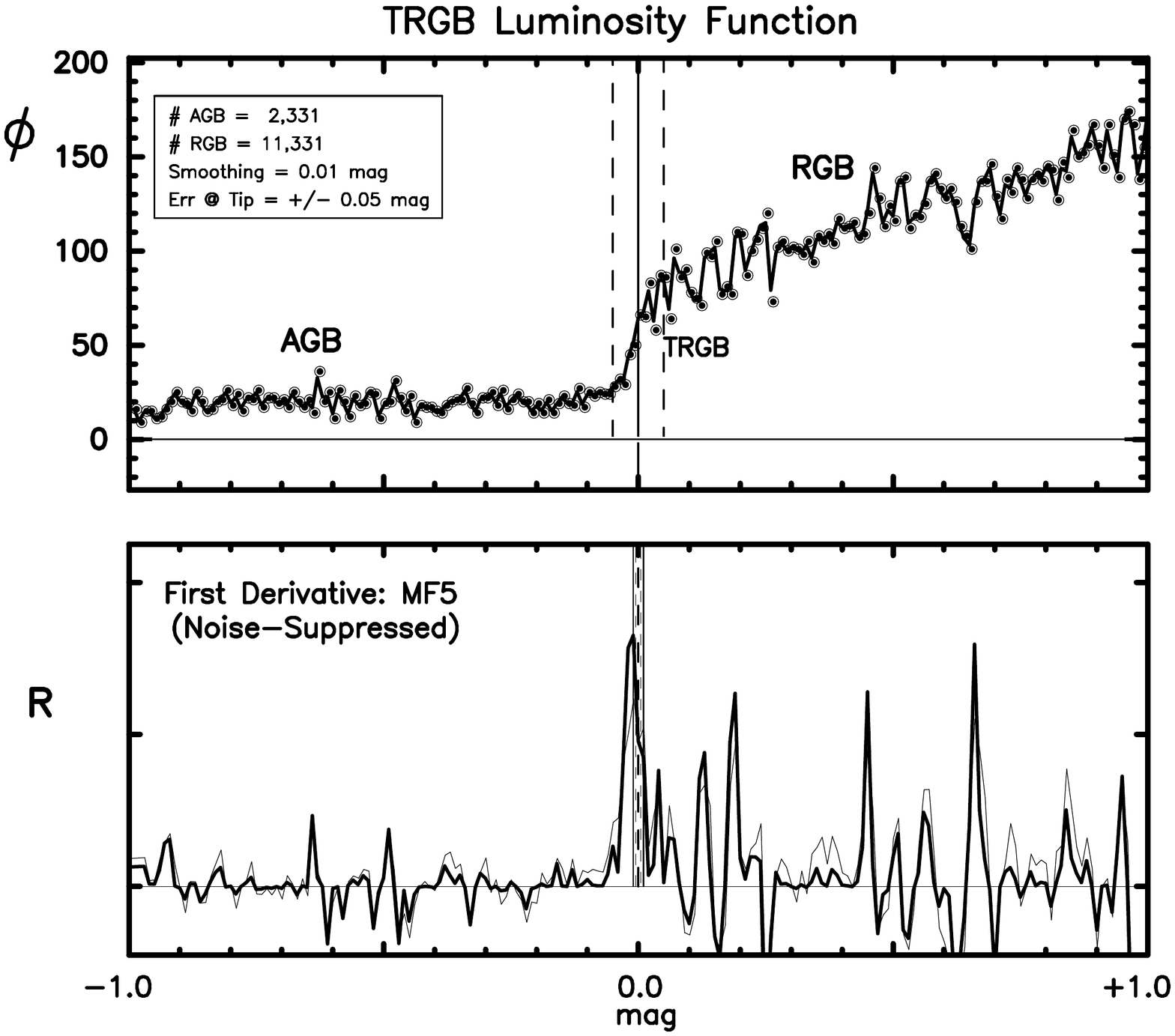}
\includegraphics[width=8.0cm,angle=-0]{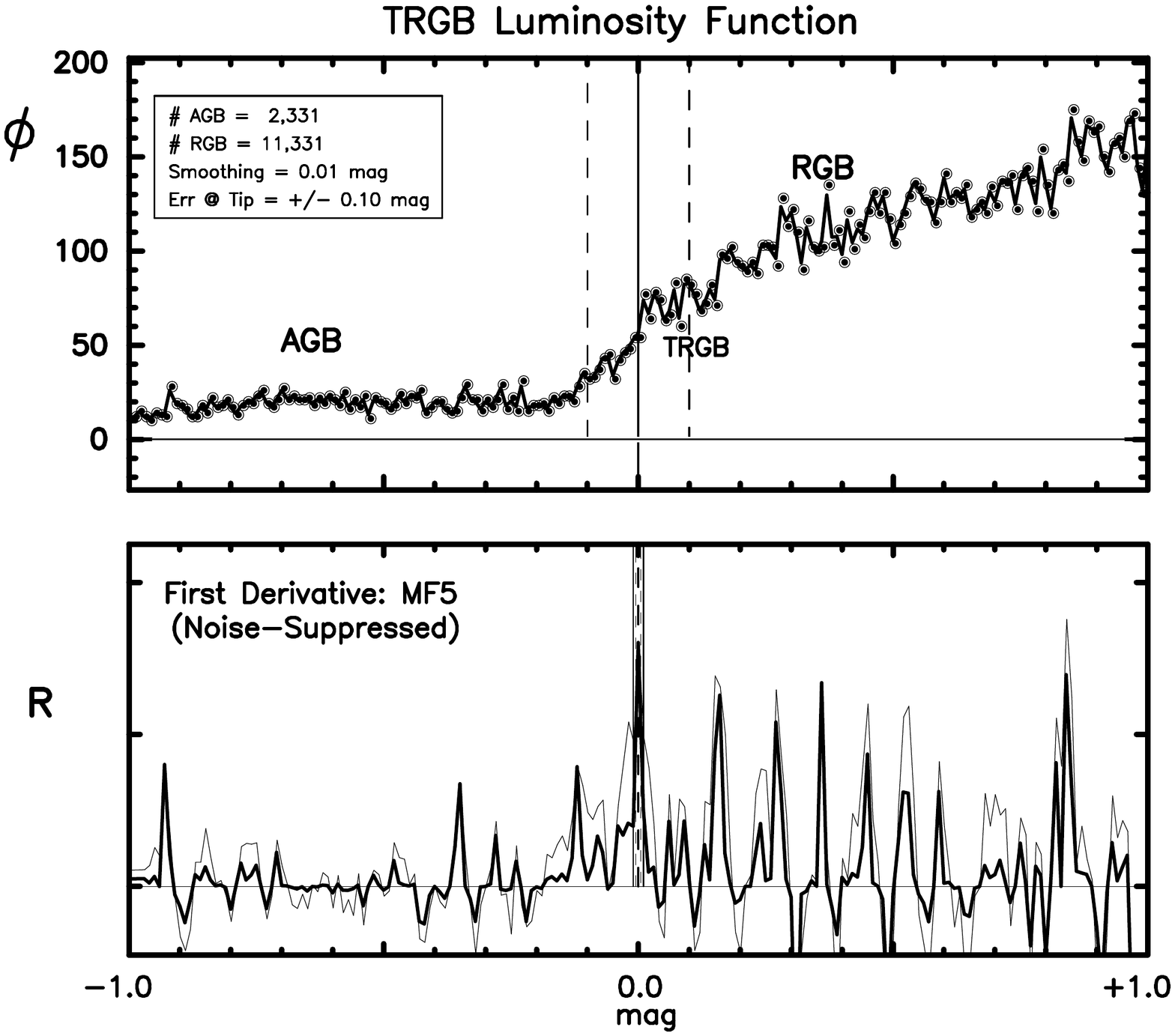}
\includegraphics[width=8.0cm,angle=-0]{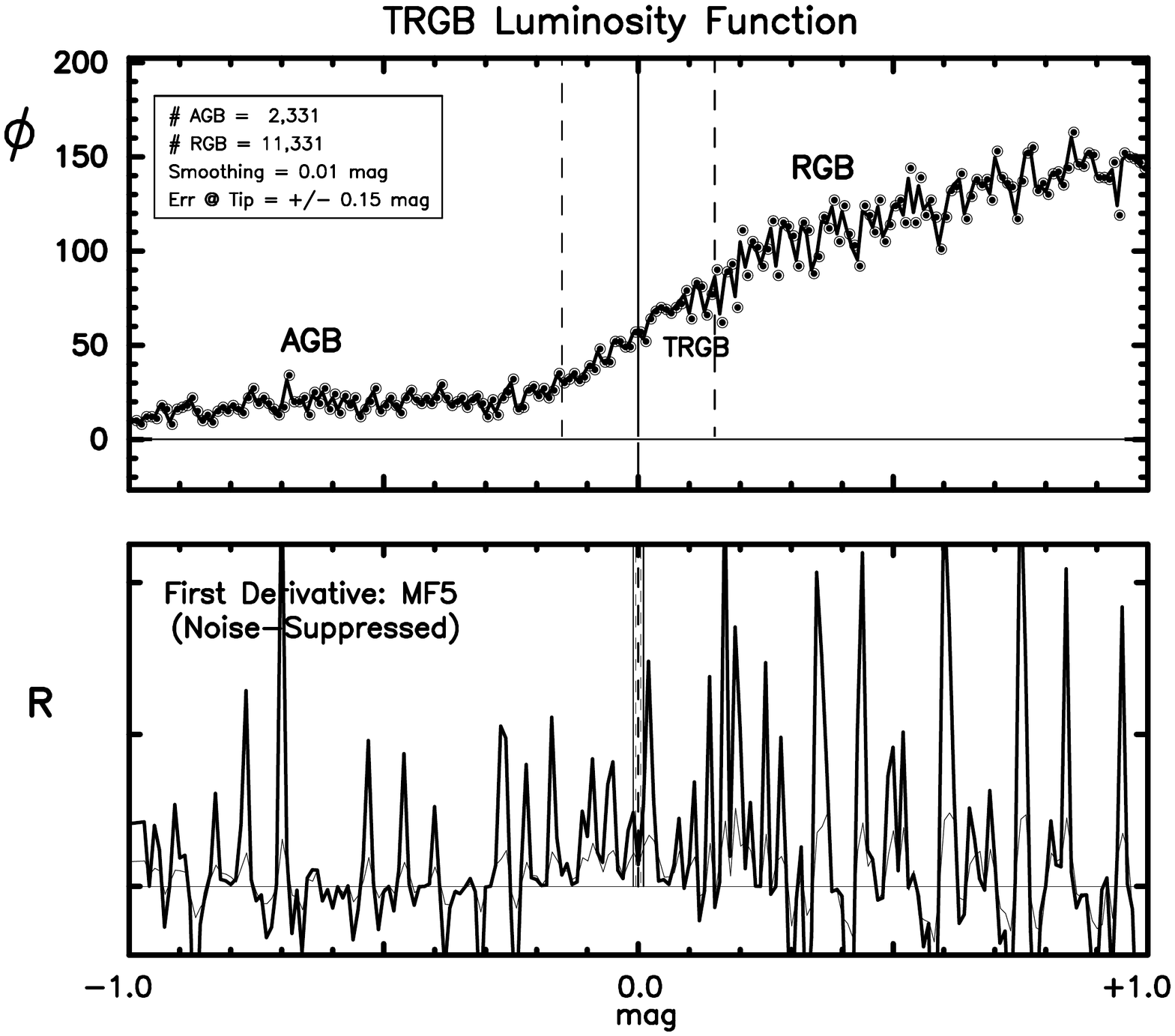}
\includegraphics[width=8.0cm,angle=-0]{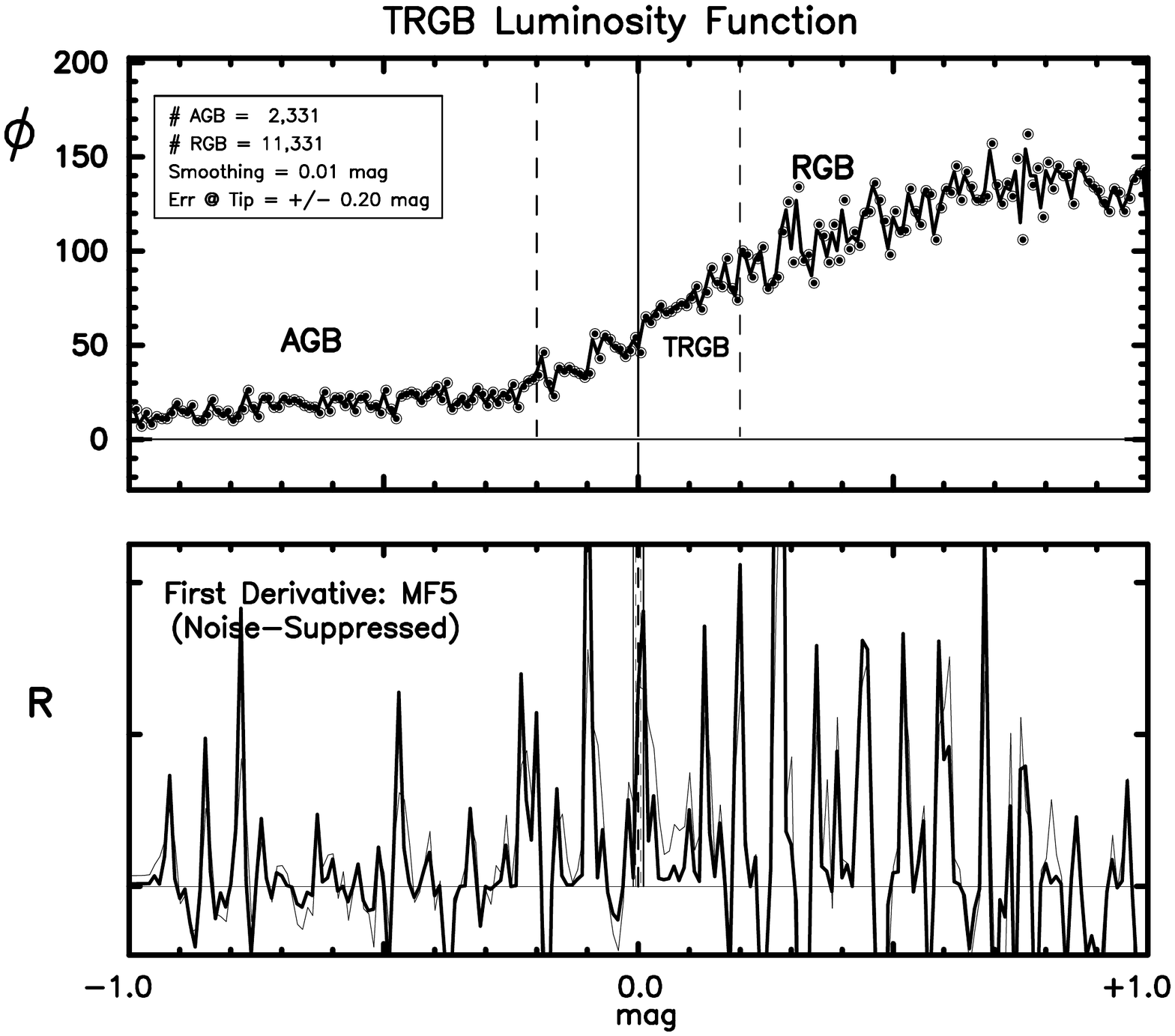} \caption{\small Six sub-panels illustrating the effect of increasing the photometric noise, (from 0.00 to $\pm$0.20~mag) at fixed smoothing ($\pm$0.01 mag) and moderately large populations of RGB stars (11,000). The lower portions of each of the six sub-panels shows the first-derivative edge-detector output in both its uncorrected (thin black lines) and it noise-weighted (thicker black line) form. See text for a detailed discussion of the trends. } \end{figure*}
\vfill\eject
\subsubsection{A Range of Photometric Errors: {\bf}{1,000 RGB Stars}, Fixed Smoothing {\bf}{$\pm$0.01~mag}}

This simulation drops the RGB population to about 1,000 stars, another factor of ten below the previous investigation.
Almost immediately, at a photometric error level of $\pm$0.02~mag, the power in the response function at the tip has dropped to a level comparable to population noise in the RGB luminosity function. 
Several `false positives' are seen (in the middle left panel of Figure 3) downsteam of the true TRGB. 
Noise spikes in the RGB magnitude range are so frequent (at this smoothing) that they can randomly appear around the tip without really being detections of the tip. 
Note the cluster of noise  spikes well below the known position of the TRGB in the lower left panel and then again a spike somewhat brighter (and certainly stronger) than the tip in the adjacent, lower right panel.

{\bf}{Summary 3--} For a population of only 1,000 RGB stars a photometric error in excess of $\pm$0.02~mag results in false positives overwhelming the tip detection, {\it in the absence of any significant smoothing} (but see Sections 4.2 and 4.3 below).

\begin{figure*} \centering
\includegraphics[width=8.0cm,angle=-0]{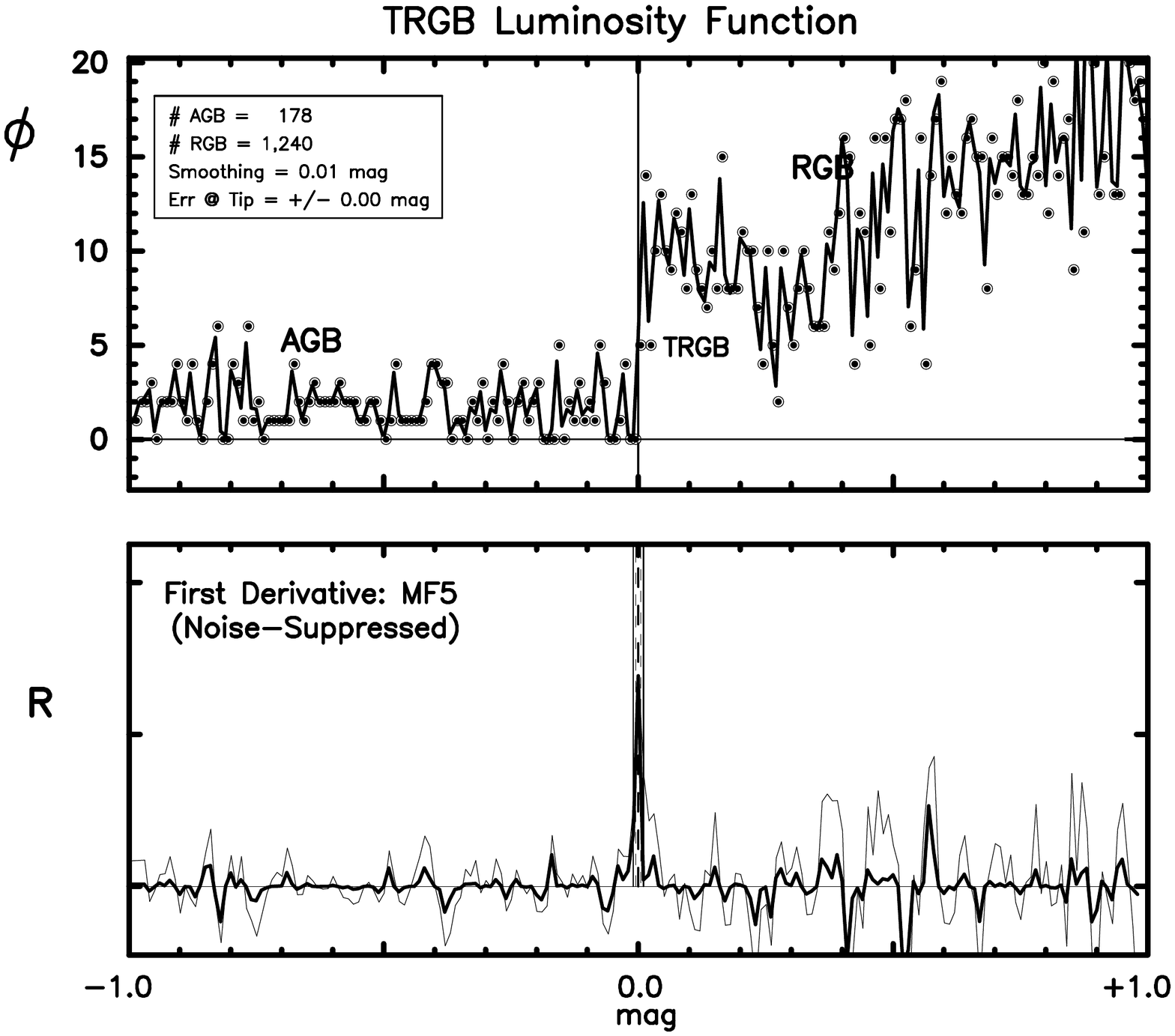}
\includegraphics[width=8.0cm,angle=-0]{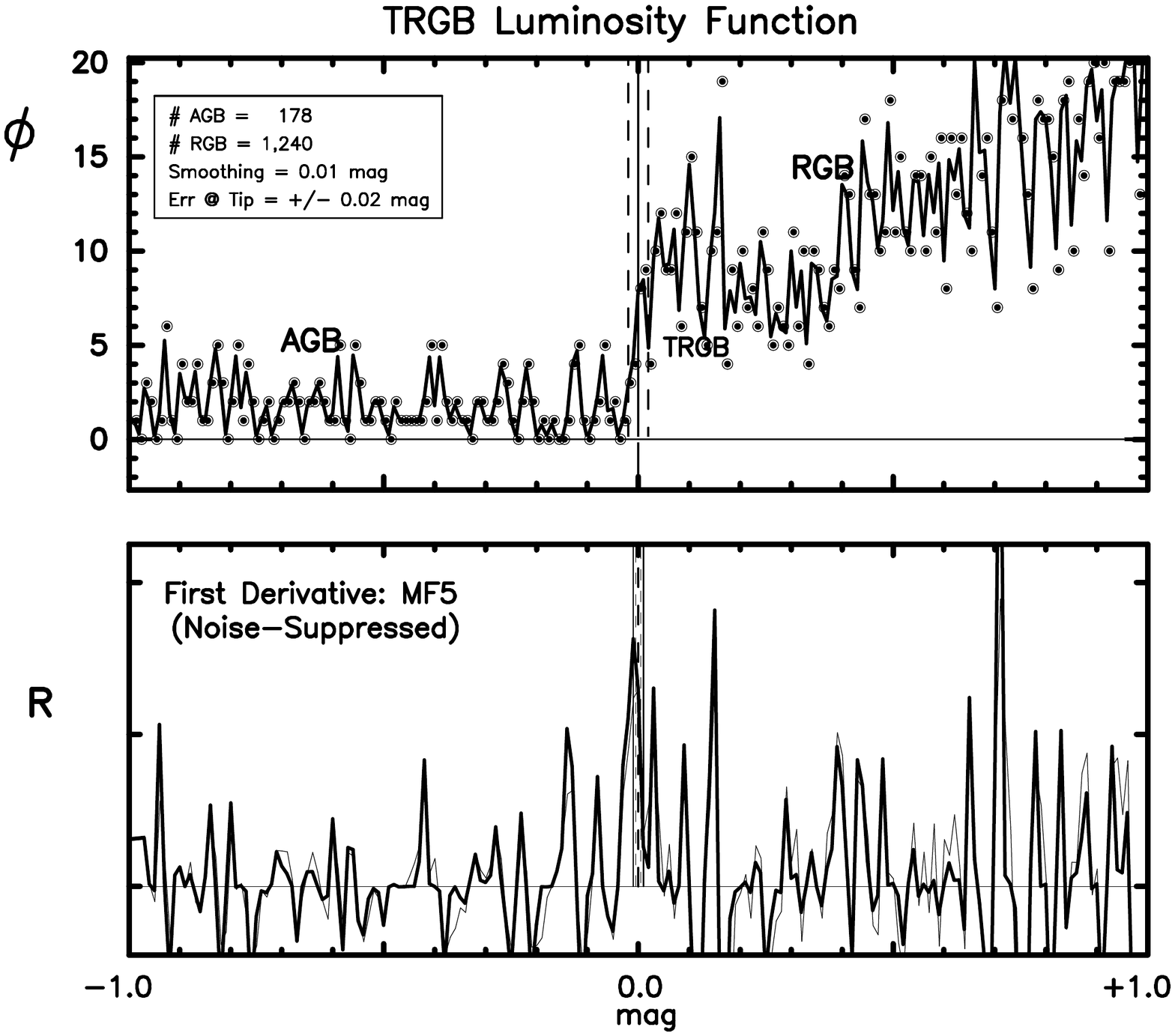}
\includegraphics[width=8.0cm,angle=-0]{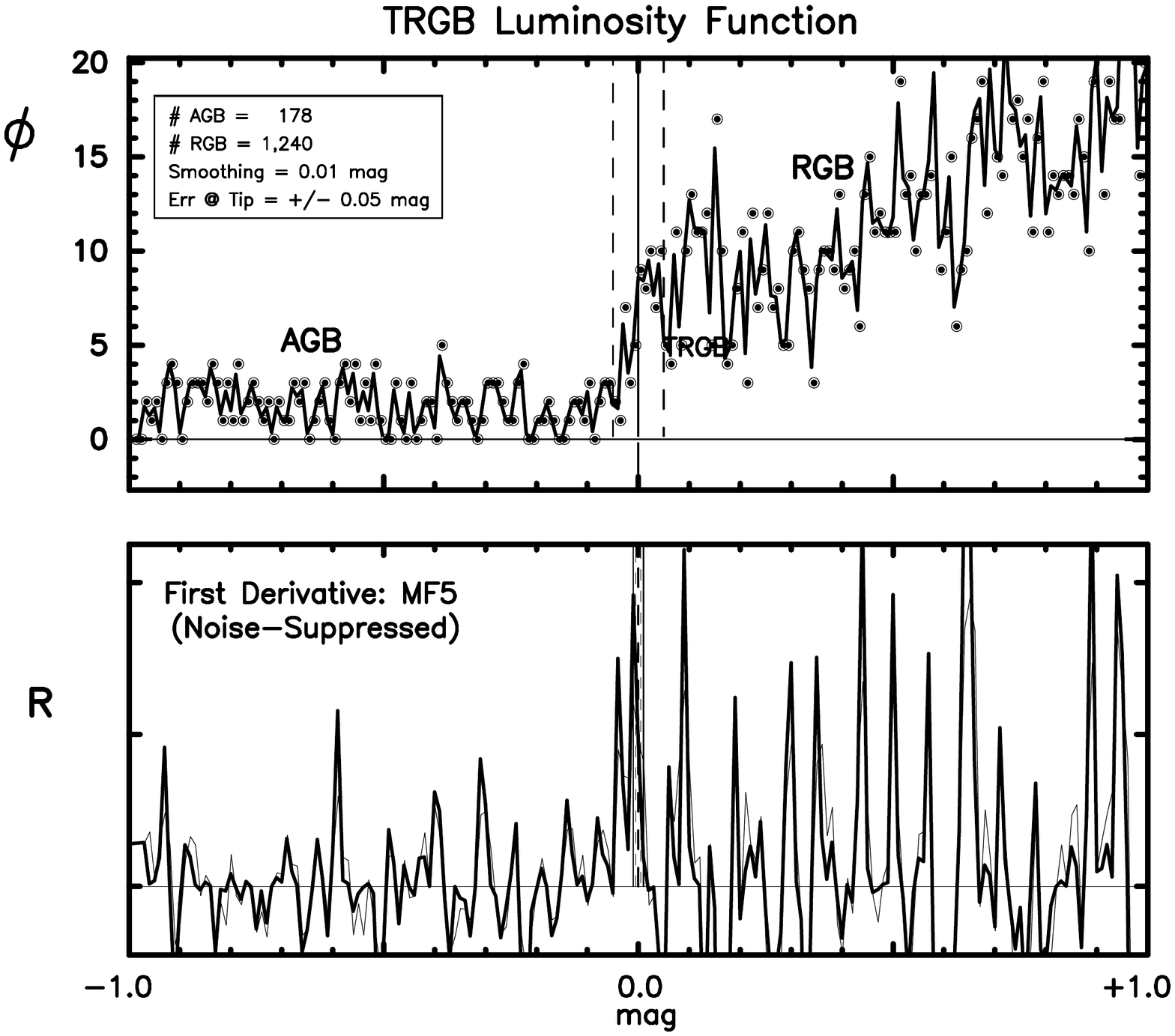}
\includegraphics[width=8.0cm,angle=-0]{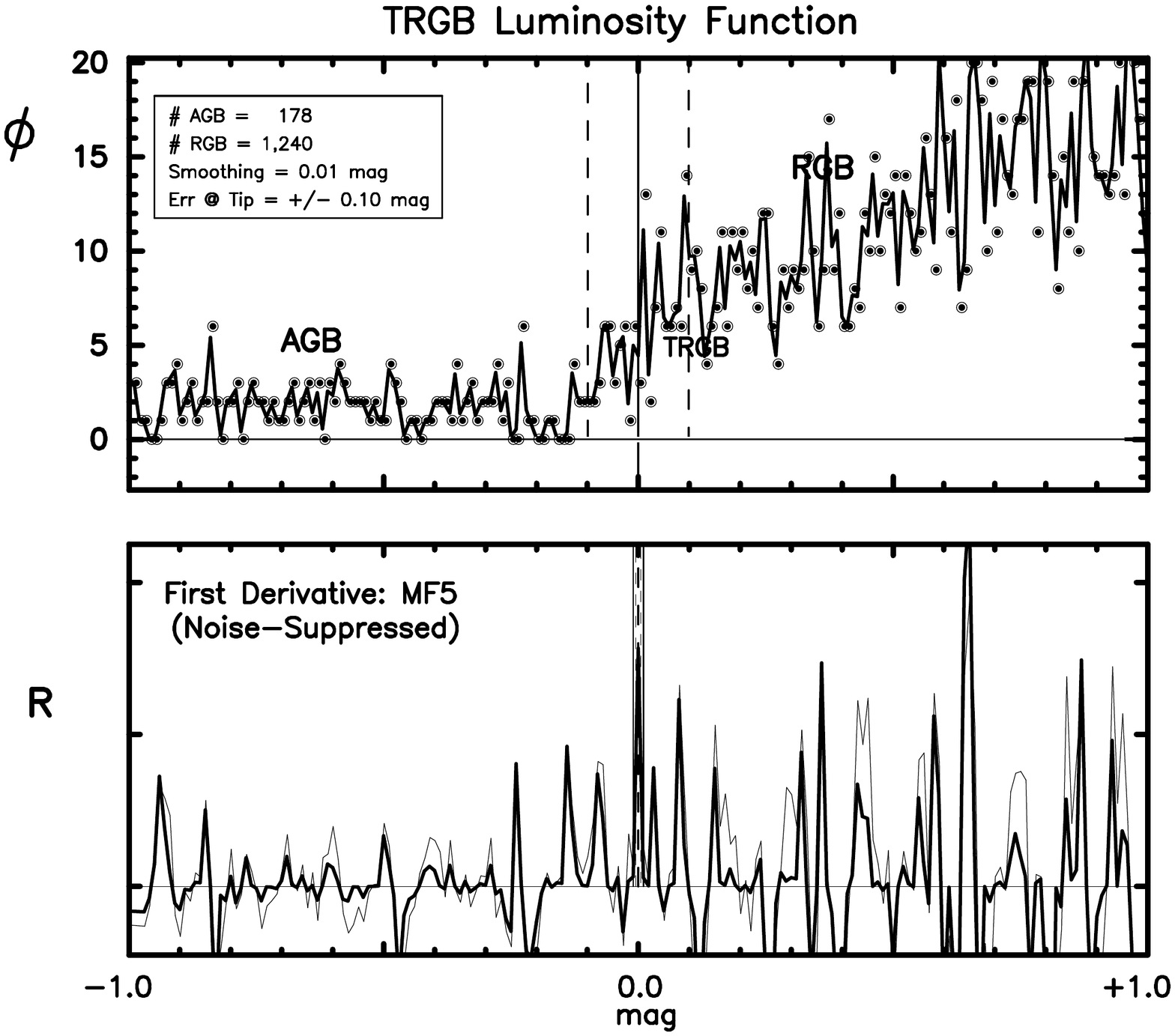}
\includegraphics[width=8.0cm,angle=-0]{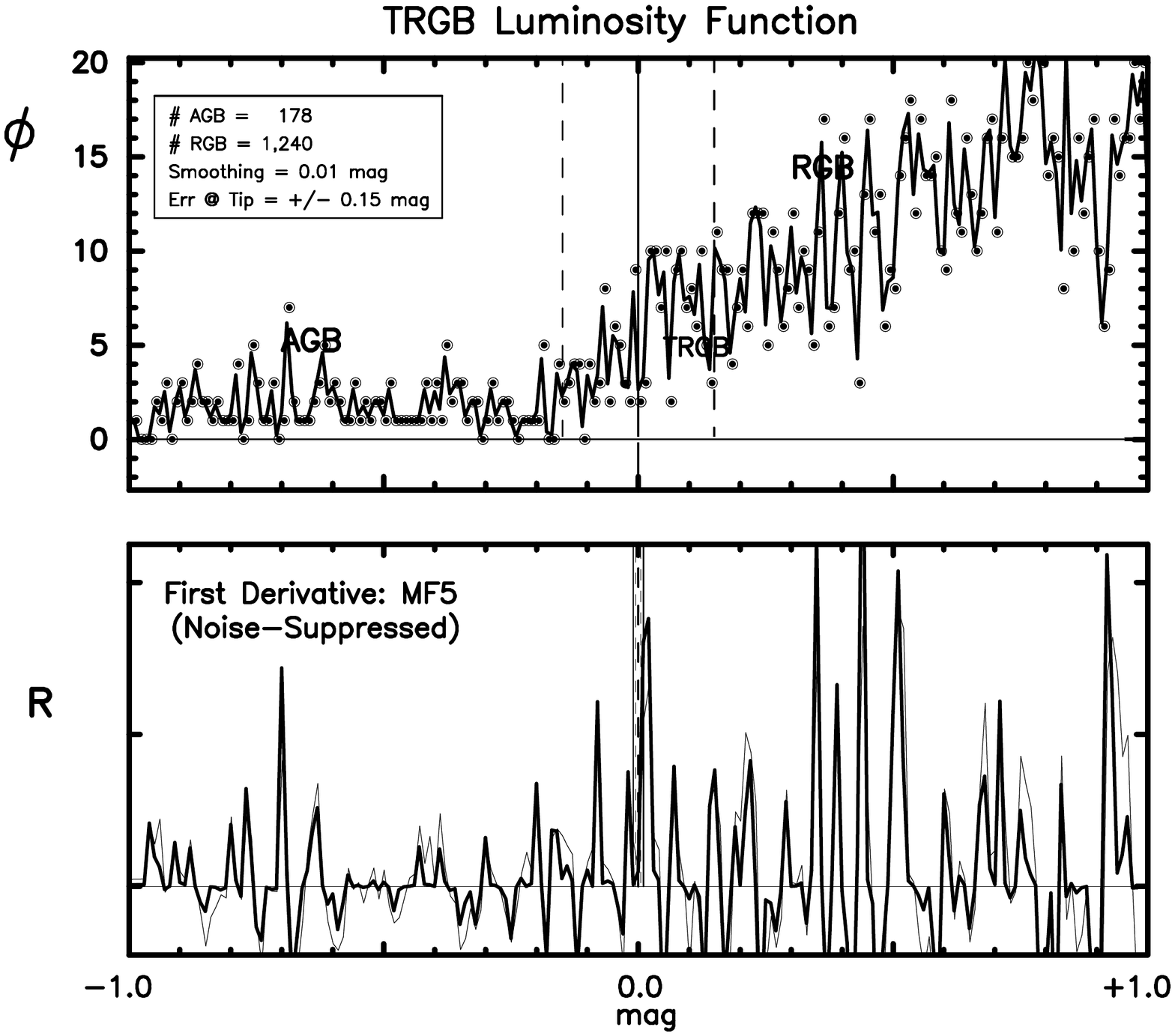}
\includegraphics[width=8.0cm,angle=-0]{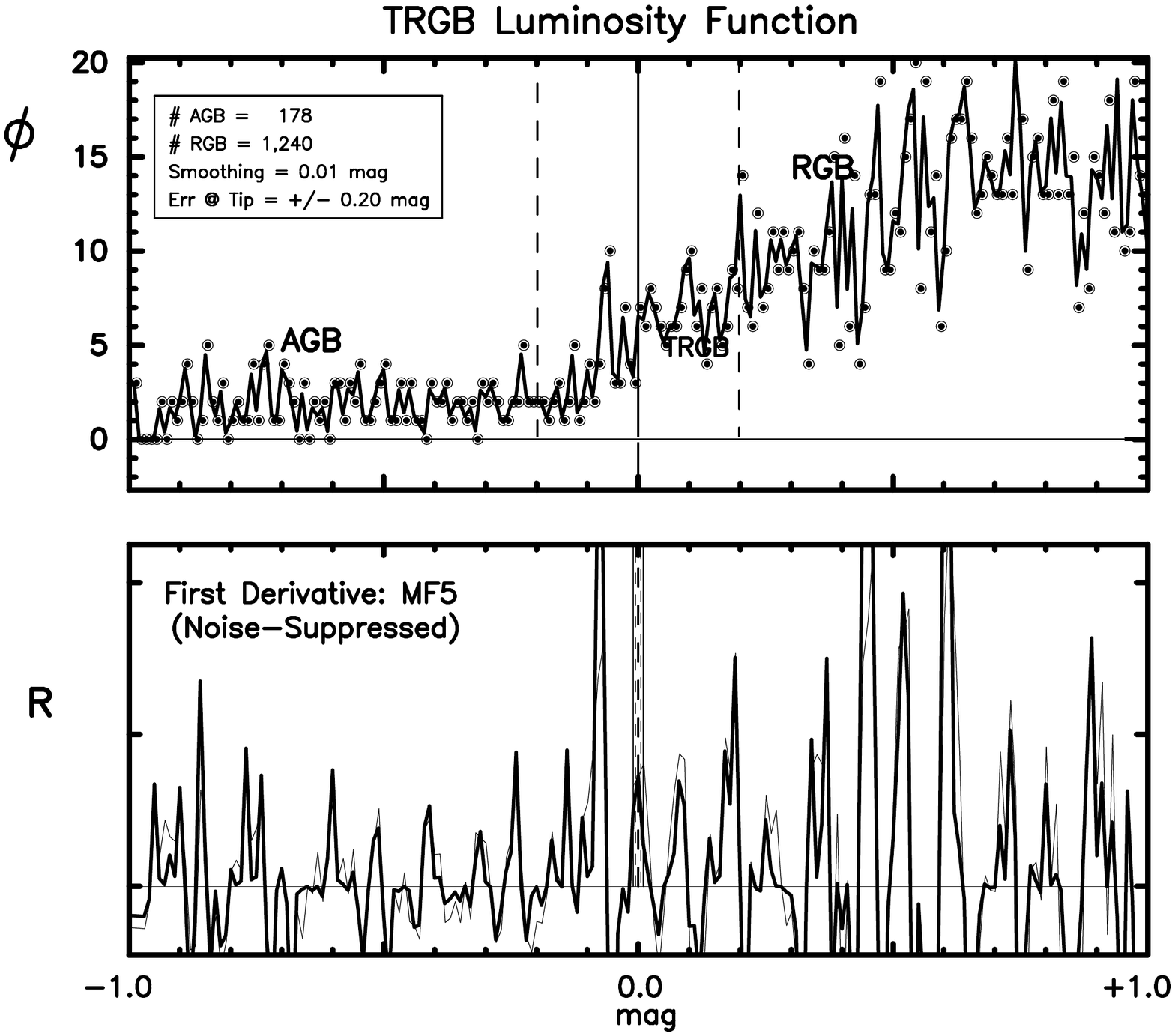} \caption{\small Six sub-panels illustrating the effect of increasing the photometric noise, (from 0.00 to $\pm$0.20~mag) at fixed smoothing ($\pm$0.01 mag) and small populations of RGB stars (1,200). The lower portions of each of the six sub-panels shows the first-derivative edge-detector output in both its uncorrected (thin black lines) and it noise-weighted (thicker black line) form. See text for a detailed discussion of the trends.} 
\end{figure*}
\vfill\eject
\subsubsection{A Range of Photometric Errors: {\bf}{124 RGB Stars}, Fixed Smoothing {\bf}{$\pm$0.01~mag}}

As may well have been anticipated by the trends already seen above in the increased number of false positives as the sample size decreased and as the photometric errors increased (at fixed smoothing), this last simulation (shown in Figure 5)  contains only 120 RGB stars, and is dominated by noise. 
While the six-to-one contrast ratio between the RGB and the AGB population still applies, the depleted populations on either side of the jump at the TRGB are so dominated by Poisson noise that (without smoothing) both the luminosity function itself and the tip-detection response function are almost indistinguishable from noise.
But with hindsight, gleaned from coming panels and figures, there is still (surprisingly perhaps) meaningful information on the position of the TRGB in all of these realizations.

{\bf}{Summary 4 --} RGB populations of this size are insufficient to provide reliable measurements of the tip magnitude, but some information can still be gained.

\begin{figure*} \centering
\includegraphics[width=8.0cm,angle=-0]{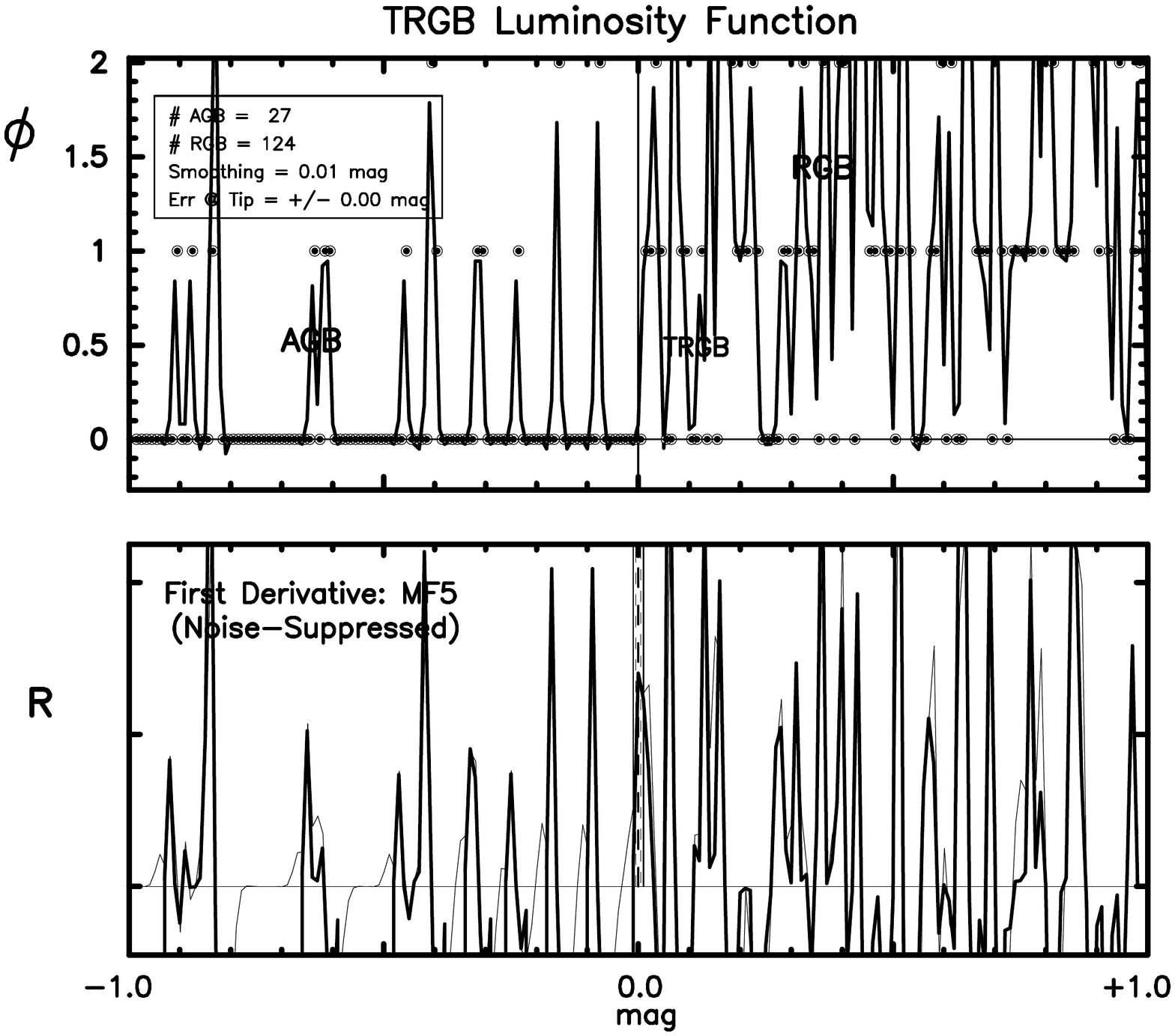}
\includegraphics[width=8.0cm,angle=-0]{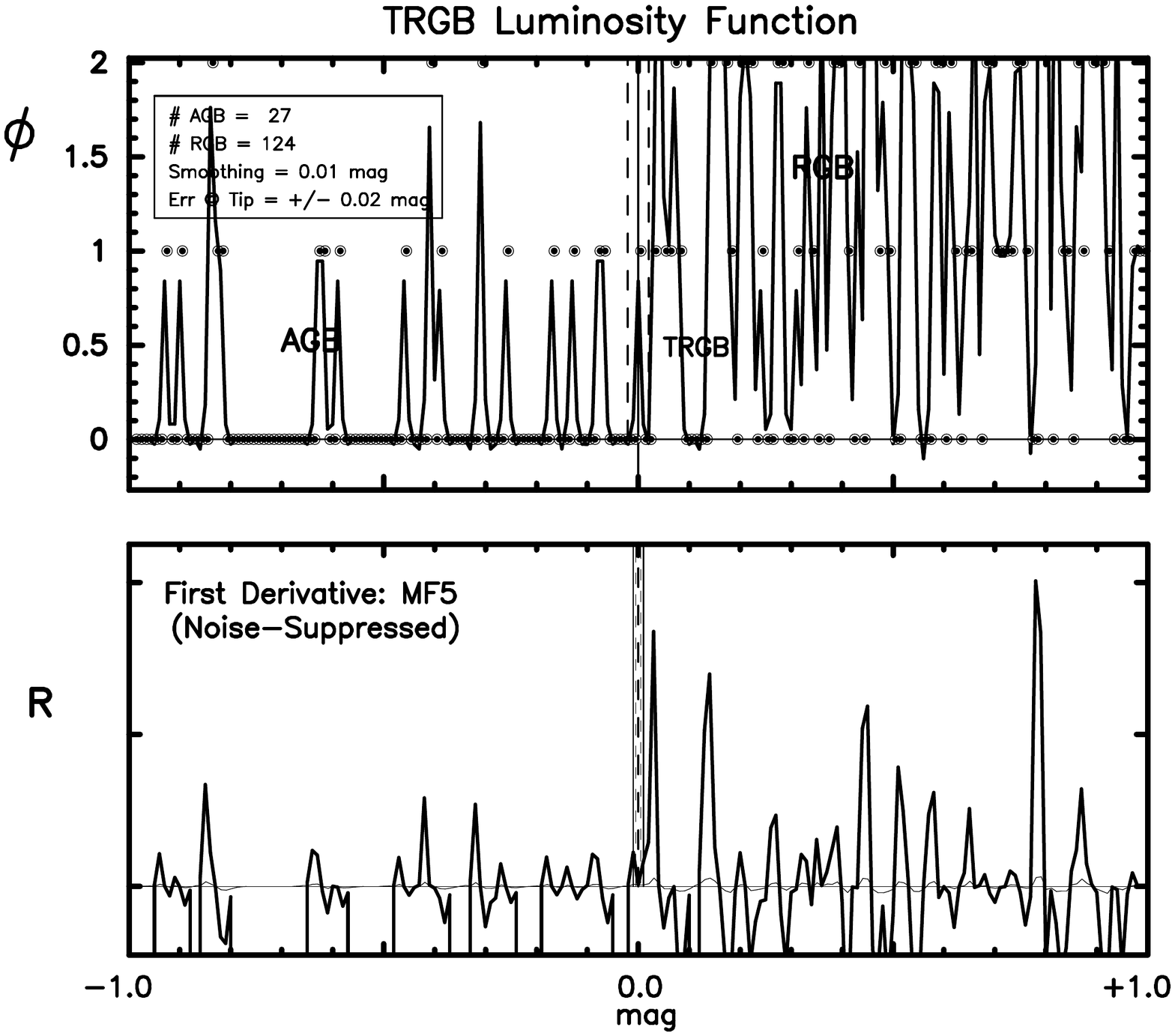}
\includegraphics[width=8.0cm,angle=-00]{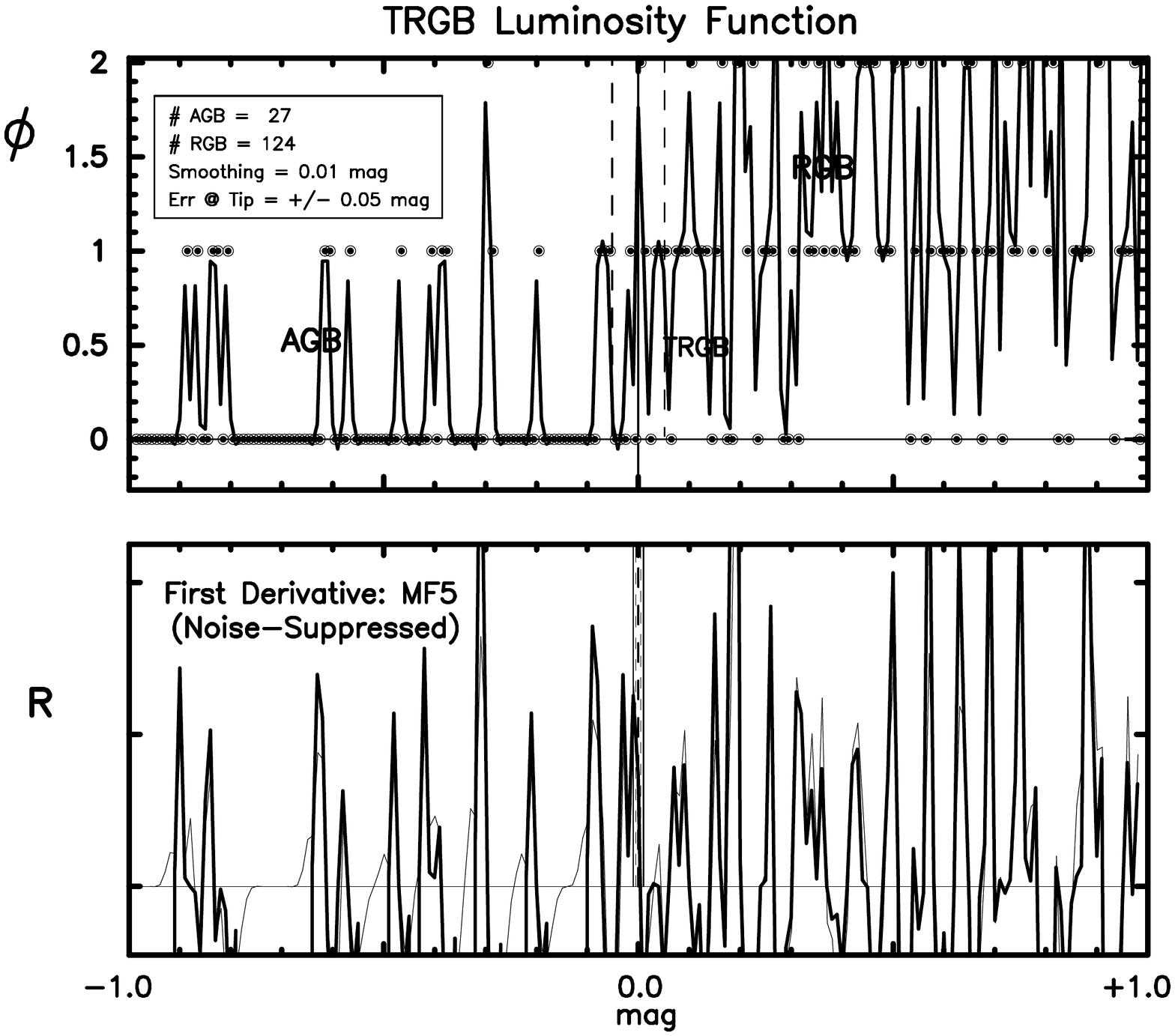}
\includegraphics[width=8.0cm,angle=-00]{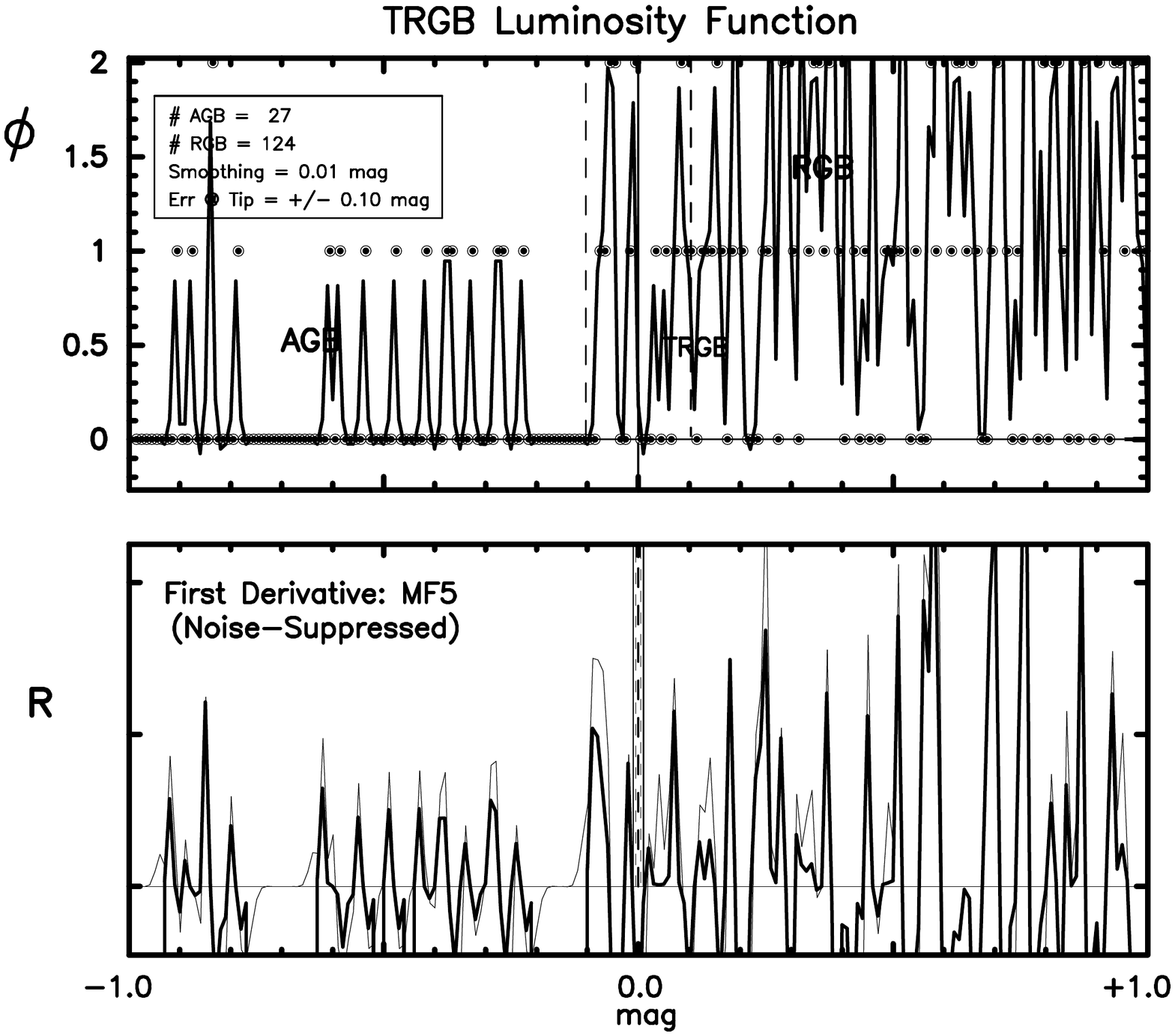}
\includegraphics[width=8.0cm,angle=-00]{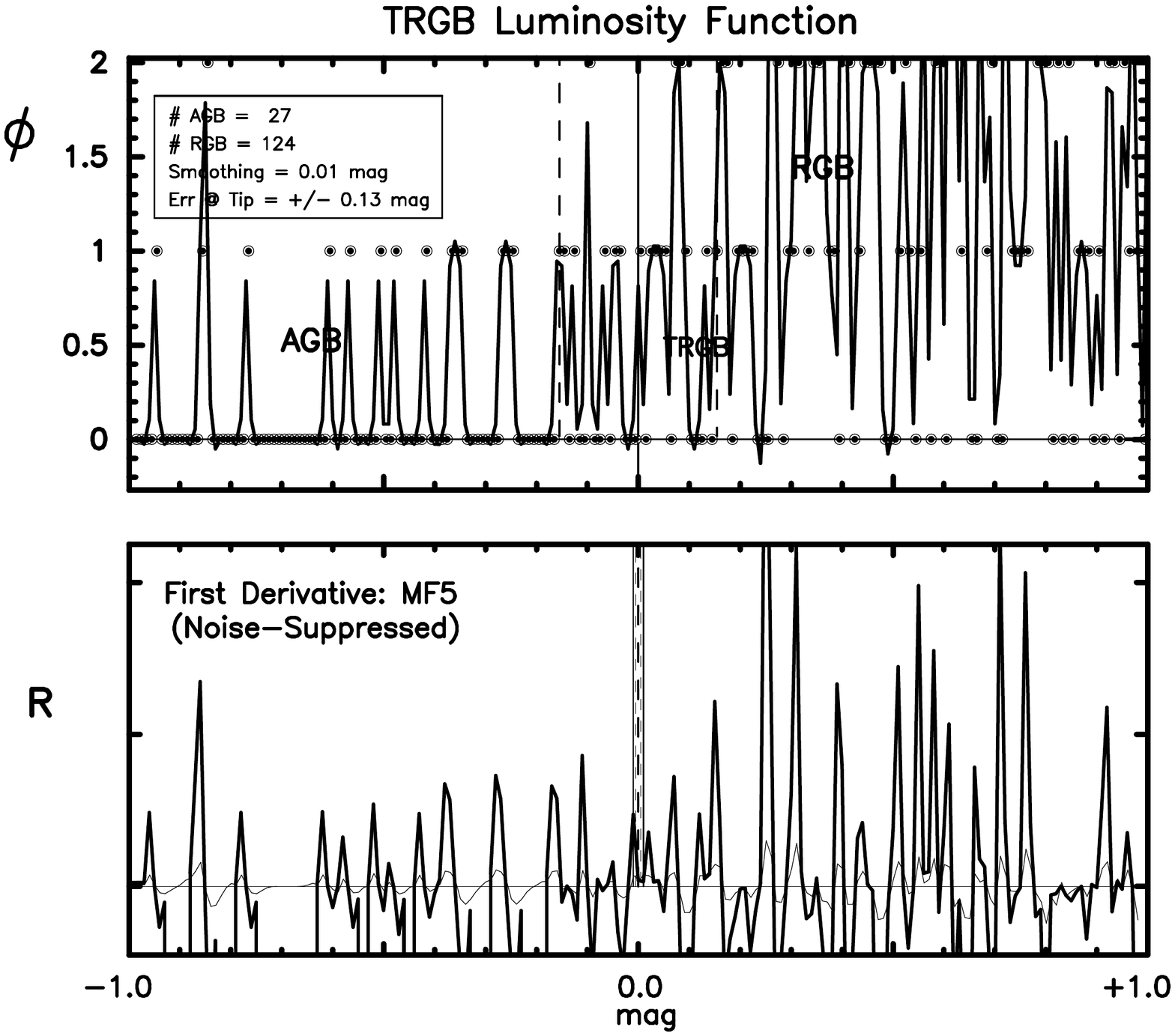}
\includegraphics[width=8.0cm,angle=-00]{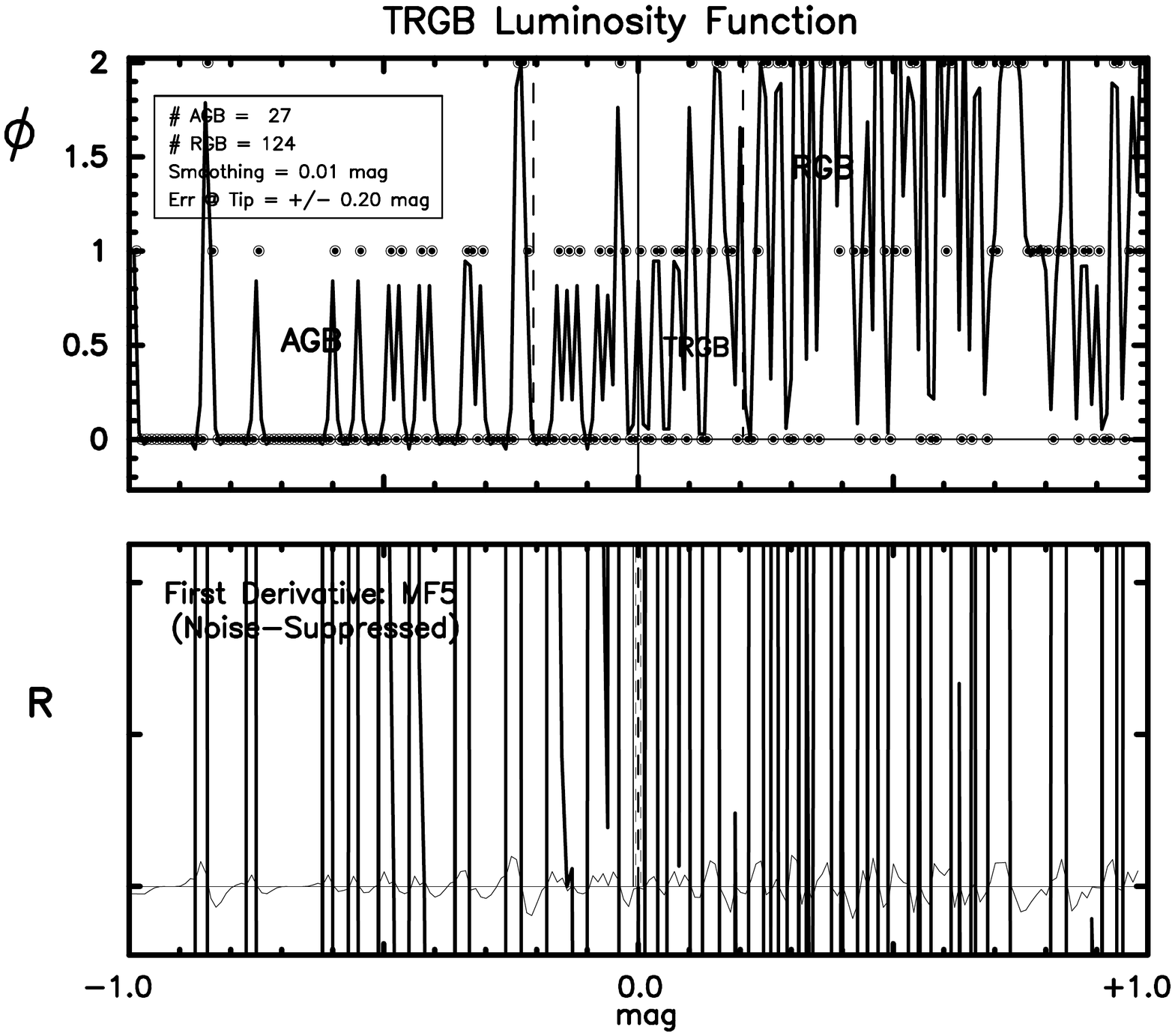} \caption{\small Six sub-panels illustrating the effect of increasing the photometric noise, (from 0.00 to $\pm$0.20~mag) at fixed smoothing ($\pm$0.01 mag) and impoverished populations of RGB stars (124). The lower portions of each of the six sub-panels shows the first-derivative edge-detector output in both its uncorrected (thin black lines) and it noise-weighted (thicker black line) form. See text for a detailed discussion of the trends.} 
\end{figure*}

\vfill\eject
\subsection{Increased Smoothing}
\subsubsection{A Range of Photometric Errors: {\bf}{120,000 RGB Stars}, Fixed Smoothing {\bf}{$\pm$0.05~mag}}

We now repeat the cycle of exploring population size effects and photometric errors, but now {\bf}{at an increased level of smoothing of the data set to $\pm$0.05 mag.}

Once again, returning to the upper left panel of Figure 6, we begin with an RGB population of 120,000 stars below the tip and a photometric error of 0.00~mag. 
At this level of precision in the data the discontinuity occurs between two bins and the smoothing is inappropriately too large, needlessly degrading the jump. 
Nevertheless the power in the first-derivative response function (bottom section of the upper right panel) is very high and well defined, as one might expect. And its width is only $\pm$0.01~mag.
Increasing the error at the tip to $\pm$0.02~mag (upper right panel) widens the discontinuity somewhat, but the smoothing of $\pm$0.05~mag is still too large. 
The output of response function itself responds to the increased photometric errors by declining in power, and widening.
In the middle left panel the smoothing and the photometric errors are identical and the fit at the tip is almost optimal. 
The response function is well centered, continues to widen with the increased errors and can be seen to be starting to develop structured wings that are due to increased, but smoothed, population noise down-stream of the TRGB.
In the final (lower right) panel the photometric errors are at their maximum for this simulation  ($\pm$0.20~mag) and the response function is widened both by the smoothing of the discontinuity in the luminosity function plane and by the encroaching population noise, smoothed out in the response function plane. It is noteworthy that throughout this simulation the mode of the response function at the true TRGB luminosity is stable at the 0.01 mag level despite the widening of the response output and the asymmetric growth of its wings.

{\bf}{Summary 5 --} A larger smoothing of 0.05 mag is too large for data with very small photometric errors. However, this smoothing becomes more appropriate when the photometric errors are comparable to the smoothing value. With a large population of RGB stars, the tip location is extremely stable in all cases.

\begin{figure*} \centering
\includegraphics[width=8.0cm,angle=-0]{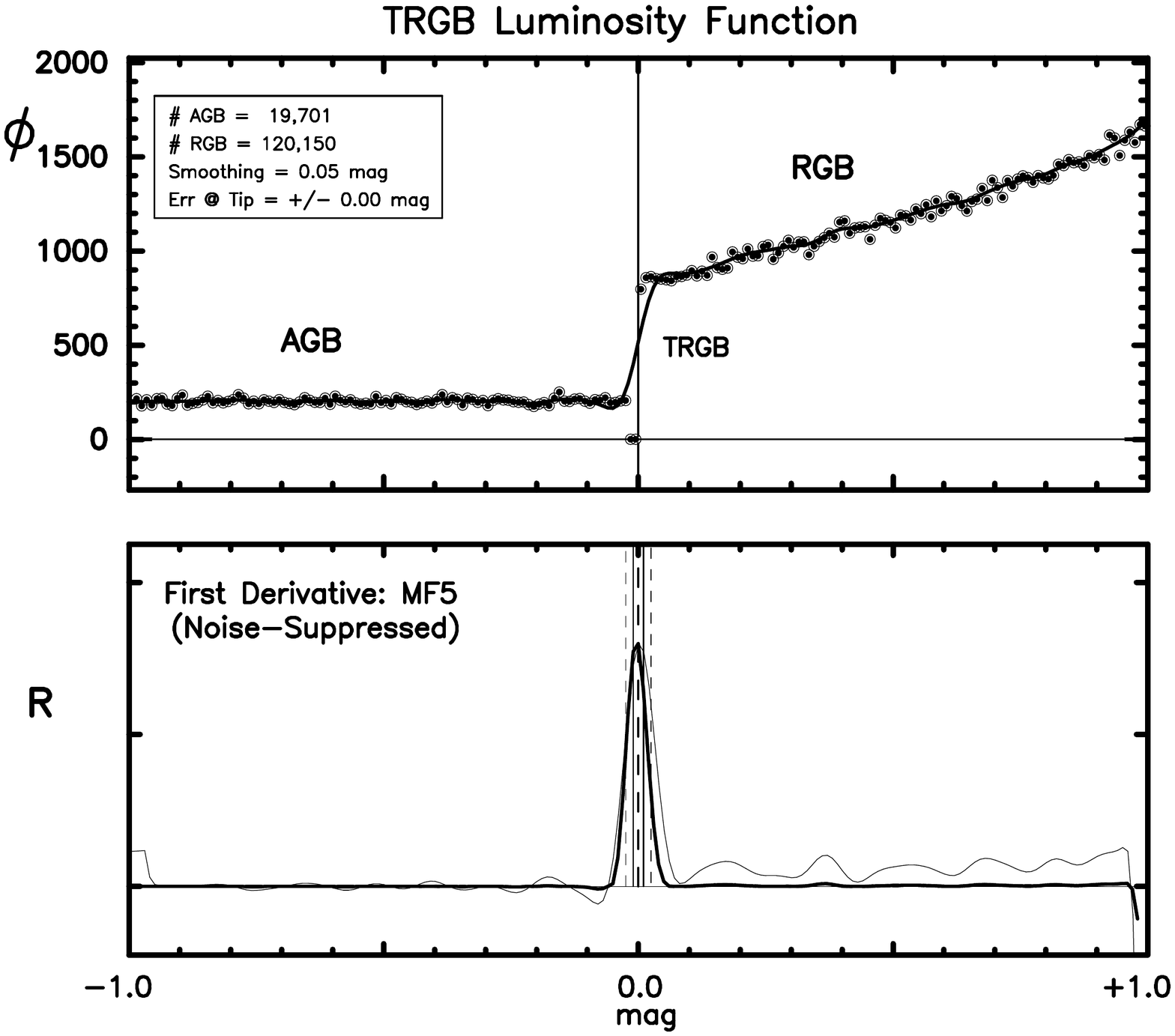}
\includegraphics[width=8.0cm,angle=-0]{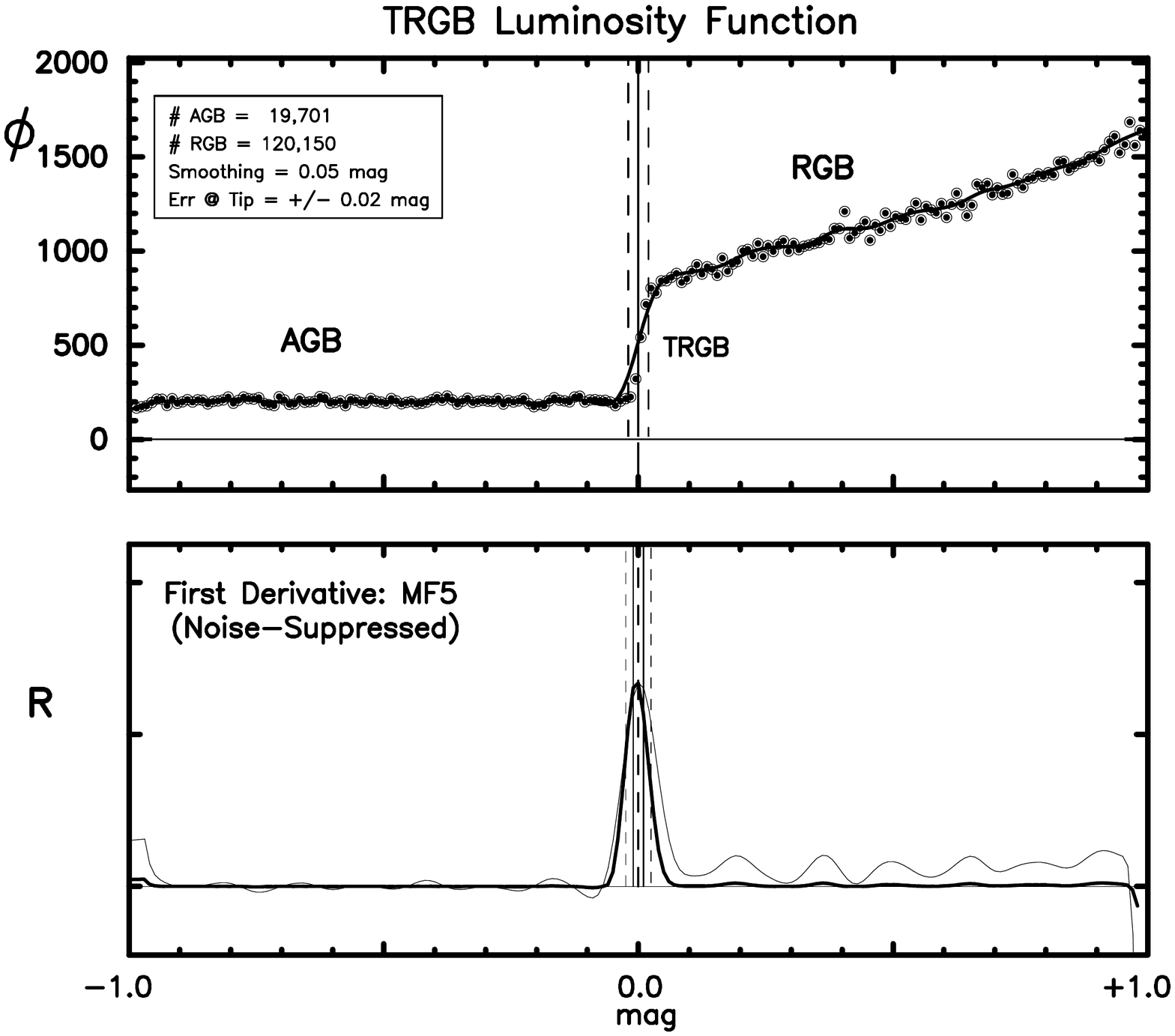}
\includegraphics[width=8.0cm,angle=-0]{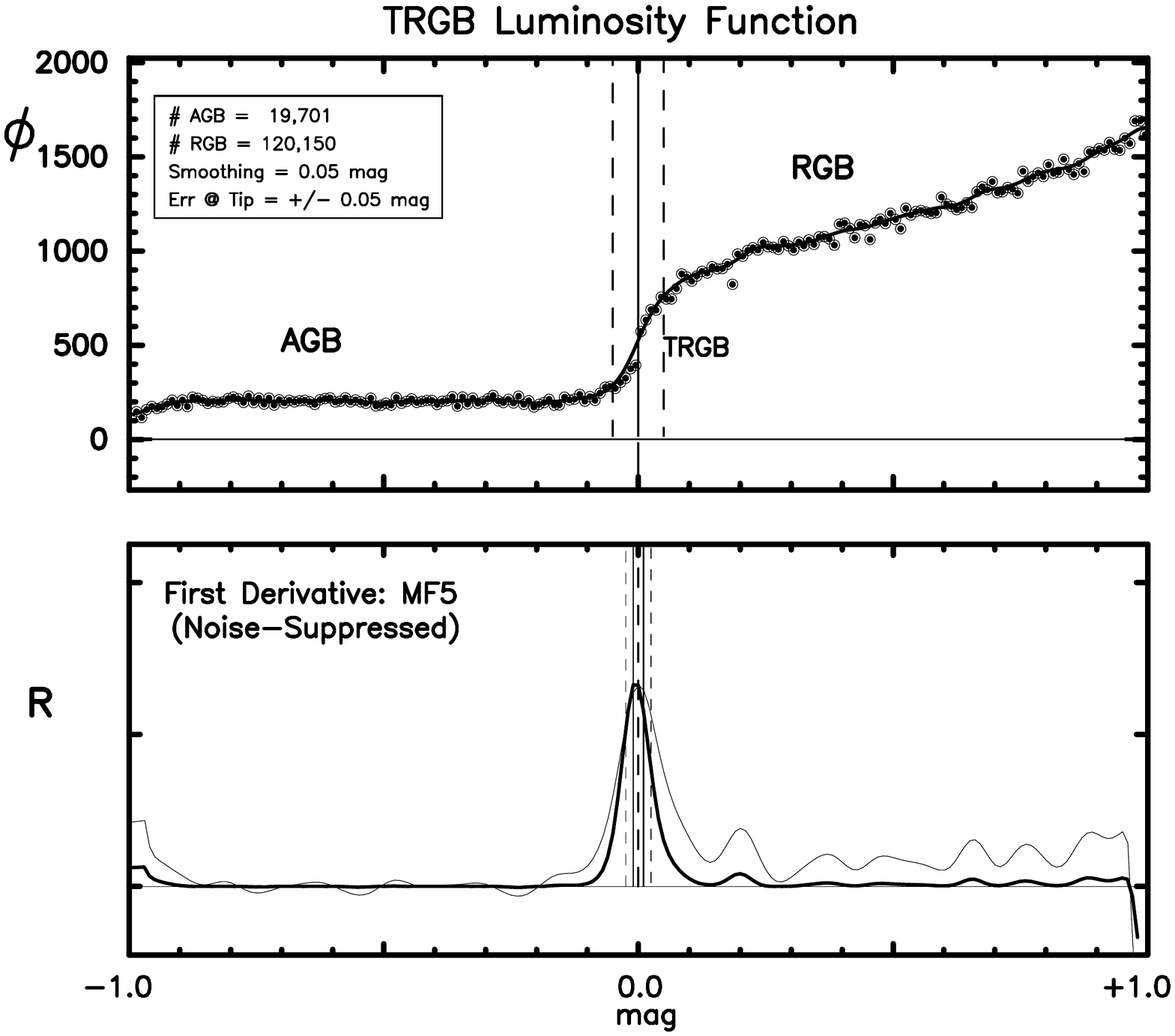}
\includegraphics[width=8.0cm,angle=-0]{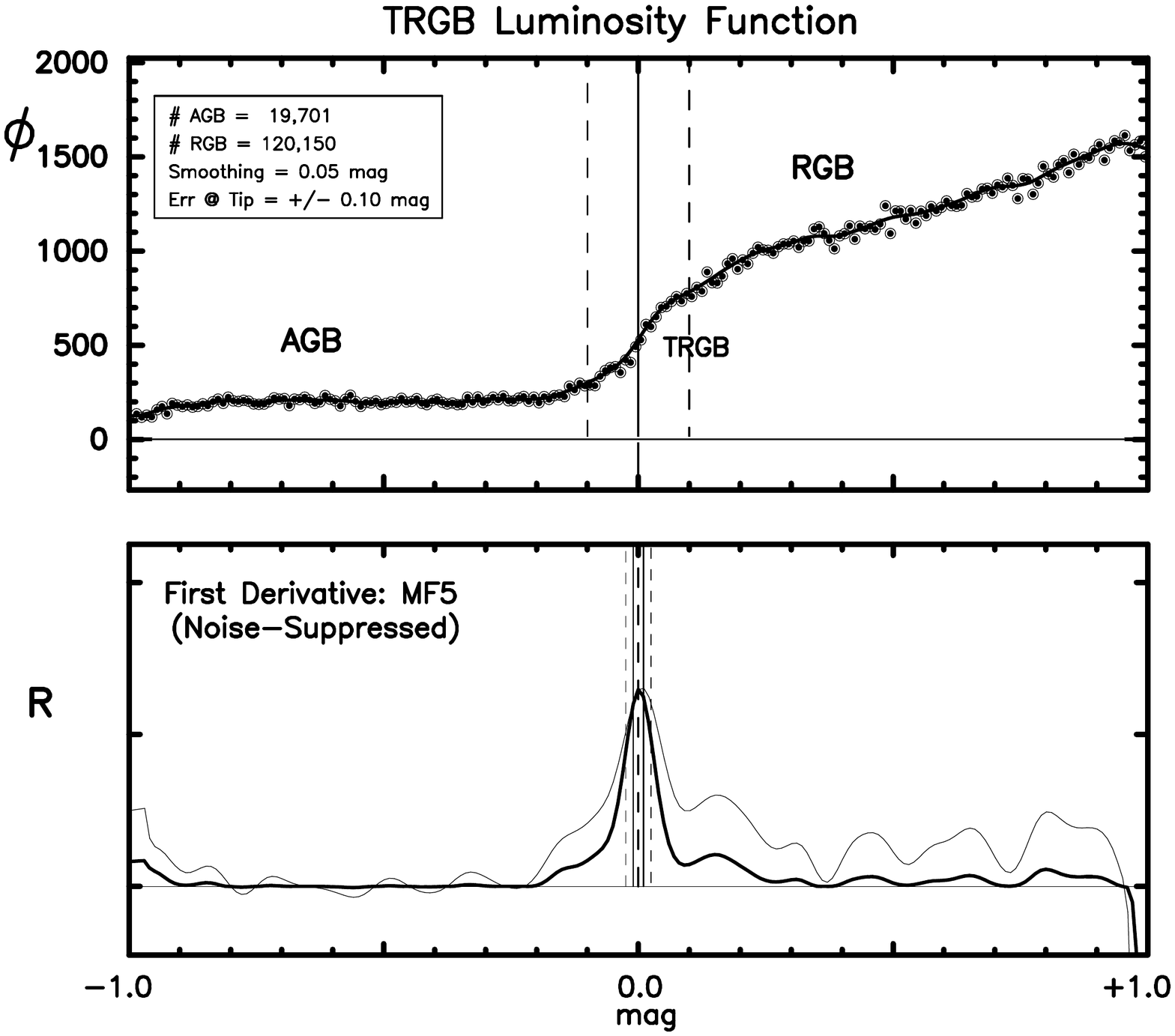}
\includegraphics[width=8.0cm,angle=-0]{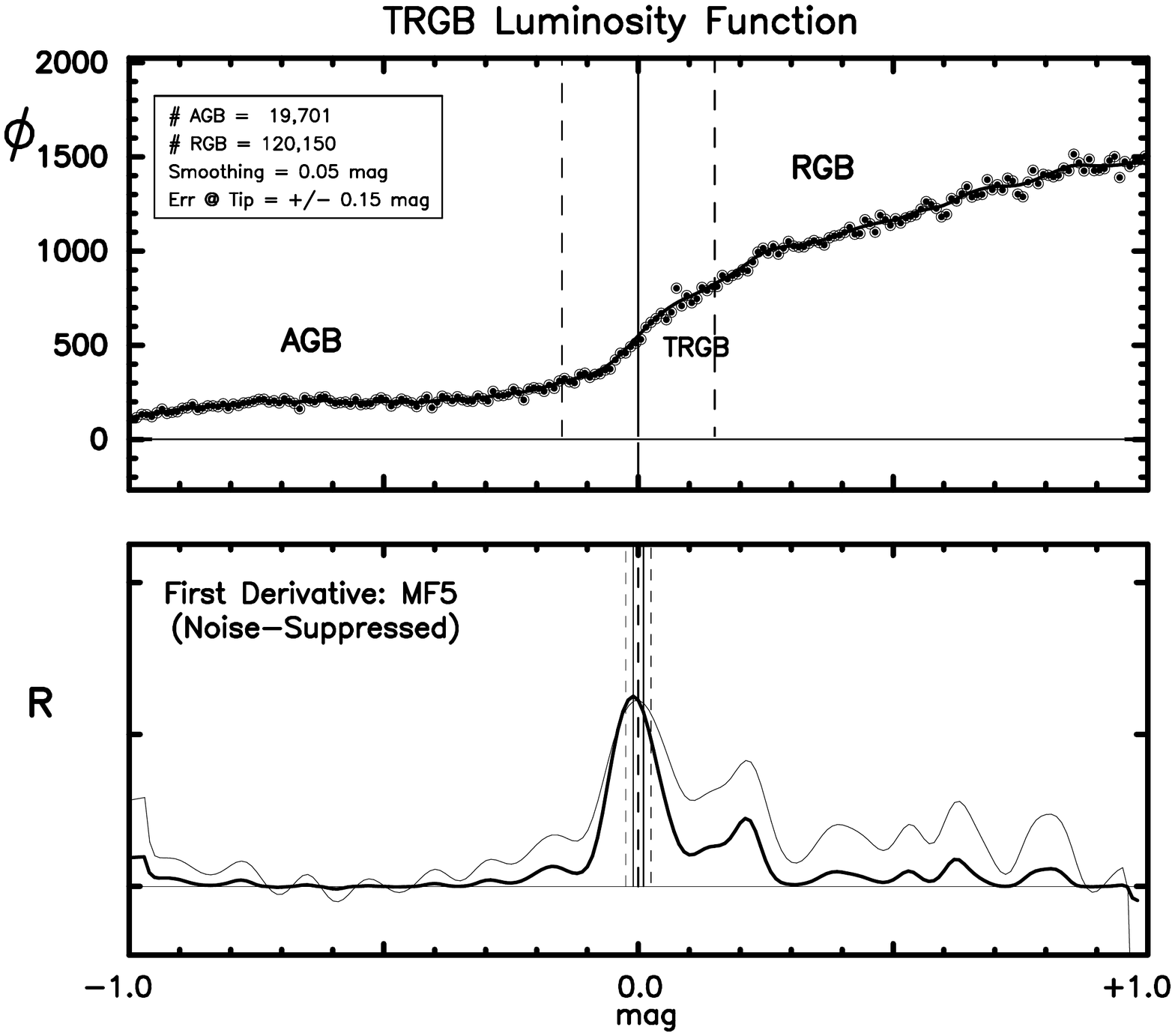}
\includegraphics[width=8.0cm,angle=-0]{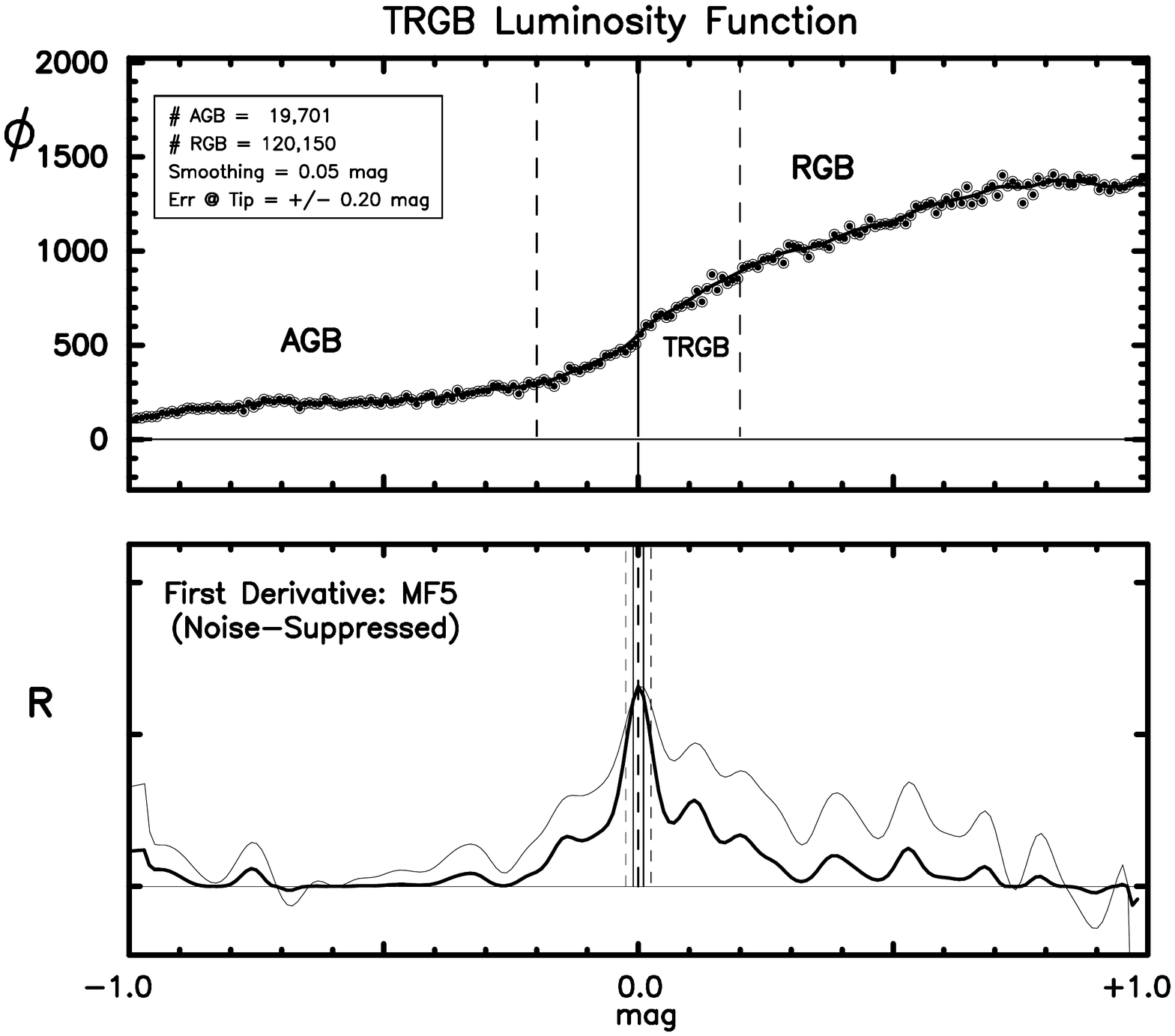} \caption{\small Six sub-panels illustrating the effect of increasing the photometric noise, (from 0.00 to 0.20~mag) at fixed, but slightly larger smoothing (0.05 mag) than previously discussed and again for very large populations of 120,000 RGB stars. The lower portions of each of the six sub-panels shows the first-derivative edge-detector output in both its uncorrected (thin black lines) and it noise-weighted (thicker black line) form. See text for a detailed discussion of the trends.} \end{figure*}

\vfill\eject
\subsubsection{A Range of Photometric Errors: {\bf}{11,000 RGB Stars}, Fixed Smoothing {\bf}{$\pm$0.05~mag}}

Despite the under-fitting of the data around the TRGB (due to the larger smoothing) the filter response is sharp, unambiguous and unbiased for photometric errors less than $\pm$0.05~mag (which coincidentally corresponds to the adopted smoothing here) for an RGB population size about 11,000 stars. 
At high photometric errors (i.e., in excess of $\pm$0.10~mag) false positives predominate up-stream (middle right panel of Figure 7) but eventually crowd around and compromise the integrity of the true tip detection, encroaching both from fainter and brighter magnitudes. For instance, the strongest peak in the lower left panels is due to a smoothed version of a clustering of random noise peaks two tenths of a magnitude below the true TRGB. The existence of a peak at the correct position in the lower right panel cannot be given much credibility given the ambient noise.

{\bf}{Summary 6 --} Photometric errors greater than  $\pm 0.10$ mag cause false positives and potential bias in the (blended) tip magnitude for populations of 11,000 RGB stars. Increased smoothing does not mitigate this effect.

\begin{figure*} \centering
\includegraphics[width=8.0cm,angle=-0]{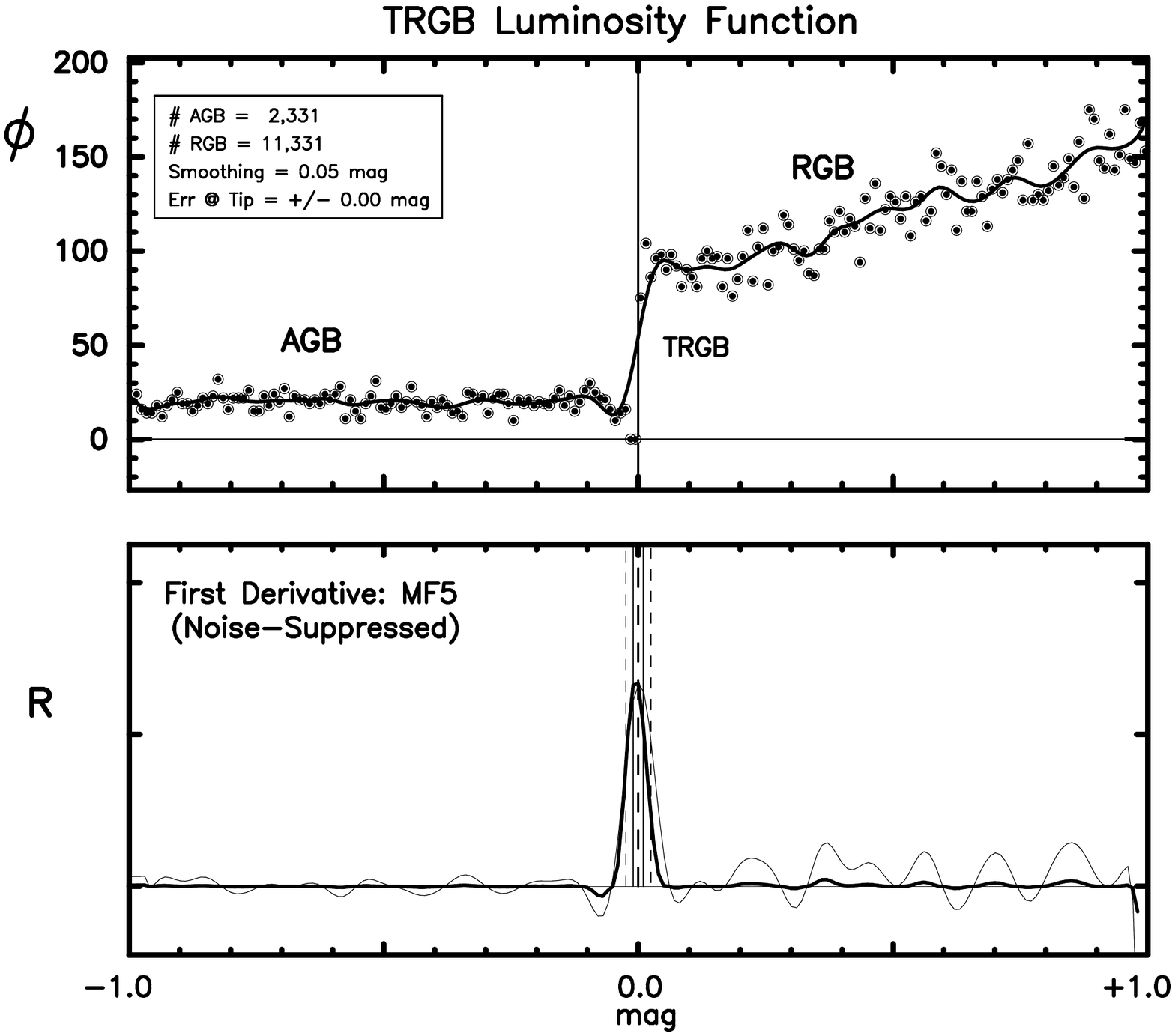}
\includegraphics[width=8.0cm,angle=-0]{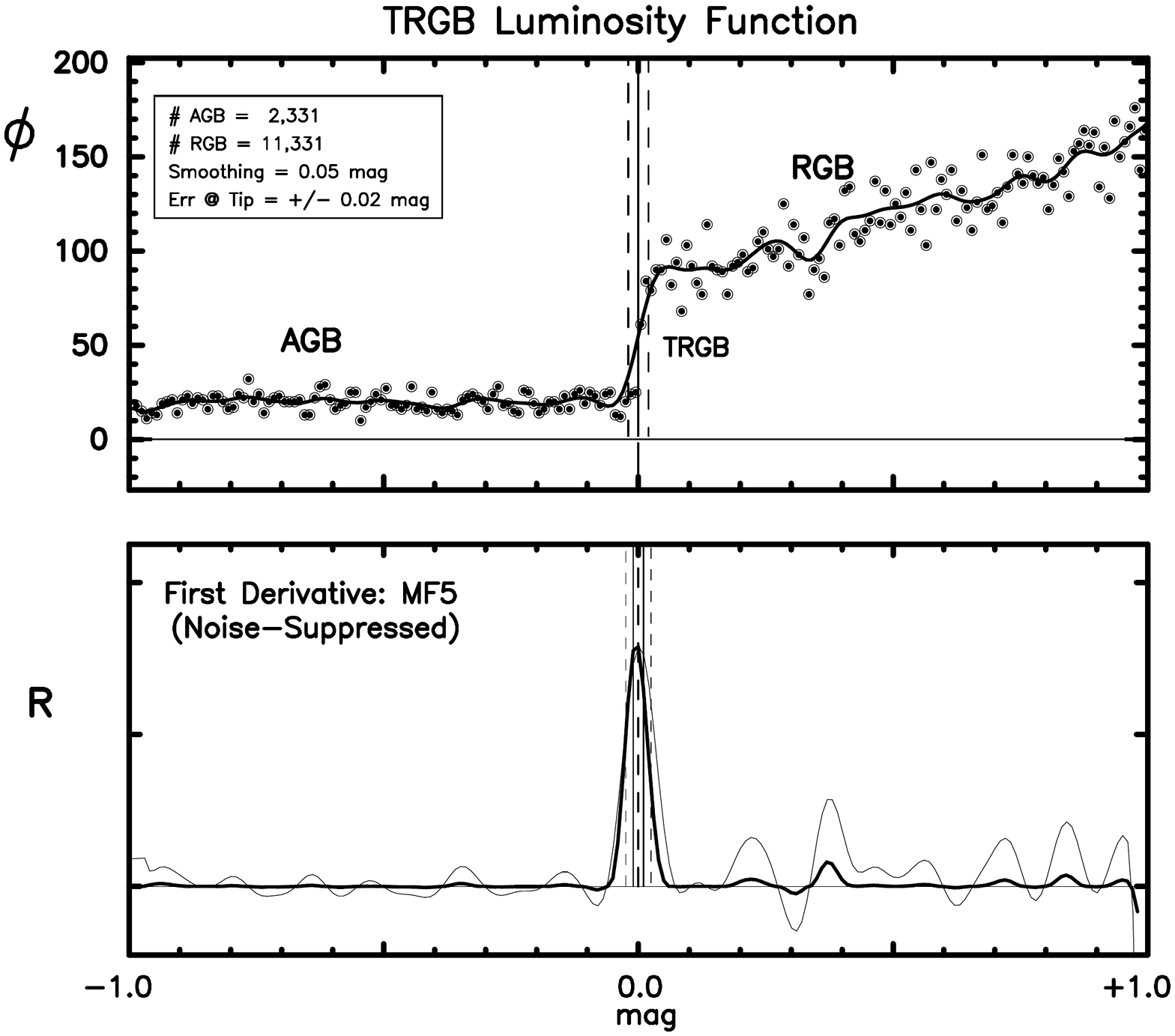}
\includegraphics[width=8.0cm,angle=-0]{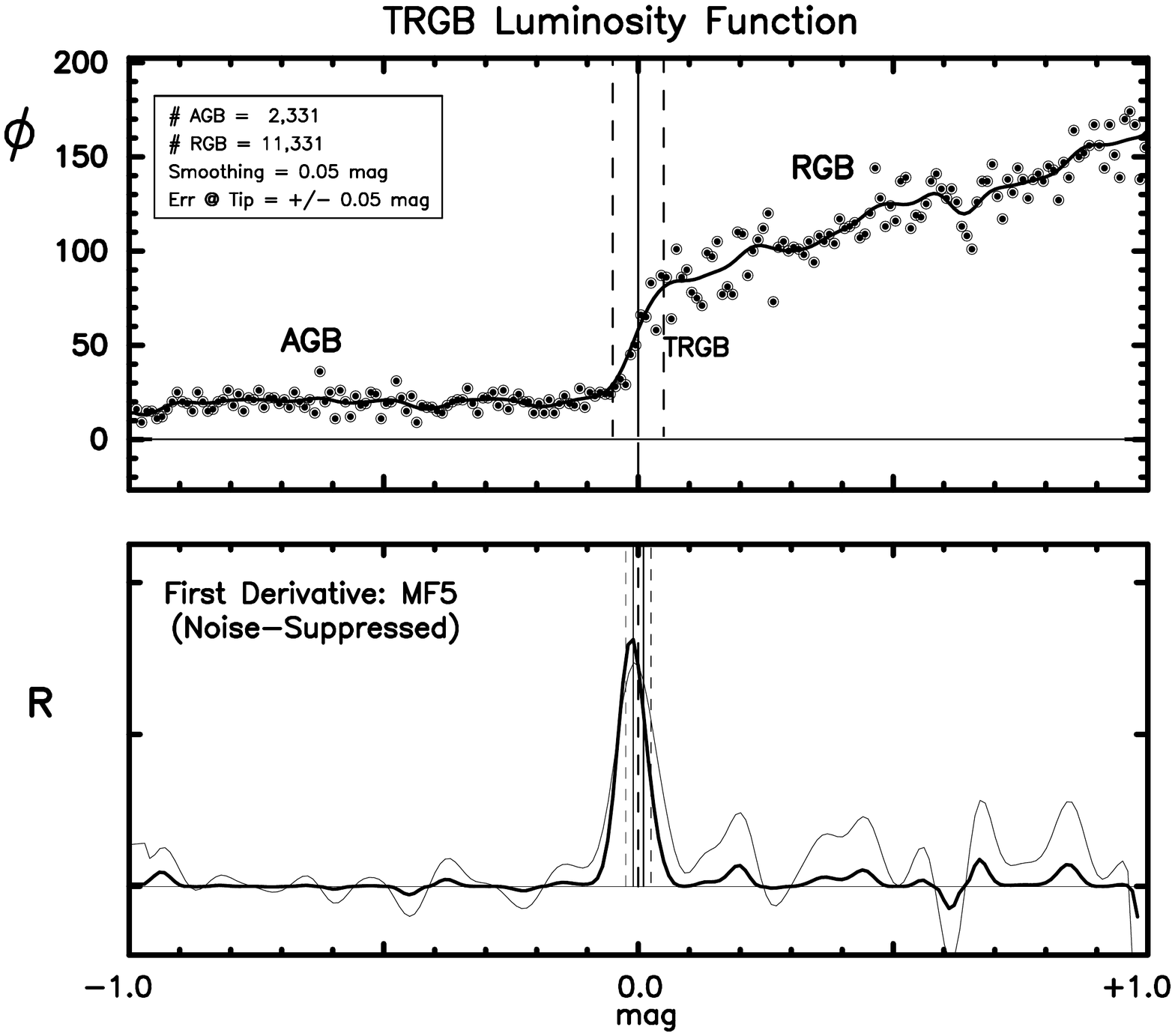}
\includegraphics[width=8.0cm,angle=-0]{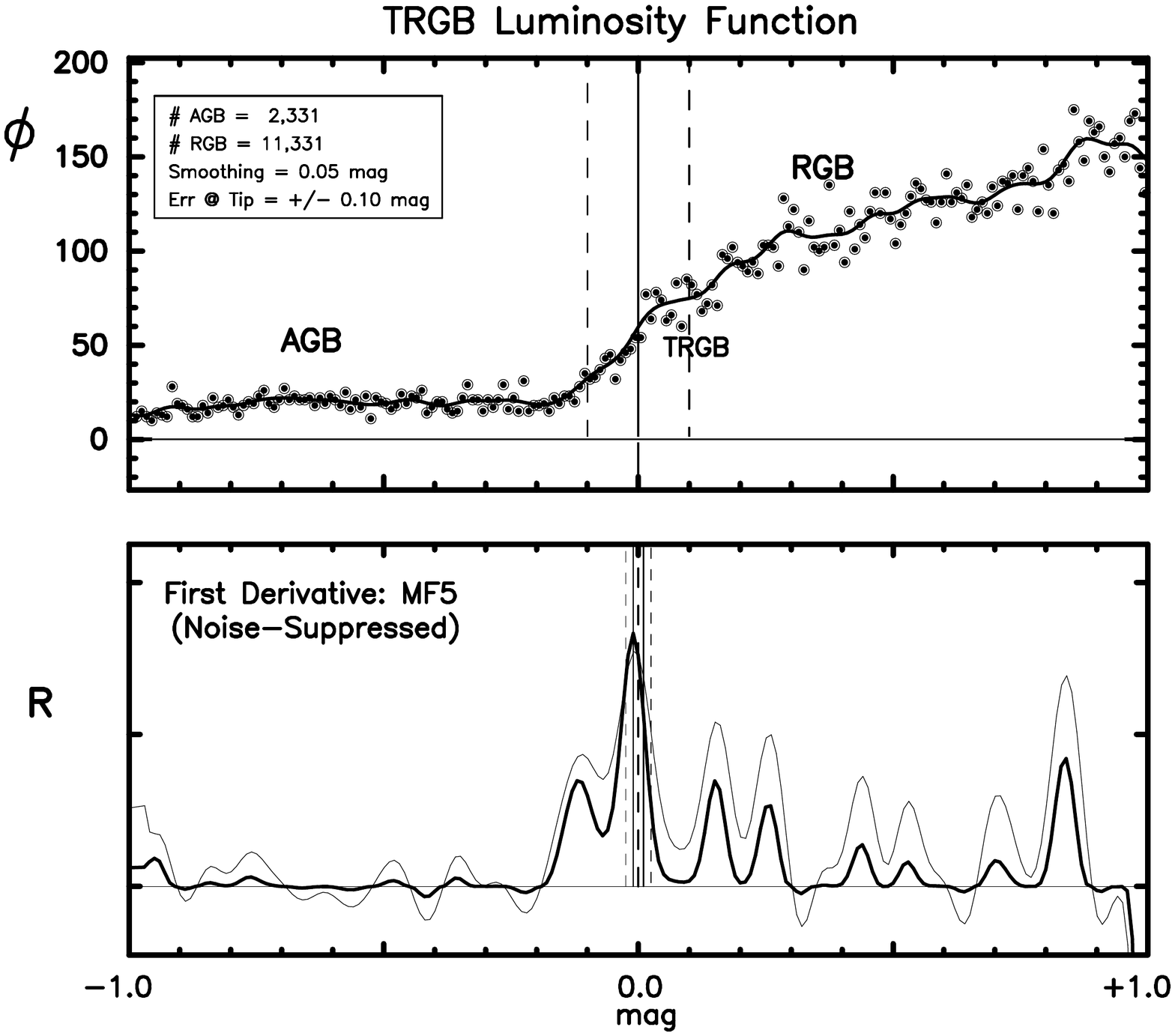}
\includegraphics[width=8.0cm,angle=-0]{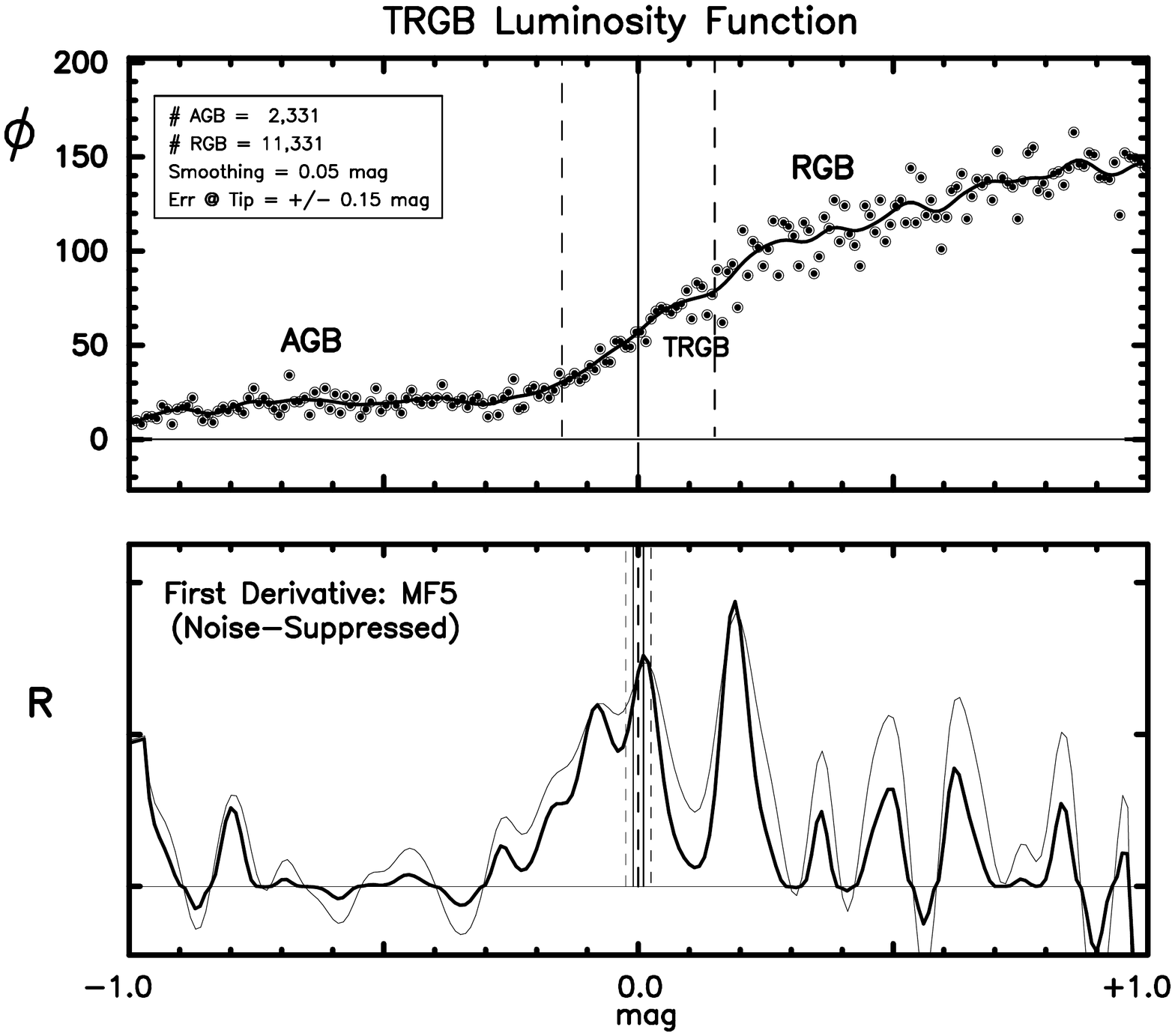}
\includegraphics[width=8.0cm,angle=-0]{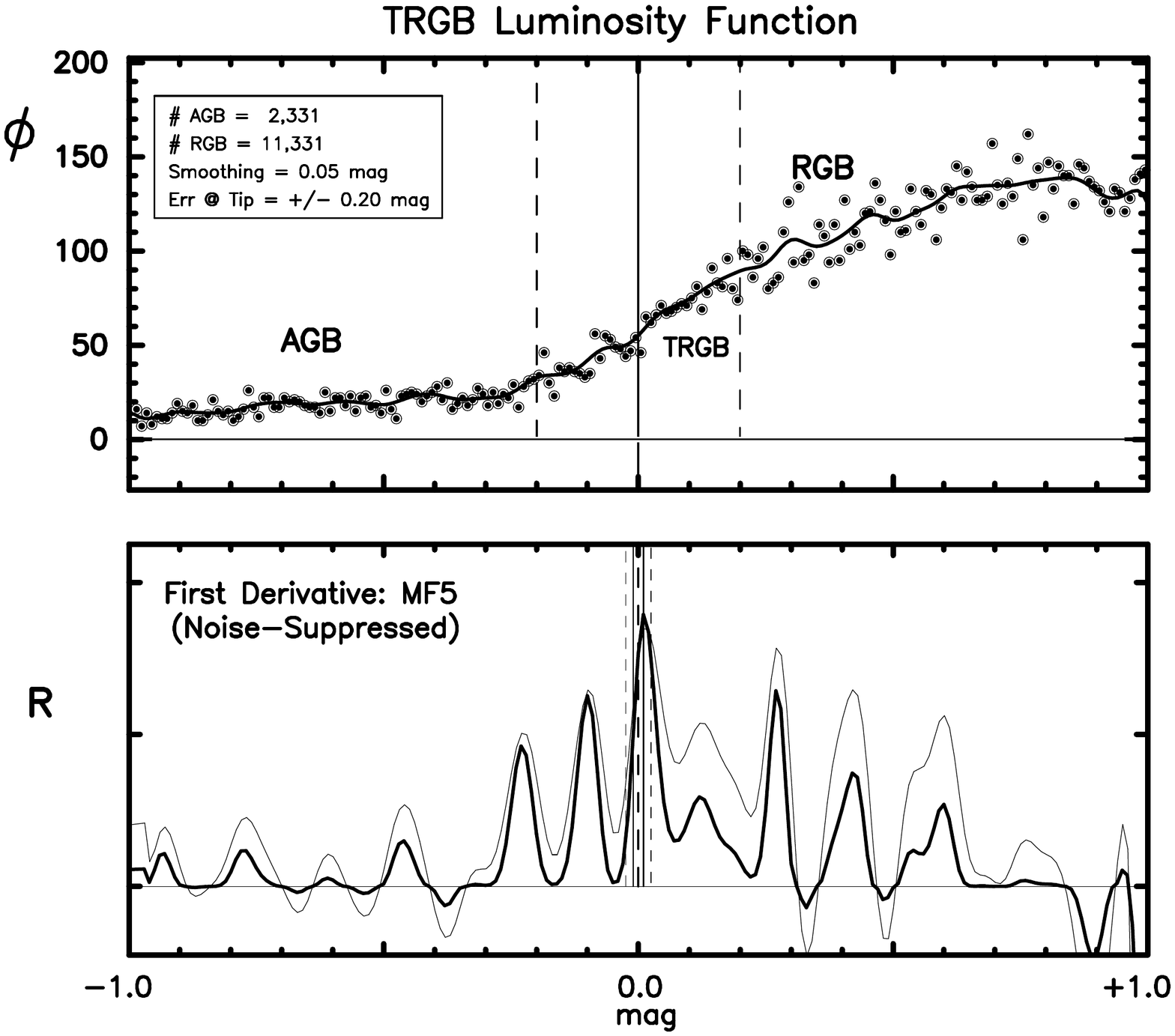} \caption{\small Six sub-panels illustrating the effect of increasing the photometric noise, (from 0.00 to 0.20~mag) at fixed, but slightly larger smoothing (0.05 mag) than previously discussed and for moderately large populations of RGB stars (11,331). The lower portions of each of the six sub-panels shows the first-derivative edge-detector output in both its uncorrected (thin black lines) and it noise-weighted (thicker black line) form. See text for a detailed discussion of the trends.} \end{figure*}

\vfill\eject
\subsubsection{A Range of Photometric Errors: {\bf}{1,200 RGB Stars}, Fixed Smoothing {\bf}{$\pm$0.05~mag}}

Dropping the sample size by another factor of ten, down to around 1,200 RGB stars below the TRGB does not substantially change the description of the situation as given in the previous section. 
The detection of the tip, as seen in in Figure 8, is relatively strong and unambiguous up to a photometric error of $\pm$0.05 mag after which competing false positives begin occurring above and below the true tip. 
\begin{figure*} \centering
\includegraphics[width=8.0cm,angle=-0]{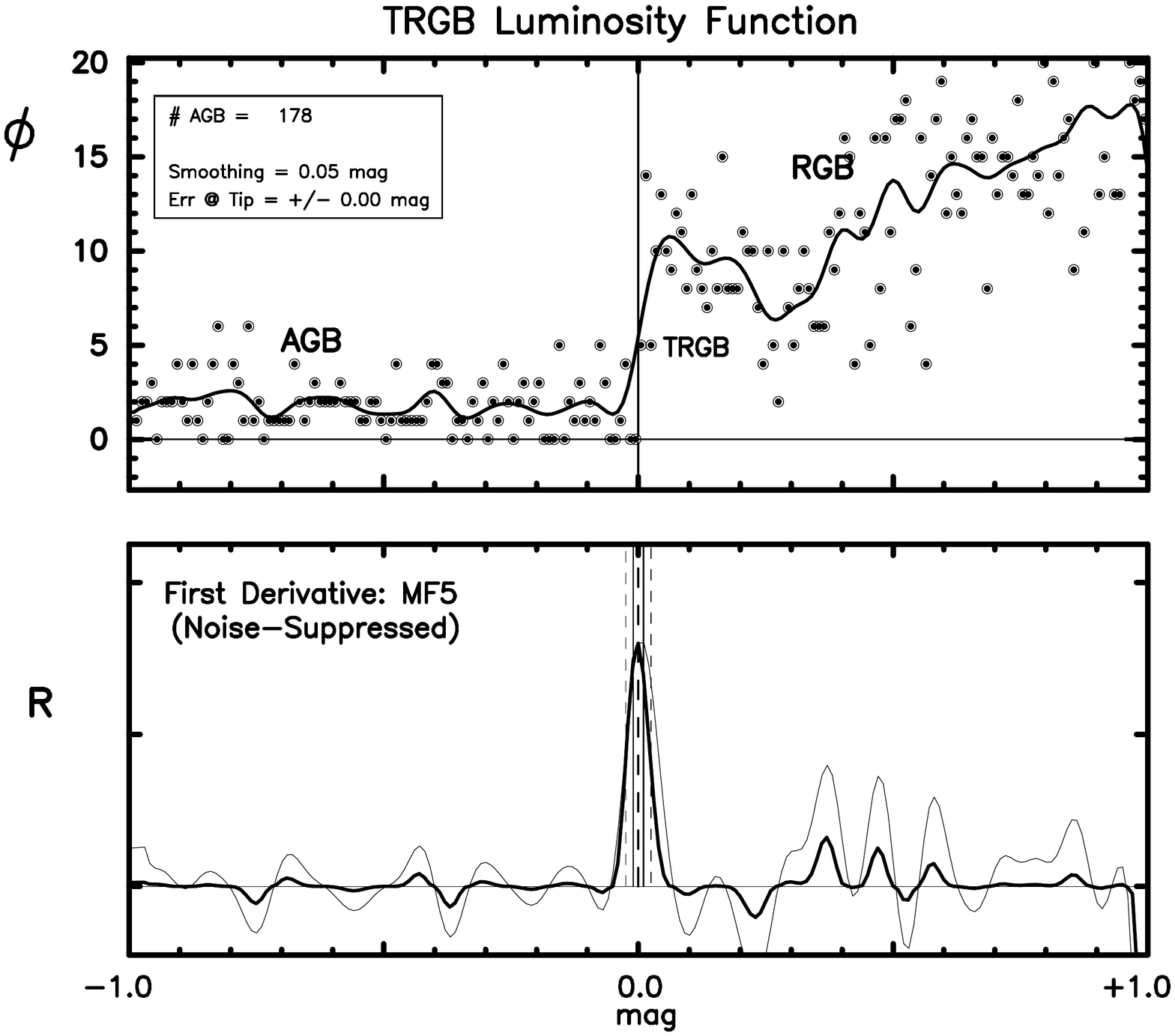}
\includegraphics[width=8.0cm,angle=-0]{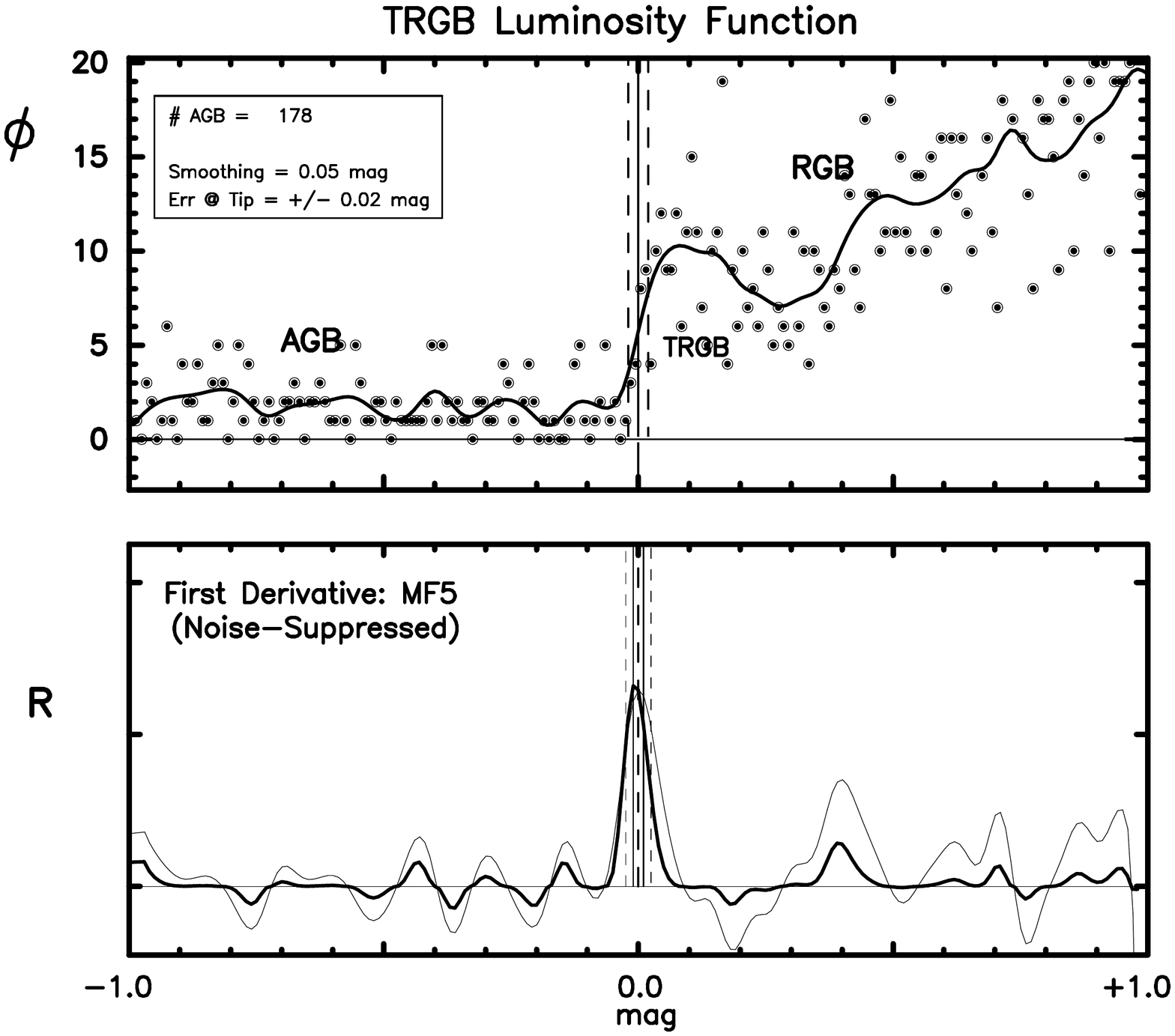}
\includegraphics[width=8.0cm,angle=-0]{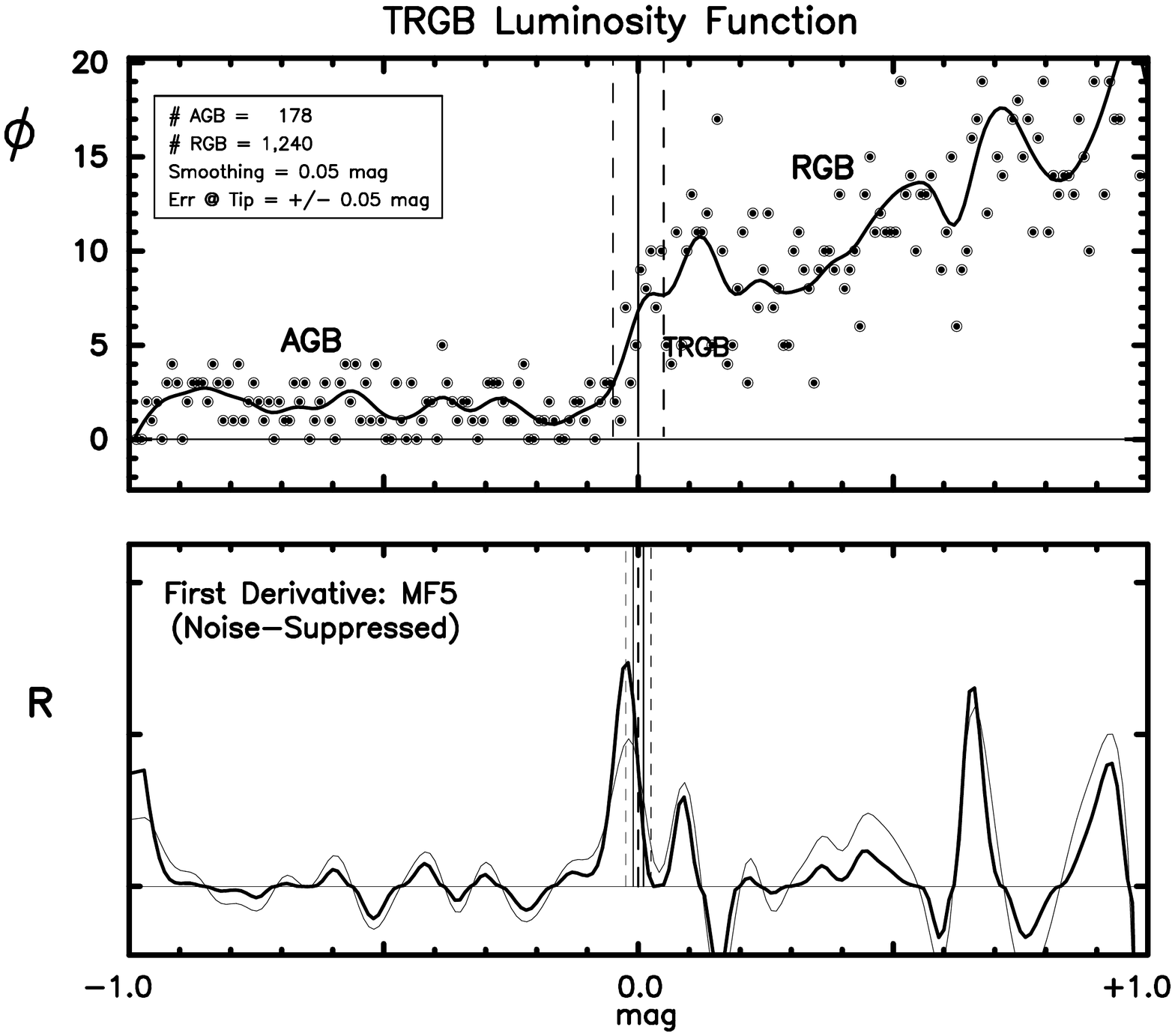}
\includegraphics[width=8.0cm,angle=-0]{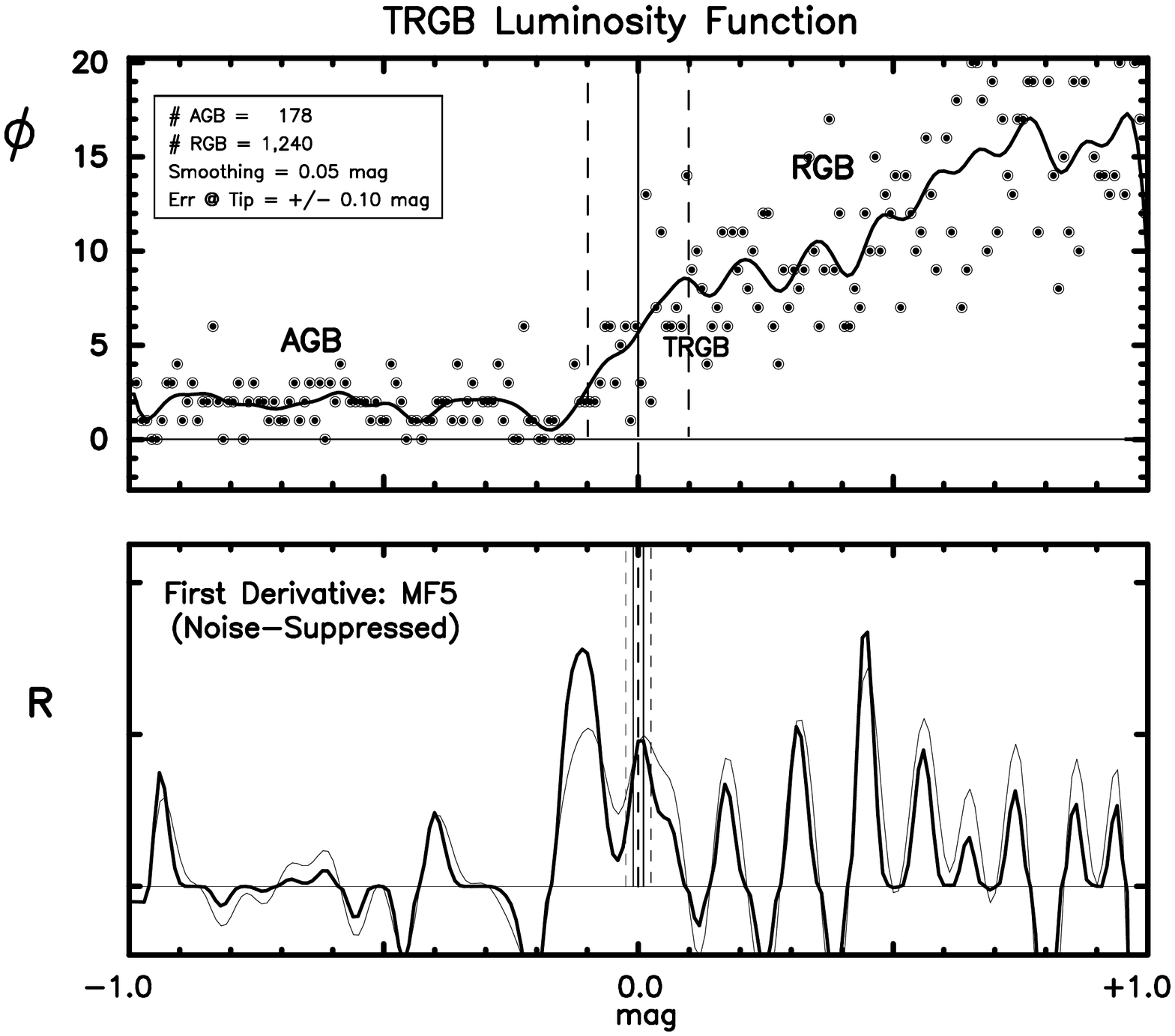}
\includegraphics[width=8.0cm,angle=-0]{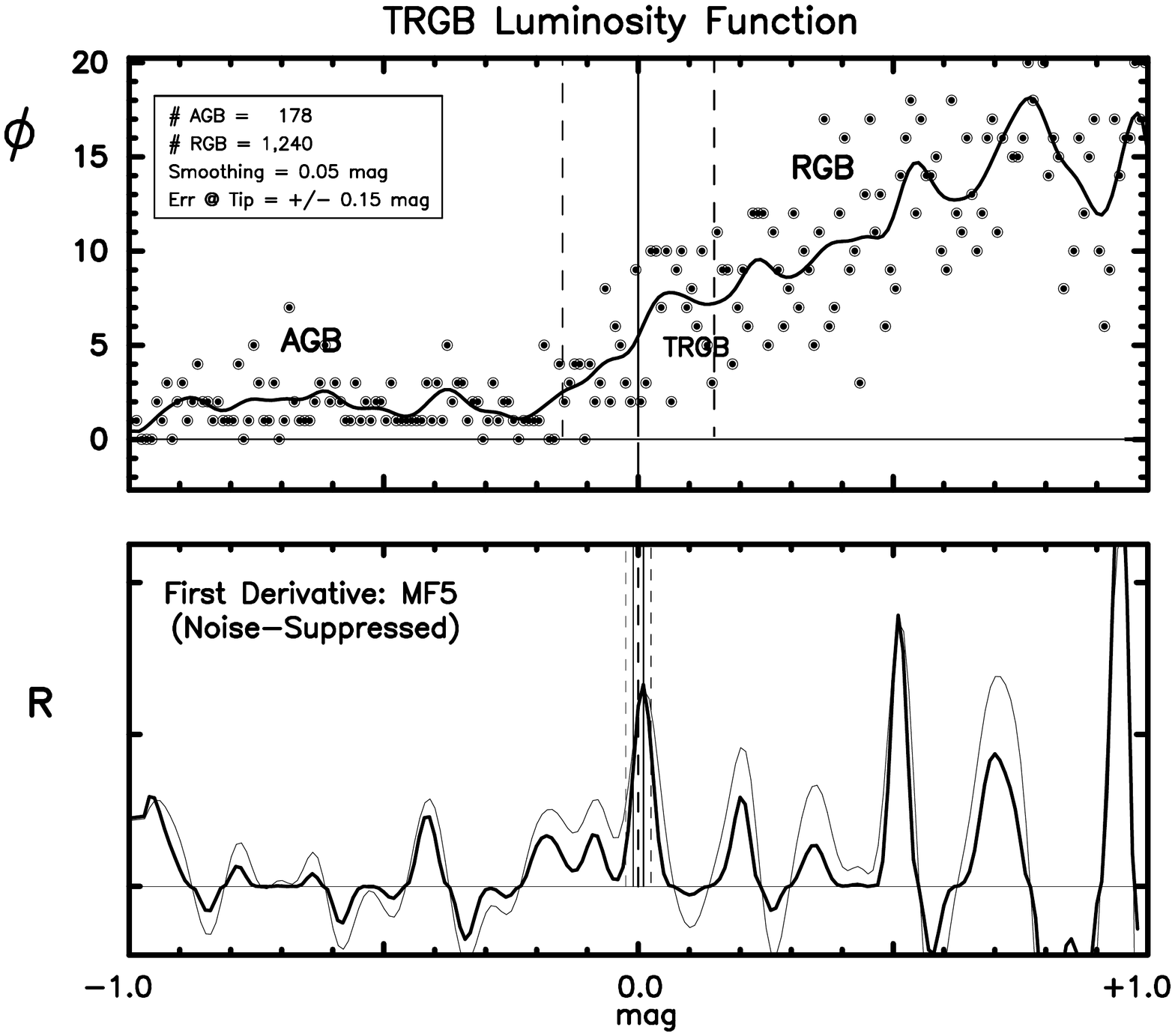}
\includegraphics[width=8.0cm,angle=-0]{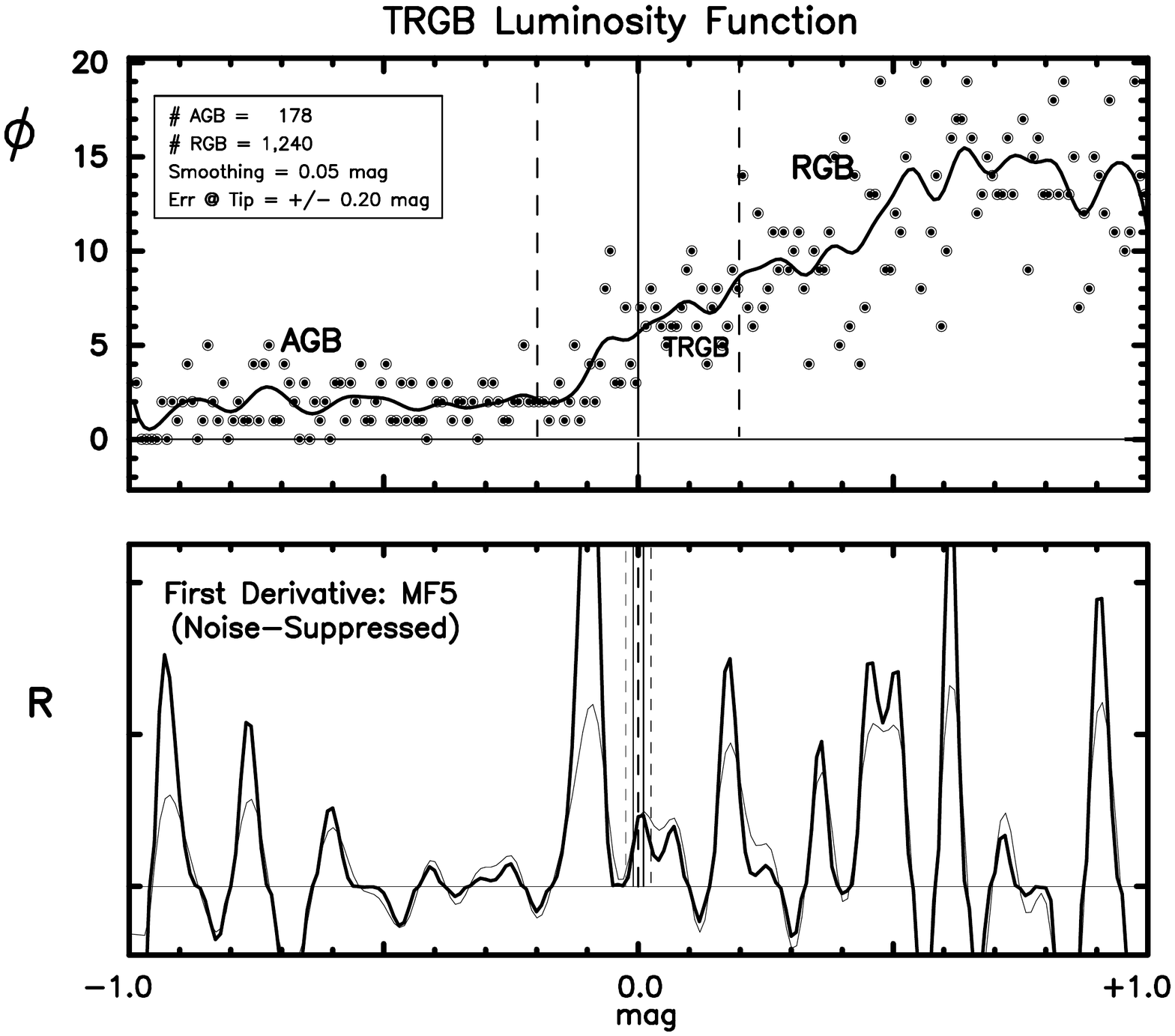} \caption{\small Six sub-panels illustrating the effect of increasing the photometric noise, (from 0.00 to 0.20~mag) at fixed, but slightly larger smoothing (0.05 mag) than previously discussed and for small populations of RGB stars (1,240). The lower portions of each of the six sub-panels shows the first-derivative edge-detector output in both its uncorrected (thin black lines) and it noise-weighted (thicker black line) form. See text for a detailed discussion of the trends.} \end{figure*}

\vfill\eject
\subsubsection{A Range of Photometric Errors: {\bf}{120 RGB Stars}, Fixed Smoothing {\bf}{$\pm$0.05~mag}}

At our smallest population size of 120 RGB stars below the tip, the advantages of smoothing 
are now becoming quite apparent in the first two panels of Figure 9,  illustrating the onset  of decreased photometric precision.  
The first detected tip is the true peak, up to an error of $\pm$0.02~mag, after which spurious noise peaks overwhelm the detection both in advance of and beyond the true position of the TRGB.
Confidently detecting the true position of the TRGB in RGB populations of around 100 stars in the upper magnitude range can only be done with high precision data and is still risky, given that noise spikes of comparable power are found systematically positioned up to $\pm$0.1 mag above and below the true tip, in virtually all of the realizations shown here. 

{\bf}{Summary 7 --} Increased smoothing can help compensate for small population sizes if the photometric quality is very good. However, using small populations is still not advisable.

\begin{figure*} \centering
\includegraphics[width=8.0cm,angle=-0]{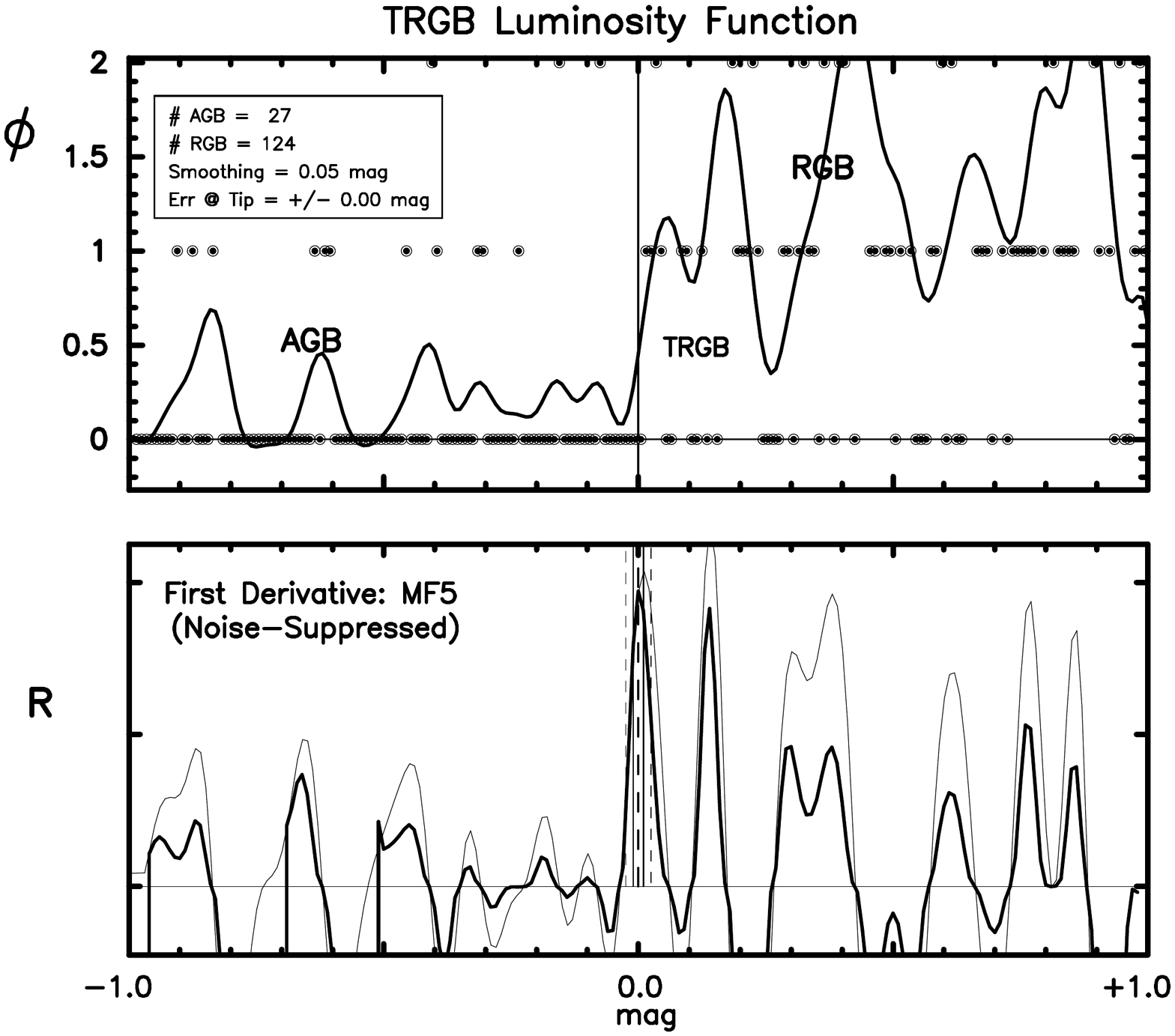}
\includegraphics[width=8.0cm,angle=-0]{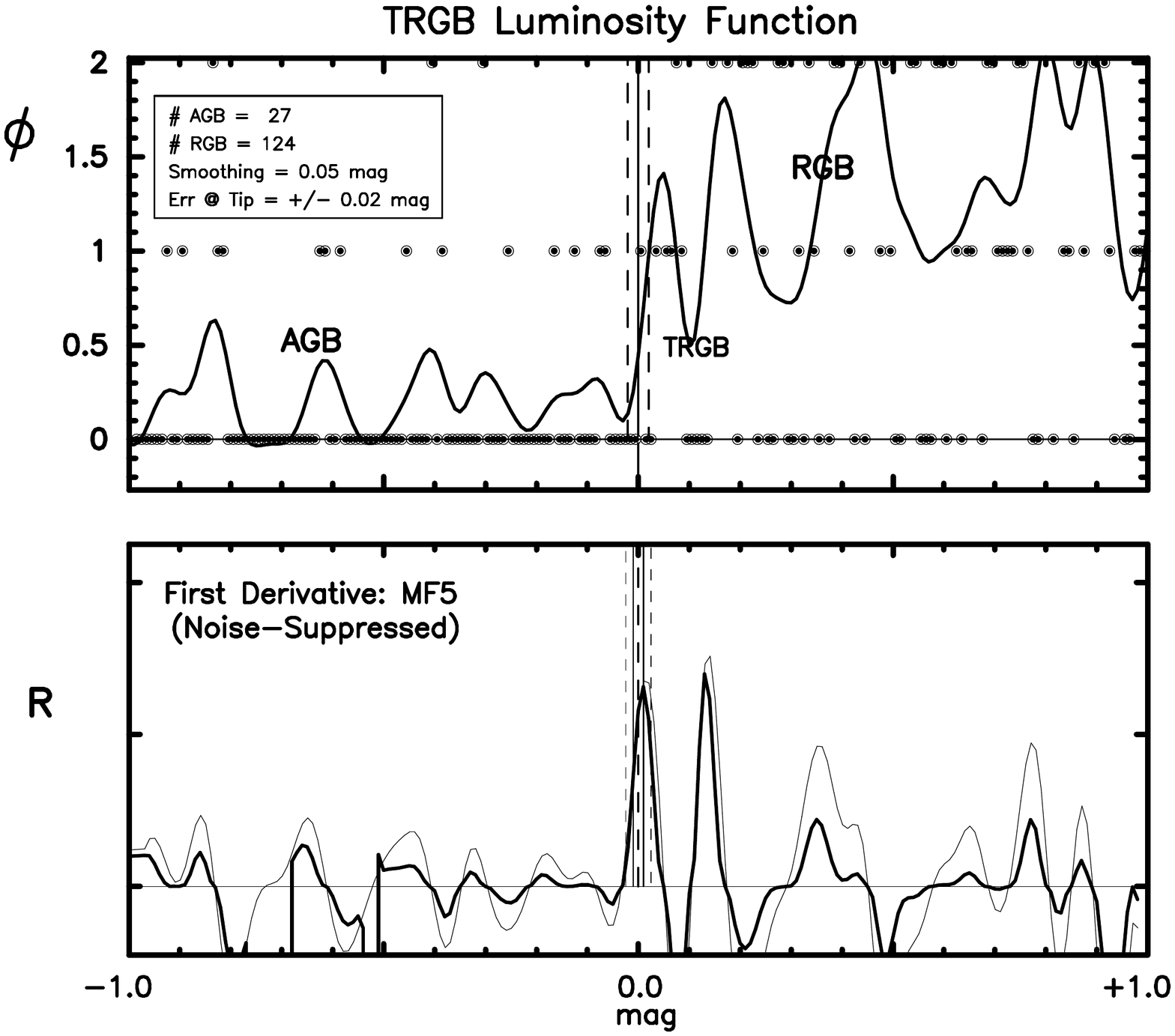}
\includegraphics[width=8.0cm,angle=-0]{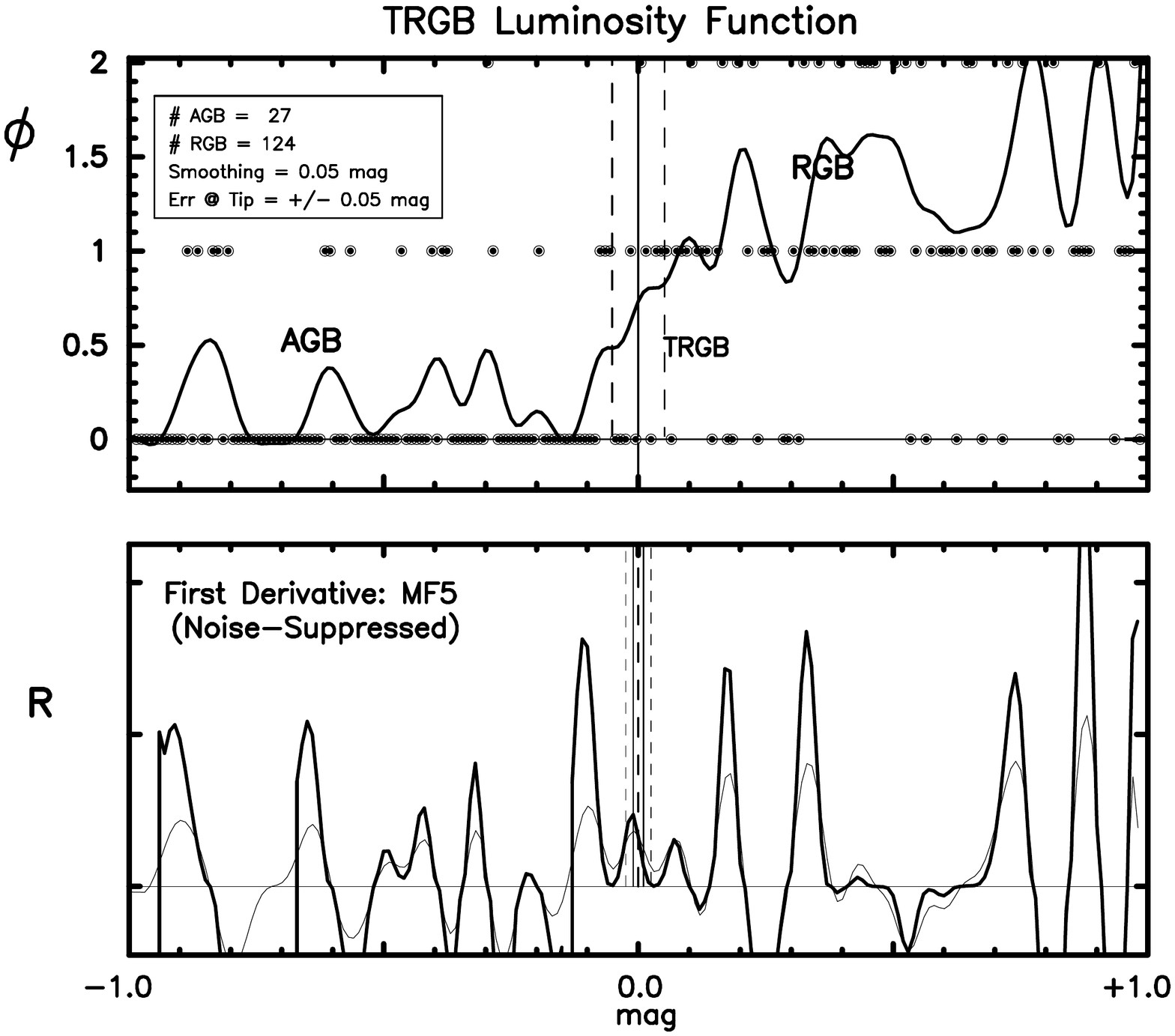}
\includegraphics[width=8.0cm,angle=-0]{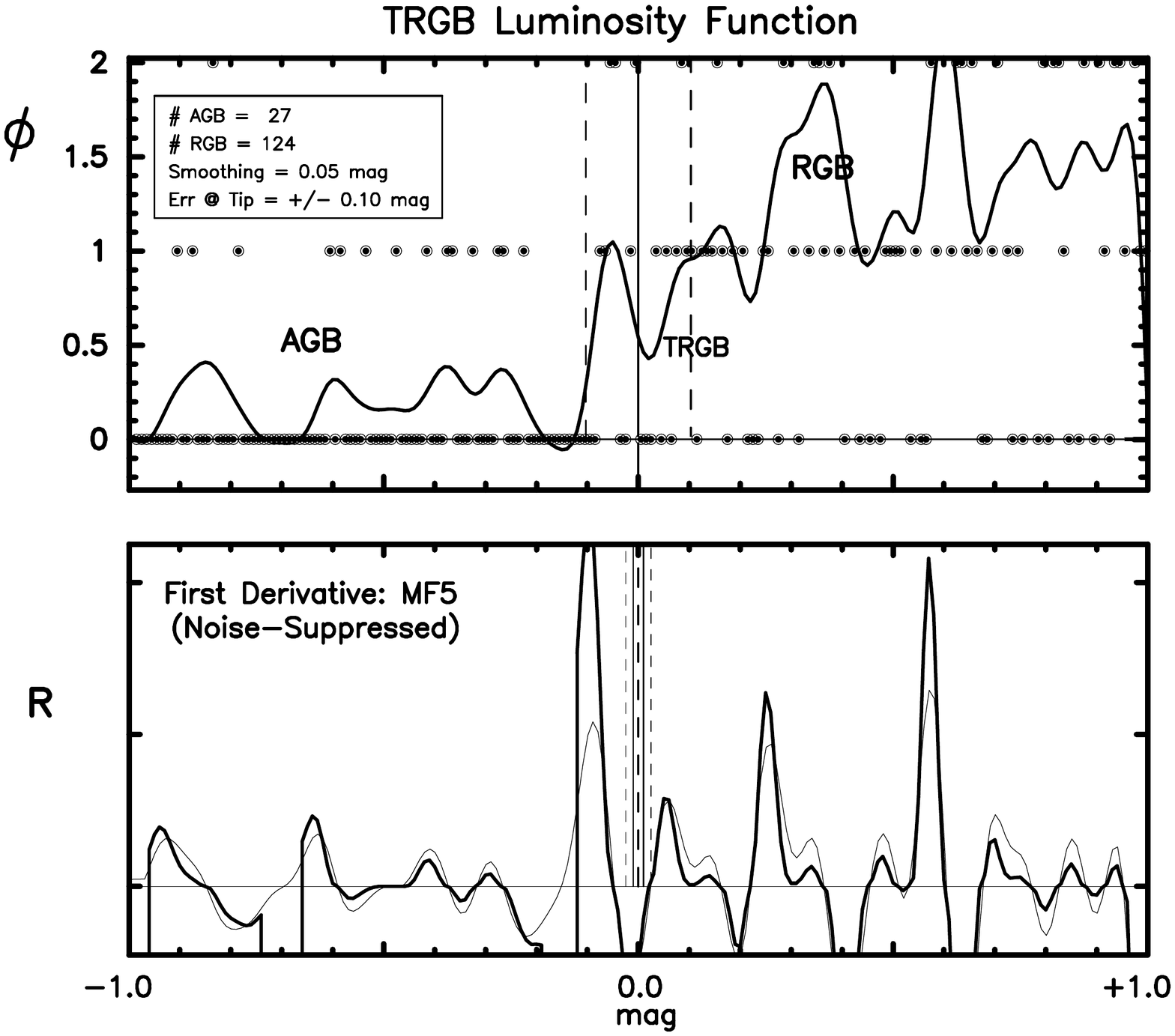}
\includegraphics[width=8.0cm,angle=-0]{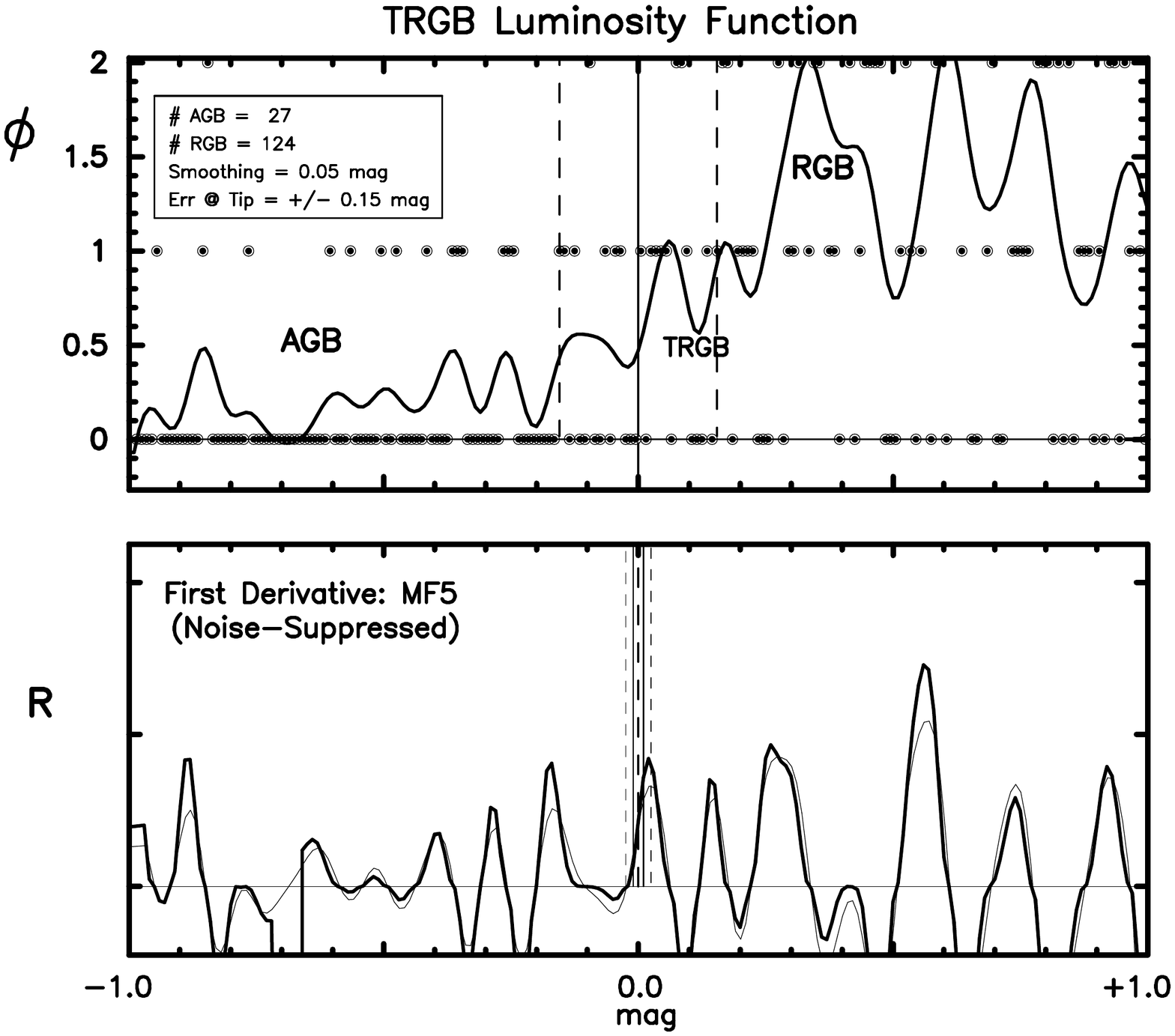}
\includegraphics[width=8.0cm,angle=-0]{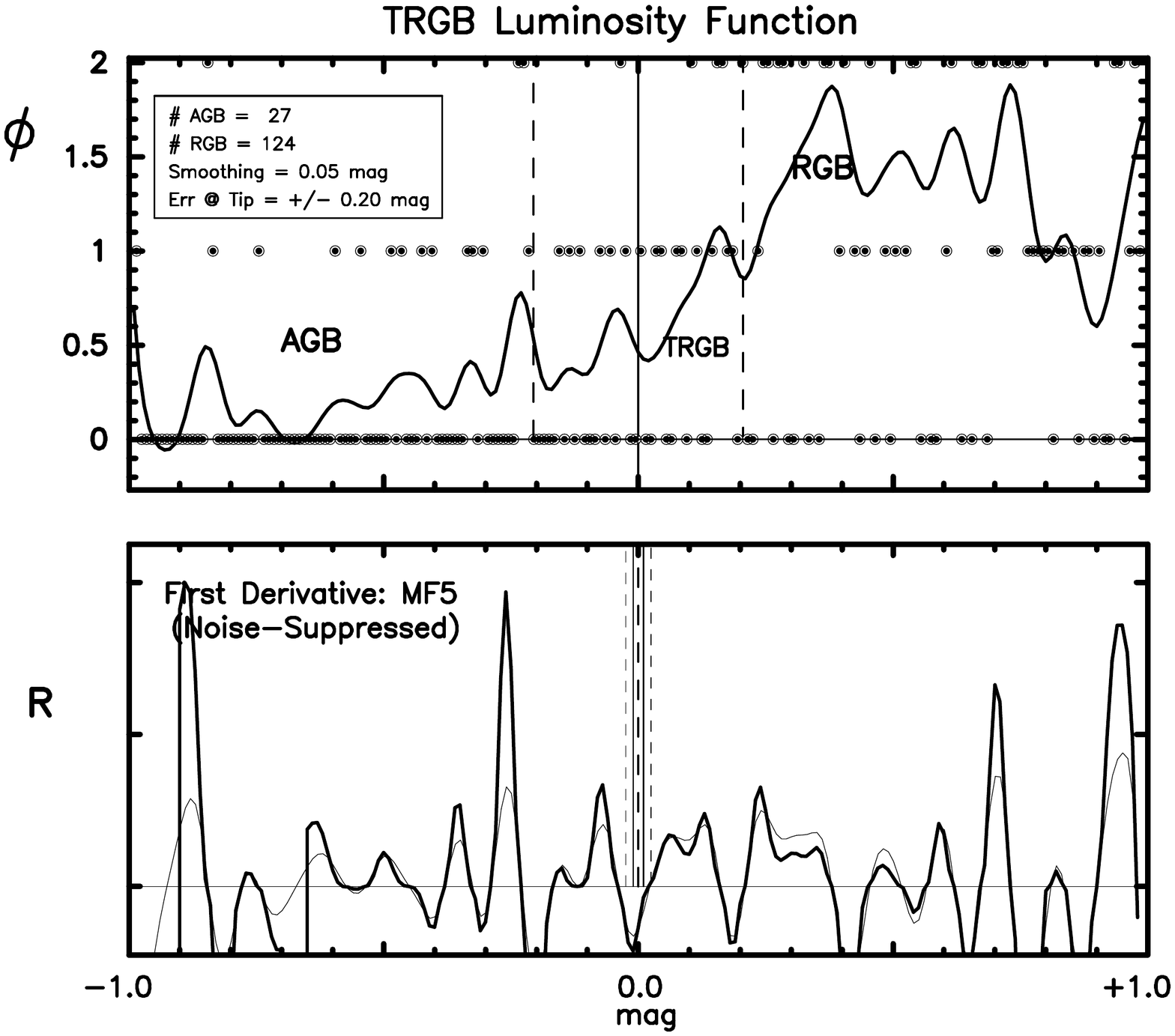} \caption{\small Six sub-panels illustrating the effect of increasing the photometric noise, (from 0.00 to 0.20~mag) at fixed, but slightly larger smoothing (0.05 mag) than previously discussed and for impoverished populations of RGB stars 124). The lower portions of each of the six sub-panels shows the first-derivative edge-detector output in both its uncorrected (thin black lines) and it noise-weighted (thicker black line) form. See text for a detailed discussion of the trends.} \end{figure*}

\vfill\eject
\subsection{Largest Smoothing Considered}
\subsubsection{A Range of Photometric Errors: {\bf}{120,000 RGB Stars}, Fixed Smoothing {\bf}{$\pm$0.10~mag}}

As we now move to overly aggressive smoothing, this large population (120,000 RGB star) simulation is clearly being over-smoothed at the tip, 
up to the point that the smoothing and the photometric error at the tip are of equal magnitude, $\pm$0.10~mag 
in this case.  As can be seen in Figure 10, the width of the response function at high signal-to-noise is controlled by the adopted 
smoothing up to the cross-over point of smoothing and photometric errors (middle right panel) after which 
the width grows with the increased photometric errors (last two panels). At high values of the photometric 
errors, the earlier-mentioned wings and low-level asymmetries are still present but obviously smoothed. Again, no bias is detected in the response function. Smoothing offers little or no advantage in the detection or measurement of the tip discontinuity in this particular scenario.

{\bf}{Summary 8 --} High levels of smoothing are not advantageous for large populations of RGB stars.

\begin{figure*} \centering
\includegraphics[width=8.0cm,angle=-0]{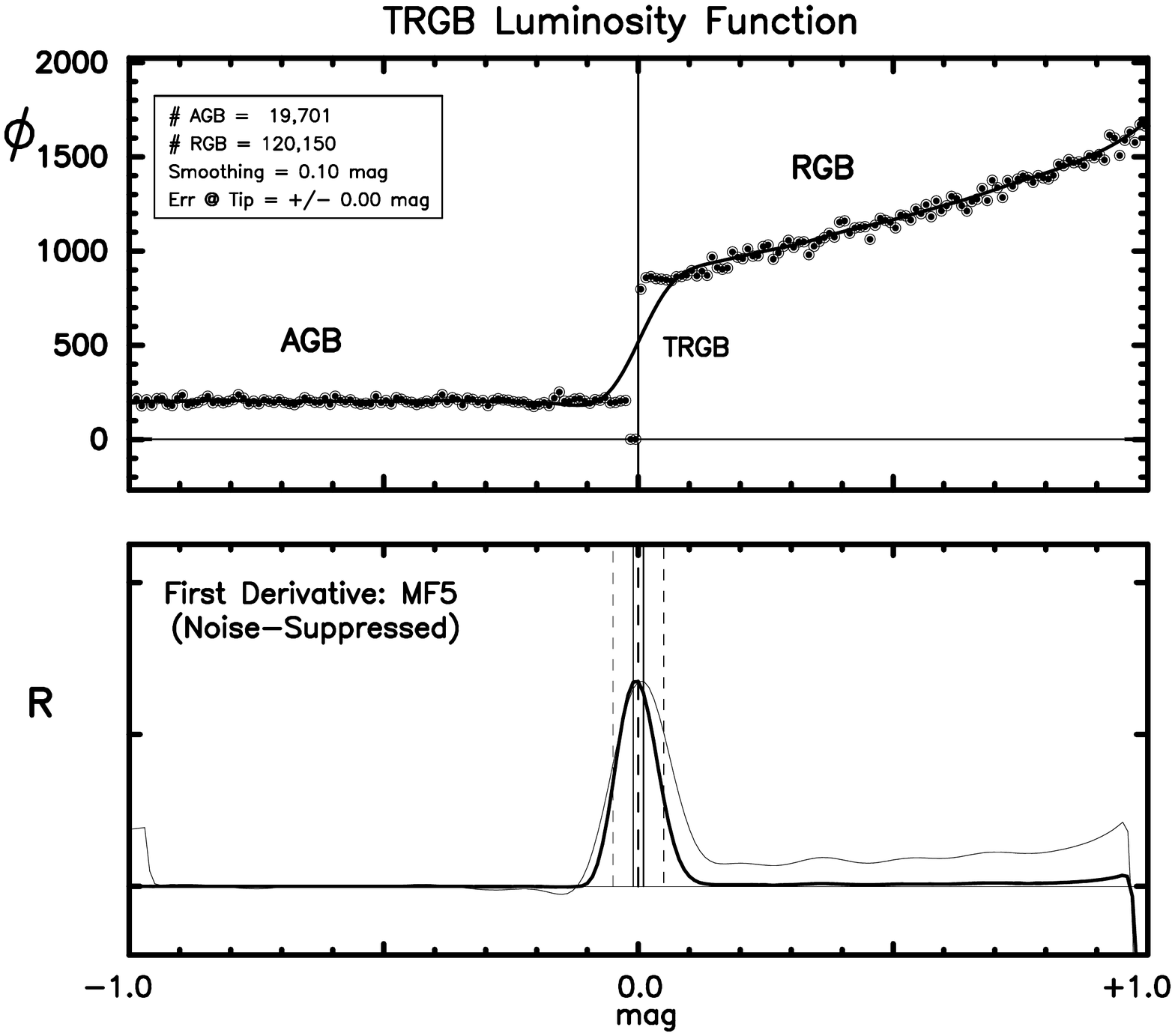}
\includegraphics[width=8.0cm,angle=-0]{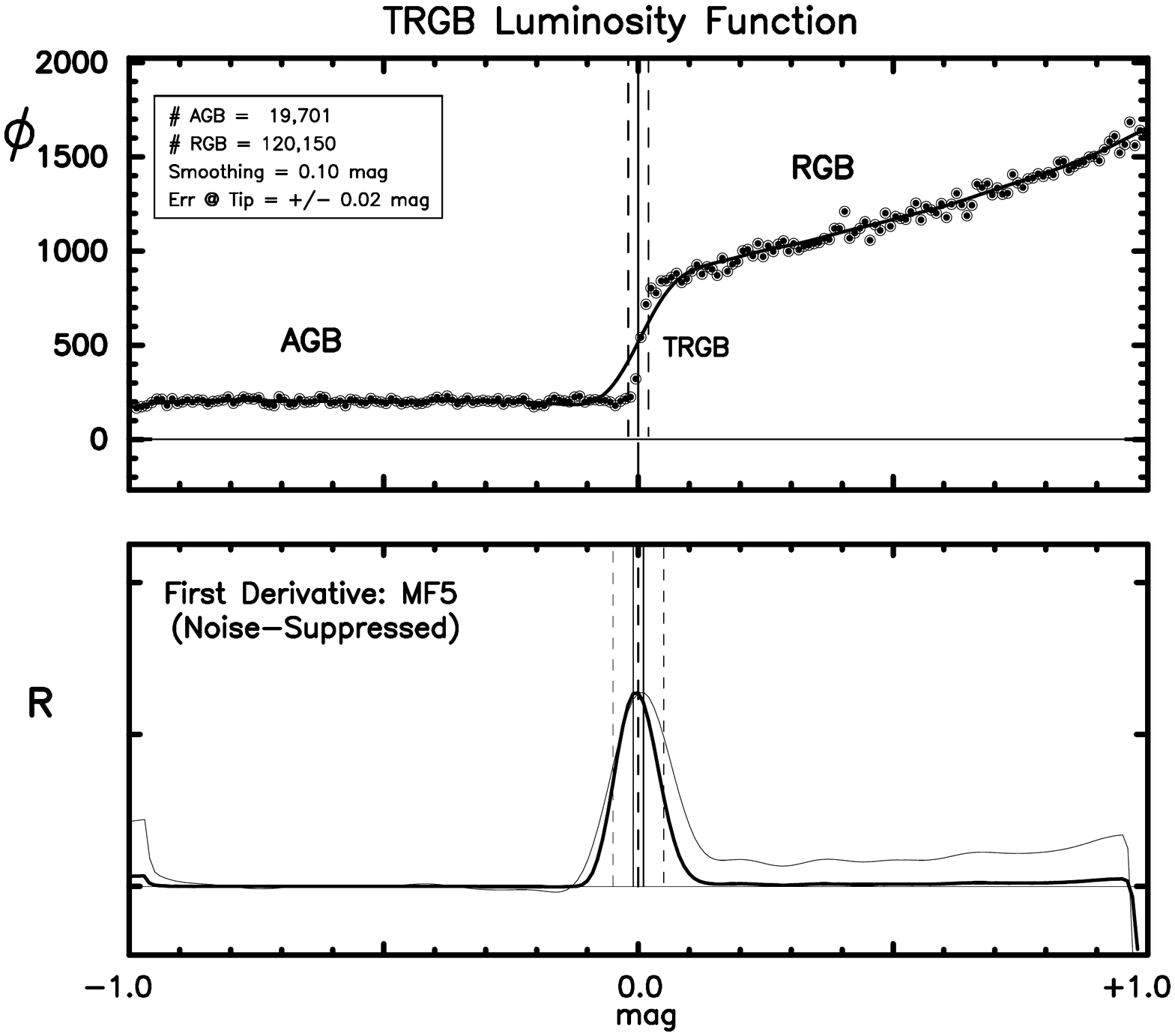}
\includegraphics[width=8.0cm,angle=-0]{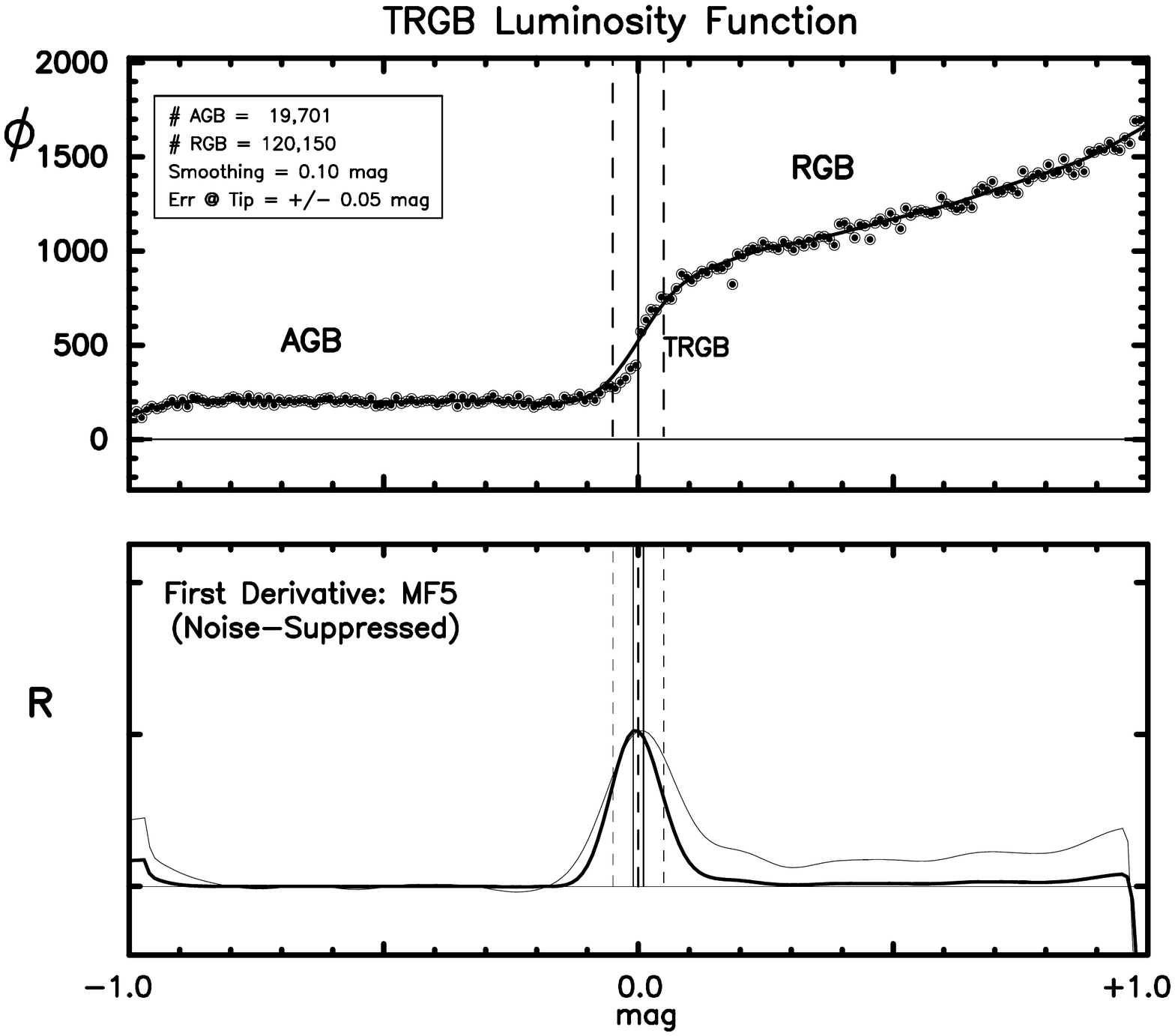}
\includegraphics[width=8.0cm,angle=-0]{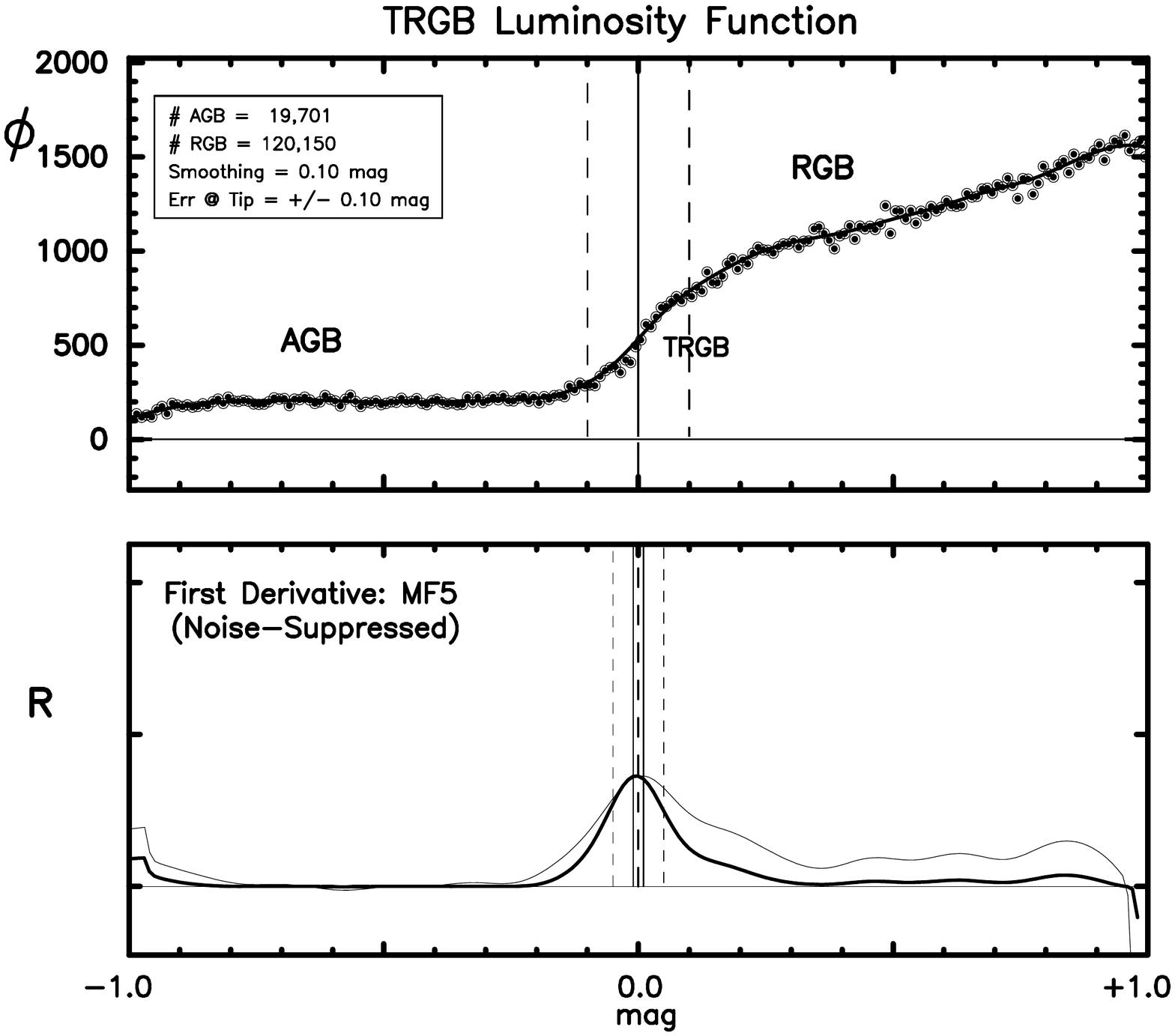}
\includegraphics[width=8.0cm,angle=-0]{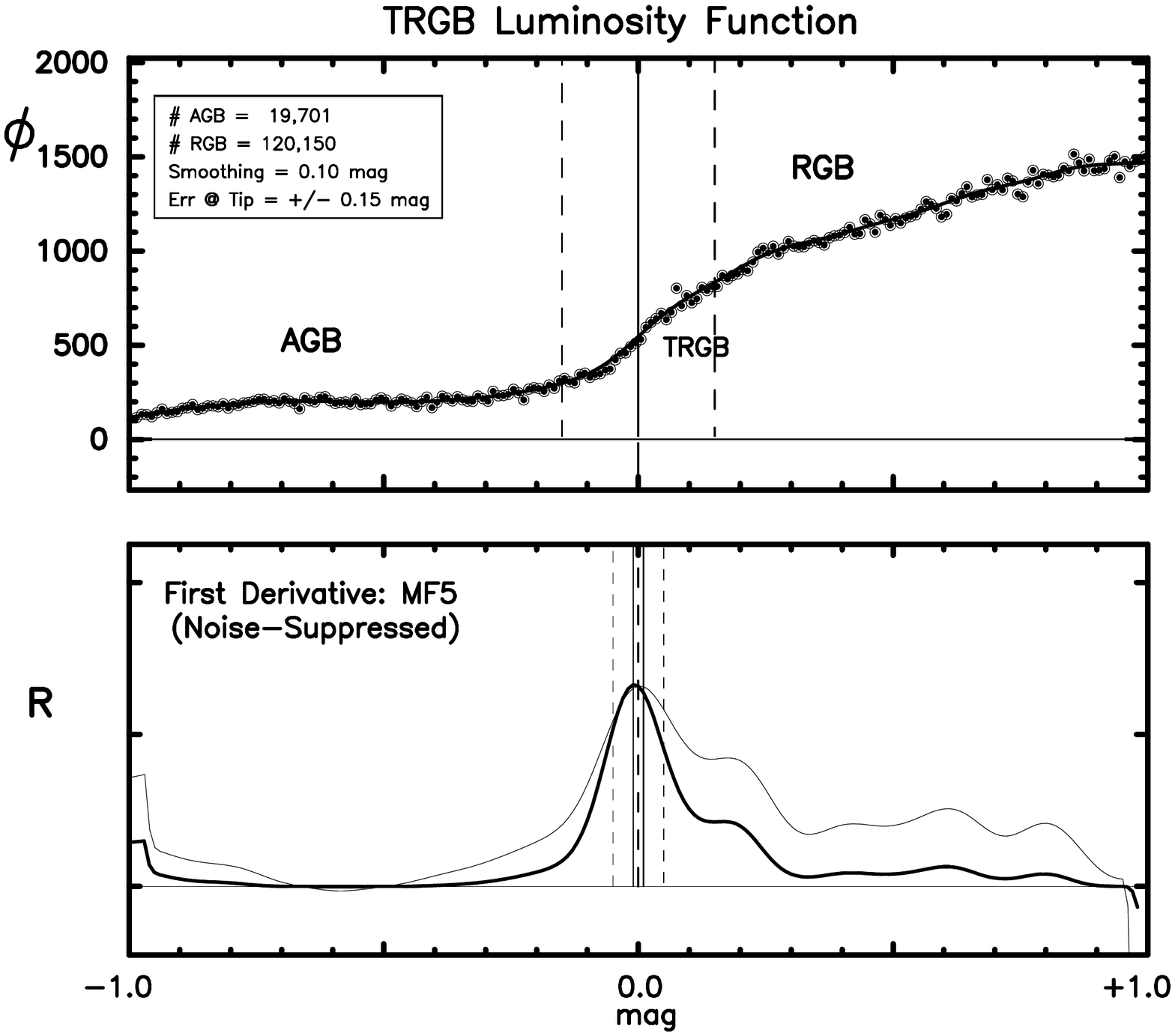}
\includegraphics[width=8.0cm,angle=-0]{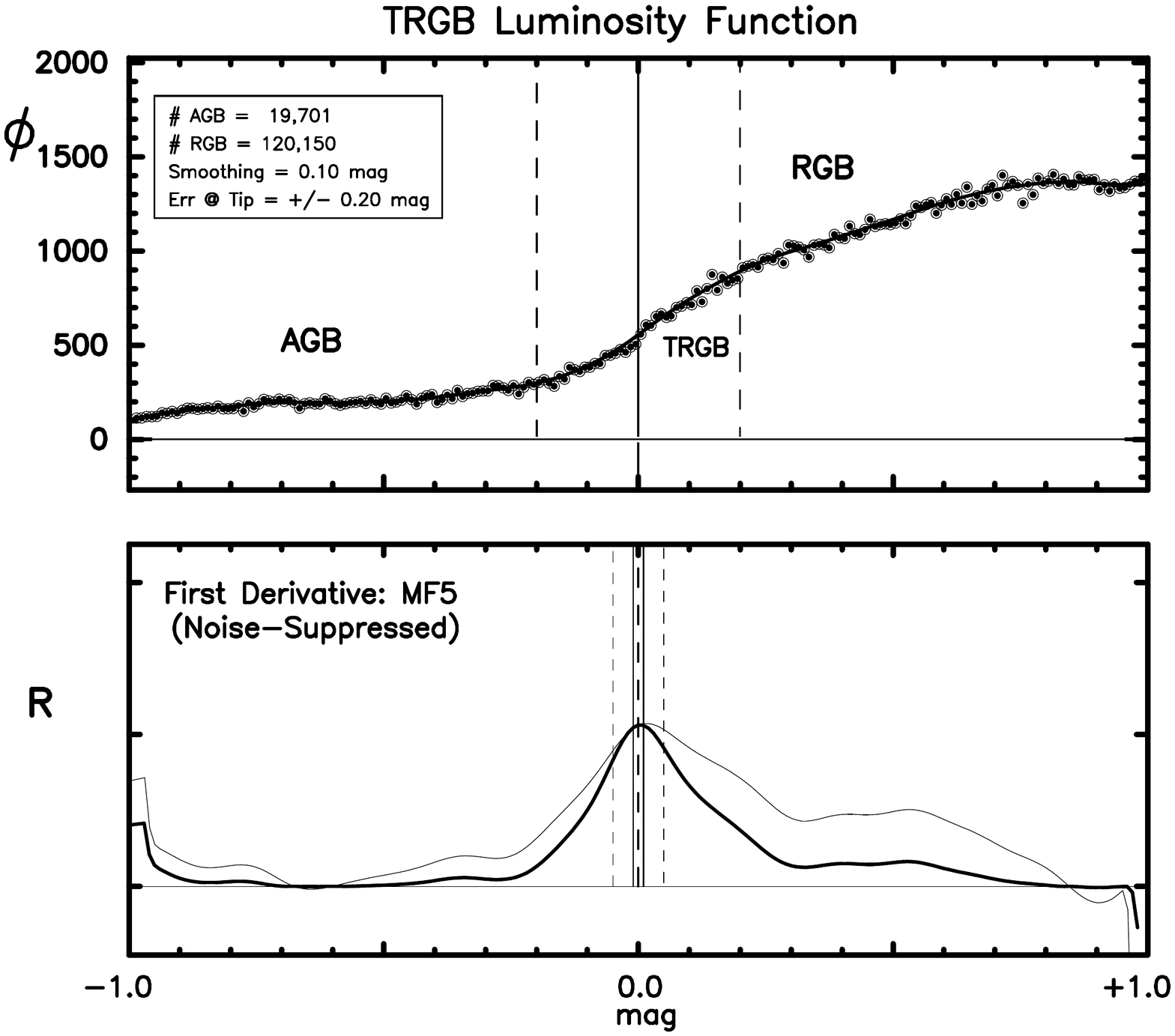} \caption{\small Six sub-panels illustrating the effect of increasing the photometric noise, (from 0.00 to $\pm$0.20~mag) at fixed, but moderate smoothing ($\pm$0.10 mag) and for very large populations of RGB stars (120,000). The lower portions of each of the six sub-panels shows the first-derivative edge-detector output in both its uncorrected (thin black lines) and it noise-weighted (thicker black line) form. See text for a detailed discussion of the trends.}\end{figure*}

\vfill\eject
\subsubsection{A Range of Photometric Errors: {\bf}{11,000 RGB Stars}, Fixed Smoothing {\bf}{$\pm$0.10~mag}}

As in the example discussed above, dropping the RGB sample to 11,000 stars in Figure 11 does not quantitatively change the description of the response function to increased photometric errors. 
However, there is now the first indication that the mode of the response function output is being drawn off center (at the $\pm$0.05~mag level) by the increased noise in the smoothed wings (last four panels). Over-smoothing should be carefully monitored. 
{\bf}{Running through a range of smoothing parameters 
can alert the user to systematic errors being introduced because of over-smoothing noise into the true peak}, as illustrated here in the last three panels.

When numerous (comparably significant) peaks are found with a low degree of smoothing, no amount of additional smoothing will reveal the true peak, but rather the resulting ``detection" will be a weighted average of the surrounding peaks which may (with enough smoothing) appear to a be a single (broad) peak, it will but probably be biased: consider smoothing the last three panels in Figure 7, as then seen in Figure 11. 
Our recommendation is that future investigators always try a number of smoothing kernels bracketing their preferred solution so as to reveal the presence (or absence) of substructure that a high degree of smoothing would otherwise gloss over.

Real world investigations into selecting an optimal smoothing have been undertaken by Beaton et al. (2019); see their Figures 5 and 8 for examples of the implementation of an iterative smoothing analysis. There one can see solutions that are over-smoothed systematically drifting from their less-smoothed solutions, being drawn away by adjacent, individually low significance, but sometimes numerous peaks.  Over smoothing in this context tends to occur when the smoothing parameter is in excess of the photometric errors at the TRGB. In addition, we note that a wide range of edge-detection methods using different smoothing kernels (and even including those using maximum-likelihood fitting techniques) were found to agree to very high (0.01 mag) precision when applied to the TRGB data for IC 1613 (Hatt et al. 2017, ApJ,  845, 146). The two papers both offer a quantitative means of selecting an optimal smoothing parameter which is the one that minimizes the quadrature sum of the random and systematic errors, generally selecting smoothing parameters that are indeed close to the measured photometric errors reported for stars at the tip.\footnote{ It should be made clear that the IC~1613 dataset and its reduction is exquisite in nature given the very high precision of the photometry and the sharpness of its tip. If similar investigations were to be shown for galaxies with lower quality data (say due to their increased distance or mixed populations), they would be unlikely to demonstrate such a high level of agreement.}

{\bf}{Summary 9 --} {\bf}{Over-smoothing can introduce a systematic bias in the presence of noise, and should be cautiously examined.}

\begin{figure*} \centering
\includegraphics[width=8.0cm,angle=-0]{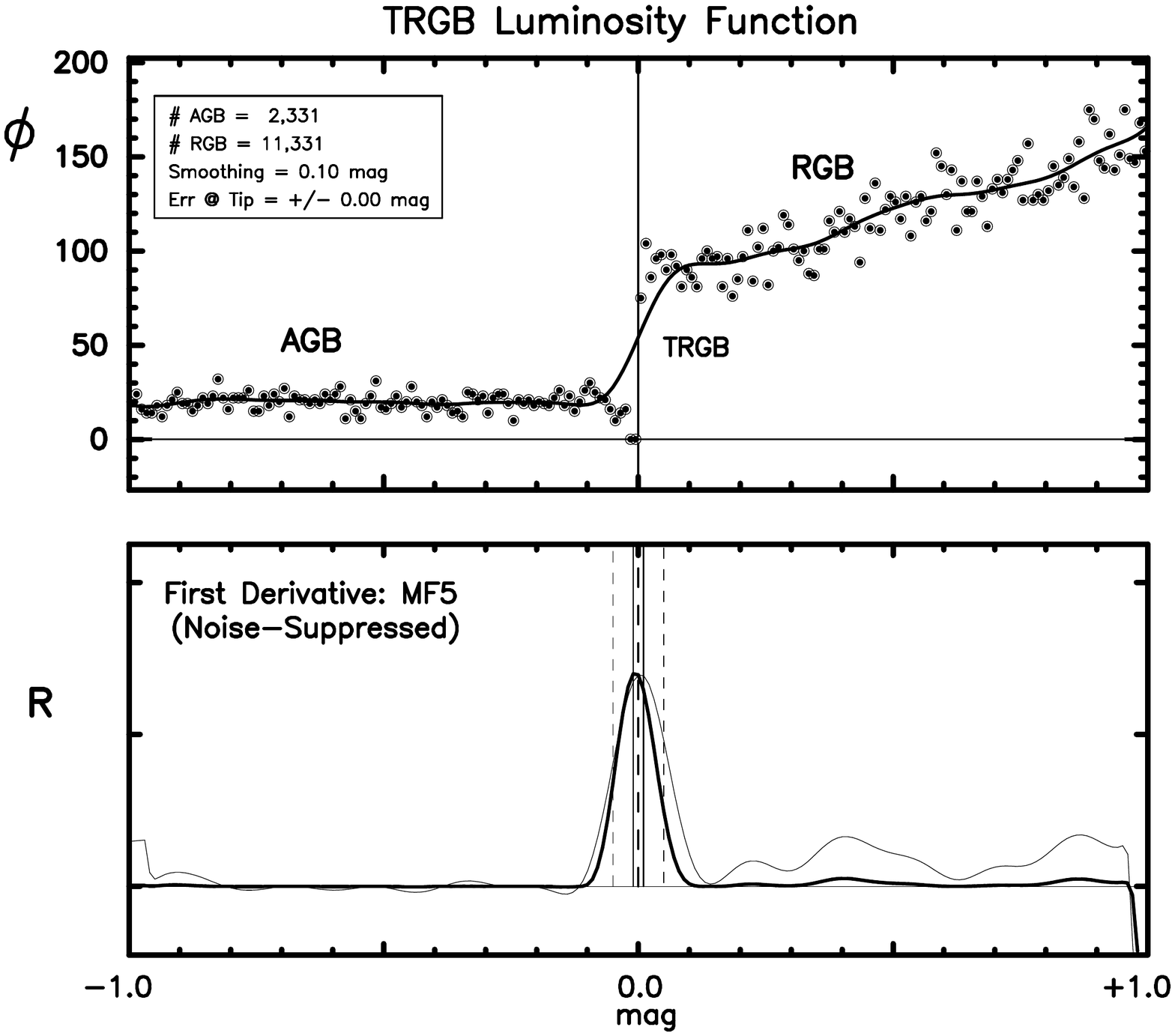}
\includegraphics[width=8.0cm,angle=-0]{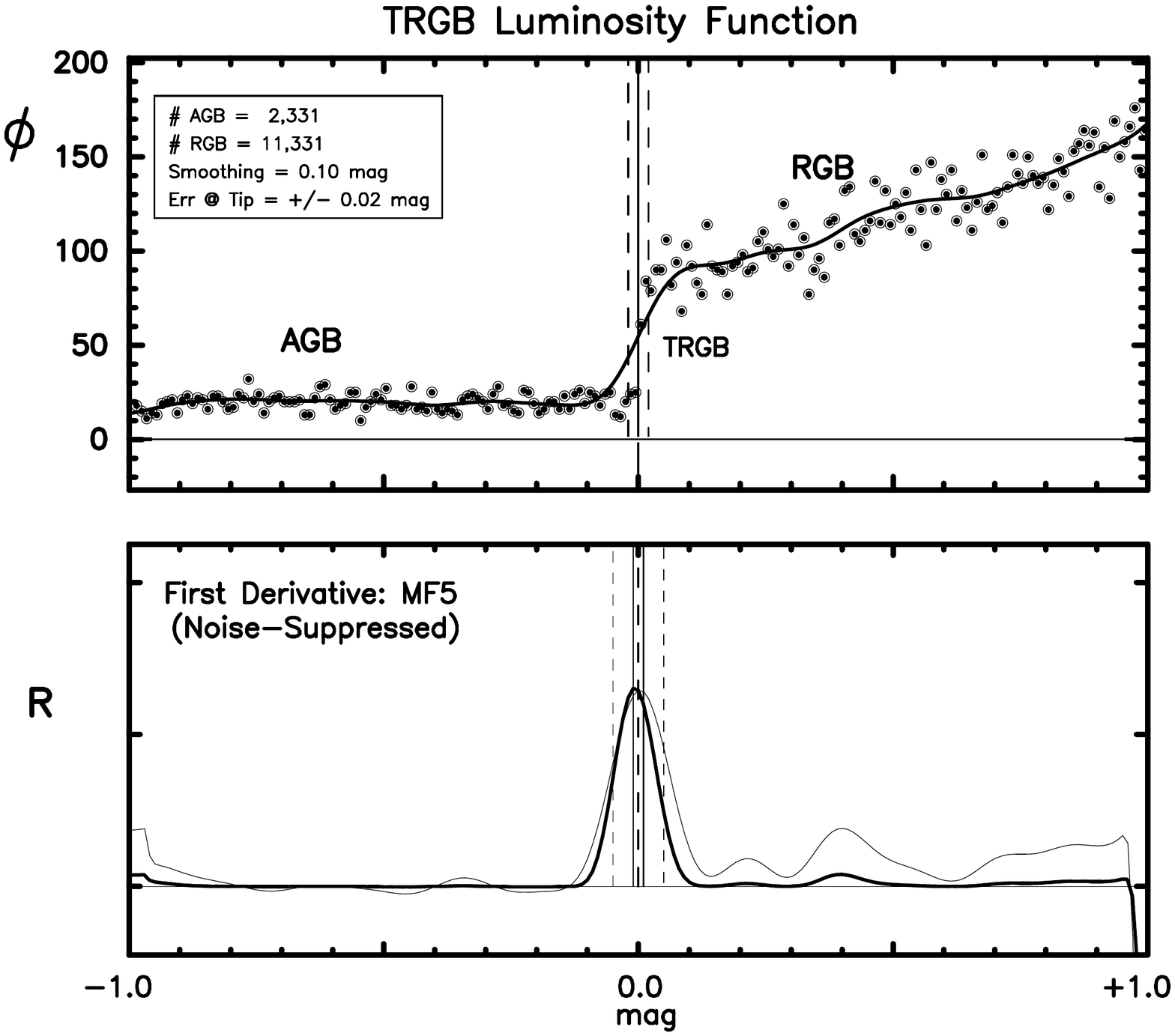}
\includegraphics[width=8.0cm,angle=-0]{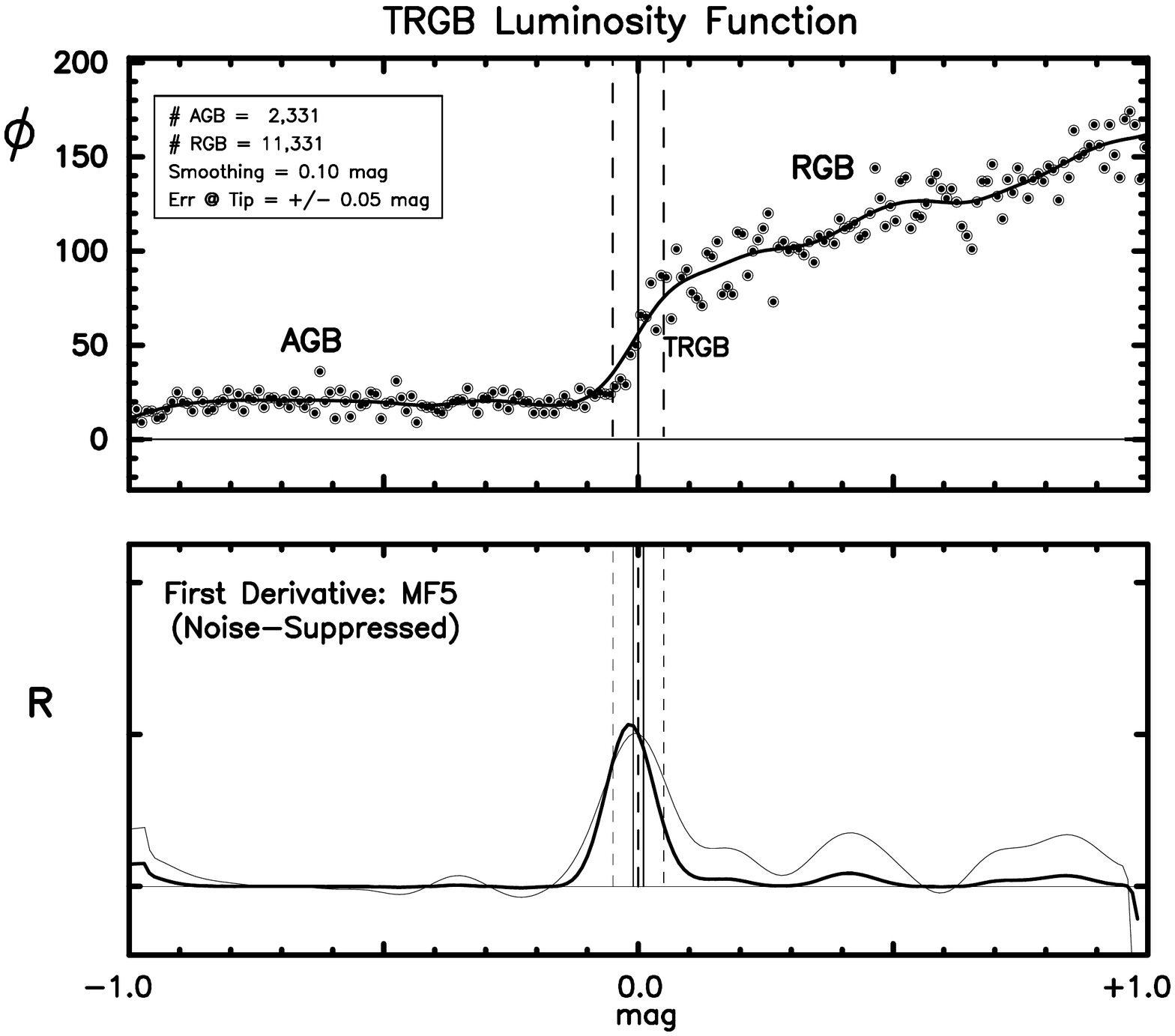}
\includegraphics[width=8.0cm,angle=-0]{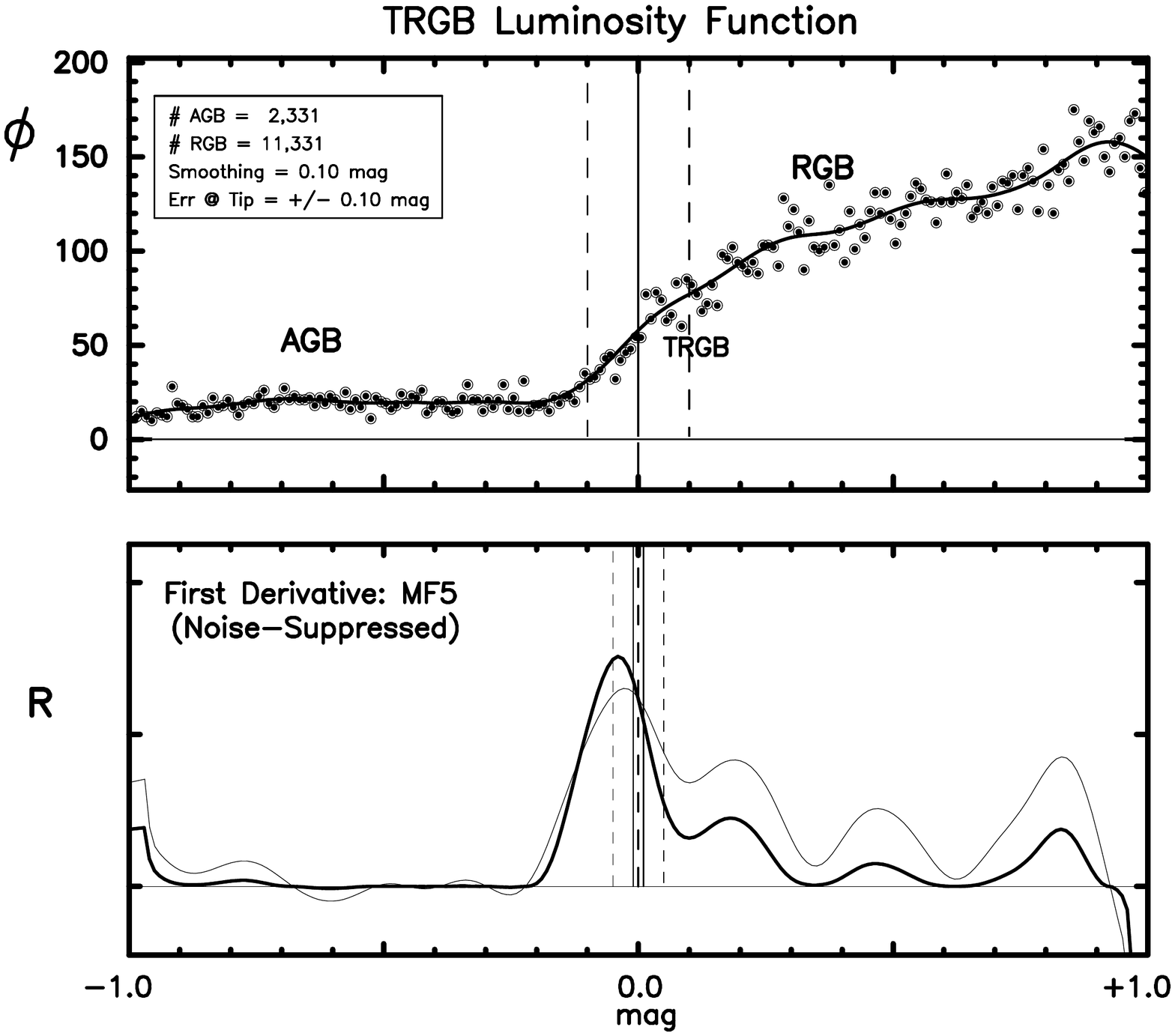}
\includegraphics[width=8.0cm,angle=-0]{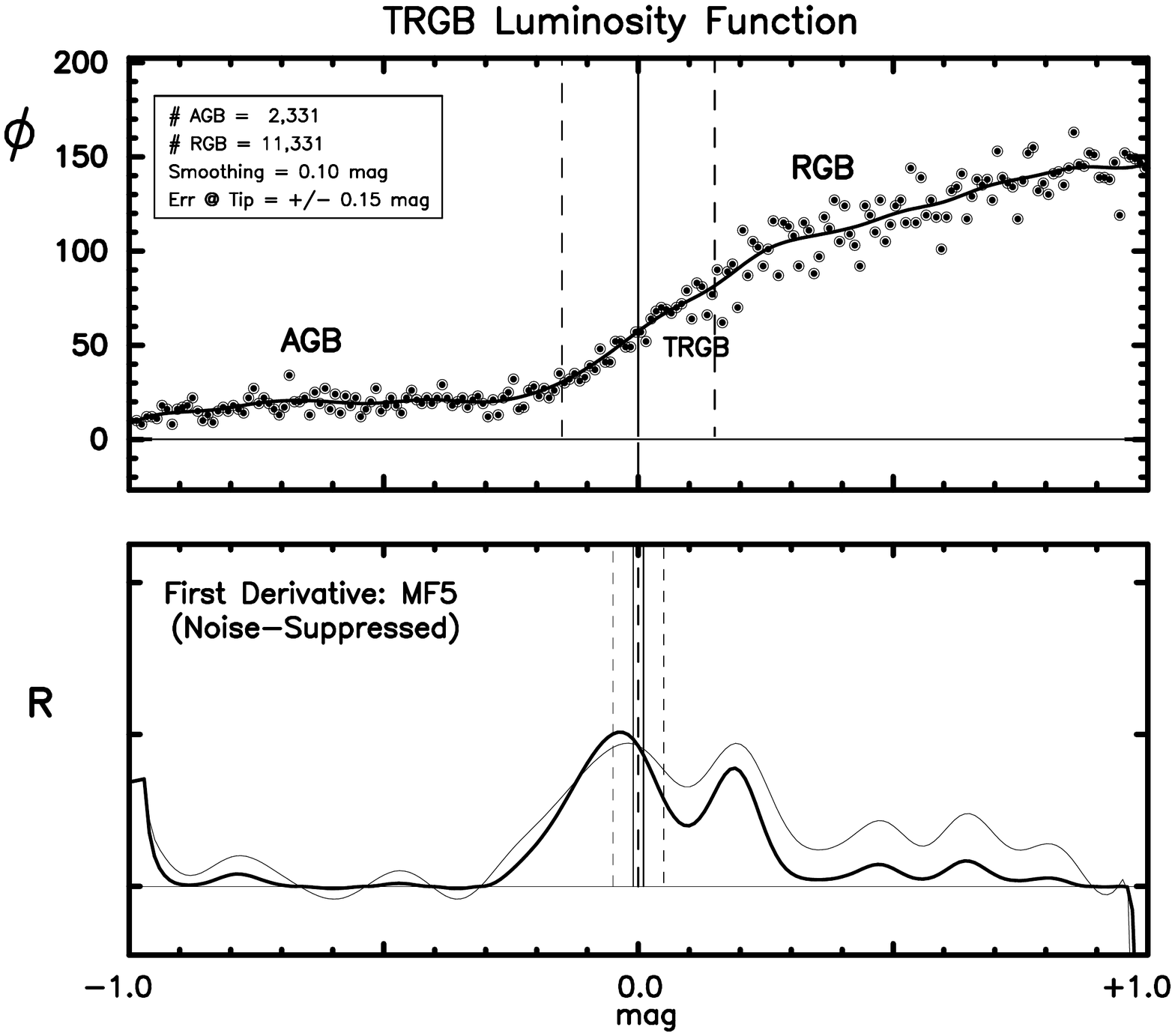}
\includegraphics[width=8.0cm,angle=-0]{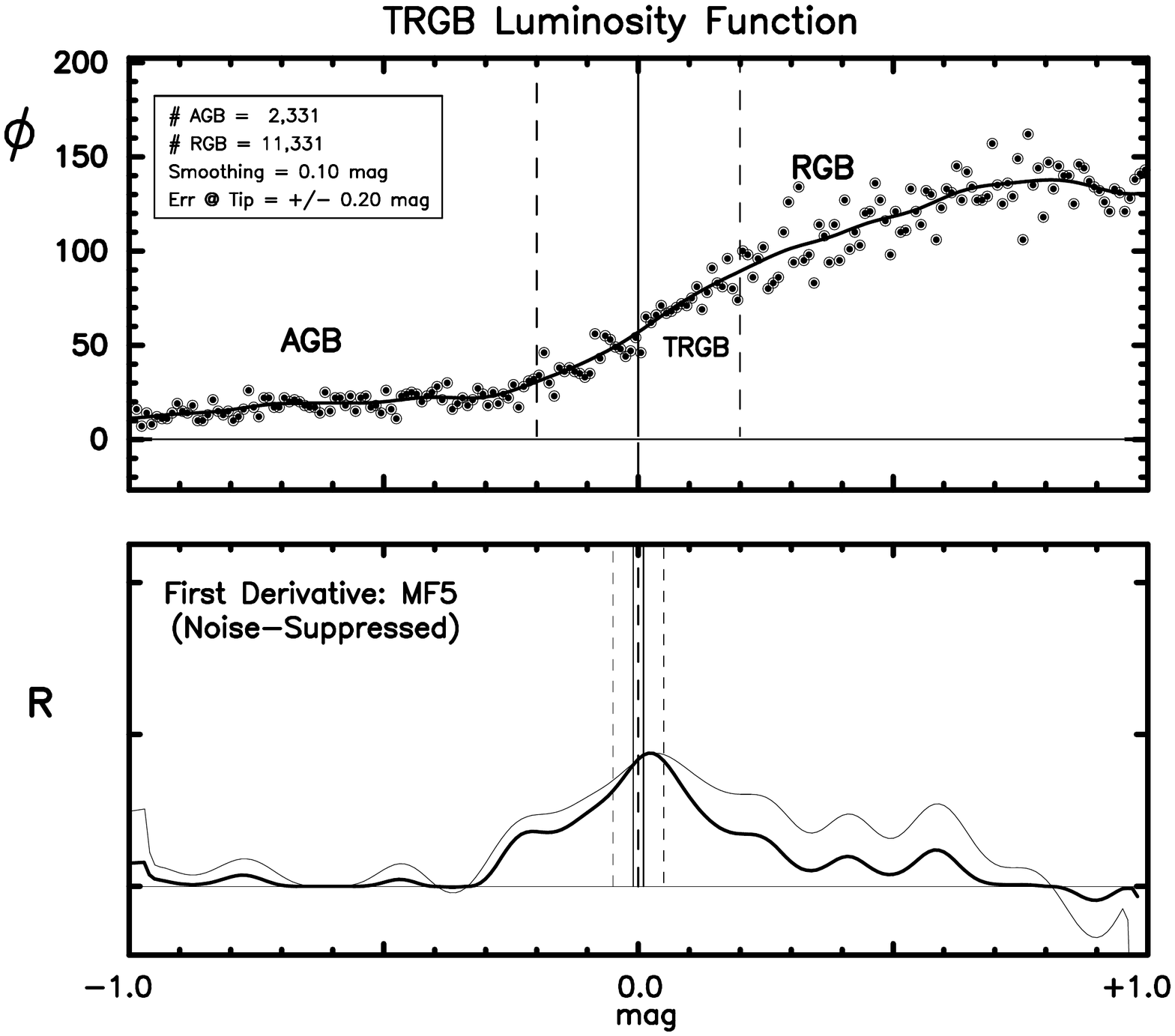} \caption{\small
Six sub-panels illustrating the effect of increasing the photometric noise, (from 0.00 to $\pm$0.20~mag) at fixed, but intermediate smoothing ($\pm$0.10 mag) of a moderately large population of RGB stars (11,000). The lower portions of each of the six sub-panels shows the first-derivative edge-detector output in both its uncorrected (thin black lines) and it noise-weighted (thicker black line) form. See text for a detailed discussion of the trends.}
\end{figure*}
\vfill\eject
\subsubsection{A Range of Photometric Errors: {\bf}{1,200 RGB Stars}, Fixed Smoothing {\bf}{$\pm$0.10~mag}}

At 1,200 stars in the RGB tip detection is unbiased and unambiguous at high signal to noise in the photometry at the tip (upper left panel of Figure 12).
At lower photometric precision adjacent noise spikes broaden and can bias the true tip detection by up to 0.1~mag (bottom two panels). 
In this realization  several of the deflections are toward brighter magnitudes, but there is no reason to believe that these are anything more than random fluctuations around the mean (see below). 
\begin{figure*} \centering
\includegraphics[width=8.0cm,angle=-0]{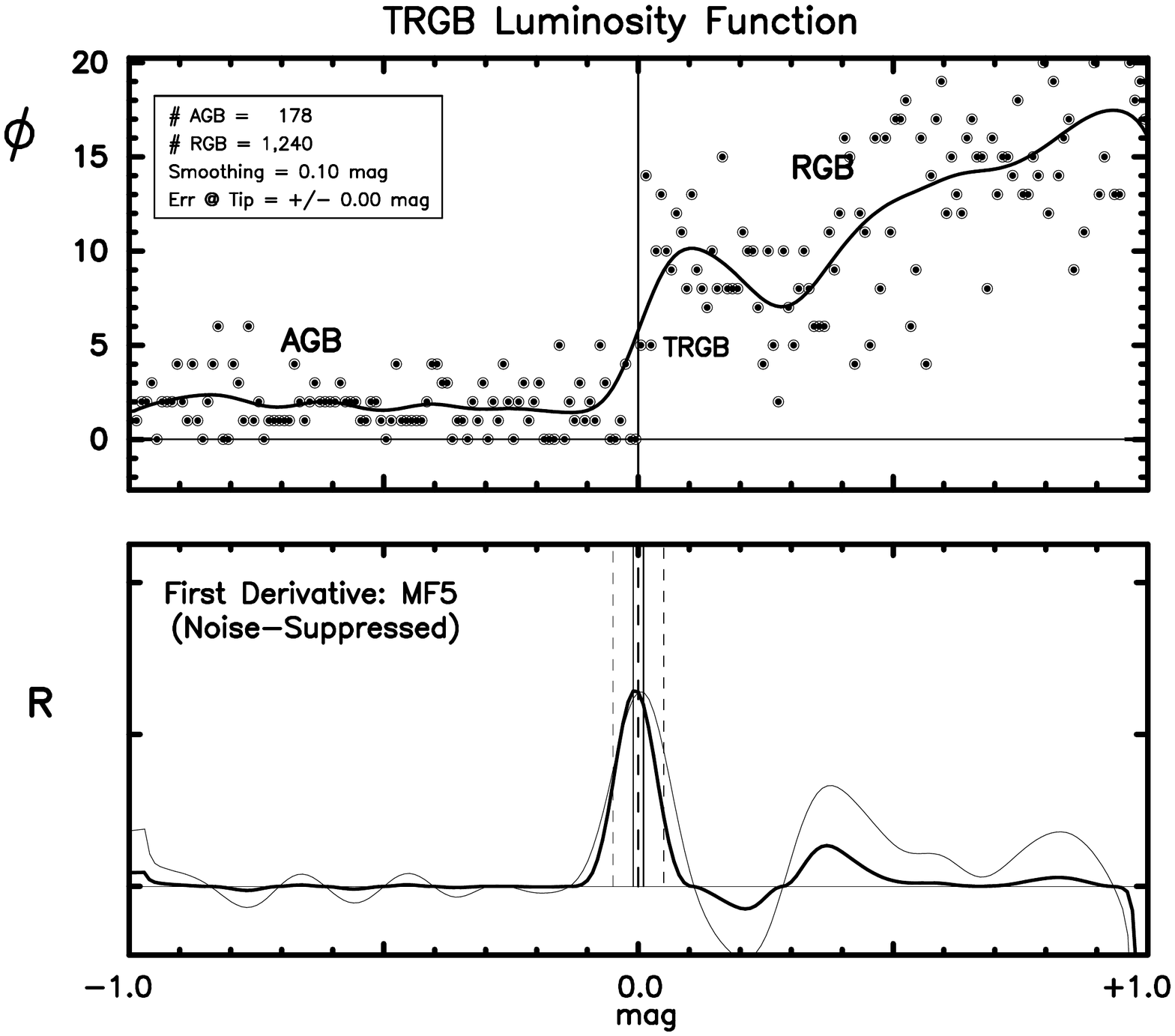}
\includegraphics[width=8.0cm,angle=-0]{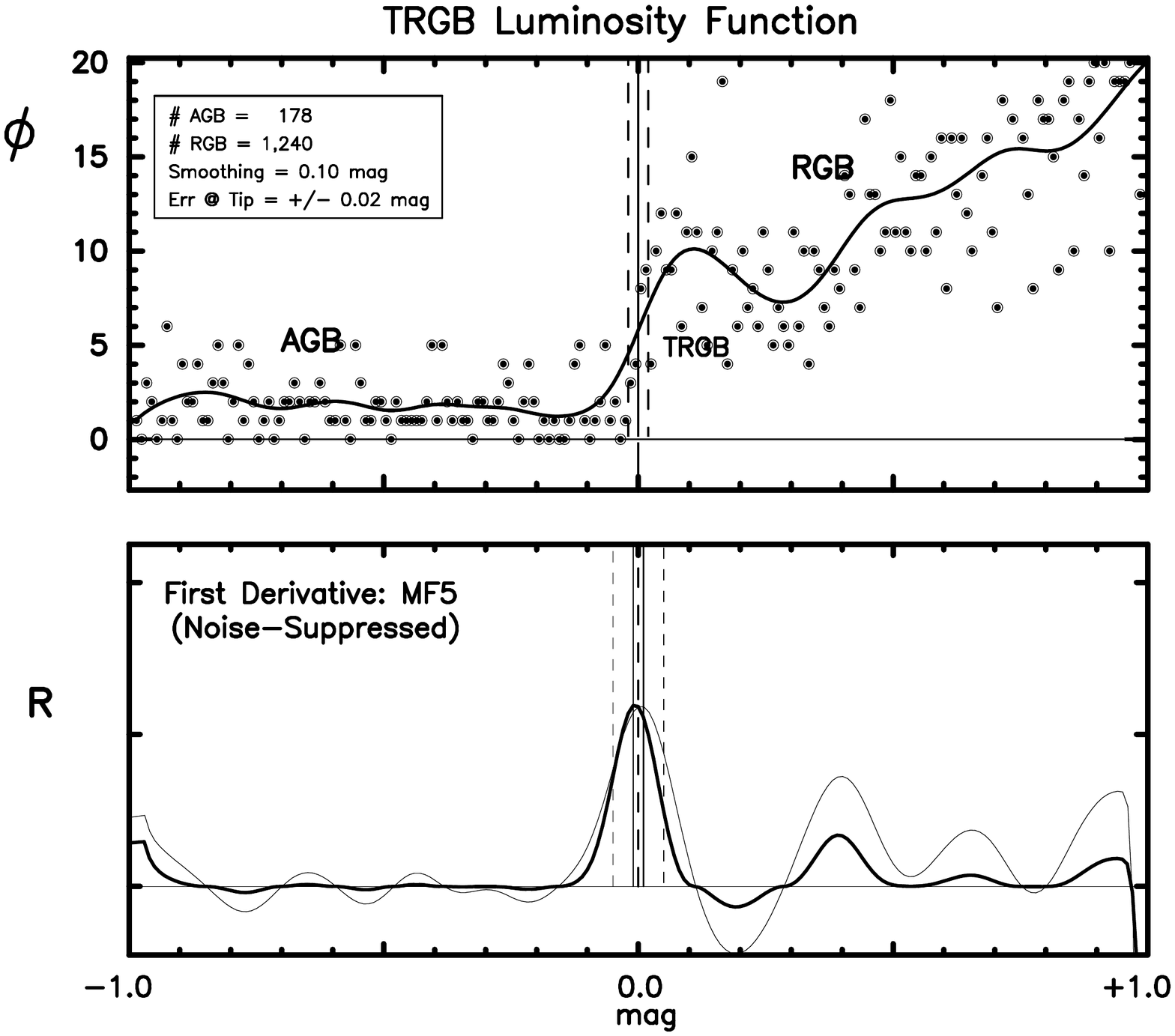}
\includegraphics[width=8.0cm,angle=-0]{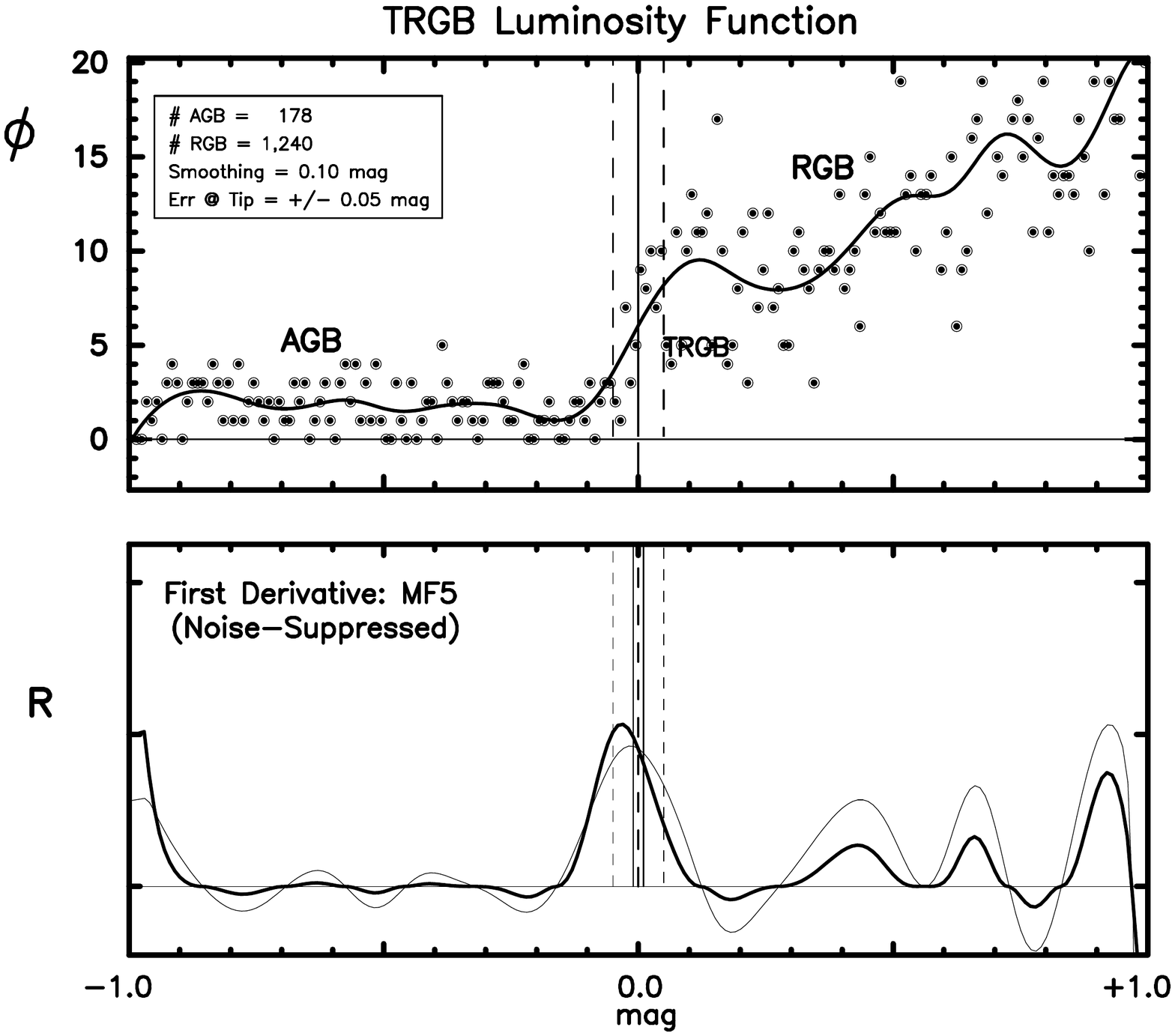}
\includegraphics[width=8.0cm,angle=-0]{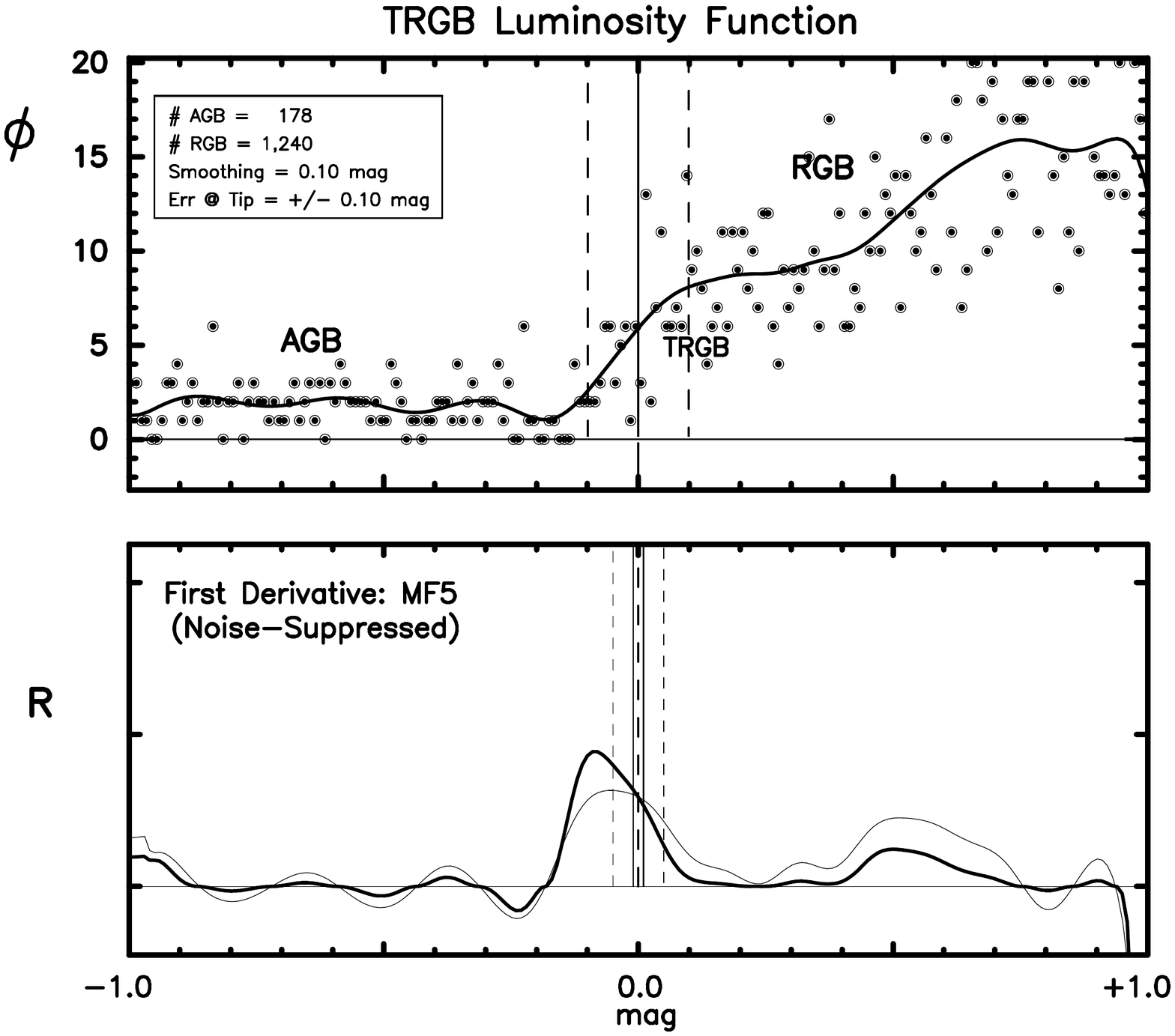}
\includegraphics[width=8.0cm,angle=-0]{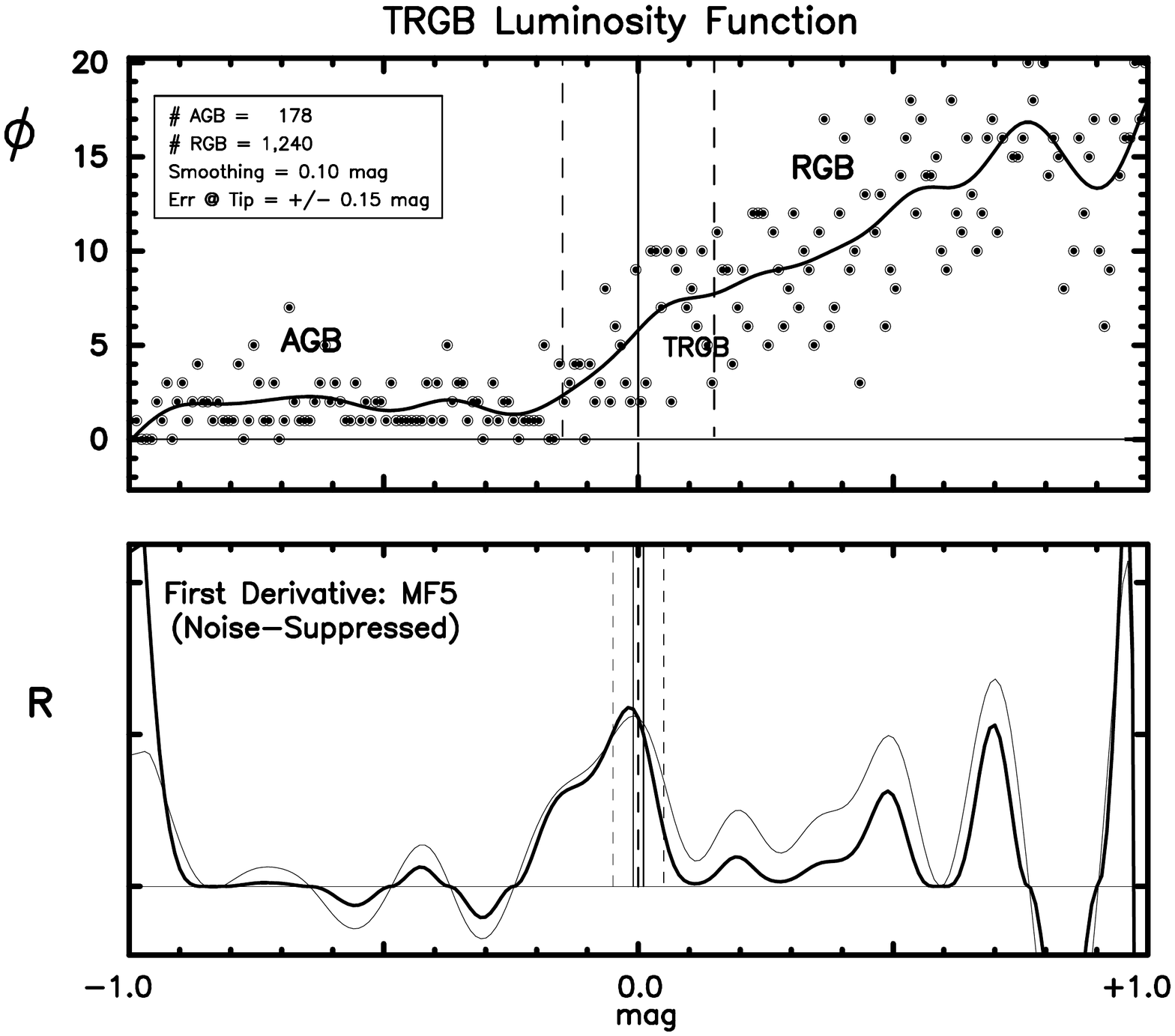}
\includegraphics[width=8.0cm,angle=-0]{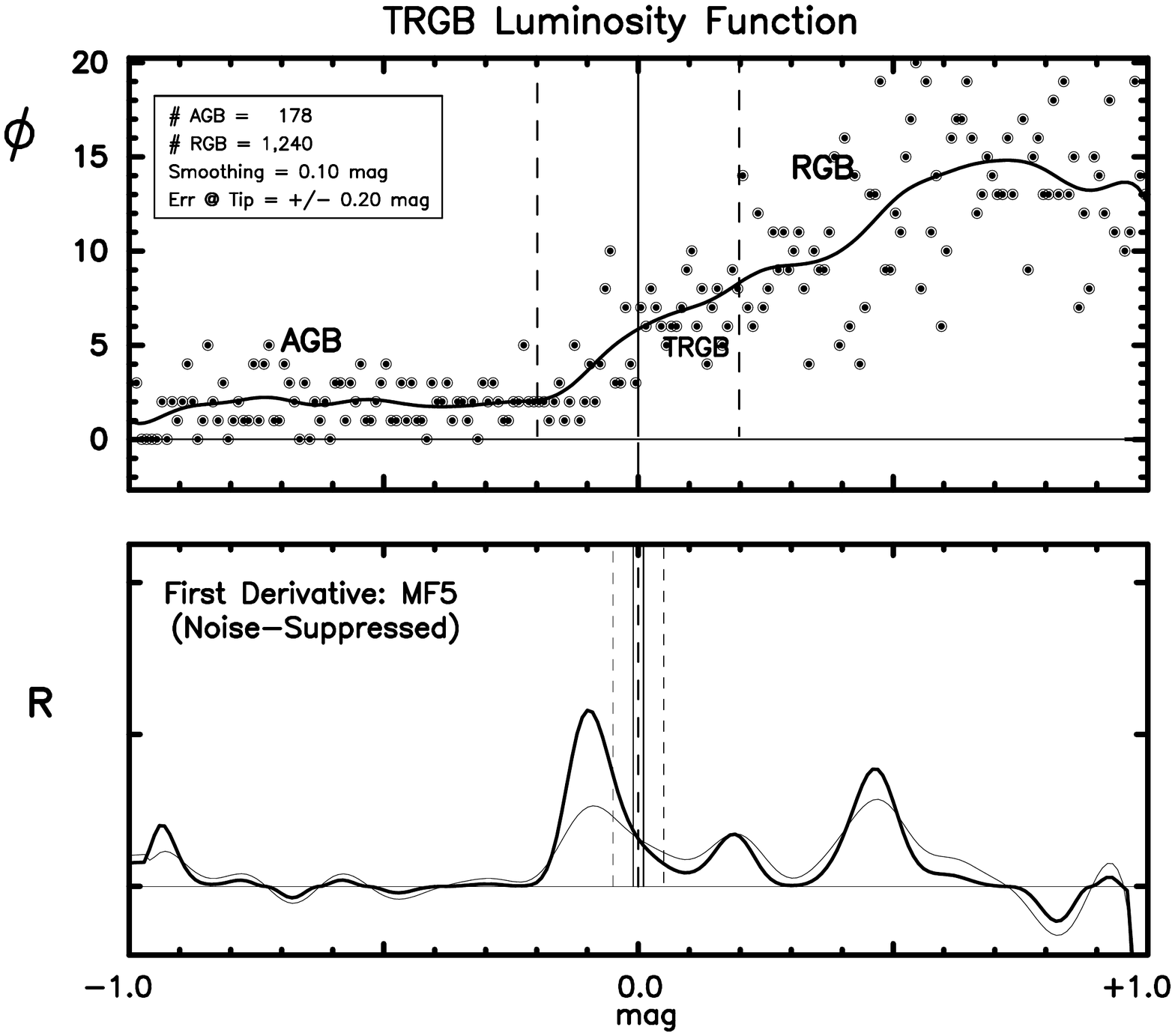} \caption{\small
Six sub-panels illustrating the effect of increasing the photometric noise, (from 0.00 to $\pm$0.20~mag) at fixed, but intermediate smoothing ($\pm$0.10 mag) of a moderately large population of RGB stars (12,000). The lower portions of each of the six sub-panels shows the first-derivative edge-detector output in both its uncorrected (thin black lines) and it noise-weighted (thicker black line) form. See text for a detailed discussion of the trends.}
\end{figure*}
\vfill\eject
\tablecaption{Guide to Simulations: Figs 2-13}
%\end{figure*}

\vfill\eject
\subsubsection{A Range of Photometric Errors: {\bf}{120 RGB Stars}, Fixed Smoothing {\bf}{$\pm$0.10~mag}}

This final realization shows the filter response to a small sample size (120 RGB stars) with large ($\pm$0.10~mag) 
smoothing applied to monotonically increasing photometric errors. At high signal-to-noise, (the top two panels of Figure 13) the true peak is properly detected, but it is 
not the highest peak over the two-magnitude interval. In this particular simulation 
the strongest (all false positive) peaks are found, four out of six times, at fainter magnitudes than the 
true tip, and by up to 0.35~mag separation. With an average of one star per RGB bin, wild statistical fluctuations, both in the luminosity function itself and in the discontinuity detector, are both to be expected and are seen. This is far from being an acceptable situation for detecting or measuring the TRGB with any degree of confidence.

{\bf}{Summary 10 --} One should not even attempt to a tip detection at low signal to noise in situations where the population size is only in the hundreds. Spurious signals will be found above and below the true tip.
\begin{figure*} \centering
\includegraphics[width=8.0cm,angle=-0]{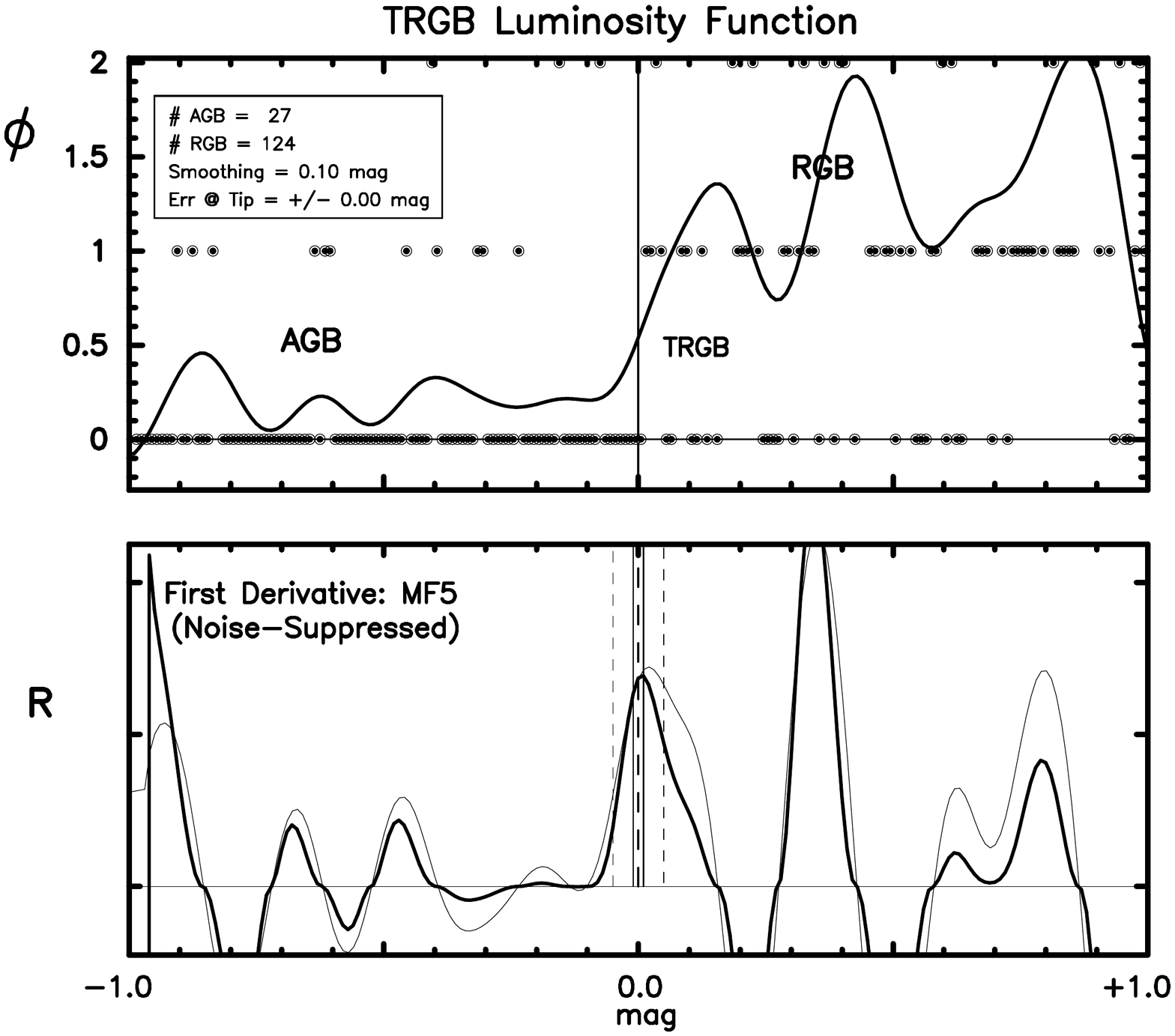}
\includegraphics[width=8.0cm,angle=-0]{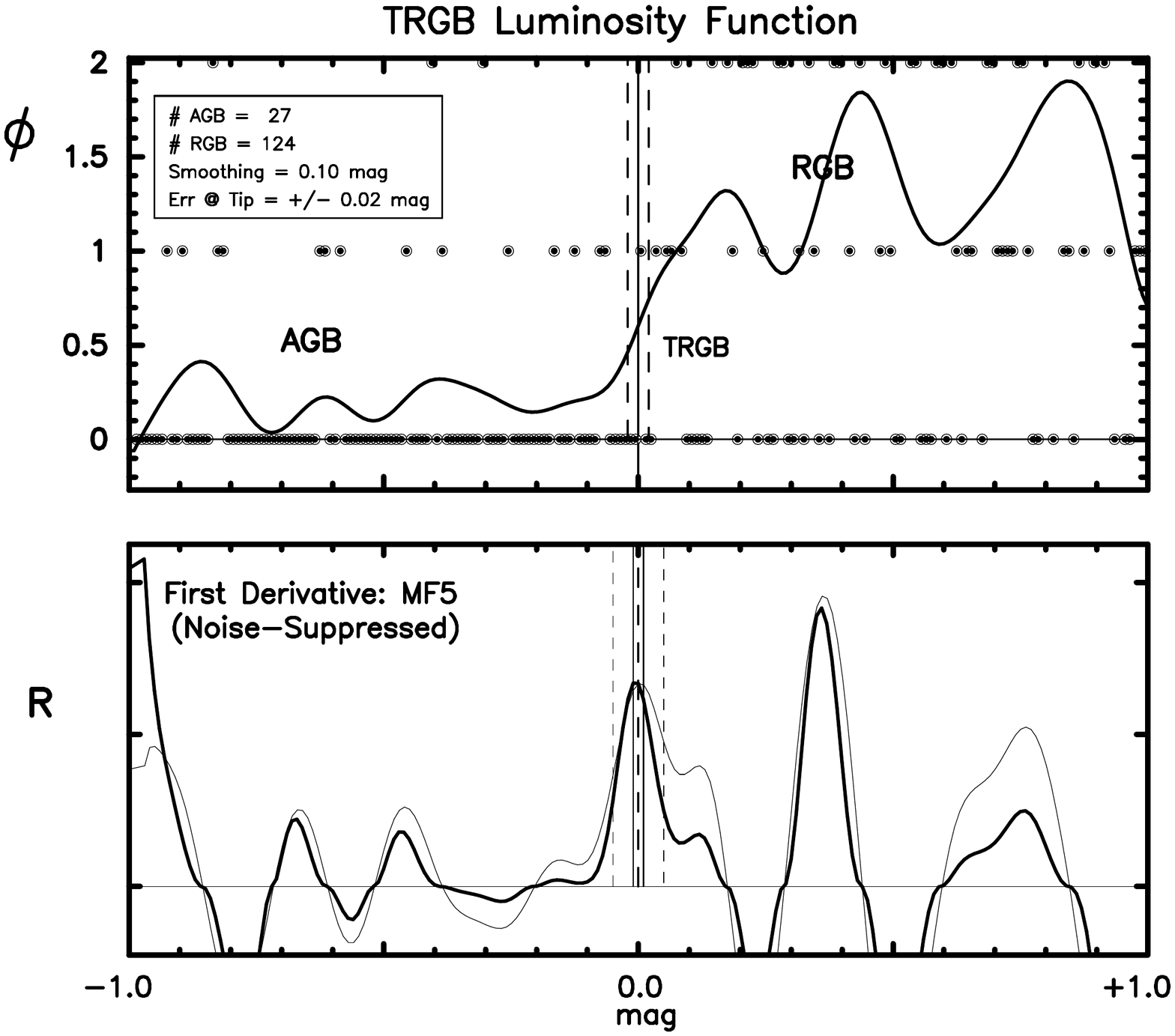}
\includegraphics[width=8.0cm,angle=-0]{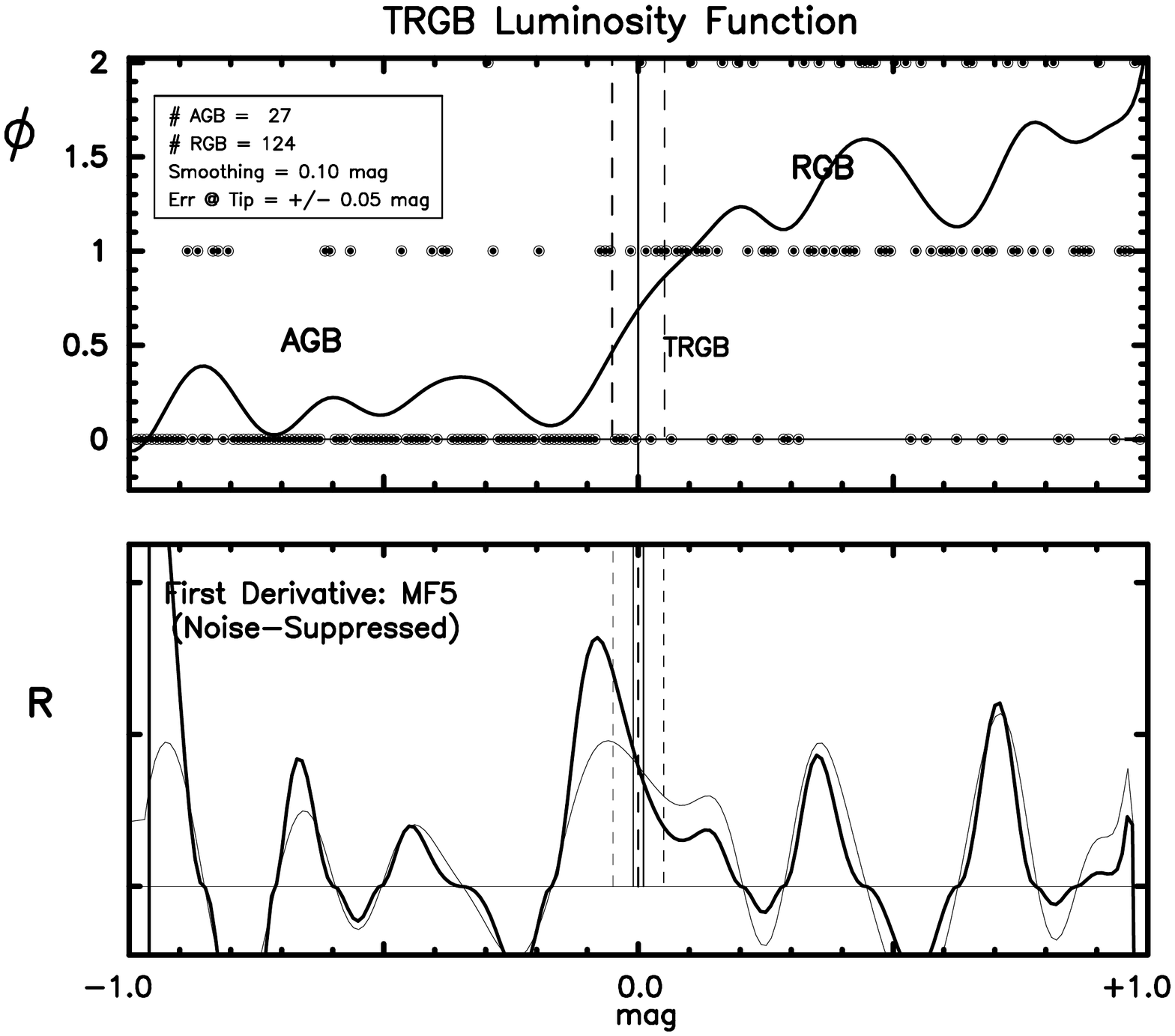}
\includegraphics[width=8.0cm,angle=-0]{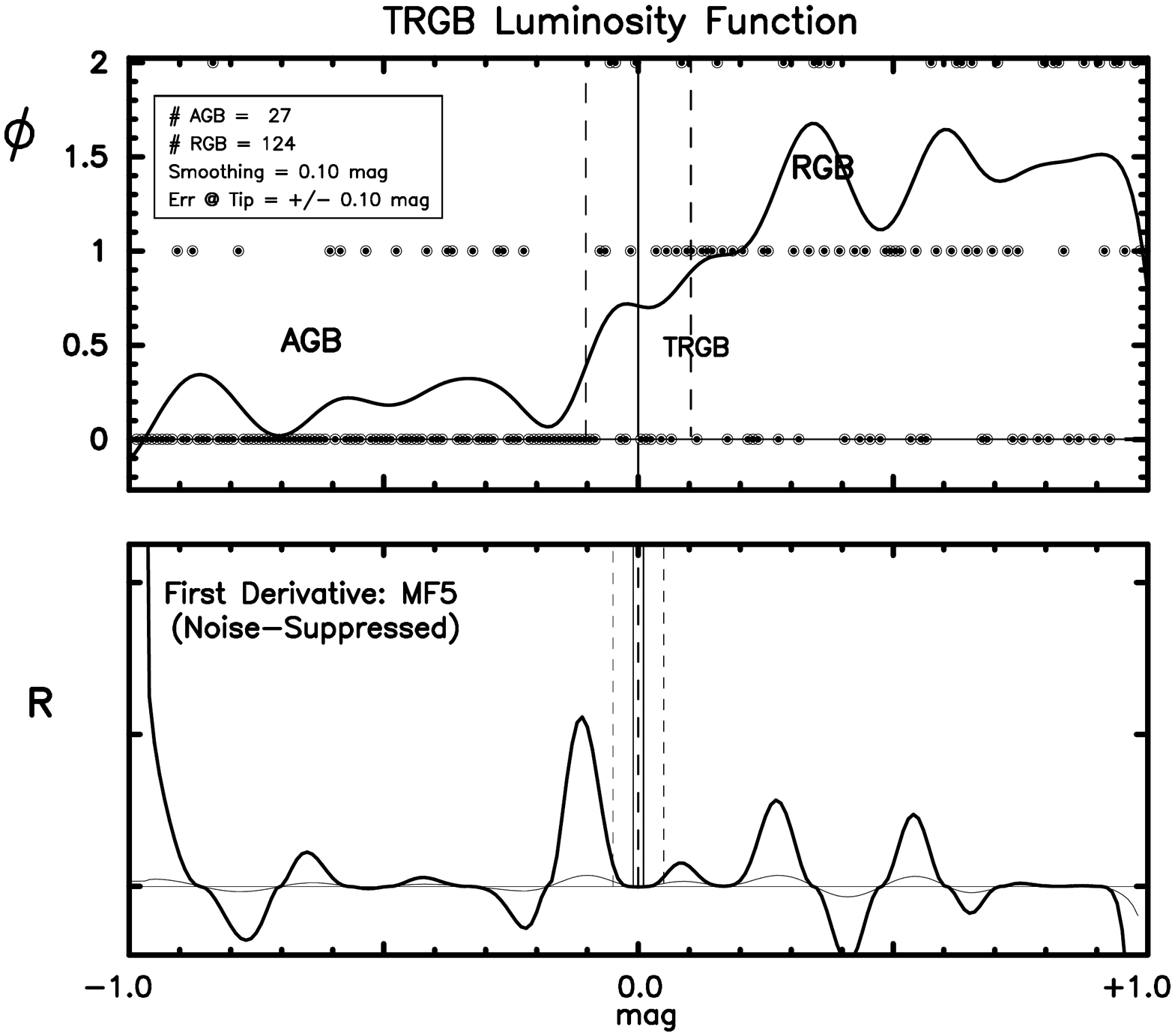}
\includegraphics[width=8.0cm,angle=-0]{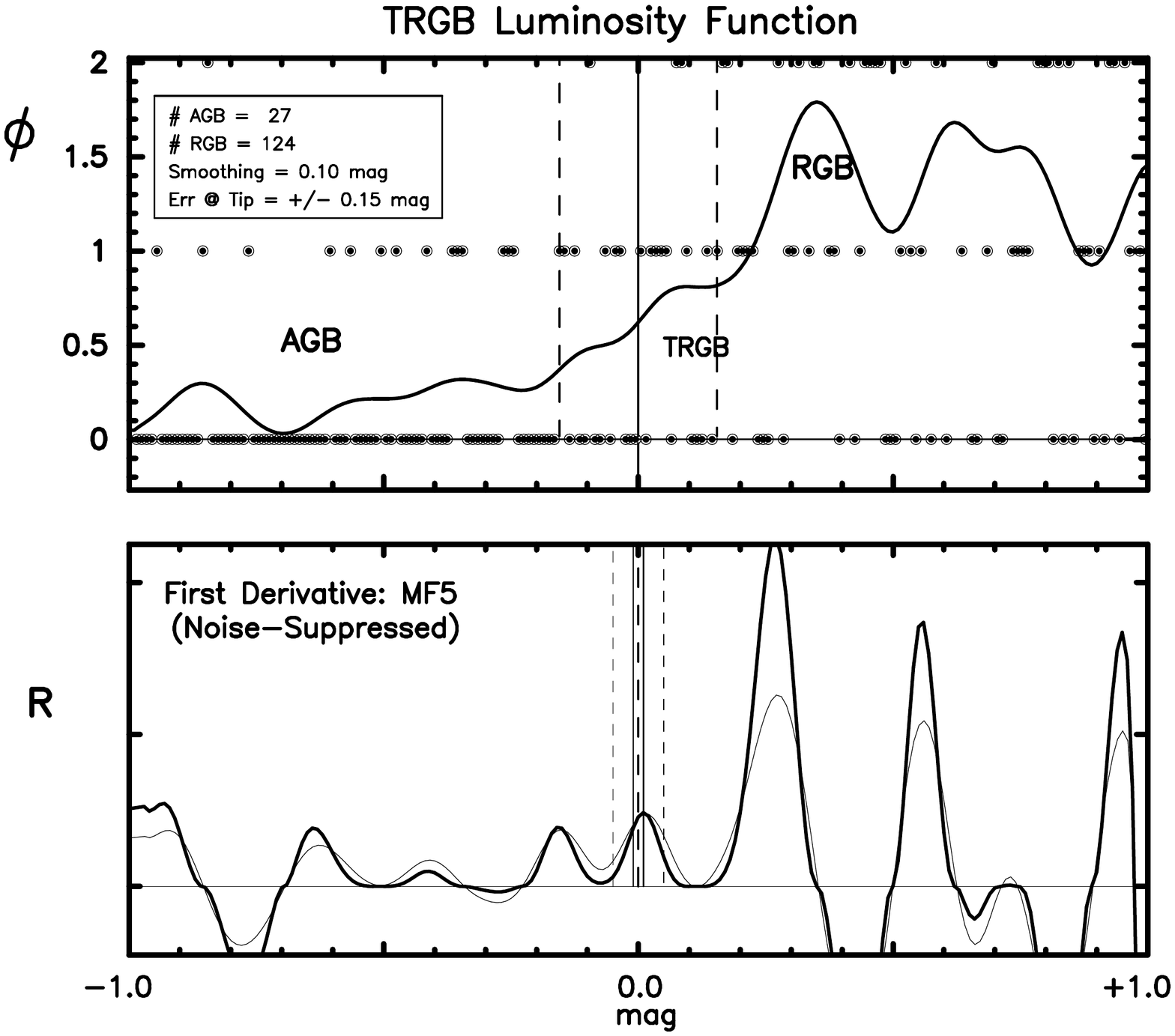}
\includegraphics[width=8.0cm,angle=-0]{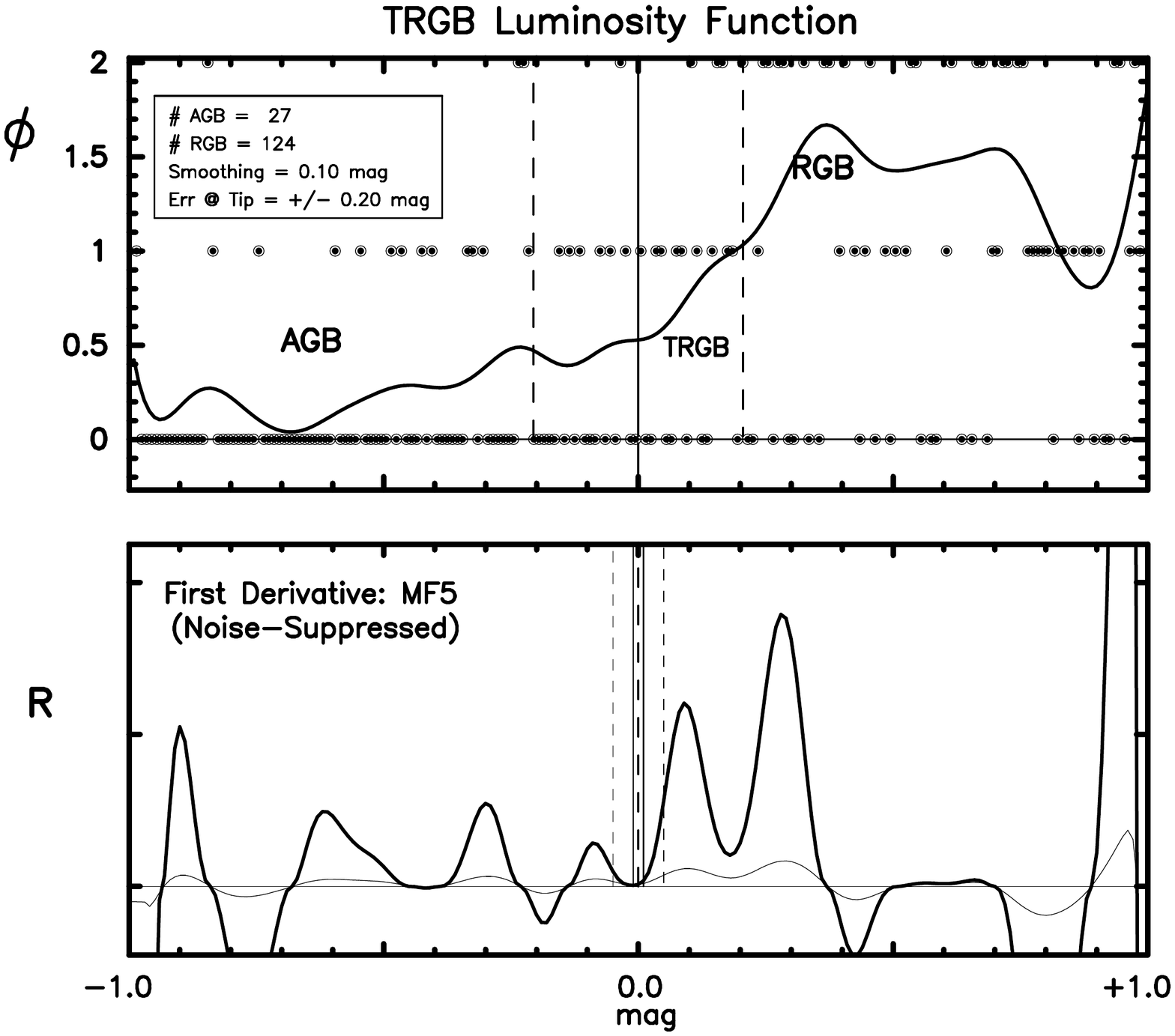} \caption{\small Six sub-panels illustrating the effect of increasing the photometric noise, (from 0.00 to $\pm$0.20~mag) at fixed, but intermediate smoothing ($\pm$0.10 mag) and an impoverished population of RGB stars (120). The lower portions of each of the six sub-panels shows the first-derivative edge-detector output in both its uncorrected (thin black lines) and it noise-weighted (thicker black line) form. See text for a detailed discussion of the trends.} \end{figure*}

\vfill\eject

\section{ Exploring Smoothing Versus Tip Uncertainty}

In Hatt et al. (2017) and in Jang et al. (2018), it has been shown that the output width of the Sobel response function (and many of its variants) is dominated by the width of the smoothing function, and it alone carries little or no quantitative information on the uncertainty in the tip measurement itself. 
This non-response of the width of the Sobel Function output to variations in smoothing 
and population size is shown in the three panels in Figure 14. 
At the bottom of each of the figures we show the Sobel filter response to the smoothed luminosity functions plotted above them. 
The two thin vertical lines centered on the response function are not determined by the the Sobel-filter response itself but rather they mark the {\it input} width ($\pm0.025$~mag) of the GLOESS smoothing function, which has nothing to do with any of the observed properties of the data. 
These lines match the observed width of the Sobel filter response function because the smoothing dominates. 
To see this, in the central panel the GLOESS smoothing width has been doubled to $\pm0.050$~mag, and the response function is seen to have exactly doubled as well; same data, same population size, but twice the width of the response. 
The final (right) panel shows the effect of reducing the luminosity function population by a factor of 10, keeping the GLOESS smoothing the same as the middle panel. 
The width of the response function is unchanged, as shown by the predicted width based on the GLOESS smoothing. For vastly different number of data points, there is no qualitative change in the width of the response function. 
We end with where we started: the width of the Sobel filter response function has little or no discernible information content on the uncertainty of the measured tip magnitude,
and it should not be used indiscriminately in any such applications.

\begin{figure*} \centering
\includegraphics[width=5.4cm,height= 8.1cm,angle=-0]{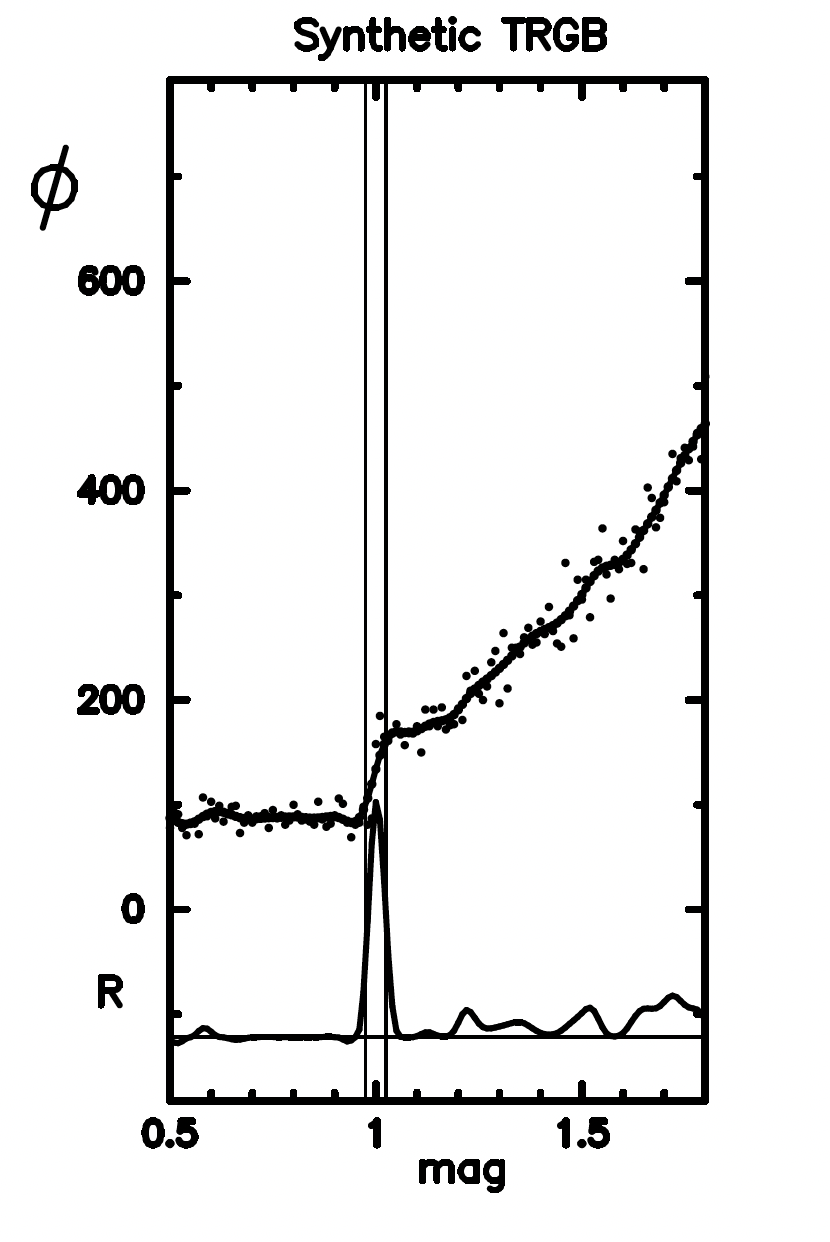}
\includegraphics[width=5.0cm,height= 8.2cm,angle=-0]{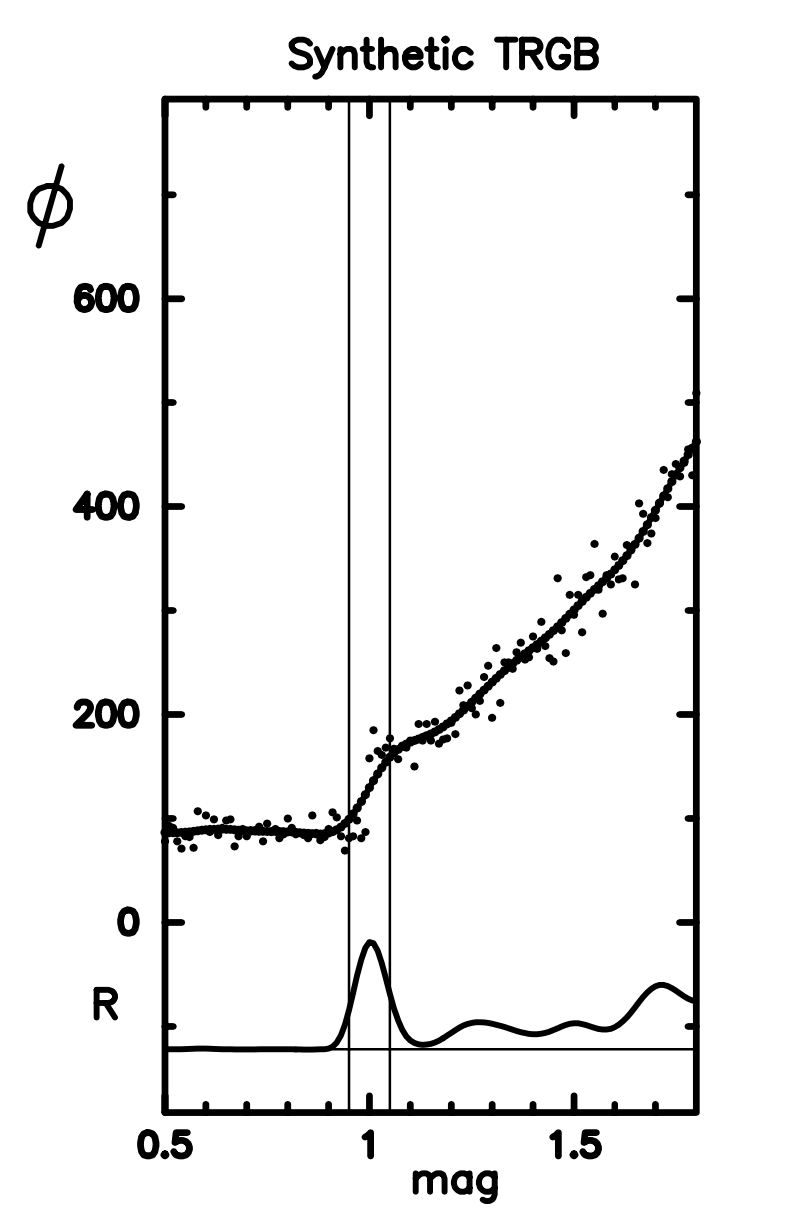}
\includegraphics[width=5.0cm,angle=-0]{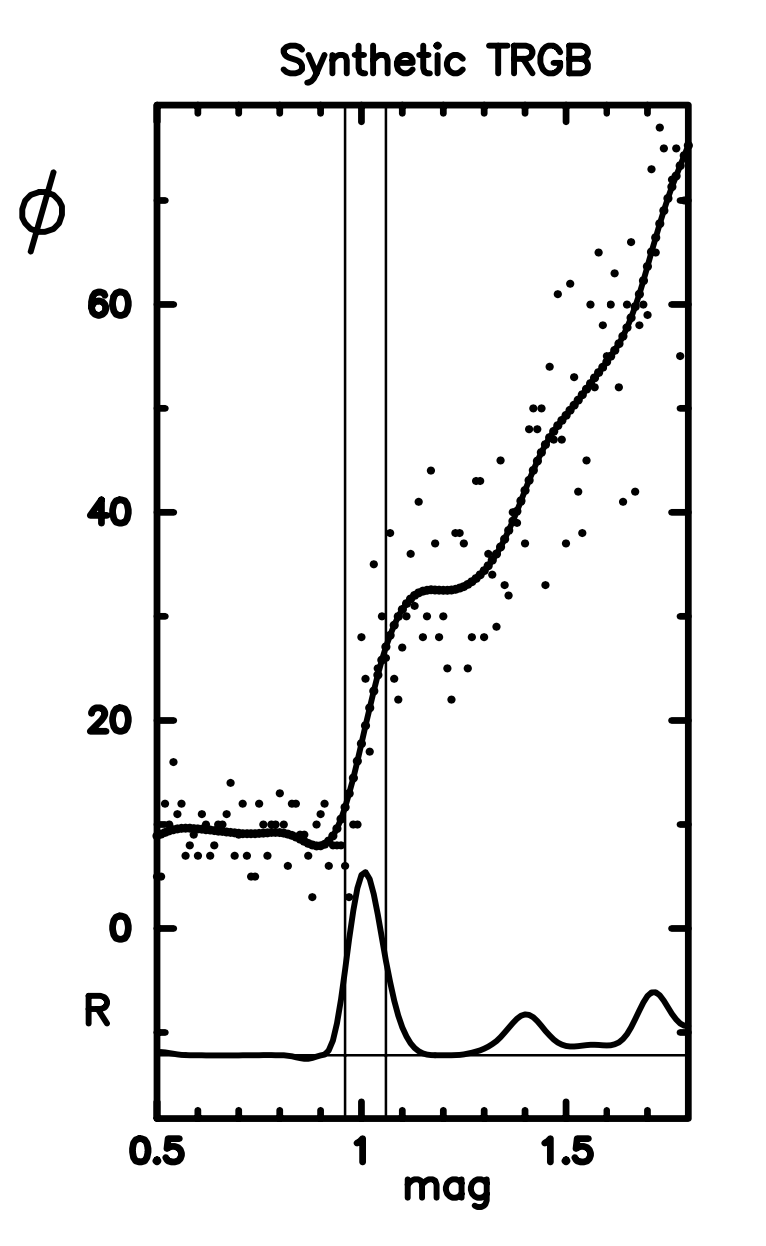} 
\caption{\small
The direct dependence of the width of the Sobel Function response to variations in smoothing (comparing the left and middle panels) and insensitivity to population size (comparing the middle and right panels). 
The left panel shows the Sobel filter tip-detection response to the TRGB luminosity function, which is shown rising diagonally across the top of the plot. 
The two thin vertical lines mark the {\it input} width ($\pm0.025$~mag) of the GLOESS smoothing function. 
They match the observed width of the Sobel filter response function. 
In the central panel the GLOESS smoothing width has been doubled to $\pm0.050$~mag, and the response function is seen to have doubled as well. 
The final (right) panel shows the effect of reducing the luminosity function population by a factor of 10, keeping the GLOESS smoothing the same as the middle panel. 
The width of the response function is unchanged, as shown by the predicted width based on the GLOESS smoothing. }
\end{figure*}

\vfill\eject
\begin{figure*} \centering 
\includegraphics[width=8.5cm, angle=-0]{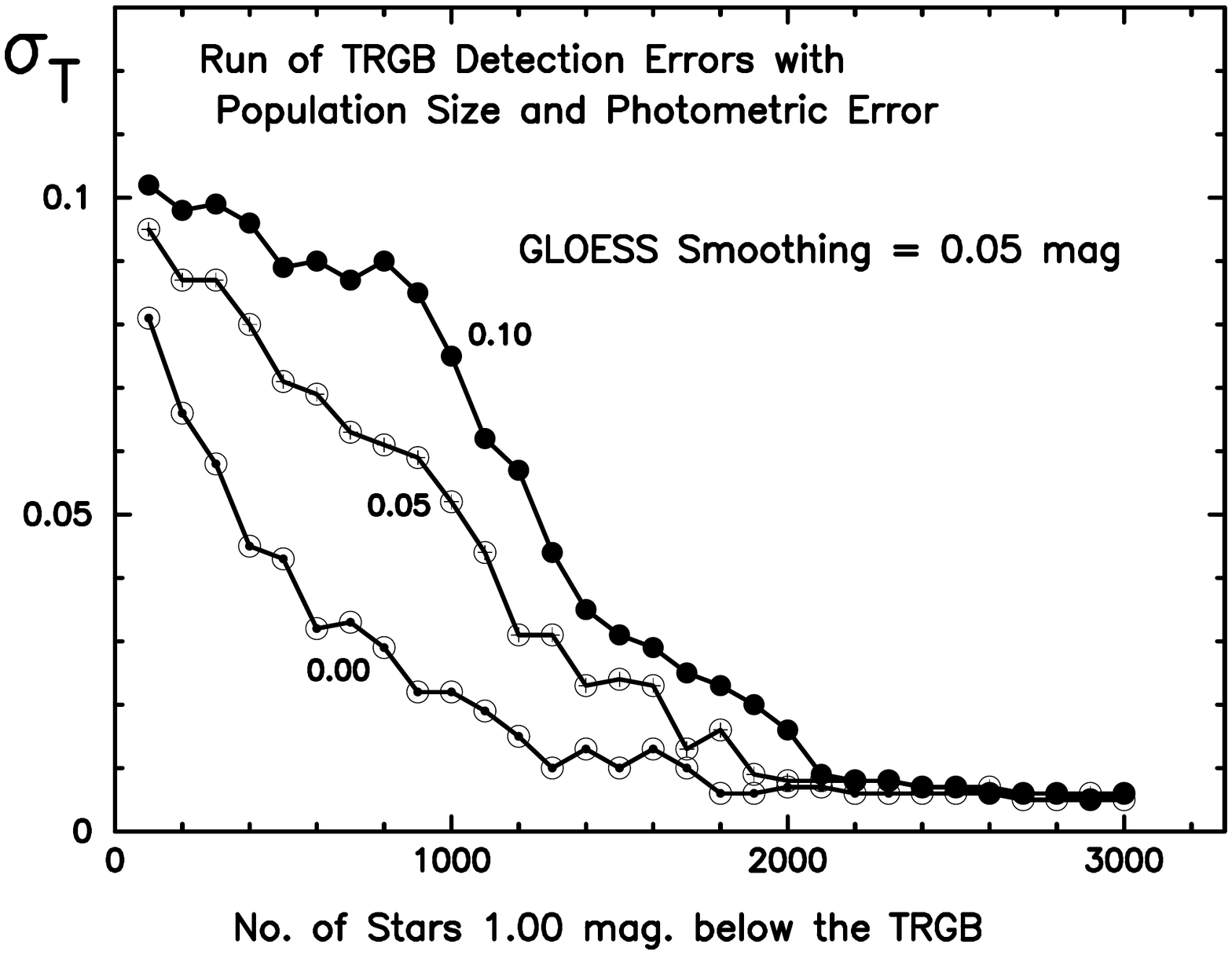} \includegraphics[width=8.5cm, angle=-0]{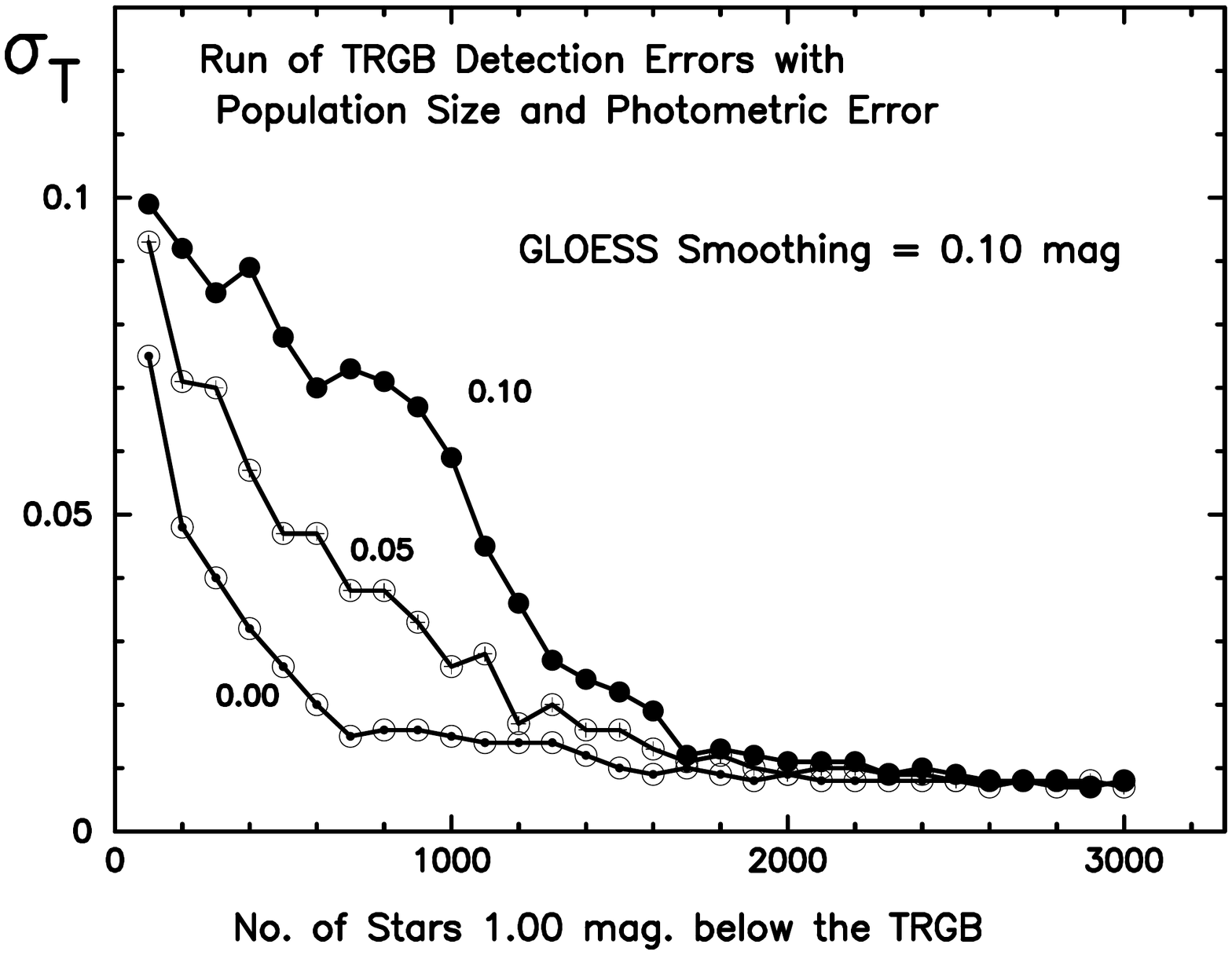}
\includegraphics[width=8.5cm, angle=-0]{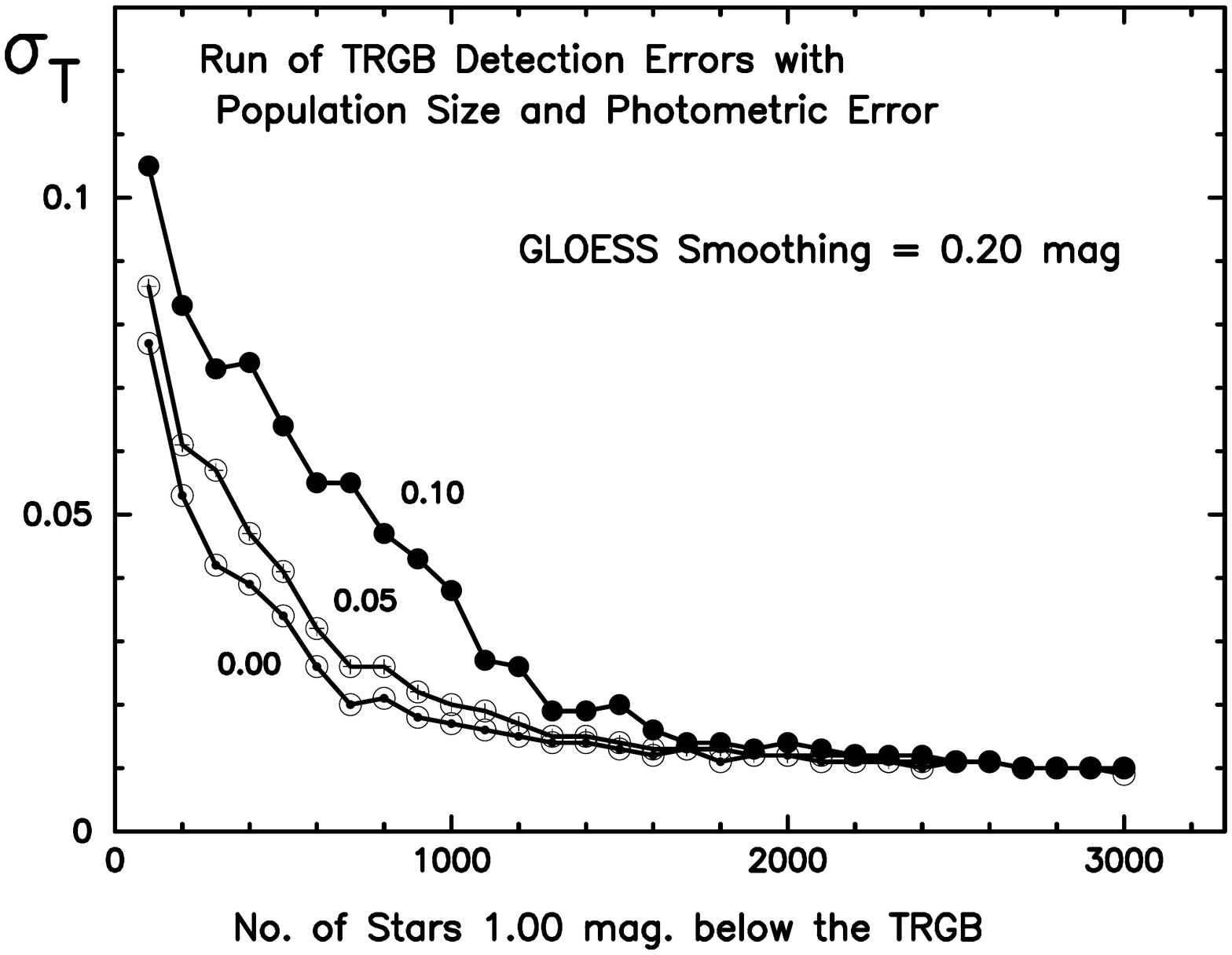}\caption{\small Variations of the run of edge-detection errors with population size and with photometric errors, as a function of the GLOESS smoothing changing from here through Figure 17. 
The vertical axis gives the  standard deviation of the measured TRGB magnitudes derived from 200 independent realizations of the luminosity function for each plotted point. 
The three line-linked symbols each track the decrease in the TRGB uncertainty as a function of increased sample size for a given mean error in the TRGB photometry. 
The horizontal axis is gives the population size normalized by the number of RGB stars in the one-magnitude interval immediately below the TRGB. 
Filled circles have photometric errors of $\pm0.10$~mag. Circled plus signs have $\pm0.05$~mag errors.  
And circled dots represent error-free photometry.  
The amount of GLOESS smoothing increases by factors of two from this to the next two figures.
Here we show the results for a  $\pm0.05$, $\pm0.10$ and $\pm 0.20$~mag smoothing. Ripples in these trends are not significant but due to small number statistics in the individual simulations.
}
\end{figure*}
However, two other observables, population size and photometric errors, do have an expected and predictable impact on the uncertainty of the TRGB measured magnitude. 
This is illustrated in the three panels in Figure 15 where the run of (statistical) edge-detection errors with population size and with photometric errors
are shown. 
In each of the panels the vertical axis gives our desired output: the standard deviation of the measured TRGB magnitudes. 
The plotted points are  derived from 200 independent realizations of the luminosity function, each smoothed and individually scanned by our edge-detection filter. 
The three line-linked symbols each track the increased precision in the TRGB measurement as a function of increasing sample size for a given mean error in the TRGB photometry. 
The horizontal axis tracks the population size of the luminosity function being scanned, defined here by the number of RGB stars in the one-magnitude interval immediately below the TRGB. 
Filled circles have photometric errors of $\pm0.10$~mag.
Circled plus signs have $\pm0.05$~mag errors.  
And circled dots represent error-free photometry. 
The amount of GLOESS smoothing increases by factors of two (across the three panels), ranging from $\pm0.05$~mag to $\pm0.20$~mag top to bottom.

Modulo the Poisson noise inherent in these finite simulations, the trends are clear: All of the determinations of the uncertainty in the TRGB measured magnitude decrease monotonically with increased sample size.
And all trend lines decrease more slowly as the photometric errors increase. 
Moving from figure to figure the effects of the GLOESS smoothing are seen to be present but sub-dominant. 
In any case, for any given observation the three parameters controlling the calculated uncertainty on the TRGB magnitude (the population size, the photometric errors and the adopted smoothing) are known and can be input into a numerical simulation tailor-made for that study giving the uncertainty for that particular tip measurement. Failing that, these plots can be interpolated for a first-order estimate of the uncertainty to be associated with the choice of smoothing, measured population size and known photometric errors (labelling each of the three realizations in each of the three smoothing plots).

One, final caveat. The simulations in this section were undertaken for situations wherein crowding is not a major source of error at the tip. However, if observations are made in regions of high surface brightness (and commensurately higher levels of crowding) then the simulations presented in the next section should take precedence. 
\medskip

\section{Crowding Simulations}
Our final set of simulations targets the important question concerning the effect of source crowding on the RGB luminosity function, especially at and around the discontinuity defining the tip.

In order to isolate and unambiguously determine the effects of crowding on the luminosity function above the TRGB we have set the AGB population to zero, which would ordinarily be found in the one-magnitude interval above the tip. We did this intentionally so as to show what the crowded population looks like unhindered by superimposing it on a true AGB population; i.e., to see the signature of crowding alone.
In each of Figures 16 through 20 we show (in the left-hand panel) the input luminosity function as a blue-shadowed white line rising in number abruptly at $M_I =$ -4.00~mag and thereafter exponentially increasing to form the Red Giant Branch luminosity function. 
The actual numbers of stars defining the simulated luminosity function are shown by the black-shadowed red line, sampled at 0.01 mag intervals. In the right panel we show the output of the edge detector in yellow. Running between the panels is a thin black (horizontal) line marking the input TRGB discontinuity. 
In the right panel the position of the discontinuity as measured by the edge-detector is shown as a dashed black line. 
Finally, in the left-hand panels in the otherwise empty space above the TRGB we have inset the simulated CCD image used in the calculation (see the captions for the key to the various symbols marking stars in the image). 

The aforementioned ``simulated CCD image" was instrumental in undertaking the crowding simulation. The image consists of a square array of 1000 by 1000 elements. As stars populated the (smooth blue line) input luminosity function seen in the left panel, they were randomly assigned a cell in the image, their magnitudes were converted to fluxes and they were added to that cell. If the cell was already occupied the flux was augmented and the cell was considered to be crowded. The process was continued until all stars were assigned to cells. The summed fluxes were then converted back to magnitudes, and these magnitudes were rebinned (at 0.01 mag resolution) thereby creating the ``crowded luminosity function" shown by the jagged red line in the luminosity function plot.

{\bf Summary 11 --} It is apparent from these simulations that self-crowding blurs the tip discontinuity. In addition, high levels of crowding can cause a bias in the measured tip magnitude. In general, it is preferred to make these measurements in low-density halo fields to avoid crowding issues.

\section{Eliminating AGB stars} 

Although AGB stars above the tip are naturally found in most halo fields they have a fairly flat luminosity function that only serves to decrease the contrast of the tip discontinuity by placing the RGB luminosity function on a slightly elevated background; but that loss of contrast does not impose a bias, only a decrease in precision. If one considers the TRGB discontinuity as a step function, then it is easy to visualize that, for reasonable high levels of the contrast ratio R $> 3 $, say, the measurement of the onset (when approached from brighter luminosities) is not influenced by level of the baseline/background upon which it is being measured. The AGB contribution, close to the tip, can be thought of as a relatively constant pedestal upon which the TRGB discontinuity is detected.

The presence of true AGB stars can be eliminated by time-domain observations of the TRGB fields. 
The Gaia Mission has shown that virtually all true AGB stars in this luminosity range are variable (see Figure 3 in Eyer et al. 2019 for the types of variables in the Gaia CMD and especially Figure 8 which gives the fraction of variables sitting at unity (red points seen directly above the downward slanting TRGB), while no RGB stars, at or below the tip, have been found to show variability greater than 0.04~mag full amplitude (black points in the aforementioned Figure 8), they would then at most contribute a $\pm$0.01~mag blurring of the tip.\footnote{The referee has correctly pointed out that Soszynski et al. (2004) have a paper entitled  ``Small Amplitude Variable Red Giants in the Magellanic Clouds". In that paper the largest amplitude variables are above the TRGB and consist of AGB stars alone (LPVs, Miras and semi-regular variables); below the tip the amplitudes systematically decrease with period (see examples in their Figure 4) and the authors believe that these fainter variables are a mix of AGB and RGB stars. They name these stars OSARGs ({\it OGLE Small Amplitude Red Giants}). In their Figure 2 the TRGB can be found at $W_I \sim $11.5~mag, and at $1.5 < (V-I) < 2.4$ in their Figure 3. In their Figure 2, cutting the lower-left panel in color restricts the RGB sub-tip population to stars that have log $P_1 < $ -1.8. Applying that cut to the Period-Amplitude plot in the panel directly above the Period-Color plot reveals that the OSARGs
below the tip have peak-to-peak amplitudes starting at 0.04 and dropping to 0.01~mag. Converting amplitudes to equivalent sigmas then suggests that these very-small-amplitude variables contribute no more scatter than $\pm$0.010 to 0.003~mag. The claim in the subsequent literature (Anderson et al. 2023) that ``Every star at the TRGB is variable" is true, at the {\it millimag} level, but it is not of concern in the context of the TRGB extragalactic distance scale.}  

Identifying and removing the variable AGB population will aid in de-blurring and  ``decontaminating'' the tip from the bright side, and increase the contrast of the discontinuity defining the TRGB as approached from either side.

\begin{figure*} \centering 
\includegraphics[width=20.0cm,angle=-0.0]{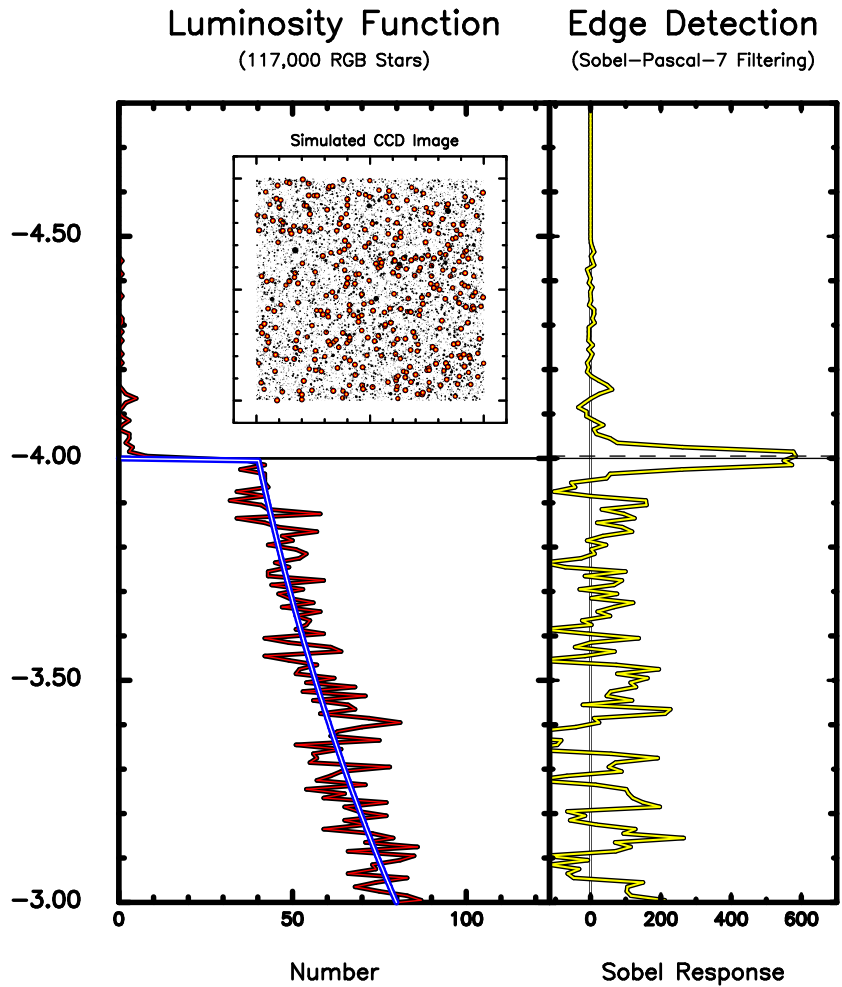} 
\caption{\small A simulation of the RGB luminosity function consisting of 117,000 stars populating the magnitude range from $M_I = $ -2.0 to a sharp cut-off at $M_I = $ -4.0~mag defining the TRGB. 
The individual stars are assumed to be error-free in their individual magnitudes. 
Noise in the luminosity function is entirely due to Poisson noise in the counts in the individual bins. 
Output from the Sobel-Pascal-7 edge detector is shown in the right panel. The black horizontal line, spanning the two panels, is the TRGB magnitude set to $M_I = $ -4.00 mag. 
The dashed black horizontal line in the right panel only is the mean value of the response function. 
The inset ``Simulated CCD Image'' in the left panel shows the positions of all stars used in the simulation. 
Large yellow dots, circled in red and black are all RGB stars from $M_I =$ -4.00 to -3.90~mag. 
Crowded stars brighter than the TRGB are shown as larger black filled circles. 
This plot is given to provide a visual impression of the self-crowding of stars near the TRGB that result in the small population of stars above the tip in this particular luminosity function.}

\end{figure*}
\vfill\eject

\begin{figure*} \centering 
\includegraphics[width=20.0cm,angle=-0.]{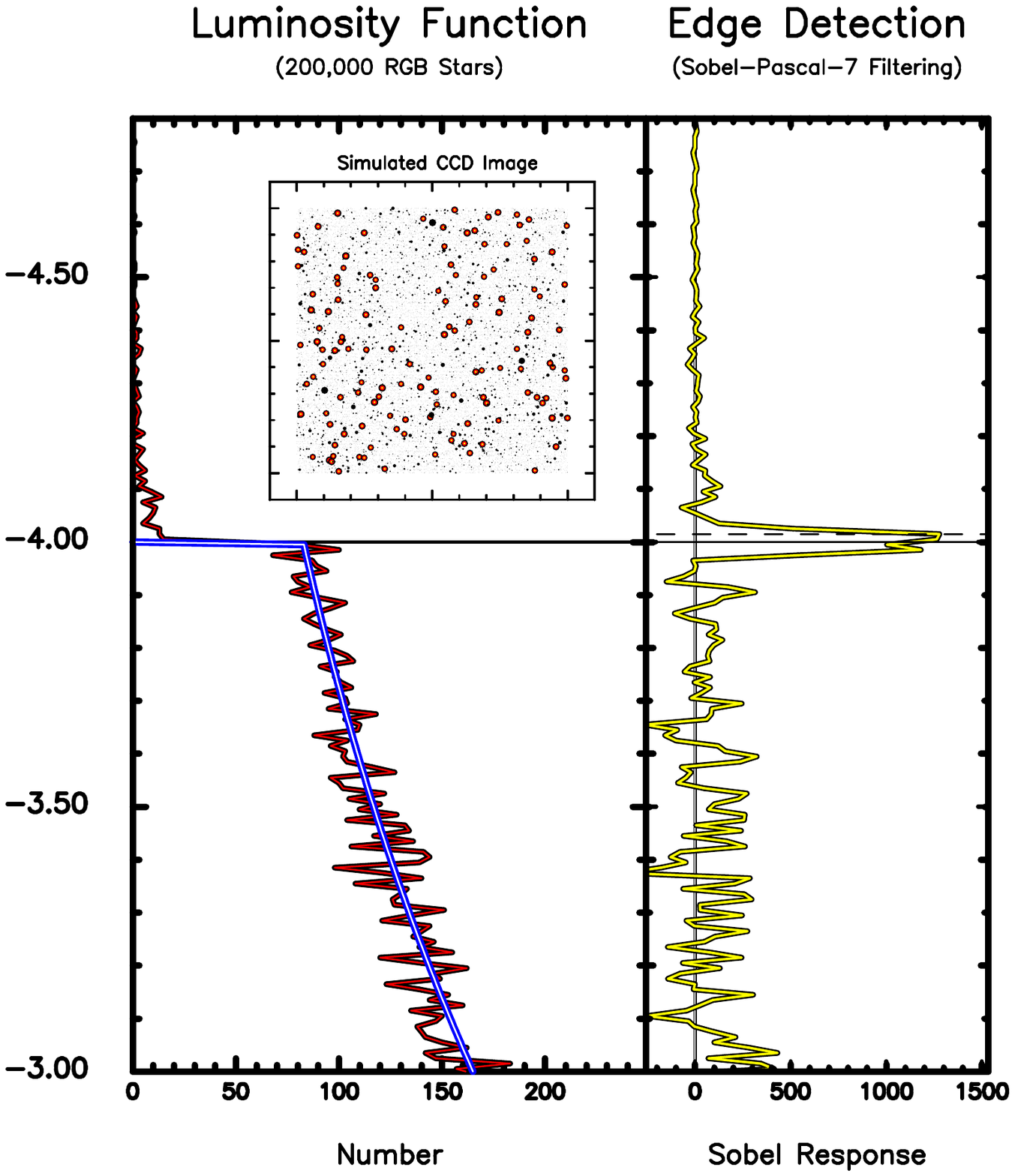} 
\caption{\small Same as Figure 16 except this simulation involve slightly over 200,000 RGB stars in the frame. Crowding is now quite apparent in the inset CCD Image, and has resulted in an appreciable population of crowded stars falling in number from the tip to 0.75~mag brighter where the self-crowding of two stars at the tip would appear as one unresolved source. For clarity, however only one star in 10 is plotted.
See text for additional details. The offset between the blue (input) line and the (red) output line in this and in the following three figures is a direct consequence of the crowding systematically producing more stars at a given magnitude by merging two or more fainter stars.
}
\end{figure*}

\begin{figure*} \centering 
\includegraphics[width=20.0cm,angle=-0.]{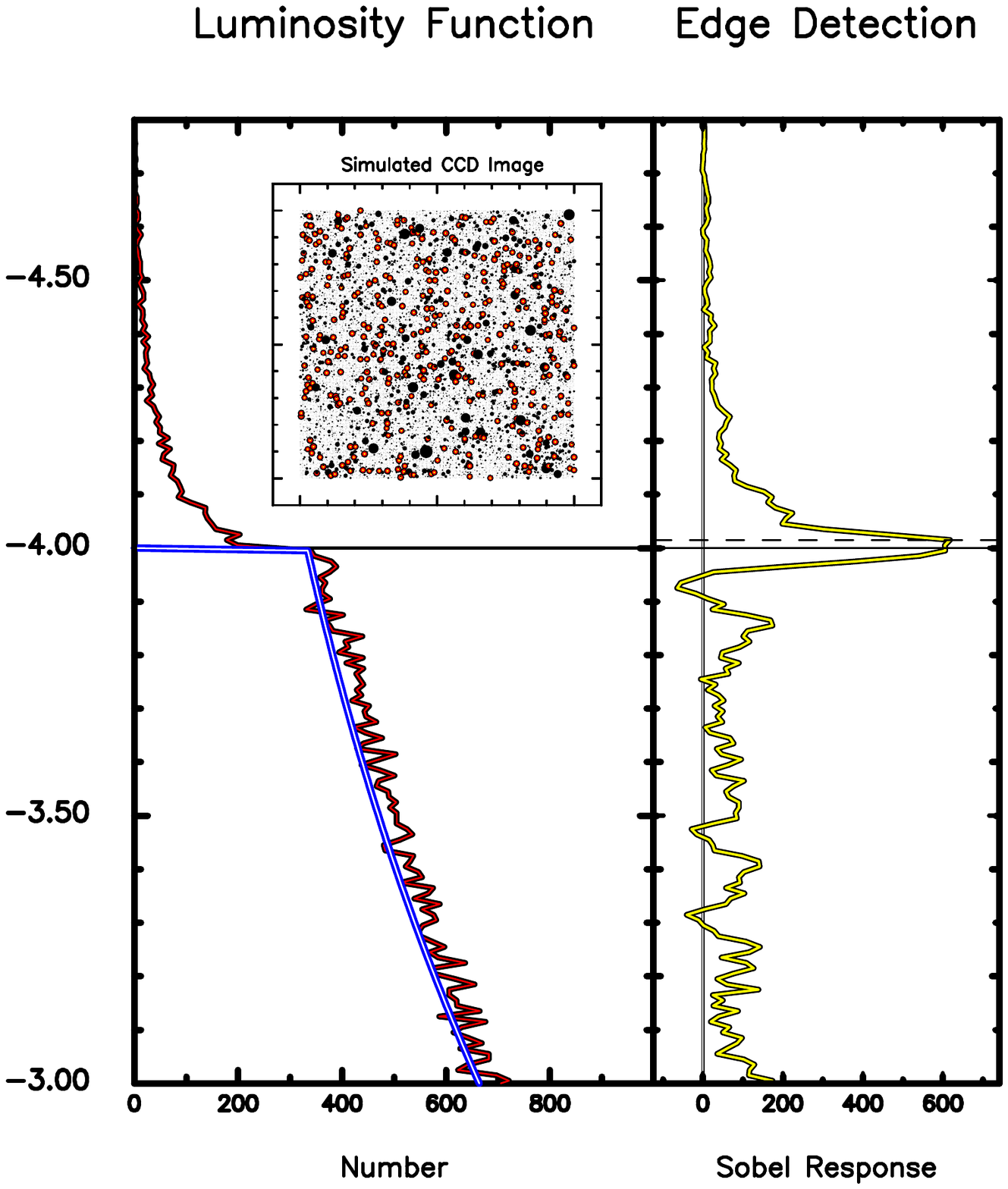} 
\caption{\small Same as Figure 16 except this simulation involves slightly over 700,000 RGB stars in the frame. Crowding is now quite apparent in the inset CCD Image, and has resulted in an appreciable population of crowded stars falling in number from the tip to 0.75~mag brighter where the self-crowding of two stars at the tip would appear as one unresolved source.
See text for additional details.
}
\end{figure*}

\begin{figure*} \centering 
\includegraphics[width=20.0cm,angle=-0.]{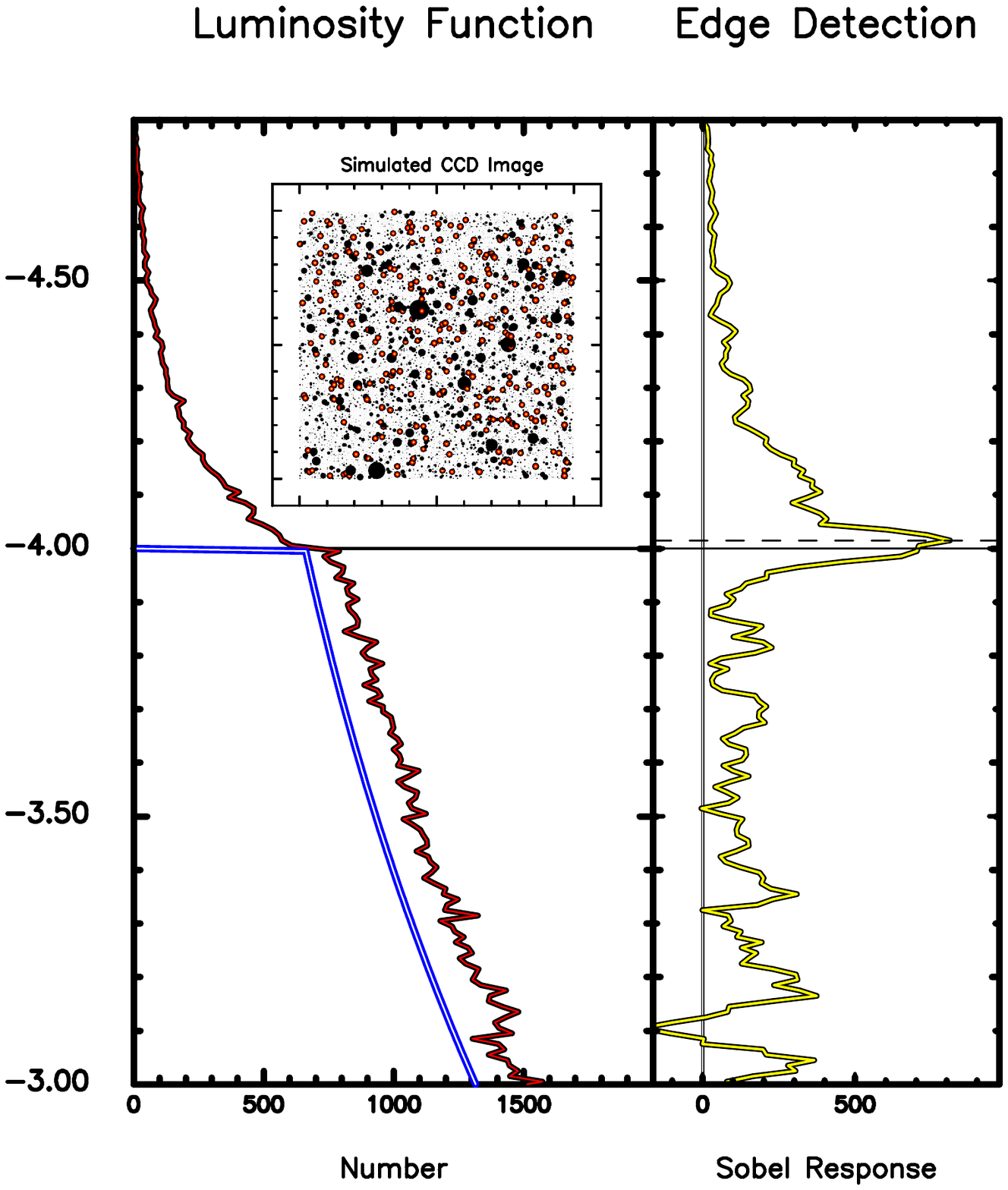} 
\caption{\small Same as Figure 16 except this simulation involves slightly over 1,370,000 RGB stars in the frame. Crowding is now quite apparent in the inset CCD Image, and has resulted in an appreciable population of crowded stars falling in number from the tip to 0.75~mag brighter where the self-crowding of two stars at the tip would appear as one unresolved source. For clarity only one star in 10 is plotted.
See text for additional details.
}
\end{figure*}

\begin{figure*} \centering 
\includegraphics[width=20.0cm,angle=-0.]{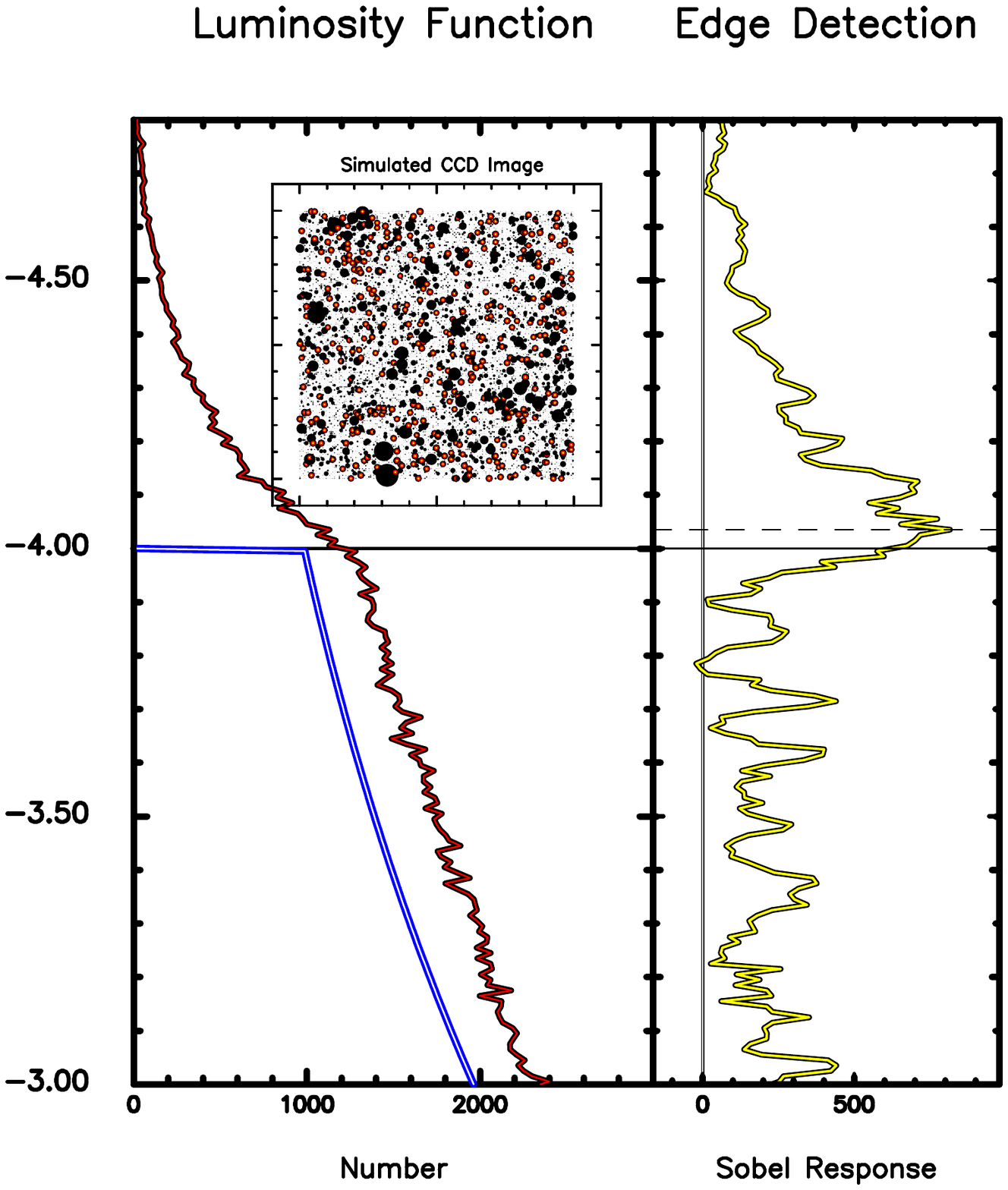} 
\caption{\small Same as Figure 16 except this simulation involves slightly over 2,000,000 RGB stars in the frame. Crowding is now quite apparent in the inset CCD Image, and has resulted in an appreciable population of crowded stars falling in number from the tip to 0.75~mag brighter where the self-crowding of two stars at the tip would appear as one unresolved source.
See text for additional details. }
\end{figure*}
\section{ Unaddressed Issues: Star Formation History of the Halo on the Position and Color Dependence of the TRGB}

In a paper by Brown et al. (2003) the case has been made for a significant population of intermediate-aged (6-8 Gyr), high-metallicity ([Fe/H] > -0.5) RGB stars being present in the halo of M31. These stars can exceed the luminosity of the old, metal-poor (standard) TRGB population, but they also ascend at a color that is far to the red of the most metal-rich, old TRGB stars (see their Figure 1f) The application of both a blue and (especially in this case) a red cut-off to the RGB stars, being used to detect the tip effectively deals with these stars. In any case, in a forthcoming paper (Freedman, Madore \& Owens, 2023) it is shown that a comparison of TRGB distances with Cepheid distances to the same galaxies give a combined scatter of only +/- 0.066 mag which {\it must bracket the total impact of all remaining random errors}, including scatter that might be introduced by star formation history difference between these halos.

Additionally, concern about the presence of young populations of red giant stars in the halos of galaxies have been raised in the literature, most recently and extensively by McQuinn et al. (2019). Their message is mostly cautionary. We agree with that stance, but offer up a number of lines of evidence arguing for optimism that the effect of younger populations (star formation history of the halo) if present, is a minor contributor to the measured scatter in the TRGB, especially in the I band.  

Some of the concern about young populations interfering with the detection and measurement of the TRGB are driven by ill-fated applications of the TRGB method too close to the disk. The following quote from the GHOSTS Team (Monachesi et al. 2014) summarizes the situation very well. ``The CMDs are mostly populated by old RGB stars (older than 1 Gyr). There are however younger populations such as blue, extended main-sequence (MS) stars ($<$ 500 Myr) or massive stars burning helium in their core (25–600 Myr old red and blue loop sequence stars). These appear primarily in the fields closer than R = 15~kpc to each galaxy, and especially along the major axis, which are dominated by disc stars.’’ 

Theory also suggests that a younger population would have little impact even if it were mixed in. From model predictions of McQuinn et al. (2019), it is expected that the F814W absolute luminosity of the TRGB should have a small dependence on stellar age of roughly 0.02 mag across an age range of 5 Gyr and 0.04 mag across 0.5 dex. Indeed, inspection of lower panel in Figure 2 of McQuinn et al. shows that the effect of age spread is indeed small, but it is also degenerate in its correlation with metallicity. Furthermore, this point has also been recently addressed in the single-authored paper by Hoyt (2023) where he states:
“A long-standing question of the TRGB concerns the extent to which age can shift the observed colors and magnitudes of TRGB stars, potentially breaking the assumption of universality in any single proposed calibration (Salaris \& Girardi 2005). Encouragingly, in this section it was shown that the Jang \& Lee (2017) QT color dependence – based on observations in the stellar halos of $L^{*}$ galaxies – describes very well the TRGB magnitude-color relation in the MCs (this study), Local Group dwarfs, and M33 (Rizzi et al. 2007). {\bf}{This consistency indicates that for RGB stars found in these environments either: the age distributions are identical, or that age-dependent variations in the I-TRGB magnitude are minimal. In either case, the I-TRGB appears well-behaved and without a measurable bias across these host environments}.”\footnote{In terms of the lack of bias due to the AGB stars
observed in these simulations we note that Hatt et al. (2017) had already remarked on this (non-effect) in their own independent simulations, stating ``... we find that the AGB component simulated here has no substantial effect on the measured TRGB magnitude. The ratio of TRGB to AGB stars near the tip is about 4:1, which might, conceivably, cause a TRGB measurement to be systematically brighter. Nonetheless, we find that the signal-to-noise of the TRGB still outweighs the noise component due to AGB stars and there are minimal systematic effects.''}
Bottom line: If you detect blue main sequence stars or red supergiant populations in your fields then you are clearly too close to the disk and any attempt to determine a TRGB distance to any such a line of sight through the galaxy is subject to systematic effects. Stay as far as possible out into the pure halo, preferably along an extension of the galaxy’s minor axis.

\section{Summary Conclusions and General Advice} 

TRGB distances have become one of the most precise and accurate means of measuring the distances to galaxies in the nearby universe (see, for instance, Dalcanton et al. 2009;  Karachentev et al. Freedman et al. 2019; Anand et al. 2022; and the cumulative holdings of TRGB distances at NED [ https://ned.ipac.caltech.edu/ ] and EDD [ https://edd.ifa.hawaii.edu/ ]) . We stress that for accuracy and precision, the method should be applied in the outer halos of galaxies, where the effects of extinction and self-crowding of TRGB stars are minimal. With these simulations, we have shown the trade-offs between a number of factors, including photometric precision, numbers of stars defining the RGB luminosity function and the effects of crowding.  These simulations can be used as a guide to optimize the choice of halo fields for accurate TRGB measurements.

The above simulations presuppose that observations of the TRGB for the purpose of extragalactic distance determinations are being made in the halos of galaxies. Thus, they are well away from disk contamination consisting of dust, gas, and stars of mixed ages, colors and spatial densities. This contamination can only degrade the TRGB detection and act (in the case of dust extinction) in biasing the apparent magnitude of the tip to fainter magnitudes. 

In a review of TRGB modeling McQuinn et al. (2019) state {\it ``Given the building histories of halos, it is reasonable to expect variations in ages and metallicities.” }
They then go on to say {\it “Assuming stellar halos are consistently metal-poor with little variation does not appear to be valid; } and add {\it ``To date, few constraints have been placed on the stellar ages.” } However, in their conclusions they more optimistically state {\it ``In the I-band equivalent HST F814W filter and JWST F090W filter, the TRGB is {\bf}{remarkably constant across all ages and metallicities probed}}''. We feel that it would be remarkable that any halo would not have a first-generation of low-metallicity red giant stars. These stars will be the brightest TRGB stars in the I band and will trigger the edge detector before any other higher-metallicity (potentially fainter) population would enter the mix. That is to say, if there is a population of fainter, high-metallicity stars in any given halo, along with the generations of lower-metallicity stars that gave rise to them, in the marginalization process undertaken before measuring the tip, the high-metallicity stars will be systematically below the first triggering of the edge detector and will simply augment the RGB luminosity function without a signature of their own (see Figure 12 in Hoyt 2023). Similar arguments can be made for the color-rectified TRGB at longer wavelengths if the curvature downward to fainter magnitudes persists at higher metallicities (which is strongly correlated with color). Finally, it needs to be emphasized that, as Figures 2 and 3 in McQuinn et al. vividly demonstrate, {\bf}{ the color and luminosity dependence of the TRGB on age is extremely small (in the I band)}, and it is largely degenerate with the color dependence of the TRGB luminosity on metallicity. The situation may be less straightforward at longer wavelengths. JWST observations will be extremely important here.

But what do the observations of the TRGB have to say on this matter? Figure 5 of Freedman et al. (2019) gives a comparison of TRGB distances with Cepheid distances to a variety of galaxies of different star formation histories (ages), different mean metallicities, different distances and different amounts of reddening. For the entire ensemble (near and far) the combined scatter is only +/- 0.11~mag, which, if equally shared between the two methods would imply that they are each good to 4\% in distance. However, if you just look at the closest sample their sigma drops to +/-0.05~mag, which means that the two methods are each good to 2\% in distance.
The take-away message is that inside of that 2\% all of the unresolved or unknown systematics are themselves contained at that same level, be it metallicity, age or biased fitting methods.  On that note we are optimistic.

Having population sizes that are sufficient to fill the RGB luminosity up to and including the tip are crucial to the extraction of an unbiased TRGB magnitude. For example, for RGB populations of less than 1000 red giant branch stars in the first magnitude interval below the true TRGB, false detections of the tip can be expected at the $\pm0.1$~mag level when the photometric precision of the data is worse than $\pm$0.05 mag  (see the lower six sub-panels in Figures 8, 9, 12 \&  13).

Degradation of the tip due to increased photometric errors can be compensated for by having increased population sizes (compare Figures 2 \& 6 with 3 \& 7).

We have demonstrated that it is best to use the least amount of smoothing possible, commensurate with the photometric errors and population sizes. When numerous (comparably significant) peaks are found with a low degree of smoothing, no amount of additional smoothing will reveal the true peak, but rather the resulting ``detection" will be a weighted average of the surrounding peaks which may (with enough smoothing) appear to a be a single (broad) peak, it will but probably be biased: consider smoothing the last three panels in Figure 6, as then seen in Figure 10. Our recommendation is that future investigators always try a number of smoothing kernels bracketing their preferred solution so as to reveal the presence (or absence) of substructure that a high degree of smoothing would otherwise gloss over. See Figure 5 and 8 of Beaton et al. (2019) for a recent implementation of this iterative smoothing analysis.

Similarly, self-crowding of RGB stars near the tip results in a population of false AGB stars which also decrease the contrast of the TRGB discontinuity, and eventually bias the tip detection (to bright magnitudes). 
However, as the simulations in Figures 16 through 20 clearly demonstrate, this effect can be predicted by the source density of RGB stars in any given field. 
Attempts to increase population statistics of the RGB by moving into higher surface brightness regions of the inner halo should be tempered because of this self-crowding effect, in addition to line-of-sight extinction issues within the disk (that are not included in this simulation). 

In the end, the characteristic (exponentially increasing) luminosity function of the {\it faux} AGB stars will betray their presence, and signal impending bias. This could, in principle be modelled away, but might best be avoided by not observing in high-surface-brightness regions to begin with. However, we do caution against smoothing data that are in the self-crowding regime. Smoothing Figures 19 or 20 would only lead to (unnecessarily) biasing the edge response to brighter magnitudes. Irreparable damage to the tip detection is seen in the highest degree of self crowding simulated in Figure 20. It too could be modelled; but, the best solution would be to re-observe the galaxy in a region of significantly lower surface density of stars.

We have simulated the CCHP adopted smoothing and filtering of the AGB/RGB luminosity function that is being used to measure the magnitude of the TRGB.   
We find that the width of the Sobel filter response function is totally dominated by the preceding GLOESS smoothing of the luminosity function. 
We have, however, calibrated the run of the uncertainty in the measured value of the TRGB as a function of (a) population size and (b) uncertainties in the stellar photometry at the tip.

For readers wishing to get a sense of the uncertainties on their distances in advance of making the observations, we suggest consulting the three panels in Figure 15. They will also be useful in validating the observed error on the tip after the population size and error at the tip have been empirically defined.

Finally, it should be noted that all of the recommendations being derived from these simulations (implicitly for the I band, where the run of TRGB magnitude is flat with color) apply equally well to those other wavelengths once the color-magnitude diagrams have been rectified.  By rotating the data using predetermined slopes of the TRGB, the respective tips will also show no trend of magnitude with color. The rectified magnitudes can then be marginalized, and an edge detection applied to the resulting color-corrected luminosity functions.
In support of this, recent articles, (purely observational and mixed with modeling), both Wu et al. (2014), their Figure 5, and Durbin et al. (2020), their Figure 3, show that in the near infrared F110W (J band) there is a clearly linear trend of TRGB luminosities with color at least over the bluest colors ranging from 0.70 $<$ (F110W – F160W) $<$ 0.95~mag. 

There are many additional sources of statistical and systematic error that these simulations have not explicitly included. These uncertainties could stem from issues in assumed PSF libraries, CTE corrections, etc.. While improving with time, some fraction of these issues still persist. And on top of this there are additional systematics when dealing with PSF photometry, such as errors in aperture corrections, or mismatching PSFs (due to telescope focus shifts, ``breathing", etc.). The list goes on, and in light of that our error budget should be viewed as a lower limit on what is occurring in the real world.
Such is the price paid  undertaking any simulation.

%As emphasized by the referee, there are many additional sources of statistical and systematic error that these simulations have not explicitly included. The referee said it well: ``These uncertainties likely stem from issues in assumed PSF libraries, CTE corrections, etc.. While improving with time, some fraction of these issues still persist. And on top of this there are additional systematics when dealing with PSF photometry, such as errors in aperture corrections, or mismatching PSFs (due to telescope focus shifts, ``breathing")’’. Thus, it is clear that our error budget should be viewed as a lower limit on what is occurring in the real world. 

\section{Epilogue}

Imagine we have two people approaching each other in the fog. They each know that there is a cliff ahead of them but it is too dark to see it. One is walking from the sea, approaching the cliff from below. The other is high above the water on a gently sloping hill approaching the cliff from above. The first may be noticing that he is walking uphill away from the water, navigating undulating sand dunes, etc. None of this topology of the local terrain can alert him to the discontinuity that he is walking towards … until he slams into it. The second adventurer notices the cracks and crevasses that he has to walk over or around, but again nothing at his feet alerts him to his pending doom … until he walks off the cliff. The AGB is the sandy seaside below. The RGB is the grassy meadow above. Neither of those features can predict what lays ahead. 

\vfill\eject

\section{References}

\medskip
\noindent
Anand, G.S., Tully, R.B., Rizzi, L., et al. 2022, ApJ, 932, 15

\noindent
Anderson, R.I., Koblischke, N.W. \& Eyer, L. 2023,  arXiv:2303.04790

\noindent
Baade, W., 1944, ApJ, 100. 137

\noindent
Beaton, R.L., Seibert, M., Hatt, D., et al. 2019, ApJ, 885, 141  

\noindent
Berendzen, R., Hart, R. \& Seeley, D. 1976, {\it Man Discovers the Galaxies}, 

Science History Publications, New York, pg. 43 

\noindent
Brown, T.M., Ferguson, H.C., Smith, E., et al. 2003, ApJL, 592, 17 

\noindent
Cassisi, S. \& Salaris, M. 2013, {\it Old Stellar Populations}, Wiley-VCH  

\noindent 
Cioni, M.-R.L., van der Marel, R.P., Loup, C., et al.
2000, A\&A, 359, 601 

\noindent
Conn, A.R., Lewis, G.F., Ibata, R.A., et al. 2011, ApJ, 740, 69

\noindent
Da Costa, G.S. \& Armandroff, T.E. 1990, AJ, 100, 162

\noindent
Dalcanton, J.J., Williams, B.F., Seth, A.C., et al. 2009, ApJS, 183, 67

\noindent
Durbin, M.J., Beaton, R.L., Dalcanton, J.J. et al. 2020, ApJ, 898, 57

\noindent
Eyer, L., Rimoldini, L., Audard, M., et al. 2019, A\&A,623, 110 

\noindent
Freedman, W.L. 1988a, AJ, 96, 1248  

\noindent
Freedman, W.L. 1988b, ApJ, 326, 691

\noindent
Freedman, W.L., Madore, B.F., Hatt, D., et al. 2019, ApJ, 882, 34

\noindent
Freedman, W.L., 2021, ApJ, 919, 16

\noindent
Frogel, J., Cohen, J. \& Persson, S.E. 1983, ApJS, 53, 713

\noindent
Gorski, M., Pietrzynski, G., Gieren, W., et al. 2018, AJ, 156, 278

\noindent
Hatt, D., Beaton, R.l., Freedman, W.L., et al. 2017, ApJ, 845, 146

\noindent  
Hoyt, T., Freedman, W.L., Madore, B.F., et al.  2018, ApJ, 858, 150 

\noindent  
Hoyt, T. 2023, Nature (accepted), arXiv:2106.13337

\noindent
Jahne, B. 1991, {\it Digital Image Processing: Concepts, Algorithms and Scientific Applications}, 

Springer-Verlag,  New York

\noindent
Jang, I.-S. \& Lee, M.-G. 2017, ApJ, 835, 28

\noindent
Jang, I.-S., Hatt, D., Beaton, R.L., et al. 2018, ApJ, 852, 60

\noindent
Karachentsev, I., Makarova, L.N., Tully, R.B., et al. 2014, MNRAS, 443, 1281

\noindent
Lamers, H.G.L.M. \& Levesque, E.M. 2017, {\it Understanding Stellar Evolution},

IOP Publishing, Bristol, UK

\noindent
Lee, M.-G. \& Jang, I.-S. 2012, ApJ, 760,14L

\noindent  
Madore, B.F. \& Freedman, W.L. 1995, AJ, 109, 1645

\noindent  
Madore, B.F., Mager, V. \& Freedman, W.L. 2009, ApJ, 690, 389

\noindent  
Madore, B.F., Freedman, W.L. et al.  2018, ApJ, 858, 11

\noindent
Madore, B.F., McAlary, C.W., McLaren, R.A. et al. 1985, ApJ, 294, 560

\noindent 
Makarov, D., Makarova, L., Rizzi, L. et al. 2006, AJ, 132, 2729

\noindent
McQuinn, K.B.W., Boyer, M., Skillman, E.D. et al. 2019, ApJ, 880, 63

\noindent 
Menendez, B., Davis, M., Moustakas, J., Newman, J., Madore, B.F. \& Freedman, W.L. 

2002, AJ, 124, 213

\noindent 
Mould, J. \& Kristian, J. 1986, ApJ, 305, 591  

\noindent 
Mould, J., Kristian, J. \& Da Costa, G.S. 1983, ApJ, 270, 471  

\noindent 
Mould, J., Kristian, J. \& Da Costa, G.S. 1984, ApJ, 278, 575 

\noindent
Nayar, S. 2022, Monograph: FPCV-2-1, Series: {\it First Principles of Computer Vision},

Computer Science, Columbia University

https://cave.cs.columbia.edu/Statics/monographs/Edge\%20Detection\%20FPCV-2-1.pdf

\noindent
Rizzi, L., Tully, R. B., Makarov, D., et al. 2007, ApJ, 661, 815

\noindent
Salaris, M. \& Cassisi, S. 2005,  {\it Evolution of Stars and Stellar Populations}, John Wiley \& Sons Ltd. 

\noindent
Sandage, A.R. \& Carlson, G. 1983, ApJL, 267, 25

\noindent
Shapley, H. 1918, ApJ, 48, 89

\noindent
Shapley, H. 1919, ApJ, 49, 311

\noindent
Shapley, H. 1930, {\it Star Clusters}, Harvard Observatory Monographs No. 2, London: McGraw Hill.

\noindent
Serenelli, A., Weiss, A., Cassisi, S., et al. 2017, A\&A, 606, 33 

\noindent
Silverman, B.W. 1986, {\it Density Estimation for Statistics and Data Analysis}, 

Vol. 26 of {\it Monographs on Statistics and Applied Probability}, 

Chapman and Hall, London

\noindent
Soszynski, I., Udalski, A., Kubiak, M., et al. 2004, Acta Astron., 54, 129

\vfill\eject

{\bf}{Appendix A: Digital Filters and Smoothing in Edge Detection}
\medskip

In most prescriptions for edge detection in digital image processing it is advised that the raw image be smoothed first to reduce the random noise in the image and then followed by an additional scan of the data using a first-derivative ``edge detector" that responds to locally-detected gradients across the image. 

Smoothing can take various forms.
They can range from simple moving rectangular averages (equally weighted) over a finite number of adjacent pixels, to a more sophisticated smoothing using triangular, biweight and/or higher-order Epanechnikov weightings, all of which symmetrically decline over a finite range of pixel support and are identically zero everywhere outside that range (see, for example, Silverman 1986 and also Jahne 1991). 
By the central limit theorem, multiple applications of any of these smoothing kernels converge on a Gaussian. A discretely-sampled (digital) Gaussian is itself, of course, a smoothing kernel (but one of infinite support, in principle). 

One of the earliest applied and certainly one of the most elementary quantized edge-detection kernels is the so-called Sobel filter. 
This filter involves the simple differencing of pixel intensity values on either side of a target position. 
The Sobel filter, in one dimension, takes the normalized 
form of [-1, 0, +1]. Indeed this is the first kernel in Figure 21 (named MF3 and shown in Panel~A) 
At the other extreme the first derivative of a Gaussian (the DoG) is also a highly effective gradient detector. 
Invoking the binomial theorem once again we 
recall that for very small samples Pascal's triangle gives the binomial terms' integer numbers of finite-support sampling (progressively approximating, and eventually converging upon a Gaussian).
That is well known. 
What is not commonly stated, but must be equally true, is that {\bf}{the 
differences between adjacent binomial terms are then discretely-sampled approximations to the first derivative of a Gaussian. }

Evaluating the location of the tip can then be done in either of two equivalent ways: (i) Find the value of x where the output of the Sobel filter is a maximum, or (ii) find the value of x where the output of the slope of Sobel filter is flat. The Sobel filter is the first derivative of the input function, the latter is the second derivative of the input function --  it is commonly known as the Laplacian. There is no difference between the two estimations of the position of the discontinuity.\footnote{
Cioni et al. (2000) were the first to suggest the use of the Laplacian as a means of locating the TRGB. Their approach differed somewhat from what we have discussed above, and they warn users about a potential bias between the Laplacian and the Sobel filter solutions. We have been in communication with Dr. Cioni, and we now all agree that when using the Laplacian, its zero crossing should be used to identify the discontinuity, and that this measure is not biased with respect to the Sobel filter.}

Pascal's Triangle of integers can also be thought of as resulting from the repeated smoothing of the initial solitary value of unit intensity by the elementary smoothing kernel [+1,+1]. 
Thus ...0, 0, 1, 0, 0, ... upon smoothing, becomes ... 0, 0, 1, 1, 0, 0, ... and then ... 0, 0, 1, 2, 1, 0, 0, ... and then ... 0, 0, 1, 3, 3, 1, 1, 0, 0, ... etc. That sequence is Pascal's Triangle.
Differentiating any row in Pascal's triangle (that is, differencing adjacent numbers in the triangle) is similarly visualized by having the row convolved by the zero-sum differencing kernel [+1,-1]. Applied to the first 
 row of Pascal's Triangle gives ... 0, 0, -1, +1, 0, 0, ... which is a very compact first derivative of adjacent pixels measured at their interface. 
Application of the differencing kernel to the second line of Pascal's triangle gives ... 0, 0, +1, 0, -1, 0, 0, ... which ushers in the appearance of Sobel filter, as mentioned above. 
Application to the third line gives ... 0, 0, -1, -1, +1, +1, ... and then the fourth line gives ... 0, 0, -1, -2, 0, +2, +1, 0, 0, ... etc.
Having said this in words, the table in Figure 22 shows the first 13 rows of Pascal's triangle, while the table in Figure 23 shows the first 13 rows of the first (digital) derivative of Pascal's triangle. 
In Figure 21 we show the first four even-numbers sets of binomial coefficients (MF3 through MF9] with a smooth Gaussian overlaid in the upper panels and symmetrically sampling the DoG, including its central point at the zero-crossing point of the kernel, shown in the lower sub-panels,

{\bf}{This immediately suggests  that, in the process of going to higher and higher approximations, the discretely-sampled Gaussian (and its derivative) each having larger support, we are spanning more and more pixels and thereby implicitly weight-smoothing the data, in addition to any previous smoothing.}
In Figure 24 we show the application of this single-step methodology to a step function.
Here it is noteworthy that the width of the response is largely independent of the order of the filter chosen.

\clearpage
{\bf}{Appendix B: A Closing Comment about the Sensitivity of the Adopted Differencing Kernels to Structure Surrounding the Discontinuity}

\medskip
It needs to be said in this closing remark that the discontinuity of the luminosity function (especially as seen directly in the I band, and in the rectified luminosity functions constructed at other wavelengths) is a very locally-defined quantity. By that we mean, that only information contained in a handful of milli-magnitude bins, ahead of and/or following the discontinuity itself, contribute to tip’s detection. Moreover, the presence or absence of stars further away from the action (i.e., from the TRGB discontinuity) can have little or no influence on the output of the edge-detection filter, since all values outside of the kernel/filter's support are set to zero. For instance, given the finite range of support adopted by the Sobel filter (the simplest example being [-1,0,+1]) only those AGB stars that have magnitudes that are within plus or minus one bin of the TRGB (AGB stars above, and RGB stars below) will have any effect on the output of the filter at the tip. The slopes of the respective AGB or RGB luminosity functions (be they positive, flat, or negative) will only produce a constant output (of the first-derivative filter) up until a  significant transition in the slope is detected (above and beyond the noise). Poisson noise at the tip will serve to smooth the tip, but it will not bias the tip detection, nor will the filter care (or know) what is happening to stars more than a tenth of a magnitude (say) above or below the tip. The presence of AGB stars immediately before the tip can only act to change the contrast in the jump by adding a pillar of counts to the difference being measured between the AGB base (seen at one side of the differencing kernel) and (the sudden) “onset” of the RGB (seen by the other side of the advancing kernel). Nothing else much matters. Everything about the TRGB is local. 

A more compact and mathematically formal way of looking at it is the fact that the derivative of a function (dF/dx)  at x(i) is found in the limit as the differencing interval (dx) goes to zero at x(i) . It does not matter what F is doing at x(i+5), or x(i+10), or x(i-5), or x(i-10), etc.

 \clearpage
\begin{figure*} \centering 
\includegraphics[width=14.0cm,angle=-0]{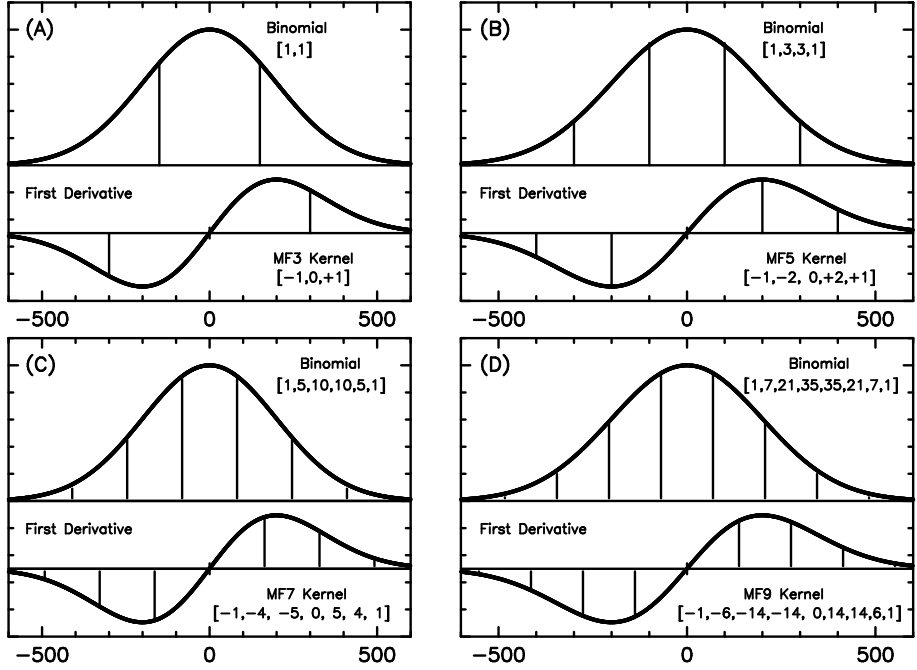} 
\caption{\small Binomial Coefficient and their First Derivatives. Each of the panels show the values of the binomial coefficients, numerically in square brackets and graphically as vertical lines in the upper half of each plot. The smooth curve is a Gaussian. In the lower half of the panel is the first derivative of the binomial kernel, again given numerically inside of square brackets, and graphically as vertical black lines bounded by the smooth black line which is the first derivative of the Gaussian.}
\end{figure*}

\begin{figure*} \centering 
\includegraphics[width=14.0cm,angle=-00]{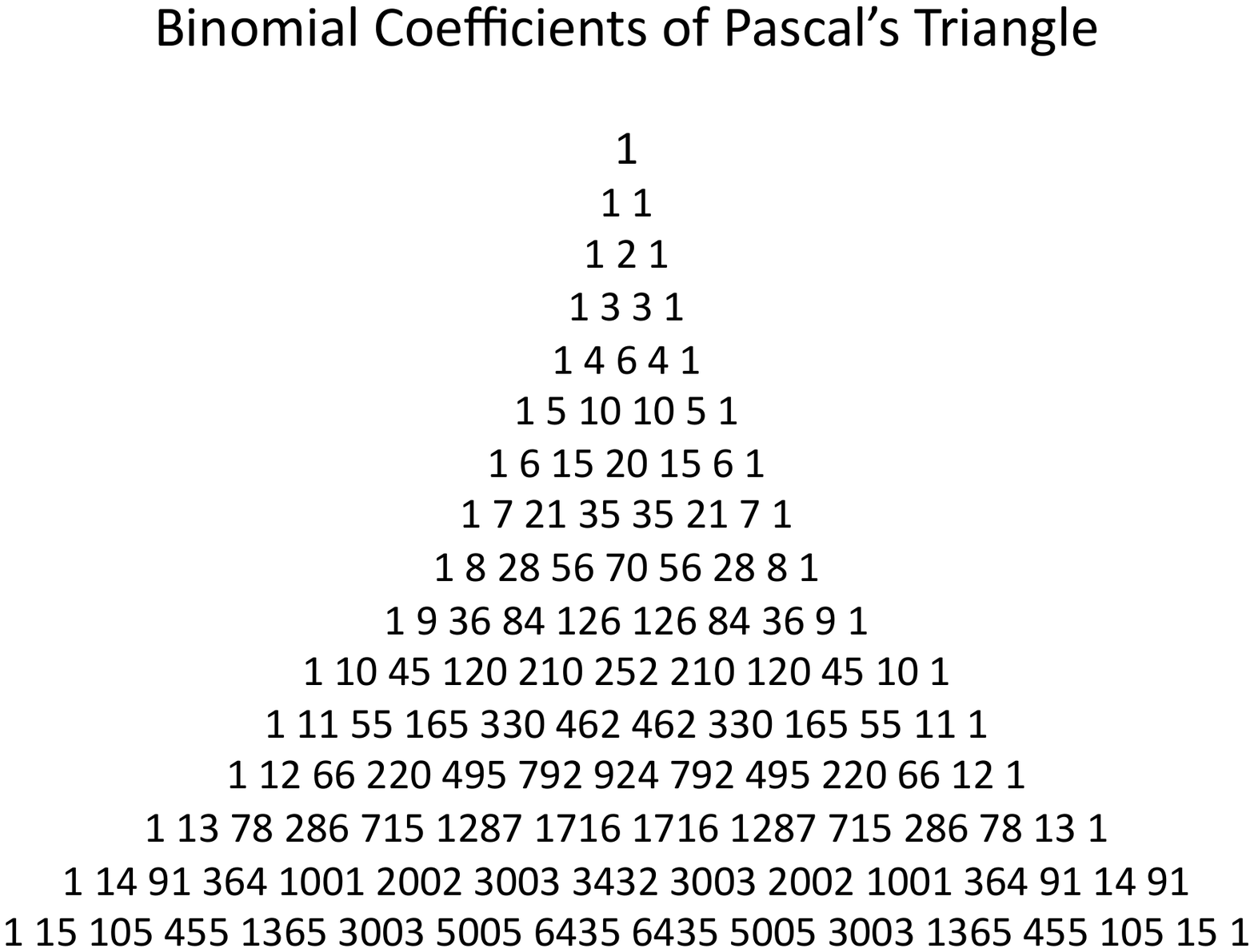} 
\caption{\small The Binomial Coefficients of Pascal's Triangle. Each line represents a digital kernel, each of which is a progressively higher fidelity approximation to a Gaussian. 
Applied to digital data these kernels act as weighted smoothing functions.} 
\end{figure*}

\begin{figure*} \centering 
\includegraphics[width=14.0cm,angle=-00]{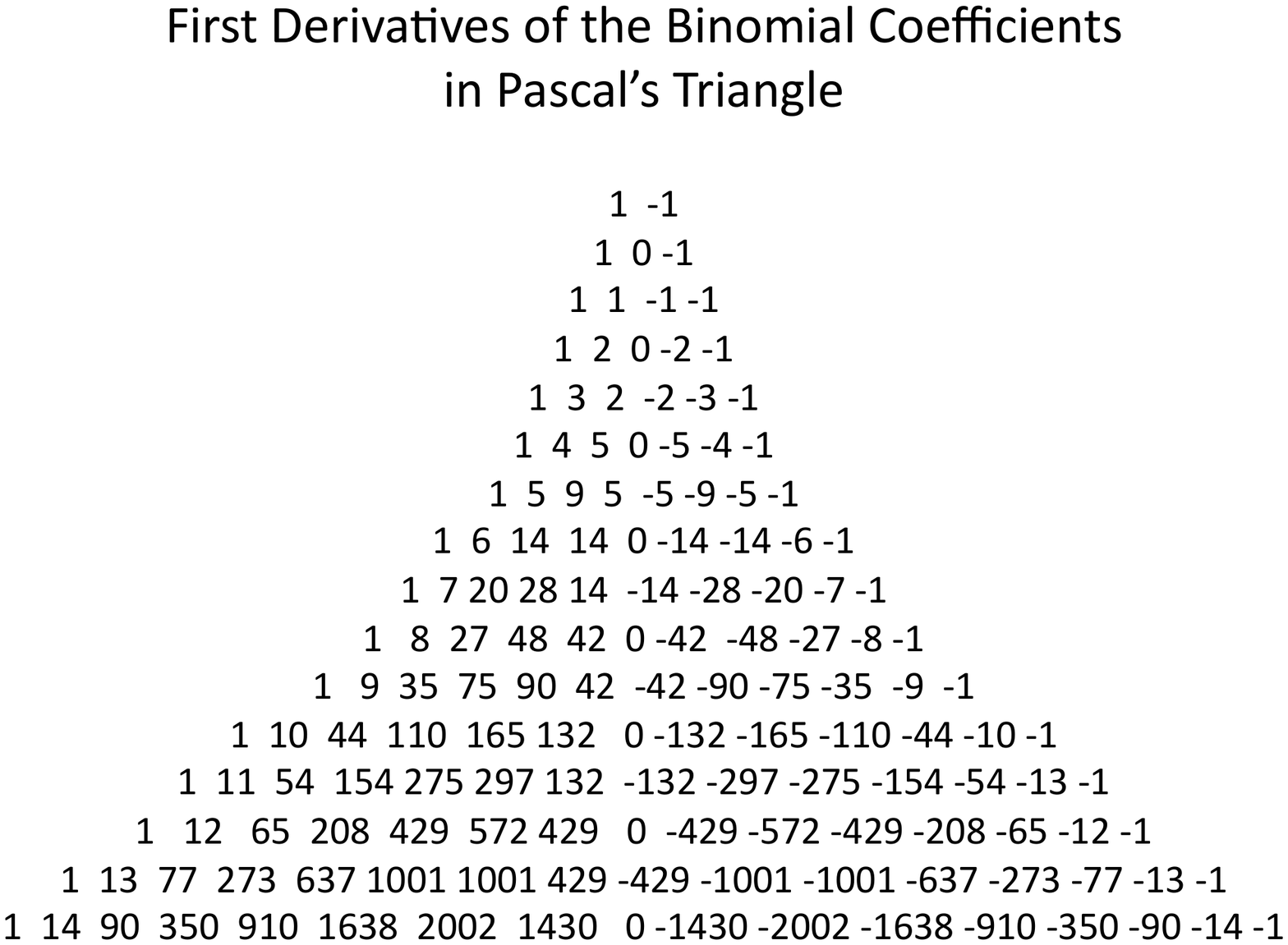} 
\caption{\small First Derivatives of the Binomial Coefficients in Pascal's Triangle, as given in Figure 23 above. The first derivative of a Gaussian (DoG) is a well-known edge detector in image processing, Each of the entries in this figure are then also edge detectors, progressively more precise approximation to the DoG. Examples of their application to the detection of a step function are shown in Figure 23 (below).} 
\end{figure*}
\begin{figure*} \centering 
\includegraphics[width=14.0cm,angle=-0]{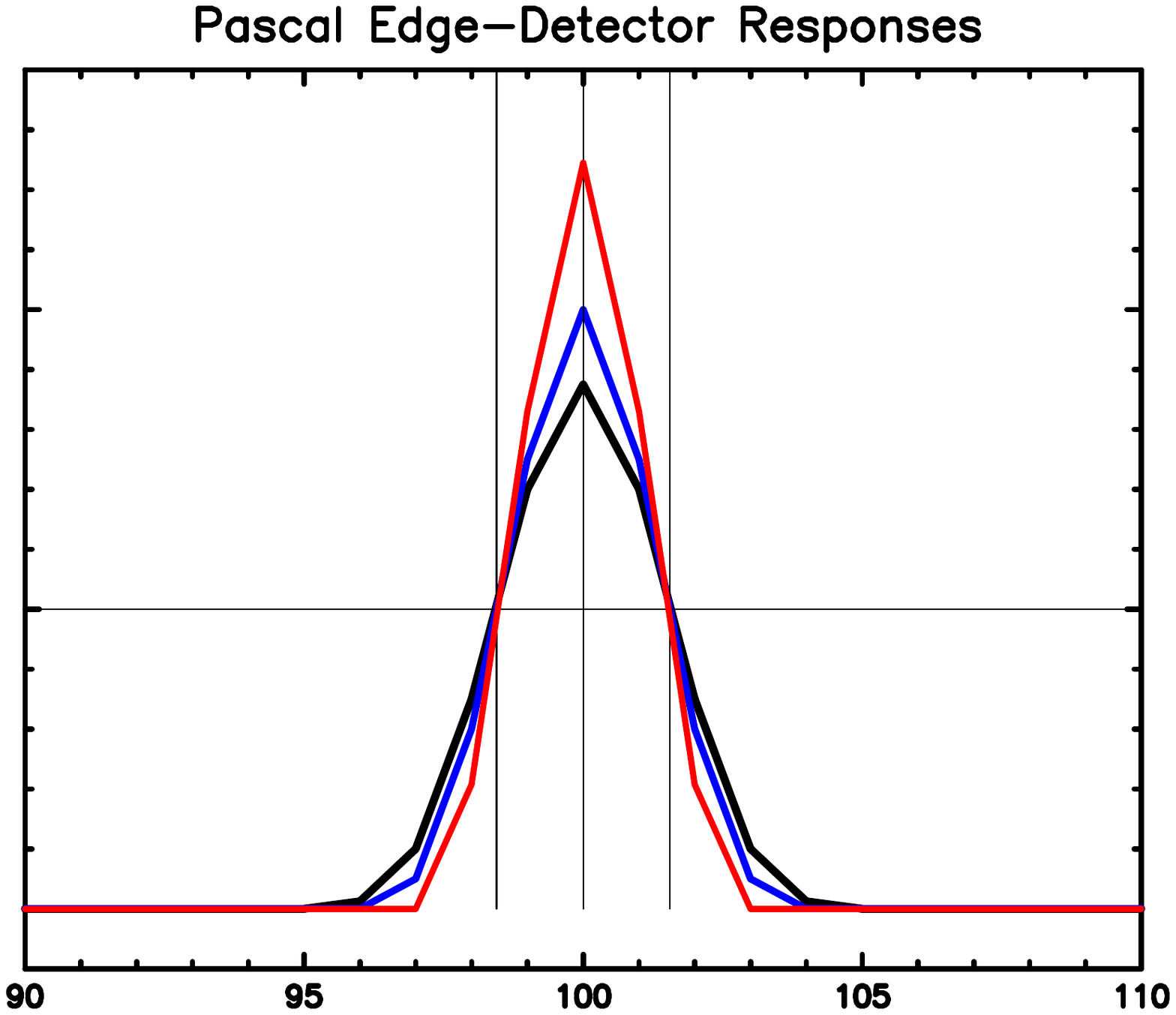} 
\caption{\small Examples of the Pascal Edge-Detector response to a step function. These are the results of applying the third, fifth and seventh (centrally-peaked) filters given in Figure 22 (above), shown as red, blue and black lines, respectively. They are scaled to equal areas, demonstrating the relatively stable width of the response functions, independent of the ``smoothing" width of the chosen edge detector.} 
\end{figure*}

\vfill\eject
{\bf}{Appendix C: Random Displacements of the Tip Revealed in Multiple Realizations of a Single Numerical Experiment}

At the urgings of the referee, we have investigated whether the mismatch between the observed and true tip magnitudes are systematic or random in nature. This experiment  has already been run and published in the study of M101 by Beaton et al. (2019) and their Figure 5 (and extended caption) and their Figure 8. The latter shows the effect of over-smoothing where highly smoothed detections drift away from their lesser smoothed versions. In addition, a wide range of edge-detection methods using different smoothing kernels and even including those using maximum-likelihood fitting techniques agreed when applied to the same dataset for IC 1613 (Hatt et al. 2017, ApJ,  845, 146),  Figure 12. The two papers both offer a quantitative means of selecting an optimal smoothing parameter which is the one that minimizes the quadrature sum of the random and systematic errors.

To shed further light on that question we offer Figure 24 which consists of nine sub-panels containing repeated runs of the simulation displayed in Sub-Panel ``e'' of Figure 11. All of the input parameters were fixed and only the random sampling of the luminosity function was allowed to change. Details are given in the extended figure caption, but our conclusions are that at this smoothing the displacements are random, but given the larger density of false (minor) peaks below the tip as compared to similar fluctuations being registered above the tip, we warn that larger amounts of smoothing will result in systematic shifts of the measured tip to fainter magnitudes.

\begin{figure*} \centering
\includegraphics[width=5.0cm,angle=-0]{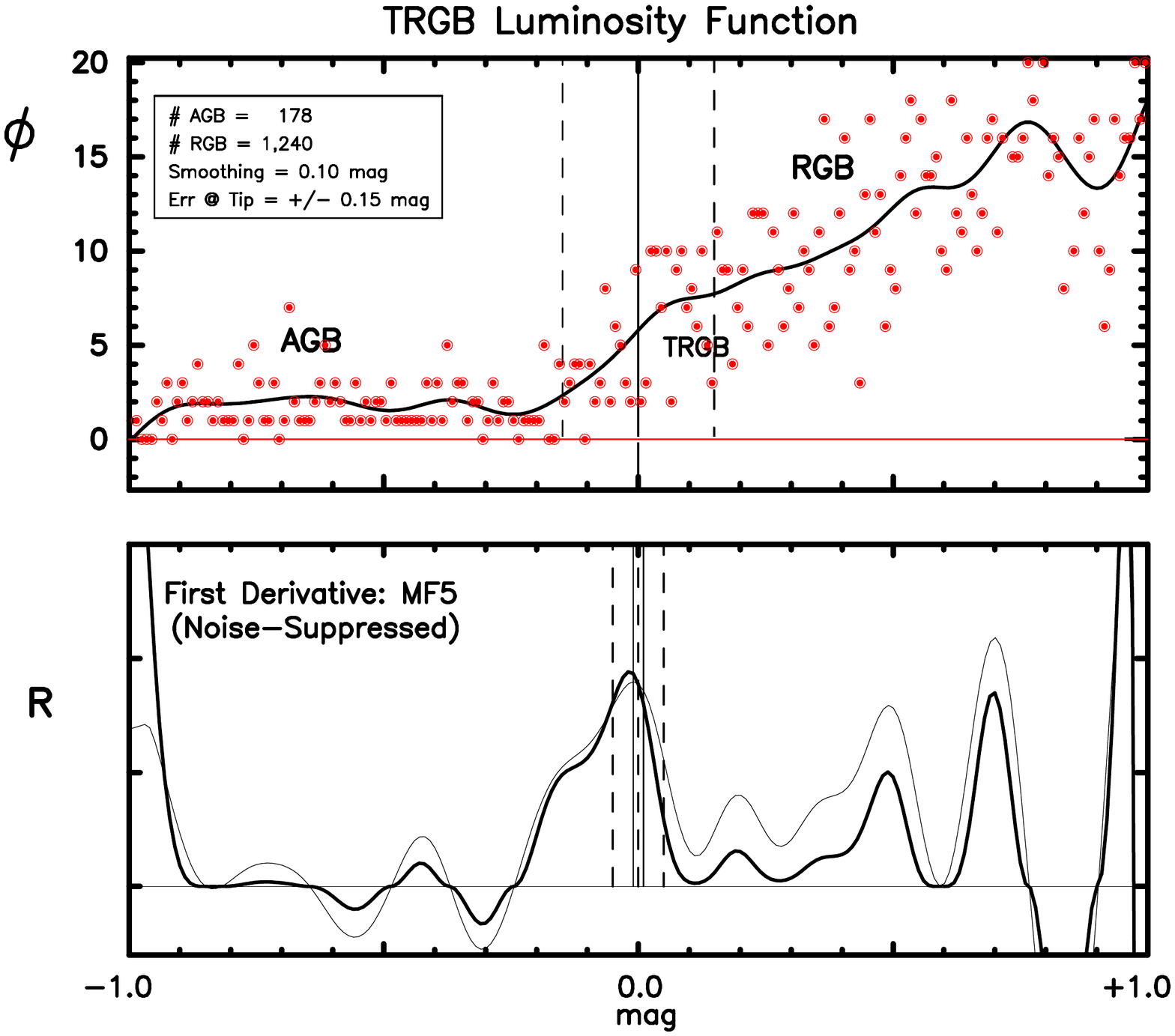}
\includegraphics[width=5.0cm,angle=-0]{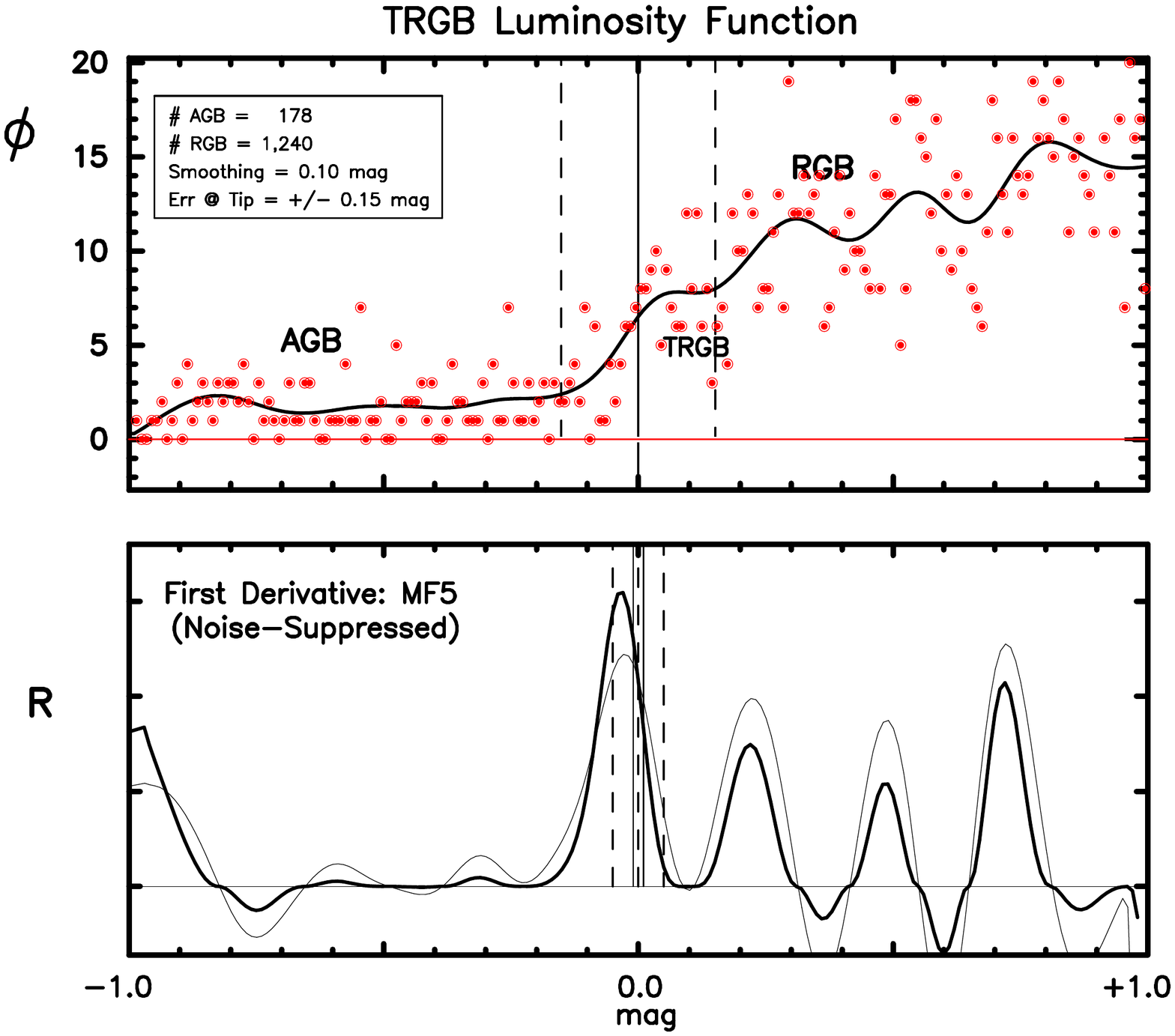}
\includegraphics[width=5.0cm,angle=-0]{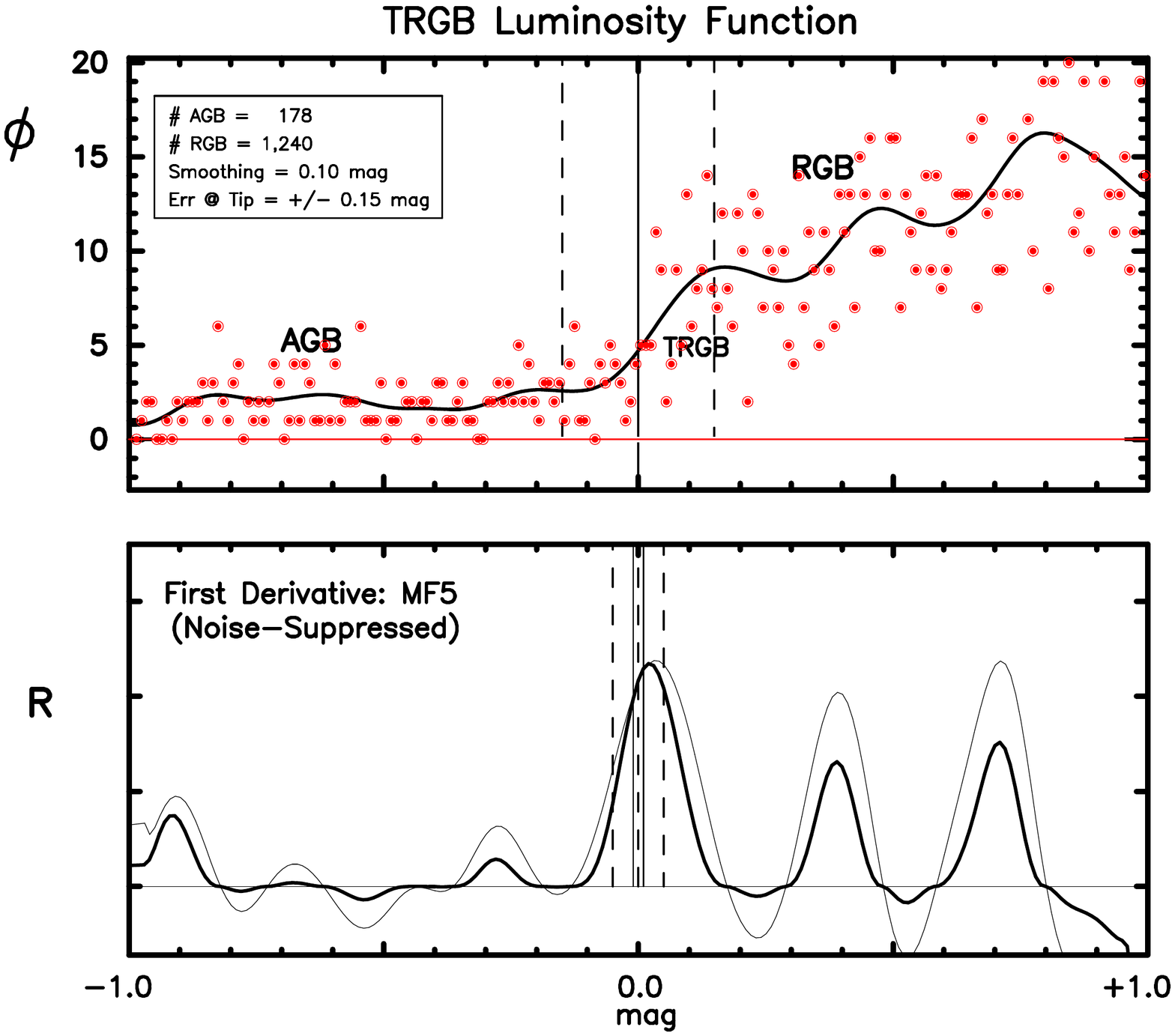}
\includegraphics[width=5.0cm,angle=-0]{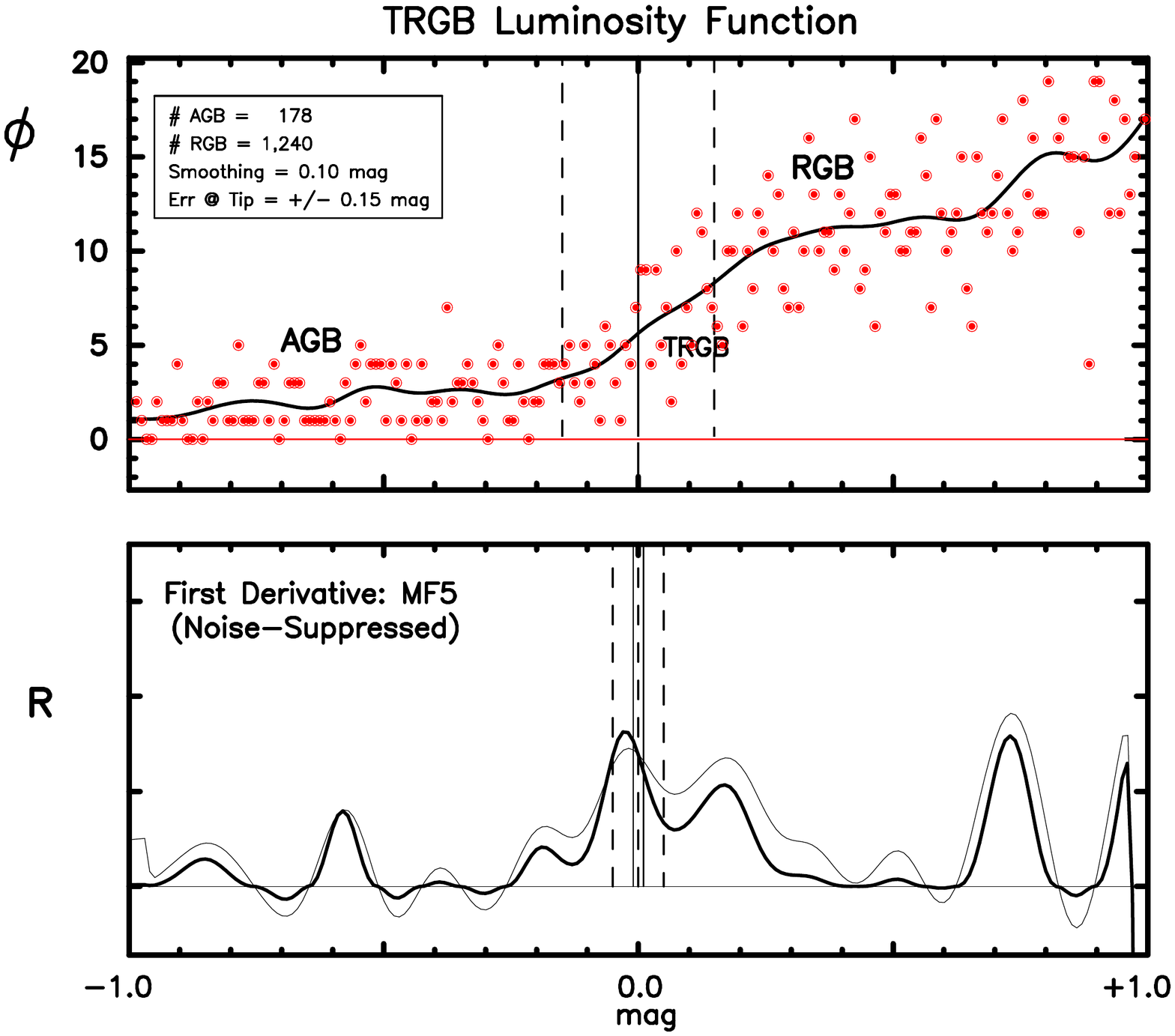}
\includegraphics[width=5.0cm,angle=-0]{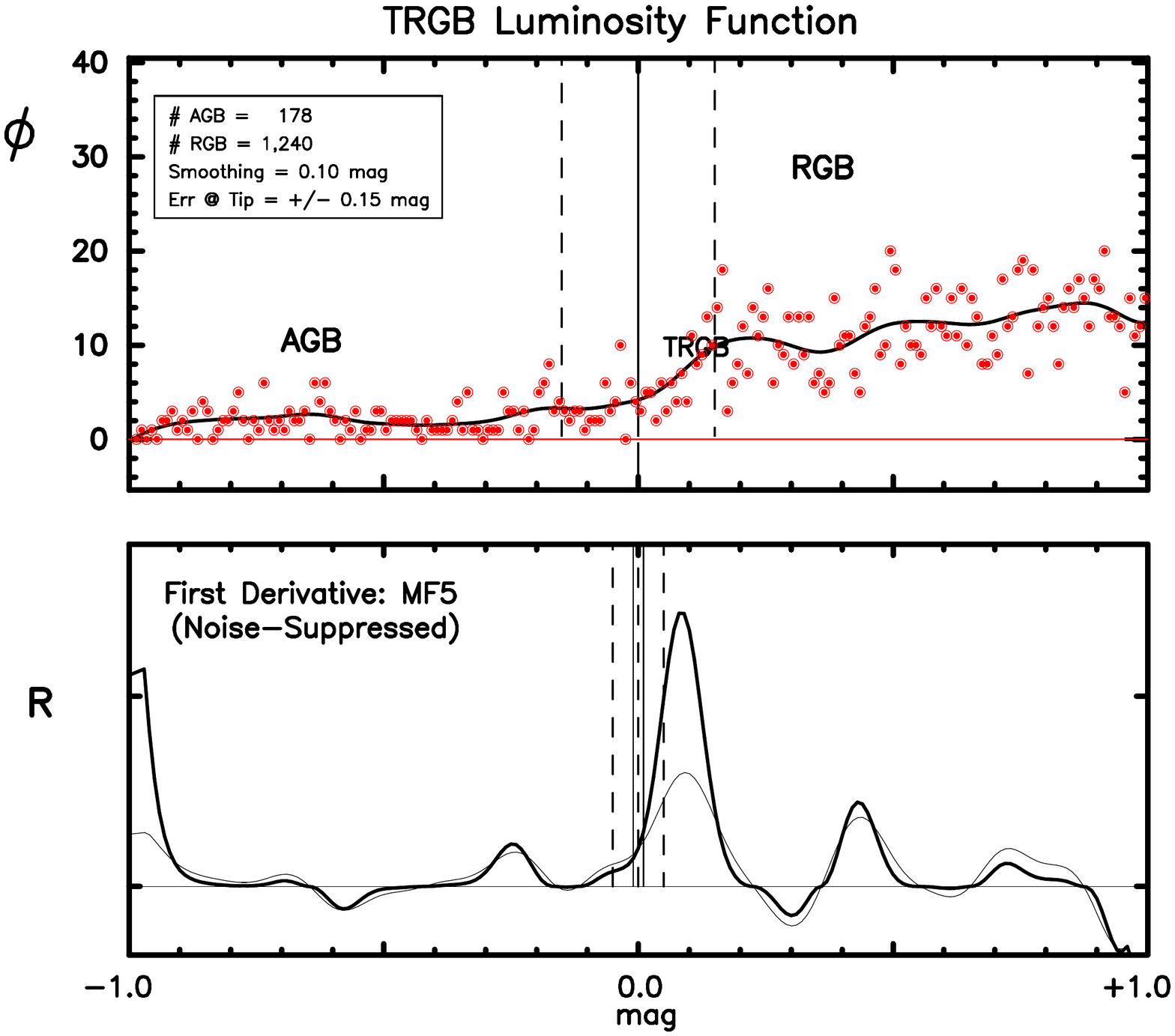}
\includegraphics[width=5.0cm,angle=-0]{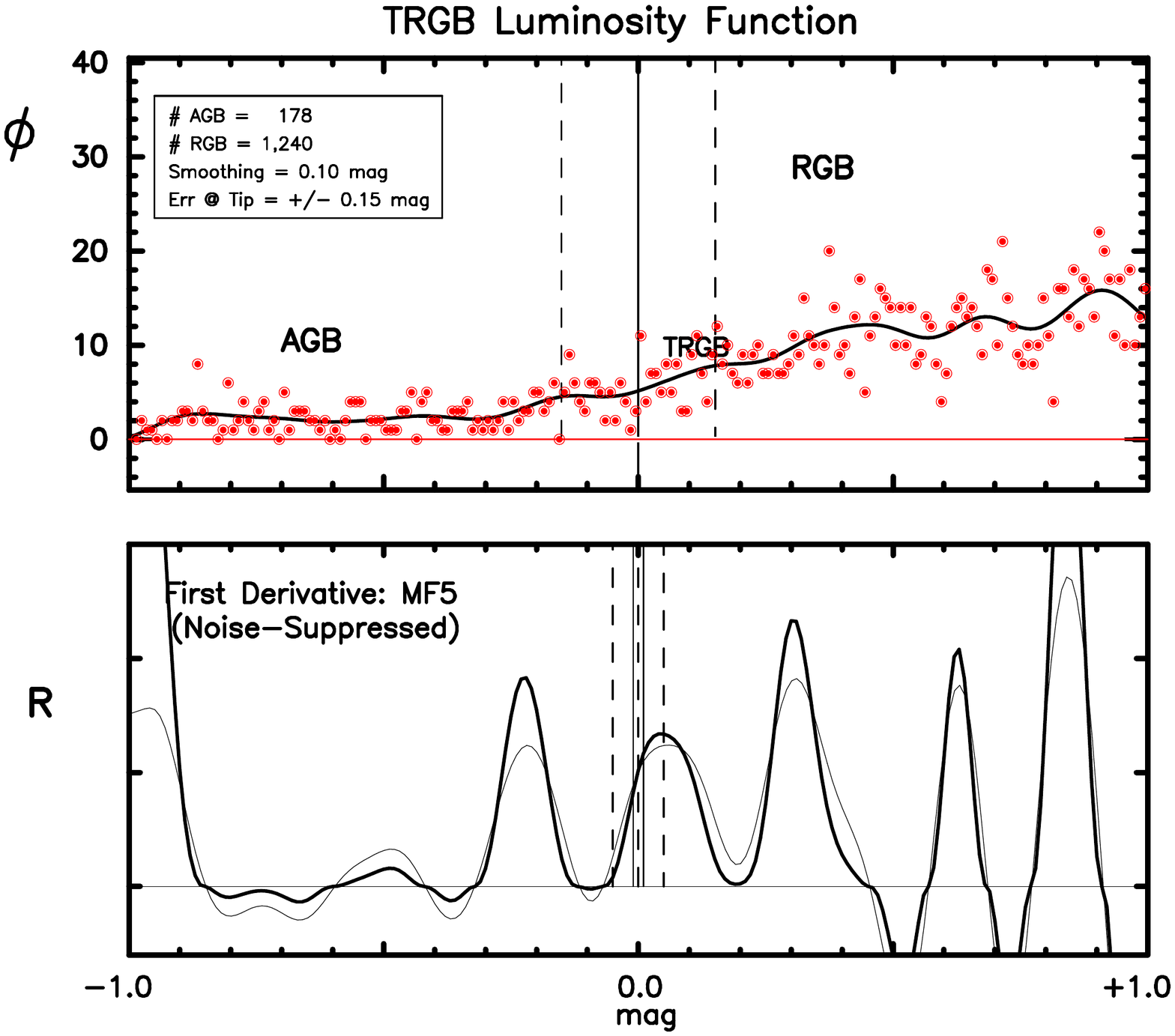} \includegraphics[width=5.0cm,angle=-0]{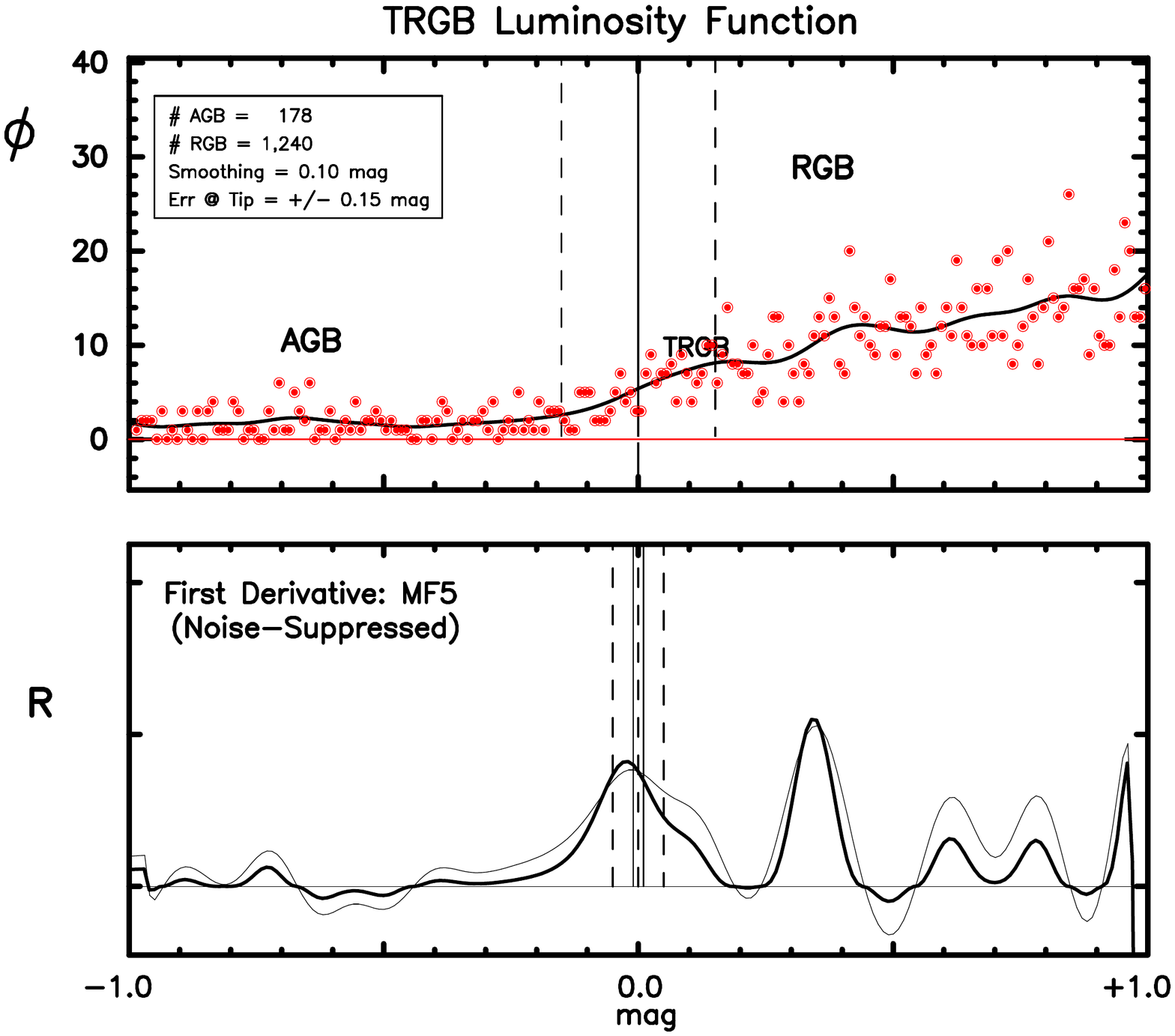}
\includegraphics[width=5.0cm,angle=-0]{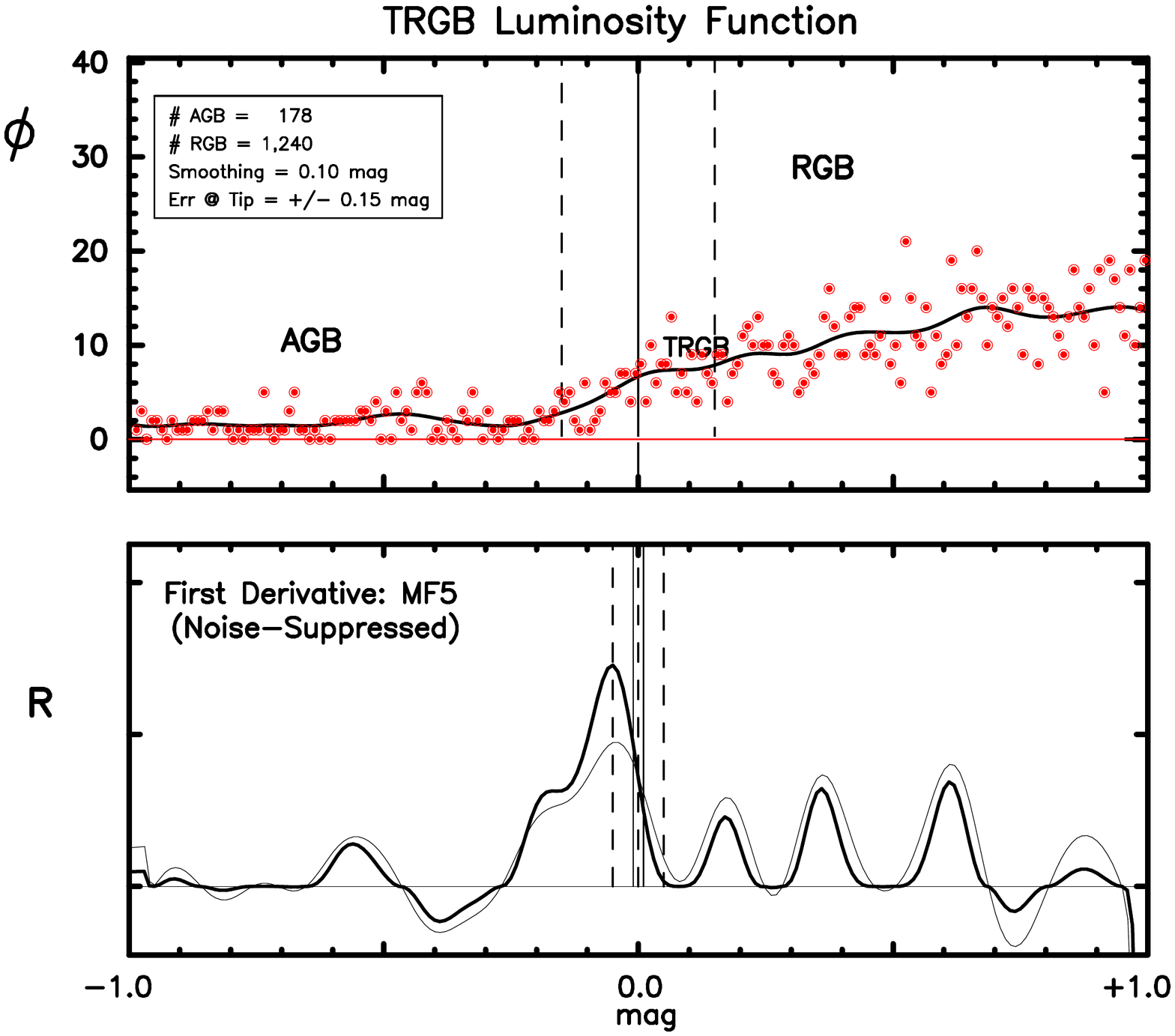}
\includegraphics[width=5.0cm,angle=-0]{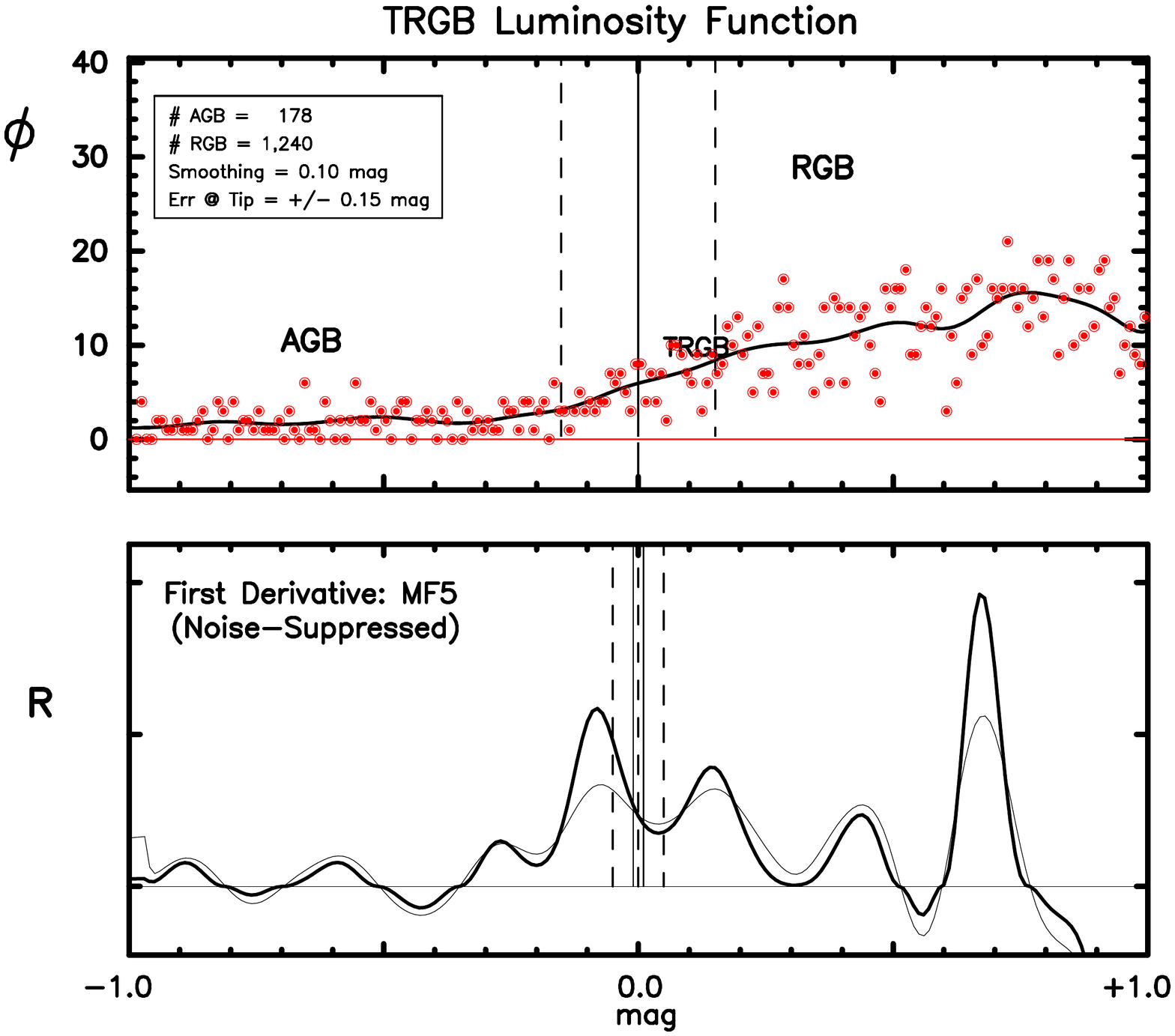} \caption{\small Nine sub-panels illustrating a random sampling of TRGB measurements for the same number of RGB stars (1,340) the same smoothing (0.10~mag) at the tip ($\pm$0.15~mag) as in Sub-Panel ``e'' in Figure 11. These independently selected examples demonstrate the random drift of the peak response of the Sobel filter around the input value shown by the solid vertical black line at 0.0~mag. One peak  (b,h and i), Read left to right and top to bottom, the peak in Sub-Panel ``i" is noticeably displaced to a brighter magnitude; the example in Sub-Panel ``e'' is displaced to fainter magnitudes. Other deflections are all within the one-sigma expected deviations shown by the vertical dashed lines; five (a, b, d, g,and h) fall to the left and two (c and f) fall to the right, although the latter is flanked by two peaks that are apparently more significant than the one found closest to the known answer. We note that more (low-level) structure in the output response is at magnitudes fainter than the true tip. Some of this structure is close enough (Sub-Panels ``d" \& ``i") that, if excessive smoothing were applied, that action would preferentially draw the measured tip magnitudes systematically towards fainter magnitudes.}
\end{figure*}
\eject
{\bf}{Appendix D. Demonstration of the Lack of Bias in the Sobel Tip Detection to 

Smoothing of a Variety of AGB Luminosity Functions and RGB stars above and

below the TRGB}

\medskip\par
In the Figures 25 through 27 we show the robust nature of the simple Sobel filter response to the the application of smoothing, and to three possible forms of the AGB luminosity function approaching the TRGB from above. Figure 25 shows a declining AGB luminosity function. Figure 26 shows an increasing AGB luminosity function and Figure 27 shows a flat AGB luminosity function as has been adopted in the simulations given in the main paper.

As expected, given the symmetric nature of the kernels being applied in the smoothing and in the tip detection, there is no resulting bias in the position reported by the Sobel filter response function. The effects of noise and the Sobel response to smoothing is also nicely discussed in Nayar (2022), especially his Figures 25, 26 and 27.

\begin{figure*} \centering
\includegraphics[width=20.0cm,angle=-0]{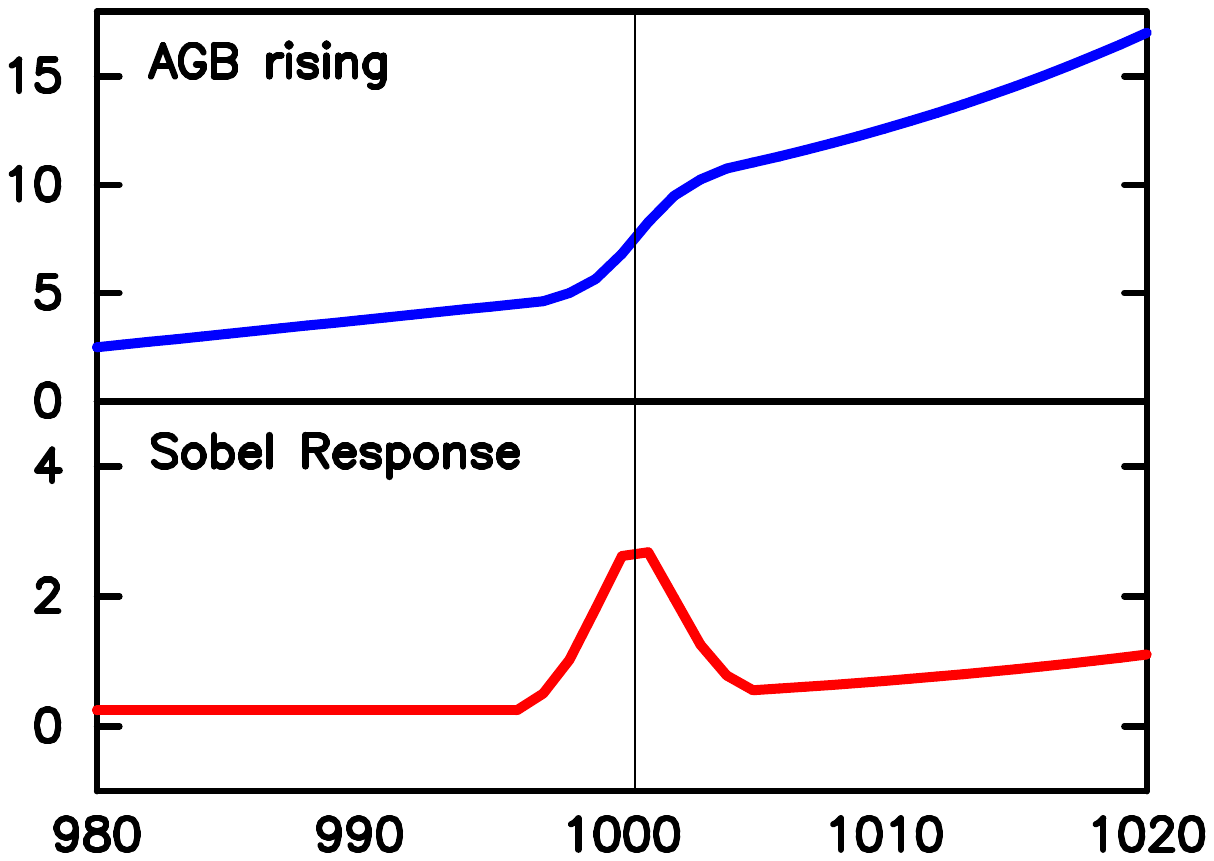}
\caption{\small Toy model of a rising AGB luminosity function + an exponential RGB luminosity function, Gaussian smoothed (upper sub-panes) and convolved with a Sobel edge-detection response filter (lower sub-panels).}  
\end{figure*}

\begin{figure*}\centering
\includegraphics[width=20.0cm,angle=-0]{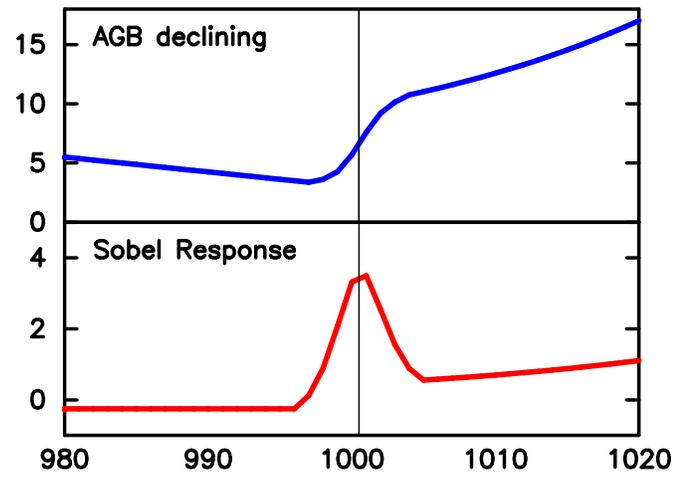}
\caption{\small Toy model of declining AGB luminosity + an exponential RGB luminosity function, Gaussian smoothed (upper sub-panes) and convolved with a Sobel edge-detection response filter (lower sub-panels).}  
\end{figure*}

\begin{figure*}\centering
\includegraphics[width=20.0cm,angle=-0]{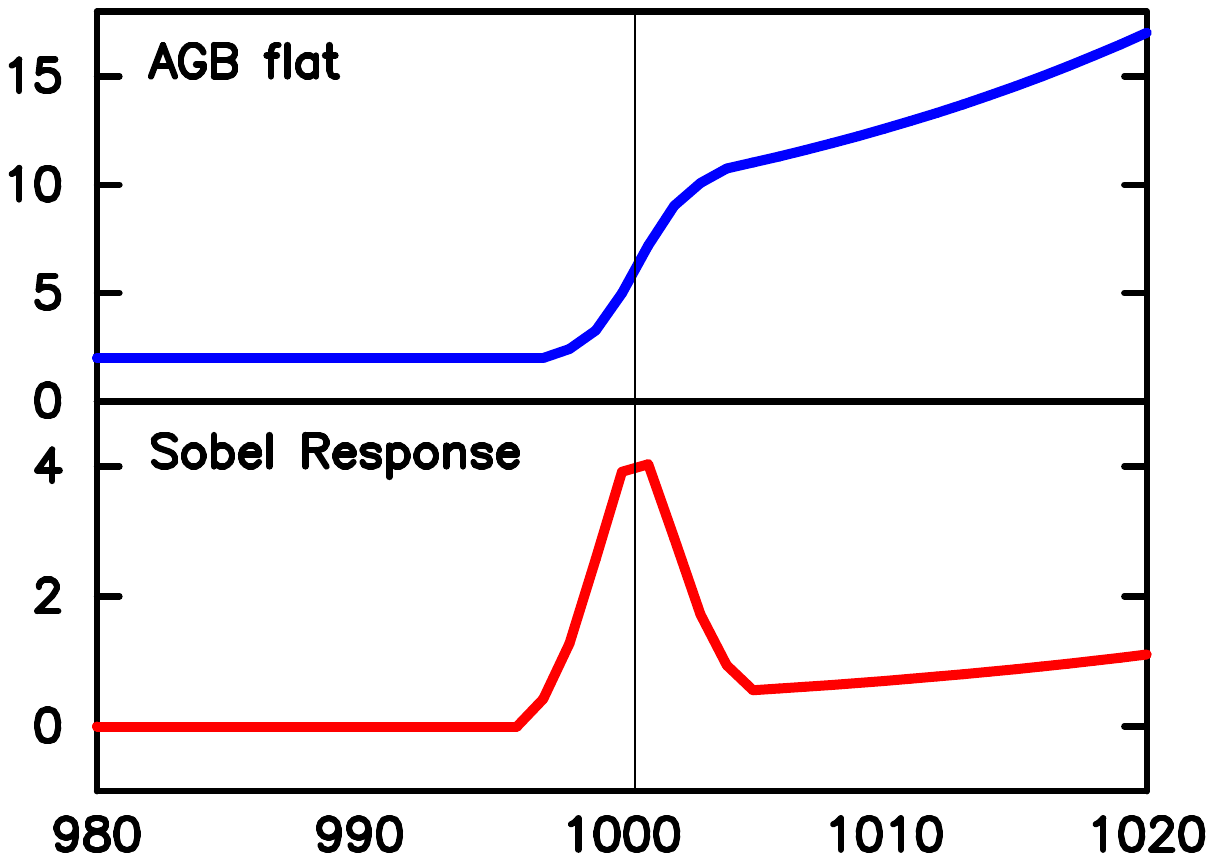}
\caption{\small Toy model of a flat AGB luminosity function + an exponential RGB luminosity function, Gaussian smoothed (upper sub-panes) and convolved with a Sobel edge-detection response filter (lower sub-panels).} 
\end{figure*}
\end{document}